\documentclass{jfm}

\usepackage{pdflscape}

\usepackage{graphicx}
\usepackage{epstopdf, epsfig}
\usepackage{amsmath}
\usepackage[title]{appendix}
\usepackage{placeins}

\usepackage{hyperref}
\hypersetup{
  colorlinks   = true, %Colours links instead of ugly boxes
  urlcolor     = blue, %Colour for external hyperlinks
  linkcolor    = blue, %Colour of internal links
  citecolor   = green %Colour of citations
}

\usepackage[normalem]{ulem}
\usepackage{color}

\usepackage{tikz}
\definecolor{magenta_1}{rgb}{0.64,0.08,0.18}
\definecolor{yellow_1}{rgb}{0.93,0.69,0.13}
\newcommand{\blueline}  {\raisebox{2pt}{\tikz{\draw[-,blue,solid,line width = 0.9pt](0,0) -- (5mm,0);}}}
\newcommand{\redline}   {\raisebox{2pt}{\tikz{\draw[-,red, solid,line width = 0.9pt](0,0) -- (5mm,0);}}}

\newcommand{\orangeline}{\raisebox{2pt}{\tikz{\draw[-,orange,solid,line width = 0.9pt](0,0) -- (5mm,0);}}}
\newcommand{\greenline} {\raisebox{2pt}{\tikz{\draw[-,green,solid,line width = 0.9pt](0,0) -- (5mm,0);}}}
\newcommand{\blackline} {\raisebox{2pt}{\tikz{\draw[-,black,solid,line width = 0.9pt](0,0) -- (5mm,0);}}}
\newcommand{\magentaline} {\raisebox{2pt}{\tikz{\draw[-,magenta_1,solid,line width = 0.9pt](0,0) -- (5mm,0);}}}
\newcommand{\yellowline} {\raisebox{2pt}{\tikz{\draw[-,yellow_1,solid,line width = 0.9pt](0,0) -- (5mm,0);}}}

\newcommand{\blackdash}{\raisebox{2pt}{\tikz{\draw[-,black,dashed,line width = 0.9pt](0,0) -- (5mm,0);}}}

\newcommand{\redcircle} {\raisebox{2pt}{\tikz{\draw [red]  (0,0) circle (2pt);}}}

\newcommand{\blackdotted}{\raisebox{2pt}{\tikz{\draw[-,black ,dotted,line width = 0.9pt](0,0) -- (5mm,0);}}}
\newcommand{\blackdashdot}{\raisebox{2pt}{\tikz{\draw[-,black ,dashdotted,line width = 0.9pt](0,0) -- (5mm,0);}}}

\newcommand{\blackSquare} {\raisebox{2pt}{\tikz{\draw[black, very thick] (0,0) rectangle (3pt,3pt);}}}

\newcommand{\magentaDiamond} {\raisebox{2pt}{\tikz{\draw[magenta, very thick] (0,0) -- (2pt, 2pt) -- (4pt,0) -- (2pt,-2pt) -- cycle ;}}}
\shorttitle{LES of a moderate-APG TBL at high $Re$}
\shortauthor{R. Pozuelo, Q. Li, P. Schlatter and R.Vinuesa}

\title{An adverse-pressure-gradient turbulent boundary layer with nearly-constant $\boldsymbol{\beta \simeq 1.4}$ up to $\boldsymbol{Re_{\theta} \simeq 8,700}$}

\author{Ram\'on Pozuelo\aff{1} \corresp{\email{ramonpr@mech.kth.se}}, 
Qiang Li\aff{1}, Philipp Schlatter\aff{1} \\ and Ricardo Vinuesa\aff{1}\corresp{\email{rvinuesa@mech.kth.se}} }

\affiliation{\aff{1}FLOW, Engineering Mechanics, KTH Royal Institute of Technology, SE-100 44 Stockholm, Sweden}

\begin{document}

\maketitle

\begin{abstract}
In this study, a new well-resolved large-eddy-simulation (LES) of an incompressible near-equilibrium adverse-pressure-gradient (APG) turbulent boundary layer (TBL) over a flat plate is presented. In this simulation, we have established a near-equilibrium APG over a wide Reynolds-number range. In this so-called region of interest (ROI), the Clauser--Rotta pressure-gradient parameter $\beta$ exhibits an approximately constant value of around 1.4, and the Reynolds number based on momentum thickness reaches $\Rey_{\theta}=8700$. To the authors’ knowledge, this is to date the highest $\Rey_{\theta}$ achieved for a near-equilibrium APG TBL under an approximately constant moderate APG.  
We evaluated the self-similarity of the outer region using two scalings, namely the Zagarola--Smits and an alternative one based on edge velocity and displacement thickness. 
Our results reveal that outer-layer similarity is achieved, and the viscous scaling collapses the near-wall region of the mean flow in agreement with classical theory.
Spectral analysis reveals that the APG displaces some small-scale energy from the near-wall to the outer region, an effect observed for all the components of the Reynolds-stress tensor, which becomes more evident at higher Reynolds numbers.
Generally, the effects of the APG are more noticeable at lower Reynolds numbers. For instance, the outer peak of turbulent-kinetic-energy (TKE) production is less prominent at higher $Re$. 
While the scale separation increases with $\Rey$ in zero-pressure-gradient (ZPG) TBLs, this effect becomes accentuated by the APG. Despite the reduction of the outer TKE production at higher Reynolds numbers, the mechanisms of energization of large scales are still present. 

\end{abstract}
% Abstract less than 250 words

\begin{keywords}
 turbulent flows, turbulence simulation
% Turbulence, turbulence simulations.
% Authors should not enter keywords on the manuscript, as these must be chosen by the author during the online submission process and will then be added during the typesetting process (see http://journals.cambridge.org/data/\linebreak[3]relatedlink/jfm-\linebreak[3]keywords.pdf for the full list)
\end{keywords}

\section{Introduction} \label{sec:Introduction}
%----------------------------------------------------
The study of boundary layers (BL) is the study of the behaviour of a fluid close to the boundary with another substance (solid, liquid or gas). We can use the knowledge of these boundary layers to predict the weather (systems atmosphere-Earth or atmosphere-sea) or even to manipulate a fluid for engineering purposes (production of electricity, mixing processes, transportation,...); note that most of these cases exhibit turbulent motions. In wall-bounded flows, as in the previous engineering examples, the optimization and control of the boundary layers is crucial for a good performance of the application under study. Some of the typical objectives are to reduce the drag in aeronautical surfaces, control the transition to turbulence or to avoid recirculation bubbles in ducts where achieving the maximum mass flow rate is important.

All the turbulent boundary layers (TBLs) in practical applications are subjected to streamwise pressure-gradients (PG), often with complex streamwise PG histories. These complicated variations of the PG can be divided into regions of adverse (APG), favourable (FPG) or zero pressure gradient (ZPG). The effects of PGs in wall-bounded turbulence are very diverse: a FPG drives the flow in a turbulent channel, while (if strong enough) it can produce relaminarization in flat-plate TBLs \citep{narasimha_sreenivasan_1973, FPG_araya2015}. On the opposite side, an APG can promote turbulence in a laminar BL and increase the turbulent fluctuations of a turbulent boundary layer; it can even produce flow separation, which is an undesirable phenomenon that reduces the performance of an aerodynamic device and can be dangerous if it happens on the wing of an airplane.
In wall-bounded flows the BL is affected by the wall geometry, the characteristics of the wall surface, the pressure-gradient distribution and the flow state beyond the BL. The effects on the wall will be seen as fluid-dynamic forces (lift and drag), but we could also be interested in effects produced after the solid, {\it i.e.} in the  the wake and its aeroacoustics properties (noise), or the wake instabilities that produce an increase in time between take-offs in airports.

Understanding the energy-transfer mechanisms within the TBL, and how they are affected by the pressure gradient, may lead to advancements in flow control and to new aerodynamic designs with higher performance.
The complexity of the problem has led to the definition of canonical TBLs in simple geometries, such as flat plates subjected to a ZPG or PG TBLs in equilibrium or near-equilibrium. 

It is important to discuss the concept of equilibrium and the effects of flow history. As discussed by \cite{Gibis2019} and \cite{Marusic_PoF_2010}, the term `equilibrium' has been used in different contexts in the literature. \cite{Clauser_1954_exp} denoted ``equilibrium profiles'' as those profiles of a boundary layer which develop maintaining a non-dimensional ``constant history'', defined as a constant ratio of the pressure-gradient and wall-shear forces, which can be written as the Clauser pressure-gradient parameter $\beta=\delta^{*}/\tau_w (\textrm{d}p/\textrm{d}x)$, where $\delta^{*}$ is the displacement thickness, $\tau_w$ is the shear stress at the wall, and $(\textrm{d}p/\textrm{d}x)$ is the pressure gradient. 
Later, the term `equilibrium' was used by \cite{rotta1962turbulent} and  \cite{townsend_1956_eqBL} as a synonym of self-preserving flow and in \cite{townsend_1961} it is used to refer to regions of the flow (`equilibrium layers') where there is a balance between rates of energy production and dissipation. In this context we will follow the same criterion as  \cite{Marusic_PoF_2010}, avoiding the use of equilibrium boundary layers and referring to near-equilibrium TBLs when the mean velocity-defect $U_{e}-U$ in the outer layer exhibits self-similarity (note that $U$ is the mean streamwise velocity and $U_e$ the velocity at the boundary-layer edge).

After these studies, many PG TBL experiments and simulations have been performed, with the aim of obtaining constant-$\beta$ distributions in order to obtain PG TBLs with a well-defined flow history, in this case with constant PG magnitude. 
If the equation of momentum in the streamwise direction is integrated across the boundary layer under several assumptions, then the momentum integral equation will relate the evolution of the momentum thickness $\theta$ with $\delta^*$ and $\beta$.

From he momentum integral equation, it is possible to derive the argument that $\beta$ must be constant to obtain self-similarity based on integral assumptions.

However, note that the ZPG TBL exhibits at least two different scalings: the near-wall region which scales properly using viscous units and the outer region, which does it with outer units (such as $\delta^*$ or the boundary layer thickness).

Therefore, the interest in establishing well-defined and close-to-constant $\beta$ distributions is not related to integral self-similarity arguments but lies in the fact that it is an integral parameter that quantifies the ratio between the pressure forces and the friction forces the boundary layer is subjected to throughout its evolution.

If this ratio had strong variations, especially at low Reynolds numbers, then it would be difficult to understand whether the observed phenomena are due to a local effect of the pressure gradient or the result of many different ratios of friction/pressure forces.

In this sense, a constant-$\beta$ configuration, in principle, allows to separate pressure-gradient and Reynolds-number effects, by comparing TBLs that only differ in the magnitude of the ratio between friction and pressure forces, since that ratio is maintained during the evolution of each BL, therefore, not mixing the contribution of flow history.

An example of constant $\beta$ are ZPG TBLs, which have been widely studied numerically e.g. by \cite{schlatter_orlu_2012, Sillero_2013pof} and experimentally by  \cite{bailey_2013_JFM, Orlu_Schlatter_exp2013, marusic_2015}. Regarding near-equilibrium APG simulations, it is important to highlight the direct numerical simulation (DNS) by \cite{Kitsios2016} with a constant $\beta=1$; the self-similar DNS TBL at the verge of separation ($\beta=39$) by \cite{Kitsios2017} or the well-resolved large-eddy simulation (LES) database comprising different PG intensities by \cite{bobke2017}. 
Some relevant experimental databases include the near-equilibrium APG TBLs by \cite{skare_krogstad_1994, MTL_expSANMIGUEL}, and the studies by \cite{MONTY2011} and \cite{harun_monty_2013}, where the flow history was not controlled.

For a complete study of PG TBLs it is necessary to obtain databases with near-equilibrium conditions extending over long streamwise regions so the effects of the PG and the Reynolds number can be clearly identified and studied. 
In this study we contribute towards that goal with a new well-resolved LES of incompressible and near-equilibrium APG TBL over a flat-plate with a nearly-constant value of $\beta \simeq 1.4$ over a large Reynolds-number range up to $Re_{\theta} \simeq 8,700$ (where $Re_{\theta}$ is the Reynolds number defined in terms of edge velocity and displacement thickness). The $\beta$ value is not constant along the streamwise development of the TBL, but its rate of reduction is small enough to be considered as nearly-constant, since the APG effects are larger for low Reynolds numbers and here we focus on a region of high Reynolds number. A comparison with experiments at similarly high $\Rey$ and $\beta$ values is carried out in this work, and even if the flow history of the experimental $\beta$ exhibits a larger variation, the profiles of the simulation and the experiment are in good agreement, indicating that these small deviations from a constant $\beta$ are not relevant in the region of high Reynolds number.

This is one of the largest simulations of a near-equilibrium APG TBL extending over a Reynolds-number range comparable to that of wind-tunnel experiments \citep{MTL_expSANMIGUEL}. 
Throughout this work, the results will be compared with the well-resolved LES ZPG TBL by \cite{E-AmorZPG}, which exhibits a similar Reynolds-number range. These data-sets allow for a proper study of APG and Reynolds-number effects in the turbulent statistics as well as in the energetic scales involved in turbulence.

The article is organized as follows: in \autoref{sec:NumSetUp} the studied databases are presented, together with the numerical setup of the new simulation. 
The turbulent 2D statistics are compared with experimental data in \autoref{sec:RS_peaks_and_exp}.

Different scalings for the statistics will be considered in \autoref{sec:Scalings} and in order to understand the energetic scales as well as their distribution, the spectral analysis of the Reynolds stresses will be presented in \autoref{sec:Spectra}. 
In \autoref{sec:Conclusions} some conclusions on APG and Reynolds-number effects will be drawn, and an outlook will also be given. In appendix \hyperlink{AppA}{A} we include for the sake of completeness other turbulent statistics such as the streamwise evolution of integral parameters and the turbulent kinetic energy (TKE) budgets as well as the other 2D statistics in outer scaling that will serve as a support material to document the phenomena seen in the Reynolds stresses and spectra seen in previous sections.
Finally, appendix \hyperlink{AppB}{B} gives a breve description of the subgrid-scale model used in the present simulation.

\section{Numerical setup} \label{sec:NumSetUp}
The objective of the present simulation setup is to obtain a TBL developing under a moderate APG over a long region of near-equilibrium flow from low to high Reynolds numbers. To achieve a high-Reynolds-number simulation starting from a laminar flow we have taken as a reference the ZPG LES simulation by \cite{E-AmorZPG} based on \cite{Schlatter_etAl_LES_2010}, which on itself was validated using DNS by \cite{schlatter_orlu_2010}. The box size in the streamwise direction ($L_x$) was chosen to be identical to that of the ZPG simulation. The effect of an APG will increase the growth rate of the boundary layer together with the size of the energetic scales, and in order to account for this we increased the box size in the wall-normal ($L_y$) and spanwise directions ($L_z$) with respect to the values used by \cite{E-AmorZPG}, as can be observed in table \ref{tab:param}. Since the code used for the simulation, SIMSON \citep{simson_techrep}, uses Fourier modes in the streamwise direction, all the flow that has exited the domain through the top boundary due to the growth of the boundary layer, needs to enter again into the domain at the end of the box through a fringe region. 
The fringe length in the APG case will have to be longer than in the ZPG configuration, since a larger amount of flow leaves the domain in the former case and it will need to return through a negative wall-normal velocity component $V$, which will increase the Courant--Friedrichs--Lewy (CFL) number. The time step of the simulation was kept constant in order to perform Fourier analysis of the temporal series of the velocity components. Therefore, in order to have a constant time step that would not be too low and maintain a stable CFL, the fringe length had to be increased so the maximum $V$ was reduced in that region. It is important to mention that the CFL also depends on the spatial discretization, and since SIMSON uses Chebyshev polynomials in the wall-normal direction $y$, the wall-normal grid spacing ($\Delta y$) at the top boundary is too fine, if this spacing could be increased, then the CFL number would be reduced and it would be possible to use a shorter fringe with a higher $V$.

\subsection{Parameters of the simulation} \label{subsec:Param_sim}
The streamwise, wall-normal and spanwise coordinates $(x,y,z)$ and other lengths or distances are non-dimensionalized with the reference length $\delta^*_0$ (which is the displacement thickness of the inflow laminar boundary layer). In some parts of the text $\delta^*_0$ is not written for the sake of simplicity. That is also the case for the velocities, which are non-dimensionalized with the inflow free stream velocity $U_{\rm ref}$.
For all of the simulations in table \ref{tab:param} the same code (SIMSON) was employed, and the well-resolved LES is based on the same sub-grid-scale (SGS) model, ADM-RT, which stands for approximate deconvolution relaxation-term \citep{Schlatter_2004}, as discussed in more detail in appendix \hyperlink{AppB}{B}. In the present simulation, denoted by b1.4, we have used a new feature in the code: the implementation of message-passing interface (MPI) communication in single precision. In this version of the code, all the computations continue to be in double precision, but the time spent in communication between processors has been reduced by casting the data into single precision and reducing by half the amount of data to communicate. 

After selecting the box size we proceed to run the simulation with ZPG boundary conditions (BC) at a low resolution. The initial conditions and inflow profile are given by a laminar Blasius boundary layer with a displacement thickness $\delta^*_0$ such that $Re_{\delta_0^*}=450$. The flow is tripped to turbulence close to the inlet at $x/\delta^*_0=10$ using a volume forcing \citep{schlatter_orlu_2012}, which is the same method as in the other databases, and with this configuration we expect the transition to turbulence to be over at $Re_{\theta}\approx 600$ \citep{E-AmorZPG, schlatter_orlu_2012}. The TBL will develop to high Reynolds numbers over a long computational domain in a similar way as it is done in a wind-tunnel experiment. This is different from setting up an auxiliary ZPG simulation and using a recycle plane for the inflow as used in \cite{Kitsios2016, Kitsios2017, Gungor_DNS_2017}, since the APG BCs downstream would affect the initial ZPG, as can be seen in figure \ref{fig:U_BCs}. At the end of the domain a fringe region \citep{SIMSON_fringe} reduces the turbulence and yields the same laminar flow as the inflow in order to obtain the periodicity required by the Fourier discretization. 
One flow-through is defined as the time required by the flow to pass through the domain once; here the mean streamwise velocity is of the order of the unity, so we will approximate the time required by the flow to cover a distance $L_x$ to be equal to that distance. When the flow is fully turbulent after $\sim3$ flow-through times we proceeded to progressively change the APG BC to the desired distribution. In the final APG configuration, a fringe length of $1500 \delta^*_0$ is needed to keep the simulation stable.
The simulation was run for enough time for the statistics to converge \citep{vinuesa2016_mec} after excluding the initial transients due to the change of resolution. 
The statistics were averaged for over 8 and 4 eddy-turnover times (ETT) at the middle and the end of the region of interest respectively (note that the region of interest ends before the fringe). In APG TBLs the eddy-turnover time varies significantly with $x$, and it is calculated as: $ {\rm ETT}(x) = \Delta T  u_{\tau}(x) / \delta_{99}(x)$, with $\Delta T$ being the period of time used for averaging and $u_{\tau}(x)=\sqrt{(\tau_w / \rho)}$ the friction velocity.
We also collected time series containing 36,000 samples over the same period. 
The time step of the simulation is held constant at $\Delta t_{\rm step}=0.25$, which ensures that in viscous units the time step along the domain is $\Delta t_{\rm step}^+ \leq 0.35$. The statistics were sampled every 4 time steps in the region of interest, which corresponds to $\Delta t_{\rm stat}^+ \leq 0.5$.
The viscous scaling (denoted by the superscript `+') for velocity, length and time scales is:  $u_{\tau}$, $l_{\tau}=\nu/u_{\tau}$ and $t_{\tau}=l_{\tau}/u_{\tau}$.

In table \ref{tab:param} we show the box size, the collocation points using the 3/2 factor for de-aliasing ($m_x, m_y, m_z$) and the resolution of the various simulations. With the previous parameters we obtain the streamwise and spanwise resolution in viscous units ($\Delta x^+ , \Delta z^+$). The maximum separation of grid points in the wall-normal direction inside the boundary layer is shown in viscous units as $\Delta y_{\rm max}^+$, and it occurs close to the boundary-layer edge at the highest Reynolds numbers. In the new simulation b1.4, for $Re_{\tau}=500$, the number of grid points below $y^+=10$ is 6, below $y^+=1$ we have 2 and the distance of the first grid point from the wall is $\Delta y_w^+=0.3$. Since the viscous length scale $l_{\tau}$ increases with the streamwise position, the resolution in viscous units close to the wall will be better at higher Reynolds numbers: for $Re_{\tau}=800$ we have 7 grid points below $y^+=10$, 3 grid points below $y^+=1$ and $\Delta y_w^+=0.2$.
The color code that will be used along this work is shown in table \ref{tab:param}.

\begin{table}
  \begin{center}
\def~{\hphantom{0}}
    \begin{tabular}{ l r  r  r  r  r  r  r  r  r  r  r r}
    Case  & $L_x/\delta_0^*$ & $L_y/\delta_0^*$ & $L_z/\delta_0^*$ & $m_{x}$ & $m_{y}$ & $m_{z}$ & $\Delta x^{+}$ & $\Delta y_{\rm max}^{+}$ & $\Delta z^{+}$ & Colour & Reference \\[3pt]
    
    b1             &    3000   &   140   &    250  &    3072 &     301 &     576 &   21.8    & 8.4 &   9.7  & \redline     & Bobke (\citeyear{bobke2017}) \\
    b2             &    3000   &   180   &    320  &    3072 &     361 &     768 &   21.6    & 7.9 &   9.2  & \greenline   & Bobke (\citeyear{bobke2017}) \\
    m16            &    3000   &   180   &    220  &    3072 &     361 &     576 &   21.2    & 7.7 &   8.3  & \blueline    & Bobke (\citeyear{Bobke_2016}) \\

    ZPG            &   13500   &   400   &    540  &   13824 &     513 &    1152 &   22.6    & 19.6 &  10.9  & \blackline  & Eitel-Amor (\citeyear{E-AmorZPG}) \\
    b1.4           &   13500   &   800   &   1080  &   13824 &     301 &    1920 &   21.7    & 30.1 &  12.5  & \orangeline & Present study \\

    \end{tabular}
  \caption{Parameters of the simulations used in this paper. The inlet displacement thickness $\delta_0^*$ corresponds to the displacement thickness of a laminar flow for a Reynolds number $Re_{\delta_0^*}=450$. The box size is $(L_x/\delta_0^*, L_y/\delta_0^* , L_z/\delta_0^*)$  and $(m_x, m_y, m_z)$ are the number of collocation points including the $3/2$ factor for de-aliasing in the Fourier directions ($x$ and $z$). The spatial resolution in viscous units $(^+)$ of the streamwise and spanwise directions $\Delta x^{+}, \Delta z^{+}$, has been calculated using $(m_x, m_z)$ at the position $x/\delta_0^*=250$, which corresponds to a friction Reynolds number $Re_{\tau}\approx 210$ for all the simulations. The maximum viscous distance between grid points in the wall-normal direction inside the boundary layer is observed close to the boundary-layer edge and for the highest Reynolds number of each simulation. }
  \label{tab:param}
  \end{center}
\end{table}

The Reynolds-number and pressure-gradient ranges are shown in table \ref{tab:ROI},together with the region of interest (ROI) for each simulation. 
The ROI in the APG cases starts at the point where the maximum $\beta$ is achieved, while in the ZPG TBL, the starting point is taken at $Re_{\tau}=500$, which approximately corresponds to $Re_{\theta}\approx 1500$, where according to \cite{E-AmorZPG, schlatter_orlu_2012} the flow is independent of the inflow conditions and the tripping.
The last point of the ROI was chosen as the position where the skin-friction coefficient exhibits a clear tendency upwards due to the effects of the fringe; note that the large scales have not been affected yet.
In the ROI we observe a slowly-decaying positive $\beta$, where $\overline{\beta}$ is the average $\beta$ along the ROI. Note that this is not the accumulated $\overline{\beta}$ defined by \cite{Vinuesa_2017}, but a simple average instead. The standard deviation is $\sigma({\beta})$ and these quantities, together with the defect shape factor  $G=(H_{12}-1)/(H_{12}\sqrt{c_f/2})$, are given in table \ref{tab:ROI}. Note that $H_{12}$ and $c_f$ are the shape factor and the skin-friction coefficient, respectively. In the ROI of the b1.4 case, $G$ lies between 10.5 and 11.6, which is in agreement with the results reported by \cite{bobke2017}: 9.8 and 12.1 for the b1 and b2 cases, respectively.
\begin{table}
  \begin{center}
\def~{\hphantom{0}}
    \begin{tabular}{ l c  c  c  c  c  c c c}
    Case  & Range of $x/\delta_0^*$ & Range of $Re_{\tau}$ &  Range of $Re_{\theta}$ & Range of $\beta$ & $\overline{\beta}$ & $\sigma({\beta}) / \overline{\beta}$ & $G$ \\[3pt]
    
    b1             &     718 --  2101   &  350 -- 750   &  1320 -- 3080  & 1.12 -- 0.85 & 0.99 &  0.08 &    9.8 -- 10   \\
    b2             &    1148 --  2001   &  480 -- 730   &  2260 -- 3530  & 2.19 -- 1.86 & 2.06 &  0.04 &   12.1 -- 12.3 \\
    m16            &     823 --  2001   &  370 -- 740   &  1790 -- 3650  &  2.8 -- 1.9  & 2.42 &  0.11 &   12.5 -- 13.5 \\

    ZPG            &    1251 -- 11666   &  500 -- 2500  &  1430 -- 8200  &  $\simeq 0$  & $\simeq 0$ & $\simeq 0$ &   7.1 -- 7.2 \\
    b1.4           &    2455 --  8117   &  800 -- 1900  &  3700 -- 8700  & 1.65 -- 1.20 & 1.41 &  0.10 &   10.5 -- 11.6  \\
    \end{tabular}
  \caption{ Flow characteristics in the ROI for the various cases.}
  \label{tab:ROI}
  \end{center}
\end{table}

\subsection{Boundary conditions} \label{subsec:BC}

These simulations are statistically homogeneous in the spanwise direction $z$, therefore, they are statistically two-dimensional in the streamwise and wall-normal directions, and have periodic boundary conditions in the streamwise and spanwise directions. 
Since no cross-flow is present in the spanwise direction the only boundary condition (BC) is imposed by the periodicity. In the streamwise direction, the periodicity forces the outflow to be the same as the inflow, and this is achieved through a fringe region as mentioned above. 
The inflow is given by a Blasius boundary layer at $Re_{\delta^*_0}=450$.
At the bottom boundary of the computational domain there is a flat plate, the BCs of which are no-slip and no transpiration. 
The PG will be imposed at the top boundary of the domain with the free-stream boundary conditions discussed next.

\subsubsection{Free-stream boundary conditions}
Three conditions have to be imposed at the top of the domain:

\begin{equation} 
    \frac{\partial W}{\partial y}=0,
\label{eq:BC_dWdy}
\end{equation}

\begin{equation} 
    \Omega_z = \frac{\partial V}{\partial x} - \frac{\partial U}{\partial y}=0,
\label{eq:BC_Vort_z}
\end{equation}

\begin{equation}
    U_{\rm top}(x)= \left\{ \begin{array}{ll}
             U_{\rm ref}, &  x \leq x_z \\
             \\ U_{\rm ref}\left( 1 + \frac{x - x_{z}}{x_{b}} \right)^{-m}, & x > x_z.
             \end{array}
   \right.
\label{eq:BC_APG_U}
\end{equation}
The first one relates to the statistical homogeneity in $z$, through the derivative in the wall-normal direction of the spanwise mean velocity as shown in (\ref{eq:BC_dWdy}).
Homogeneity in $z$ implies that $W$ is statistically zero, therefore its wall-normal derivative should also be zero.
Homogeneity also implies that the mean derivatives in $z$ are statistically zero, a fact that has an implication in the mean streamwise vorticity, {\it i.e.}  $\Omega_x = \partial W/ \partial y - \partial V / \partial z = 0$.
This could be implemented in the simulation in multiple ways, but since the perturbations in the farfield are negligible in our case, the easiest way is to set $\partial \widetilde{w} / \partial y = 0$ for each time step, where ($\widetilde{ ~~ }$) denotes instantaneous quantities.
The second BC is zero spanwise vorticity, given by (\ref{eq:BC_Vort_z}) as in \cite{Kitsios2016, Abe_2019};
finally, the third BC will introduce the adverse pressure gradient through a decaying mean streamwise velocity $U_{\rm top}(x)$ using a power-law distribution. First, a constant velocity is used to provide an initial ZPG development and at $x=x_z=350$ the power law is applied as indicated in (\ref{eq:BC_APG_U}). At the end of the domain the fringe raises the velocity to the initial $U_{\rm ref}$ values. The power law was also used as in \cite{Bobke_2016}, based on the theoretical studies by \cite{Townsend_1956_structure} and \cite{mellor_gibson_1966} in order to obtain a near-equilibrium TBL.
The use of (\ref{eq:BC_APG_U}) together with a fringe region may produce instabilities in the code where the fringe region starts, which were damped by the LES filter. The parameters of the power law were chosen as in the m16 simulation \citep{Bobke_2016}, {\it i.e.},  $x_z=350$,  $x_b=60$ and $m=0.16$.

\begin{figure} 
\centering
    \includegraphics[width=0.6\textwidth]{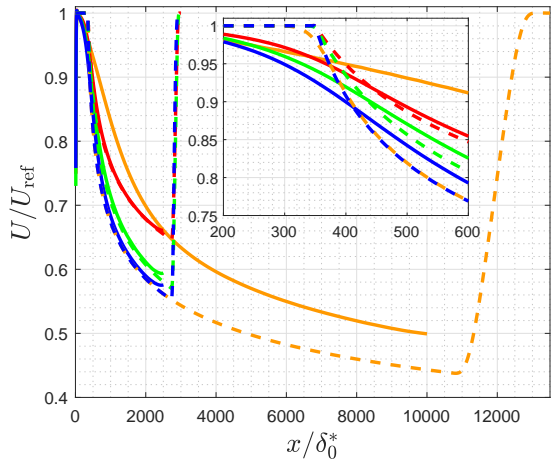}
  \caption{Streamwise evolution of (dashed) velocity at the top of the domain $U_{\rm top}(x)$ and (solid) velocity at the boundary-layer edge $U_e(x)$. Colors: (\protect\orangeline) b1.4; (\protect\redline) b1; (\protect\greenline) b2; (\protect\blueline) m16.}
%   Colors as in table \ref{tab:param}.}
\label{fig:U_BCs}
\end{figure}

 In figure \ref{fig:U_BCs} we show the streamwise evolution of $U_{\rm top}$ for the various cases under study, and it can be observed that this quantity is constant up to $x=350$, while it decays following a power law with different exponents $m$ for each APG \citep{bobke2017}. Note that the later rise of $U_{\rm top}$ is produced by the fringe.
 Since the BC is applied at the top of the domain and not at the edge of the BL, the velocity at the boundary-layer edge $U_e$ is not the same as $U_{\rm top}$, a fact that has been observed in multiple experiments and simulations of PG TBL and is still discussed in the scientific community as part of the problem of determining the edge of the TBL \citep{diagnostic_Vinuesa, d99_determination_2020}. The top boundary condition is far from the boundary-layer edge where the TBL  starts to develop, however, when the TBL starts to grow, this distance is reduced and the curves for $U_{\rm top}$ and $U_{e}$ come close to each other. This effect can be seen for the APGs by Bobke since the height of the domain was lower than in the larger b1.4 simulation.
 The fact that the the mean streamwise velocity $U$ exhibits a gradient ${\partial U}/{\partial y} \neq 0$ (which is a consequence of the streamwise PG) makes it harder to impose a specific velocity distribution at the edge of the BL, specially for the larger high-Reynolds simulations that require a taller computational domain for the TBL to grow. This effect could be reduced if the simulation could be performed in a domain with a variable height and not in a box of constant height.
 As a result of this effect, the imposed ZPG at $y=L_y$ is not perceived as a pure ZPG at the boundary layer edge, where the velocity instead of being constant, is slightly decaying as in an APG. As stated above, in the case where an auxiliary ZPG simulation gives the inflow, this effect of upstream influence of the APG will not be seen.
 The parameters for $U_{\rm top}$ are the same in the b1.4 and m16 cases, as can be observed in figure \ref{fig:U_BCs}, but the resulting $U_{e}$ is higher in the b1.4 than in the m16 case, which means that the decay of the velocity is not as steep in the former as in the latter and it will result in a smaller $\beta$. This shows that it is important to take into account the value of $L_y$ when setting the $U_{\rm top}(x)$ distribution to achieve a certain $U_{e}(x)$. 
 
Since we use a zero-spanwise vorticity as a BC, it is possible to rewrite the Reynolds-averaged Navier--Stokes (RANS) equations for the momentum in $x$ and $y$ in terms of the mean spanwise vorticity and its derivatives.
Being outside of the TBL means that the Reynolds stresses can be neglected if the turbulence is confined in the BL.
The first derivatives in $x$ and $y$ of $U$ and $V$ are present in the continuity equation and the mean spanwise vorticity $\Omega_z$. If those two equations are derived in $x$ and $y$ then it is possible to obtain relationships between the first derivatives of $\Omega_z$ with the second derivatives of $U$ and $V$ (present in the viscous terms of the RANS equations). Substituting the spanwise vorticity and its first derivatives in the convective and viscous terms we get:

\begin{equation}\label{eq:RANS_x_BC_OMEGA}
    U \frac {\partial U} {\partial x} + V \frac {\partial V} {\partial x} +
      \frac {1} {\rho} \frac {\partial P} {\partial x} = 
      \nu \left( \frac {\partial \Omega_z} {\partial y} \right) + V \Omega_z,
\end{equation}
\begin{equation}\label{eq:RANS_Y_BC_OMEGA}
    U \frac {\partial U} {\partial y} + V \frac {\partial V} {\partial y} +
      \frac {1} {\rho} \frac {\partial P} {\partial y} = 
      \nu \left( \frac {\partial \Omega_z} {\partial x} \right) + U \Omega_z,
\end{equation}
where $P$ is the pressure. In these equations we can see the effects of using a zero-spanwise vorticity or also making its derivatives zero. The convective terms on the left-hand side can be written as the gradient of a total pressure   $P_T / \rho = P/\rho + (U^2+V^2)/2$ caused by the effects of non-zero spanwise vorticity. Even if the spanwise vorticity is set to zero at the top of the domain, this does not guarantee that it will remain zero in all the domain outside of the TBL.
The spanwise vorticity outside of the BL is related to the curvature of $P_T$ outside of the BL due to the growth of the BL.
 
\section{ Turbulence statistics at moderately-high Reynolds numbers} \label{sec:RS_peaks_and_exp}

The streamwise development of the different simulations as well as statistics in different wall-normal profiles are given in appendix \hyperlink{AppA}{A}, where similar conclusions as the ones observed in \cite{bobke2017} for near-equilibrium flows at lower Reynolds numbers are observed and extended to higher $\Rey$ numbers. 

In this section we will present the Reynolds stresses obtained at different streamwise positions since later in section \ref{sec:Spectra} they will be decomposed in their spectral components.

\begin{figure}
\centering
\includegraphics[width=0.6\textwidth]{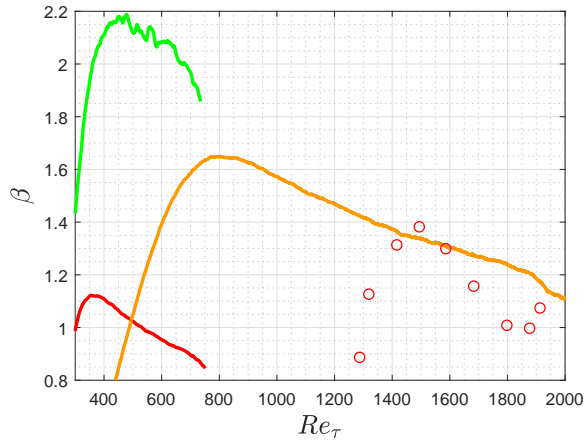}
  \caption{Evolution of the Clauser pressure-gradient parameter $\beta$ as a function of the friction Reynolds number $Re_{\tau}$ for three of the simulations Colors: (\protect\orangeline) b1.4; (\protect\redline) b1; (\protect\greenline) b2; (\protect\redcircle) experiments by \cite{MTL_expSANMIGUEL}.}
%   (colors as in table \ref{tab:param}) and the experiments by \cite{MTL_expSANMIGUEL} (represented by red circles).}
\label{fig:beta}
\end{figure}

In figure \ref{fig:beta} the Clauser pressure-gradient parameter $\beta=(\delta^*/\tau_w)  (\partial P/\partial x)_{e} $ is shown for the nearly-constant-$\beta$ simulations by \cite{bobke2017}, the current simulation and data obtained in experiments \citep{MTL_expSANMIGUEL} for a similar range of $\Rey_{\tau}-\beta$. Here, $\Rey_{\tau}=u_{\tau}\delta_{99}/\nu$ is the Reynolds number based on friction velocity and $\delta_{99}$ is the $99\%$ boundary-layer thickness, which was calculated by means of the method proposed by \cite{diagnostic_Vinuesa}.

For the following figures, the $Re_{\tau}=500$ profiles (outside the ROI) will be considered to observe effects of different $\beta$, comparing b1.4 with the other near-equilibrium APG simulations at lower $\Rey$. Three additional profiles within the ROI, at $\Rey_{\tau}=\{500, 1000, 1500\}$, will be used to determine the effects of moderate APG at higher $\Rey$ through a comparison with the high-$\Rey$ ZPG. The mean velocity profiles of these cases are shown in figure~\ref{fig:meanU} from Appendix~\hyperlink{AppA}{A}.

The inner-scaled Reynolds stresses are shown in figure \ref{fig:RSinner}, and the corresponding profiles in outer scaling can be observed in appendix \hyperlink{AppA}{A}, figure \ref{fig:RSouter}. 
The most noticeable characteristic of these TBLs is that the wall affects each component of the Reynolds-stress (RS) tensor differently. The streamwise component has in general a larger value than the other terms, making it the leading term of the turbulent kinetic energy (TKE). Since at the wall the no-slip condition makes the velocities zero, the RSs also start from a zero level. In the viscous sub-layer the mean velocity gradually increases, together with the velocity fluctuations. 
In the near-wall region, {\it i.e.} at around $y^+\simeq 15$, the streamwise Reynolds stress exhibits the well-known inner peak, while the other fluctuating components are moderately affected by the strong TKE production in this region.

The APG significantly affects the inner peak of $\overline{u^2}^+$, and when this peak is scaled in outer units its magnitude decreases with APG magnitude (see figure \ref{fig:RSouter}a)). Interestingly, the near-wall fluctuations increase slightly with $\beta$ in the other velocity components when scaling in outer units (figure~\ref{fig:RSouter}), a result which is more prominent in the case of $\overline{w^2}$.

\begin{figure}
\includegraphics[width=0.49\textwidth]{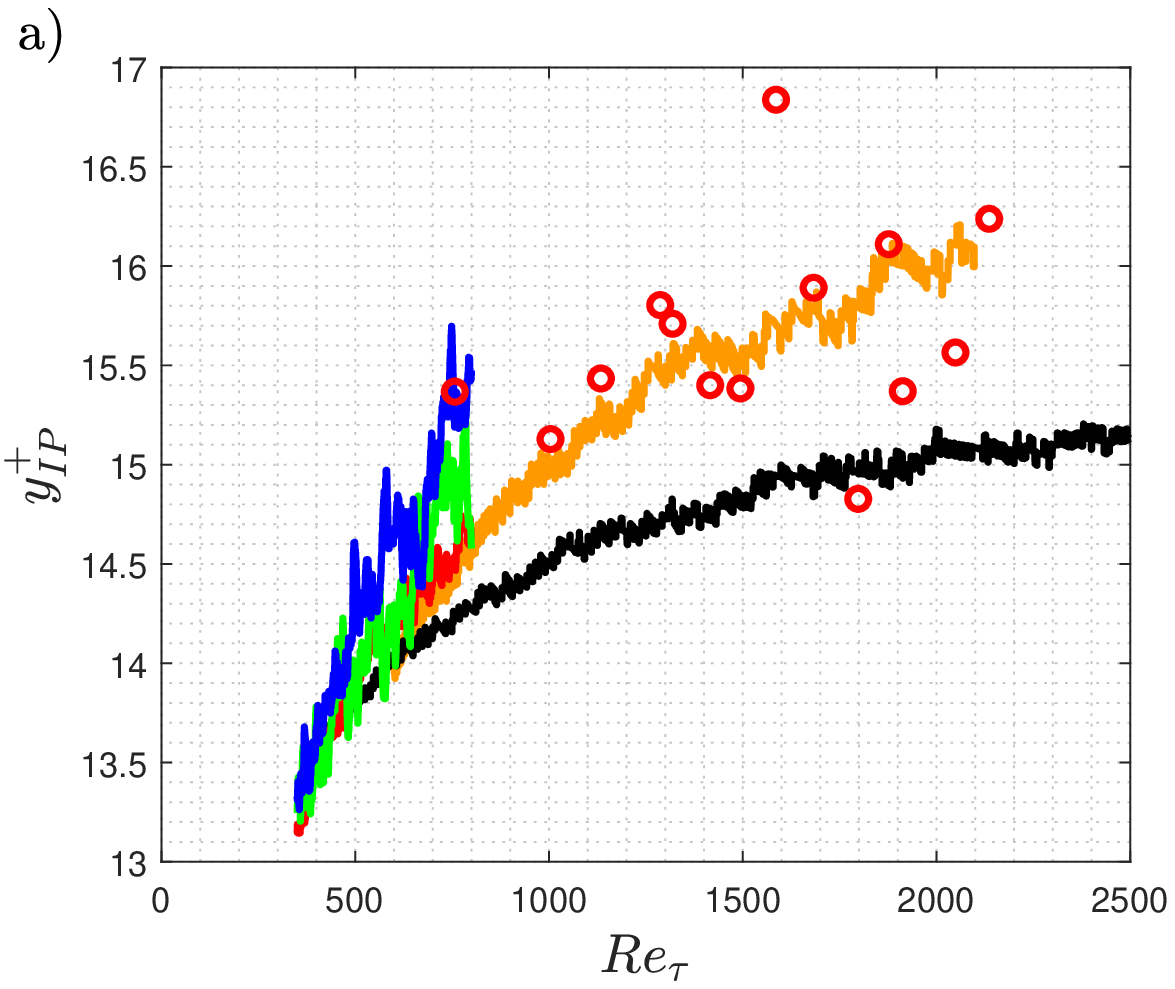}
\includegraphics[width=0.49\textwidth]{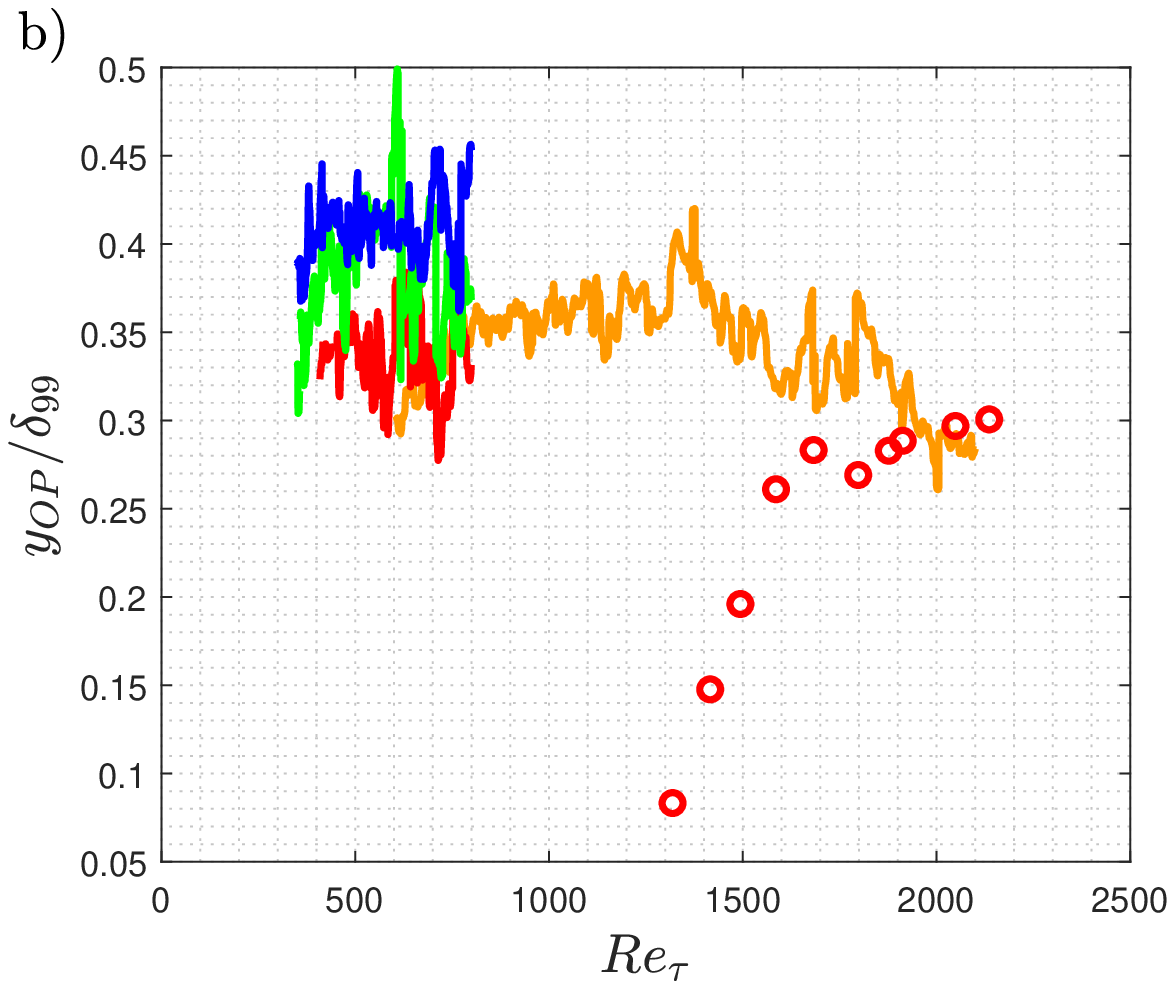} \\
\includegraphics[width=0.49\textwidth]{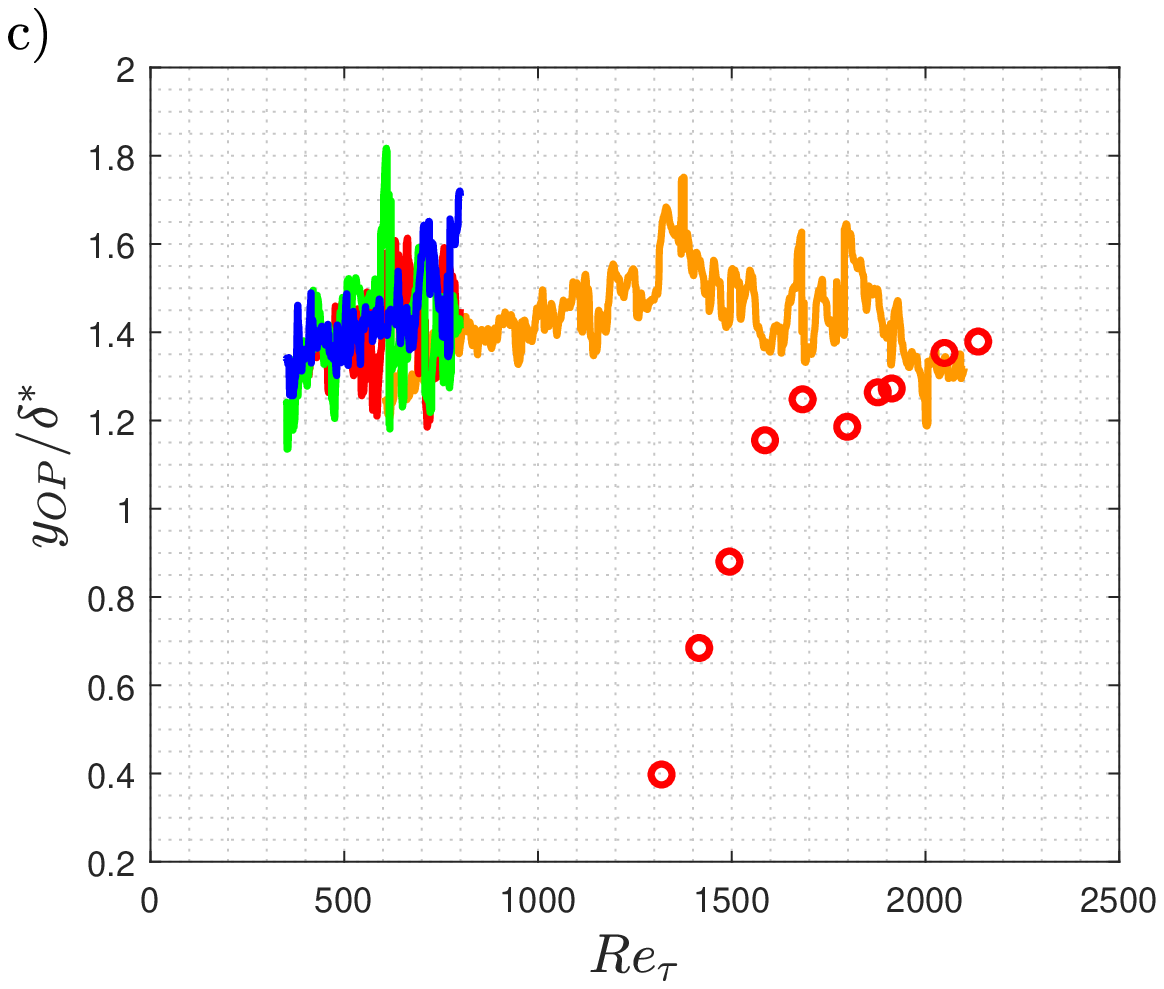}
\includegraphics[width=0.49\textwidth]{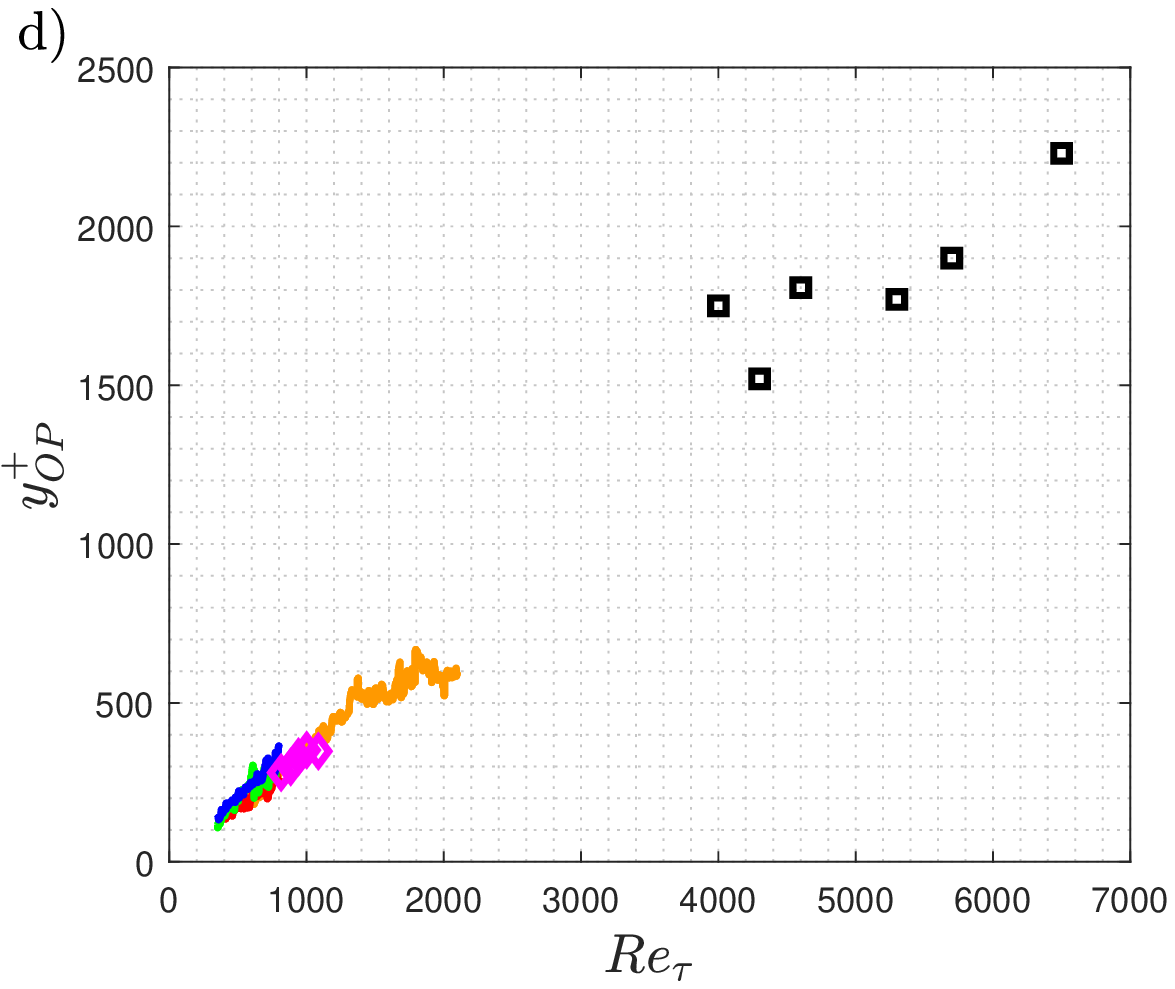}
\caption{Streamwise evolution of the wall-normal location of the inner and outer peaks of the streamwise Reynolds stress profiles. a) Inner-scaled position of the inner peak $y_{IP}^+$; b) and c) outer-peak location ($y_{OP}$) scaled with $\delta_{99}$ and $\delta^*$, respectively; d) inner-scaled outer-peak location $y^{+}_{OP}$. Colors: (\protect\blackline) ZPG; (\protect\orangeline) b1.4; (\protect\redline) b1; (\protect\greenline) b2; (\protect\blueline) m16; (\protect\magentaDiamond) $\beta=1$ DNS data from \cite{Kitsios2016}; (\protect\redcircle) experiments by \cite{MTL_expSANMIGUEL}; (\protect\blackSquare) experiments by \cite{skare_krogstad_1994}.}
\label{fig:uupeaks_loc}
\end{figure}

In the inner-scaled Reynolds stresses shown in figure \ref{fig:RSinner}, the influence of the APG can be observed in both $\overline{u^2}^+$ and $\overline{w^2}^+$, especially on the latter, from $y^+ \approx 2$ onward. However, the components containing the wall-normal velocity fluctuation are affected farther from the wall, starting at $y^+ \approx 10$ for $\Rey_{\tau}=500$ or even $y^+ \approx 20$ for higher $\Rey$. This behaviour supports the attached-eddy hypothesis on the differing contributions to the Reynolds stresses close to the wall \citep{Townsend_1976, deshpande_2021}, however, the trends are modified by the APG farther from the wall. 
The viscous scaling is appropriate for regions close to the wall, since it properly scales the mean streamwise velocity and the viscous length locates the inner peak of $\overline{u^2}^+$ at $y^+\approx 15$ (see figure \ref{fig:uupeaks_loc}a)). Furthermore, the friction velocity $u_{\tau}$ leads to more similar inner-peak magnitudes from different $\beta$ values than what is obtained using the outer velocity scale $U_{e}$. It is important to recall that the friction velocity is computed from ${\rm d}U / {\rm d} y$, which is the largest term of the near-wall TKE production also in APGs, closely connected with the formation of the inner peak in $\overline{u^2}^+$. The other components of the Reynolds stresses exhibit a better scaling using outer units even close to the wall, as shown in figure \ref{fig:RSouter} from Appendix A. 

 % Reynolds stress in inner units
\begin{figure}
\includegraphics[width=0.49\textwidth]{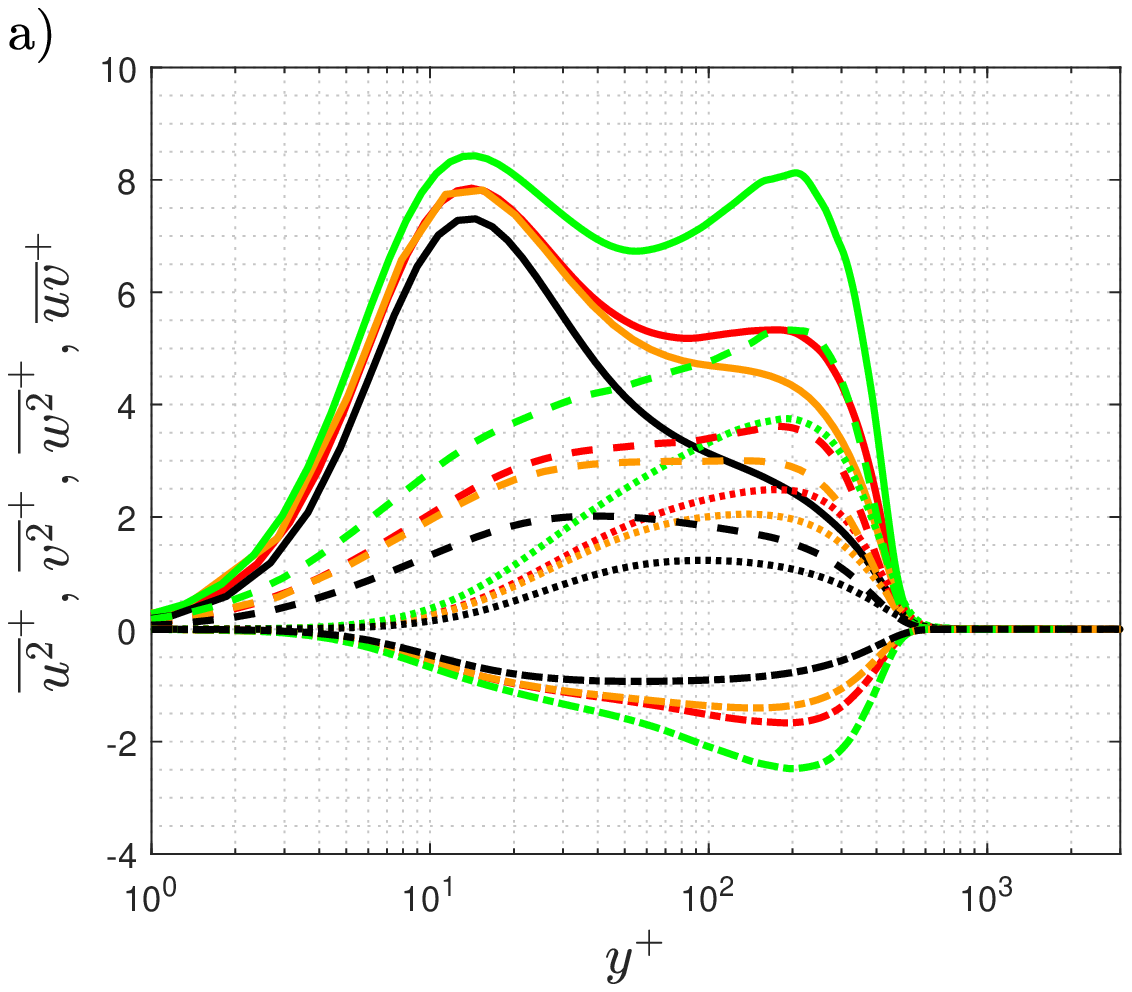}
\includegraphics[width=0.49\textwidth]{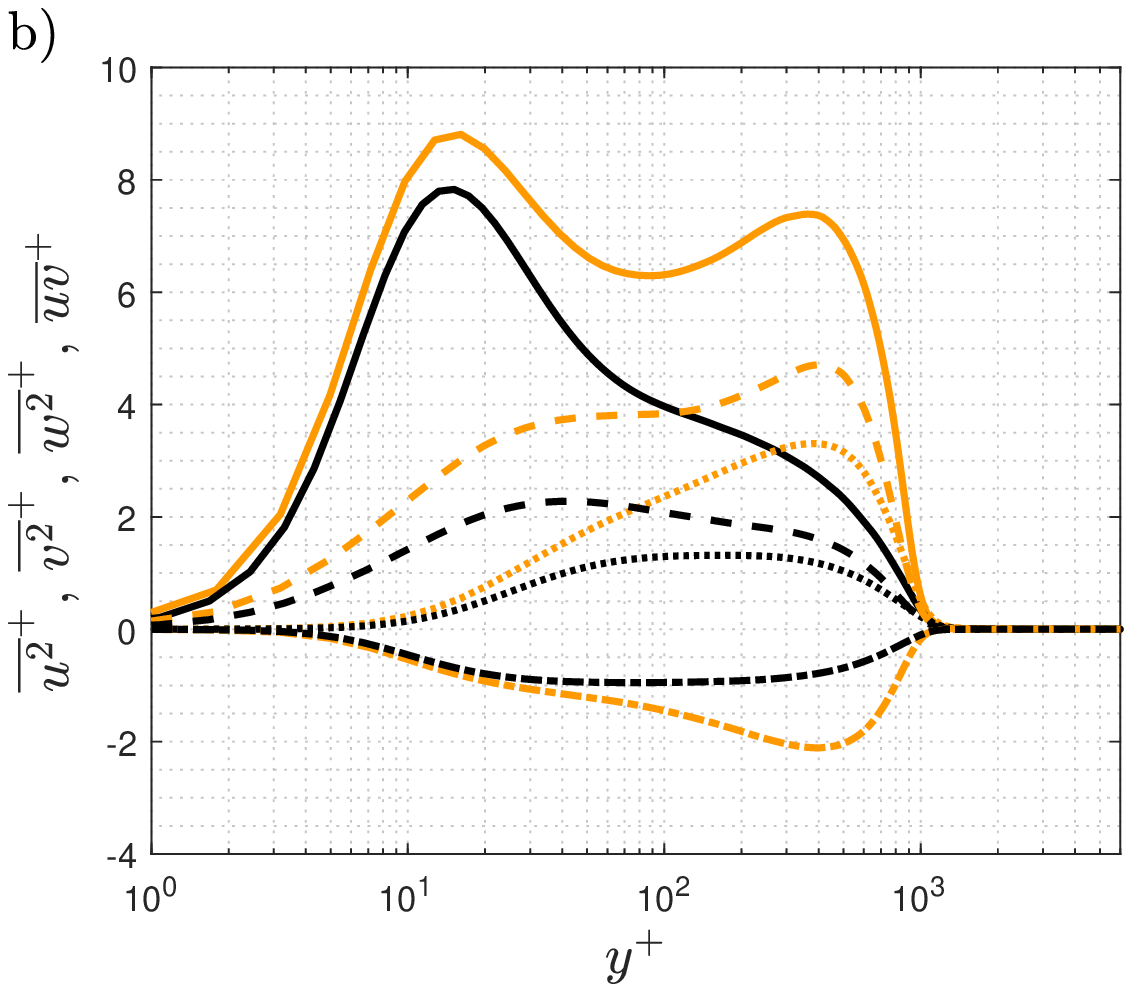} \\
\includegraphics[width=0.49\textwidth]{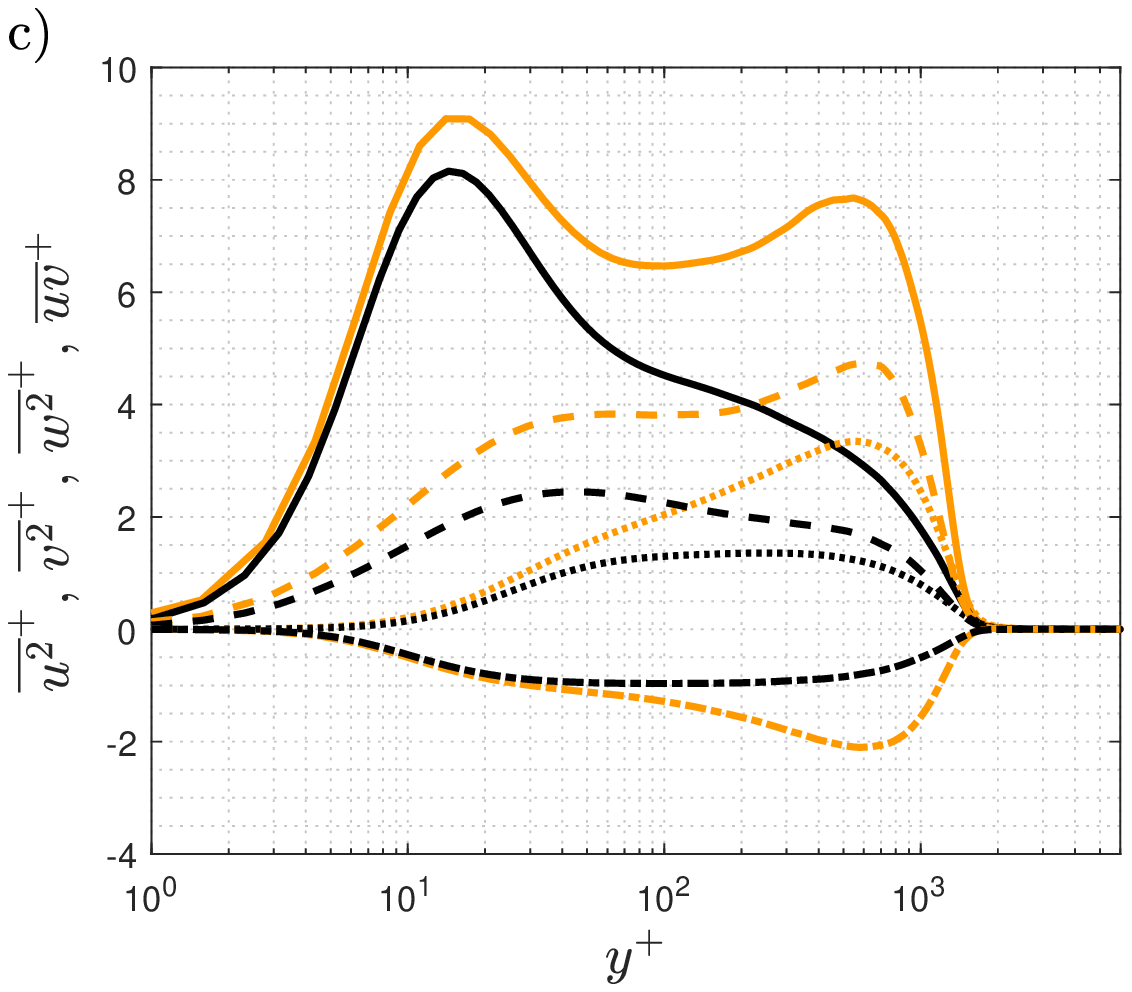}
\includegraphics[width=0.49\textwidth]{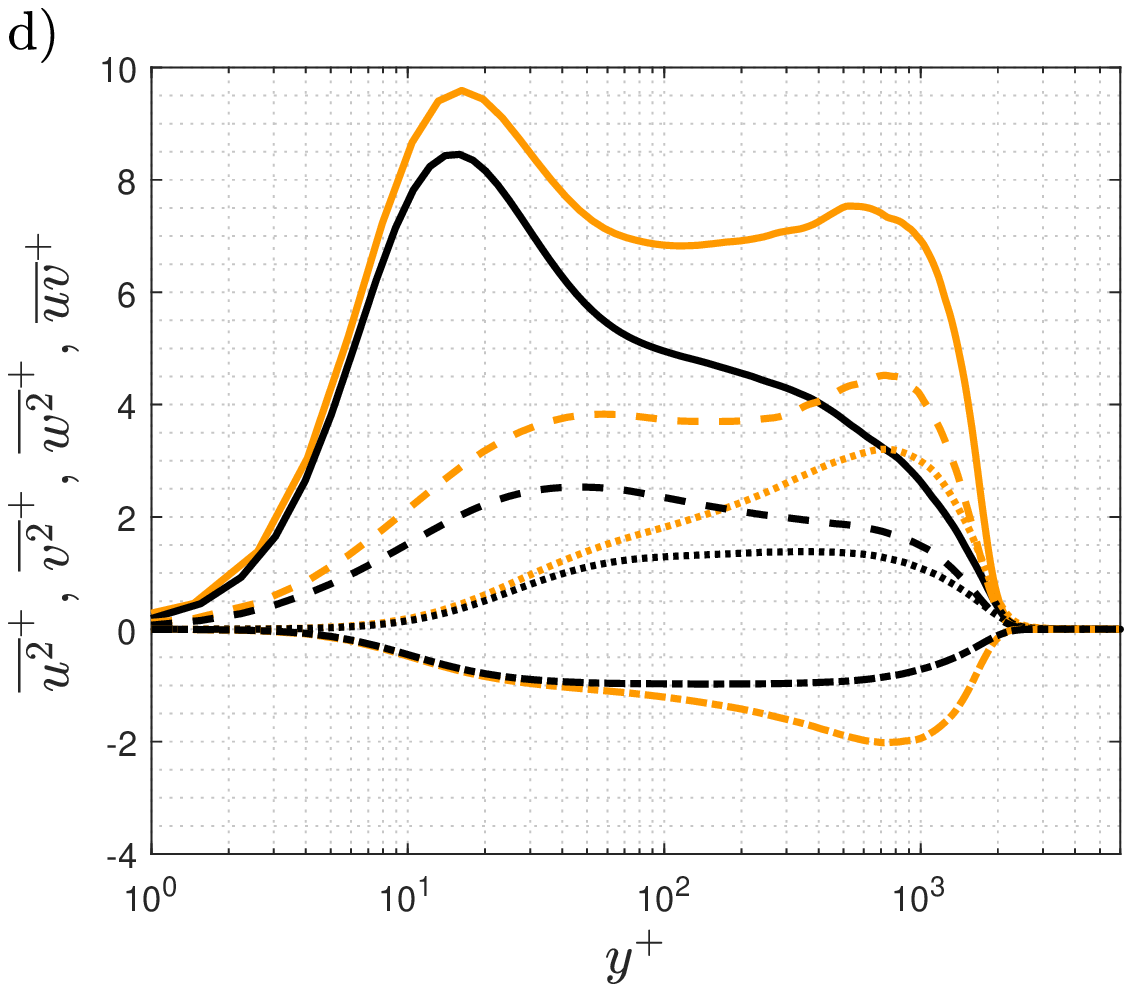}
  \caption{Inner-scaled Reynolds stresses scaled with the friction velocity $u_{\tau}$ at various matched $Re_{\tau}$: a) $Re_{\tau}=500$ where $\beta(Re_{\tau})$ intersects for the simulations b1 and b1.4; b) $Re_{\tau}=1000$ ; c) $Re_{\tau}=1500$ ; d) $Re_{\tau}=2000$. Symbols: (\protect\blackline) $\overline{u^2}^+$; (\protect\blackdotted) $\overline{v^2}^+$; (\protect\blackdash) $\overline{w^2}^+$; (\protect\blackdashdot) $\overline{uv}^+$. Colors and symbols: (\protect\blackline) ZPG; (\protect\redline) b1; (\protect\orangeline) b1.4; (\protect\greenline) b2 as in table \ref{tab:param}.}
\label{fig:RSinner}
\end{figure}

To further study the impact of APG and $\Rey$ on the near-wall and outer peaks, it is important to have fine resolutions around $y^+=15$ and well-converged statistics in the outer region. 

Since the experimental database exhibits some noise, a curve-fit approach was considered around the inner and outer peaks of the streamwise RS to determine their locations, while a simple spline interpolation was employed for the numerical data.

In figure \ref{fig:uupeaks_loc} we show the streamwise evolution of the peak locations, where a clear influence of the APG is observed. Based on our results, an increasing APG magnitude leads to a larger wall-normal location of the inner peak $y^+_{IP}$, a phenomenon which is also observed for higher $\Rey$ at a given constant $\beta$. While for the ZPG case the near-wall-peak location reaches an asymptotic value of around $y^+ \simeq 15$ for $Re_{\tau} >2,000$, the trend in the b1.4 case is currently inconclusive.

\begin{figure}
\includegraphics[width=0.49\textwidth]{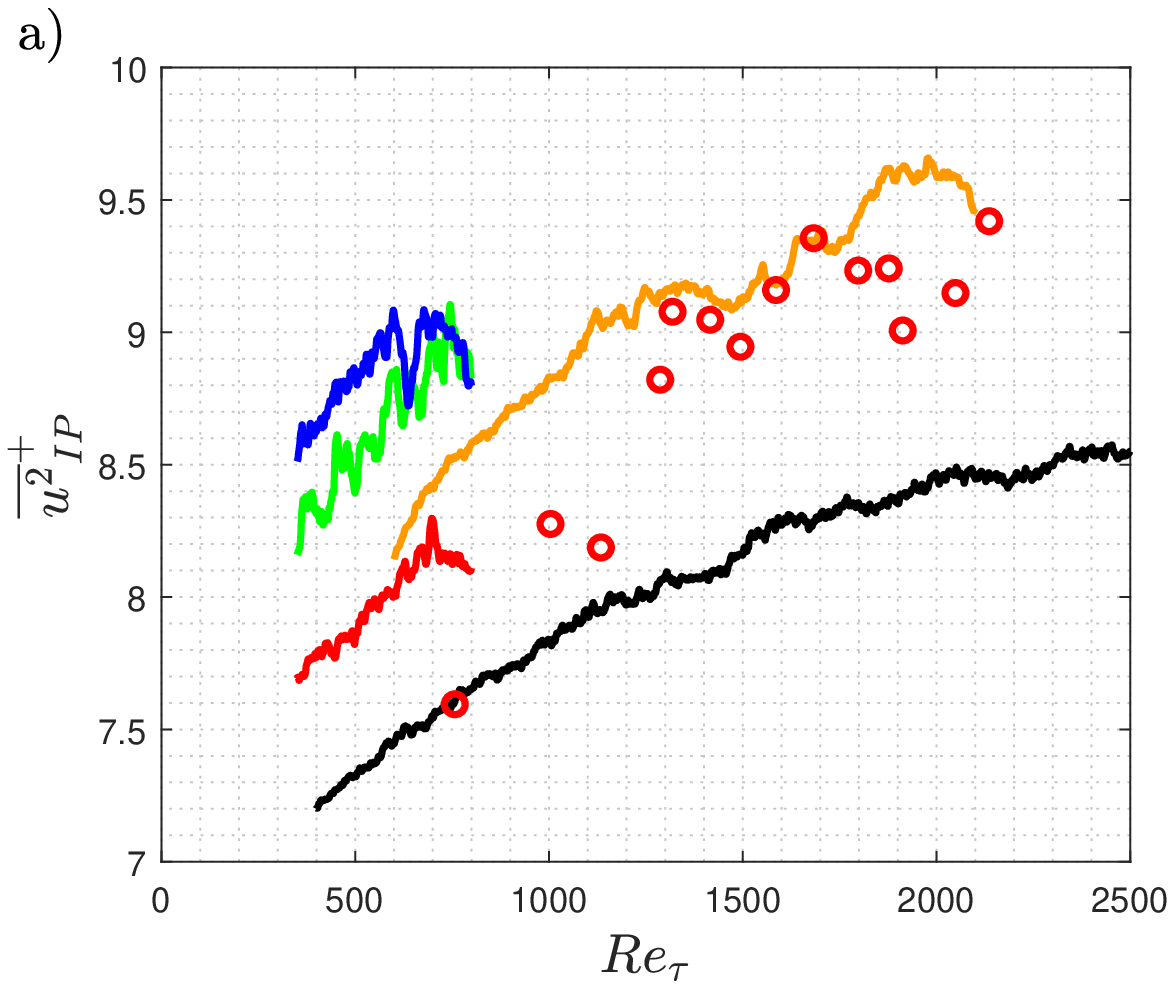}
\includegraphics[width=0.49\textwidth]{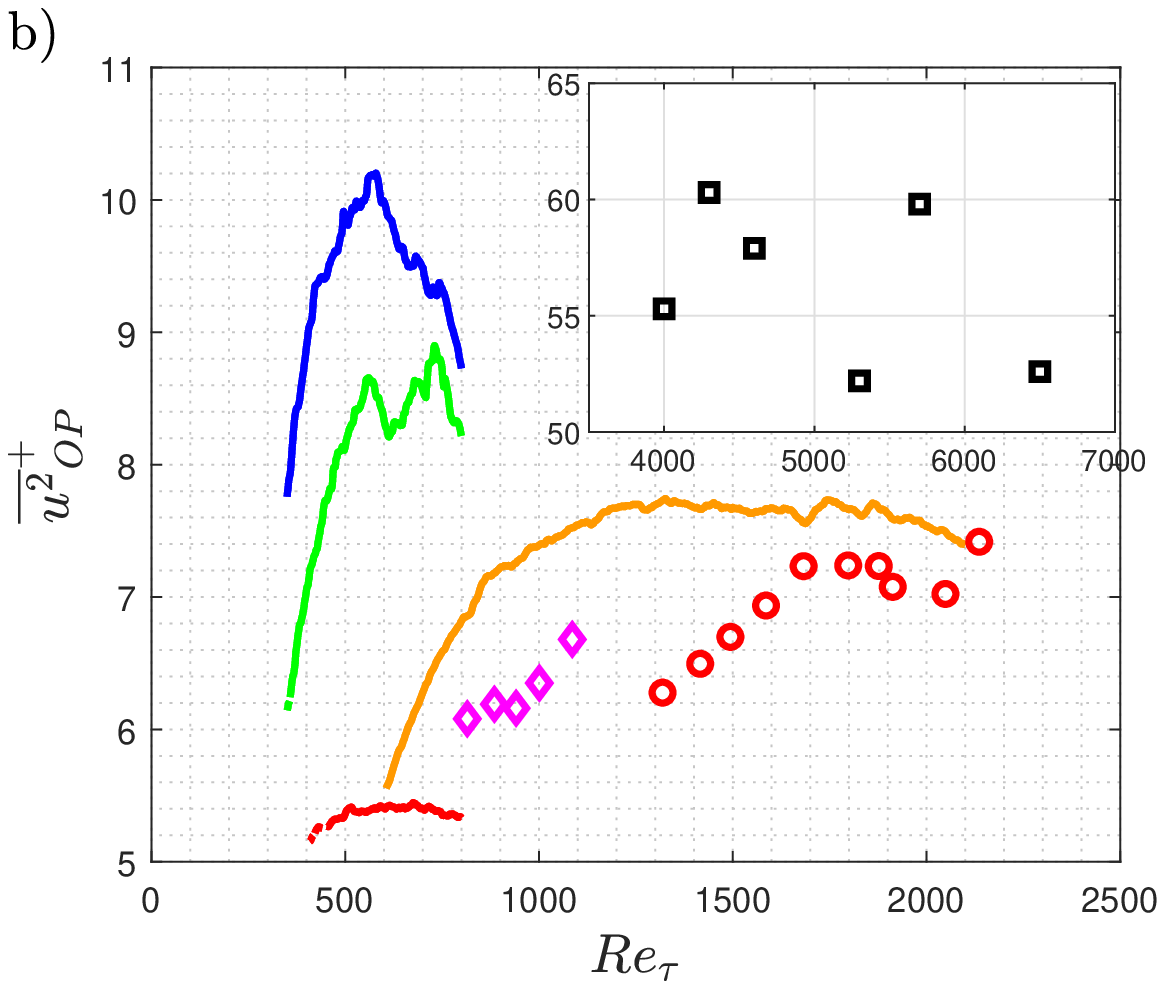}\\
\includegraphics[width=0.49\textwidth]{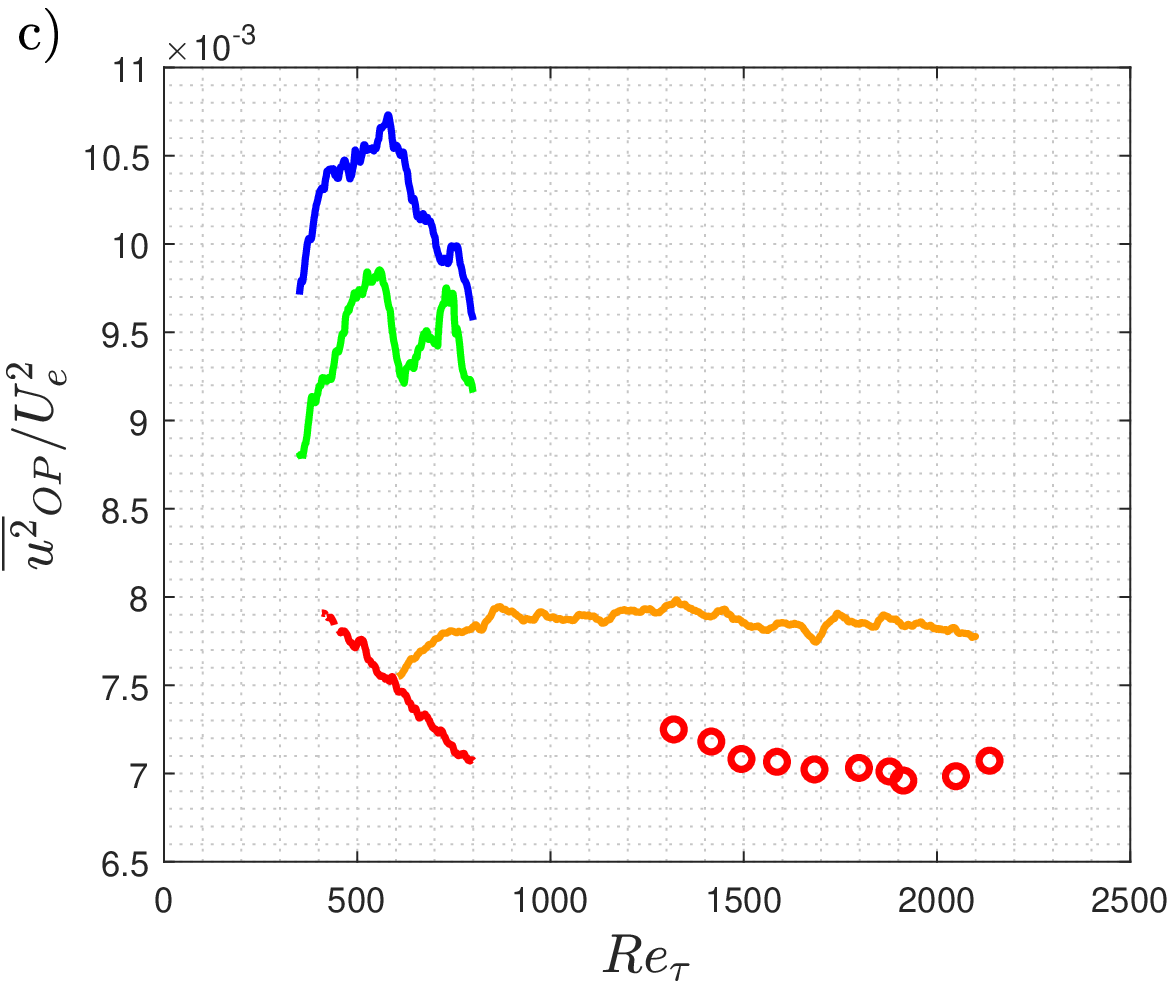}
  \caption{Streamwise evolution of the magnitude of the inner and outer peaks of the streamwise Reynolds stress profiles. a) Inner-scaled magnitude of the inner peak $\overline{u^2}^+_{IP}$; b) and c) outer-peak magnitude ($\overline{u^2}_{OP}$) scaled in inner and outer units, respectively. Colors: (\protect\blackline) ZPG; (\protect\orangeline) b1.4; (\protect\redline) b1; (\protect\greenline) b2; (\protect\blueline) m16; (\protect\magentaDiamond) $\beta=1$ DNS data from \cite{Kitsios2016}; (\protect\redcircle) experiments by \cite{MTL_expSANMIGUEL}; (\protect\blackSquare) experiments by \cite{skare_krogstad_1994}. }
\label{fig:uupeaks_val}
\end{figure}

Figure \ref{fig:uupeaks_loc}b) shows the outer-peak location $y_{OP}$ of $\overline{u^2}$, scaled with the $99\%$ boundary-layer thickness, and our results indicate that the curve reaches a slightly decaying trend with $\Rey$ in the b1.4 case. 
Furthermore, comparison with the other cases indicates that the values of $y_{OP}/\delta_{99}$ are clearly affected by $\beta$, with stronger APGs leading to larger values of the outer-peak location. Note that although the value of $\beta$ is approximately the same in the b1 simulation and in the experiment, the former is at a much lower Reynolds number, and therefore this case is expected to perceive a more intense effect of the APG. This would explain that the outer-peak location is slightly farther away from the wall in the b1 case than in the experiment.
Figure \ref{fig:uupeaks_loc}c) shows the outer-peak location of $\overline{u^2}$ scaled with the displacement thickness $\delta^*$. The low-$\Rey$ simulations exhibit a similar slowly-growing trend for the various $\beta$ cases, with an average value of around $y_{OP}/\delta^* = 1.4$. The b1.4 case appears to reach an approximately constant state at higher $\Rey$, also around 1.4, a result consistent with that reported by \cite{Sanmiguel_PRF} for the experimental data (despite the noise present in the measurements).

The inner-scaled location of the outer peak present in the DNS by \cite{Kitsios2016} and the experiments by \cite{skare_krogstad_1994} are also shown in figure \ref{fig:uupeaks_loc}d). The outer-peak location of the $\beta=1$ DNS by \cite{Kitsios2016} continues the trend defined by the b1 LES at lower $\Rey$ and lies below the line of the b1.4 LES. The values of the near-equilibrium experimental data by \cite{skare_krogstad_1994} at a much higher $\Rey$ and $\beta \approx 20$ appear to be consistent with the linear trend of $y_{OP}^+$ established by the lower-$\Rey$ data.

The magnitudes of the inner ($\overline{u^2}_{IP}$) and outer ($\overline{u^2}_{OP}$) peaks of the streamwise RS are shown in the different panels of figure \ref{fig:uupeaks_val}.
In panel a) it is possible to see the Reynolds-number evolution of the inner-scaled inner peak ($\overline{u^2}_{IP}^+$) for the various APGs, as well as that of the ZPG TBL, which is well documented in the literature \citep{E-AmorZPG, Marusic_1997_ZPG}. While it is unclear what the behavior will be for the APG cases at higher $\Rey$, our data indicates that the inner-scaled near-wall peak increases with APG magnitude. The trend from the experiment exhibits more scatter, but it appears to be in qualitative agreement with that of the b1.4 case. Note that, although the inner peak in inner scaling increases with $\beta$, it actually decreases in outer scaling for stronger APGs, as can be observed in figure \ref{fig:RSouter}.
The outer-peak value increases with $\beta$ using inner scaling, as shown in figure \ref{fig:uupeaks_val}b), and also in outer scaling with $U_e$, as illustrated in figure \ref{fig:uupeaks_val}c). In both panels, the trends are approximately constant with $\Rey_{\tau}$, where $U_{e}$ yields a reasonably flat curve even at lower Reynolds numbers.

In figure \ref{fig:uupeaks_val}b) the location of $\overline{u^2}^+$ for the $\beta=1$ DNS by \cite{Kitsios2017} and the b1 LES exhibit small differences probably associated with the different flow histories at low Reynolds numbers.
The experimental data by \cite{skare_krogstad_1994} exhibits a larger uncertainty and dispersion of the values, but an approximately-constant trend in $\overline{u^2}^+$ can be observed.

An unexpected trend is observed in figure \ref{fig:uupeaks_val}c) for the b1 simulation, a fact that could be attributed to the relatively low outer-peak values in this simulation.

%*********************************************************************************
% Experiments in MTL
%*********************************************************************************
\subsection{Comparison with experiments} 

The range of Reynolds numbers achieved in the b1.4 case allows for a direct comparison of the statistics with the experimental results by \cite{MTL_expSANMIGUEL}, where we selected two cases with matching $\beta$ and $\Rey_{\tau}$ conditions. As shown in  figure \ref{fig:beta}, the flow history of this database differs from that of the b1.4 case, a fact that will be taken into account when comparing the data.

\begin{figure}
\includegraphics[width=0.49\textwidth]{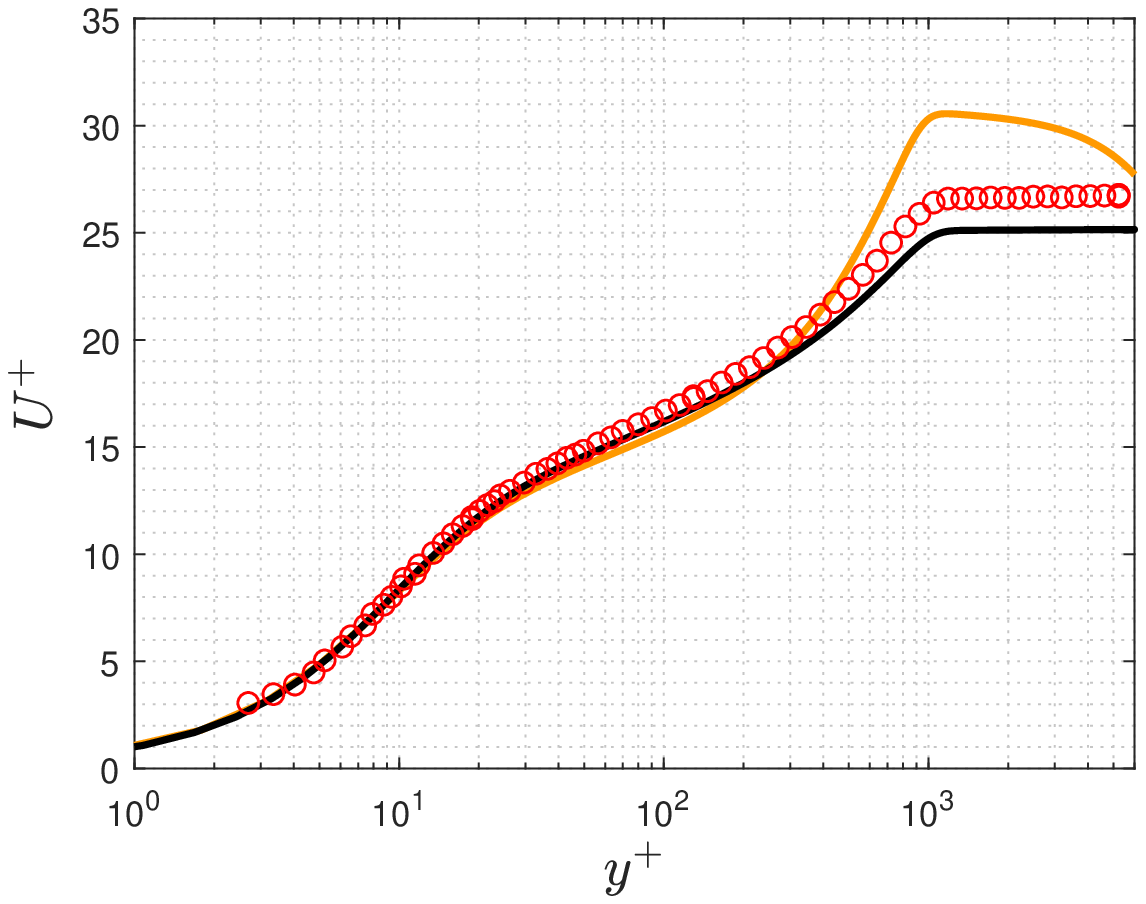}
\includegraphics[width=0.49\textwidth]{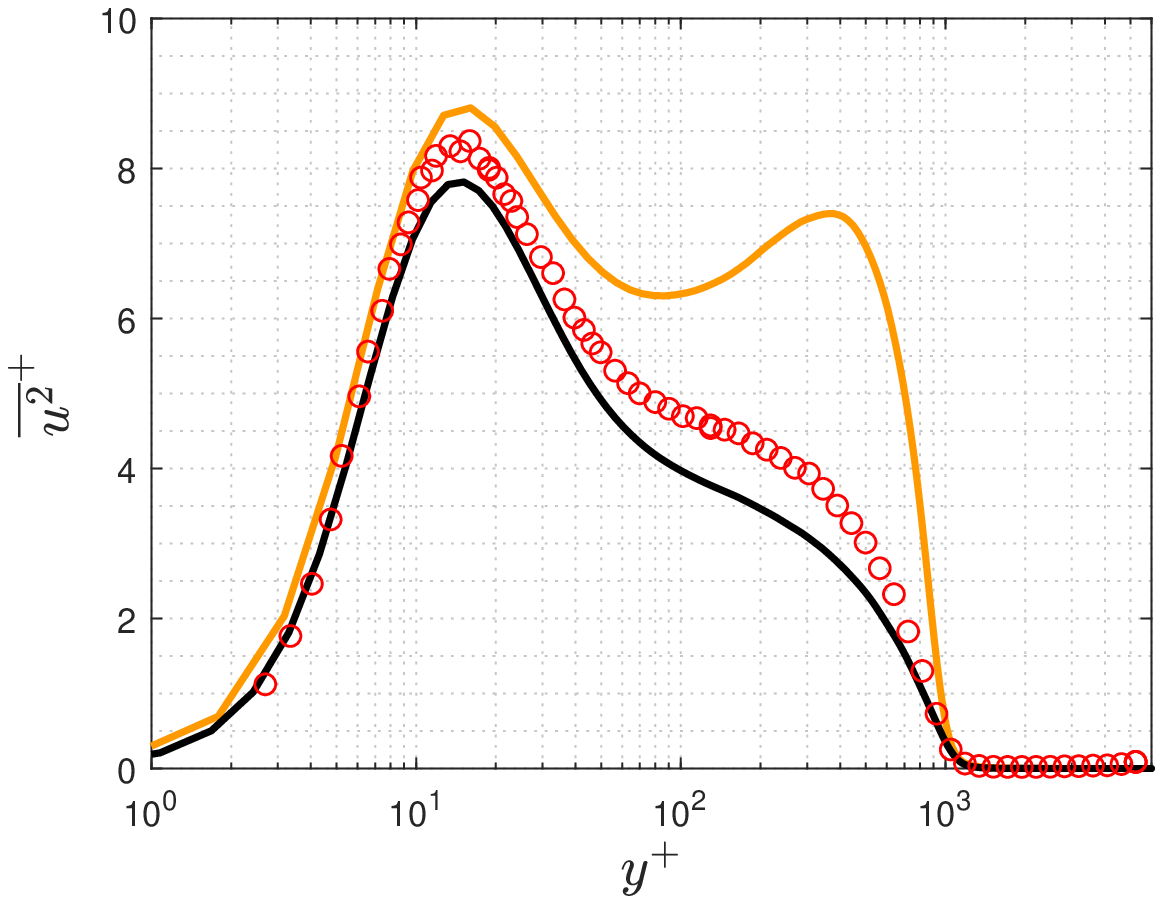}
\includegraphics[width=0.49\textwidth]{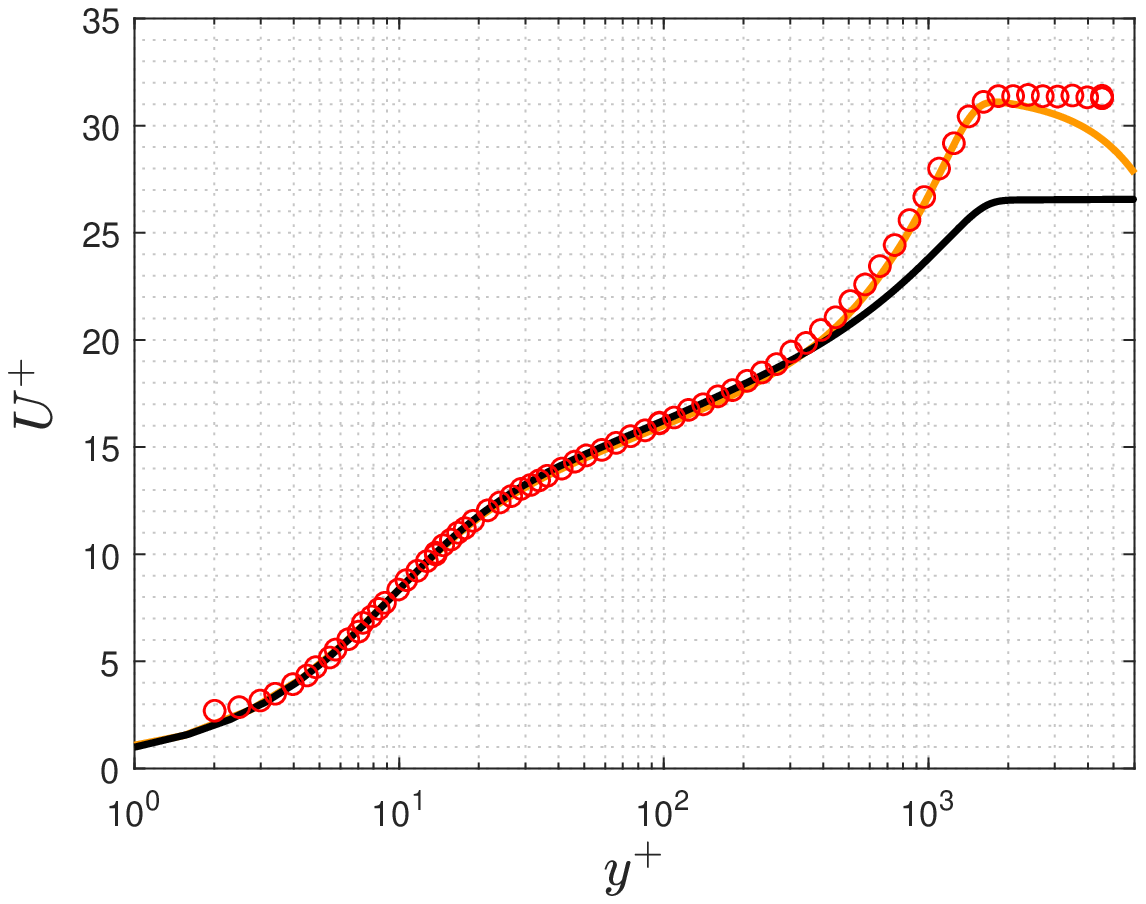}
\includegraphics[width=0.49\textwidth]{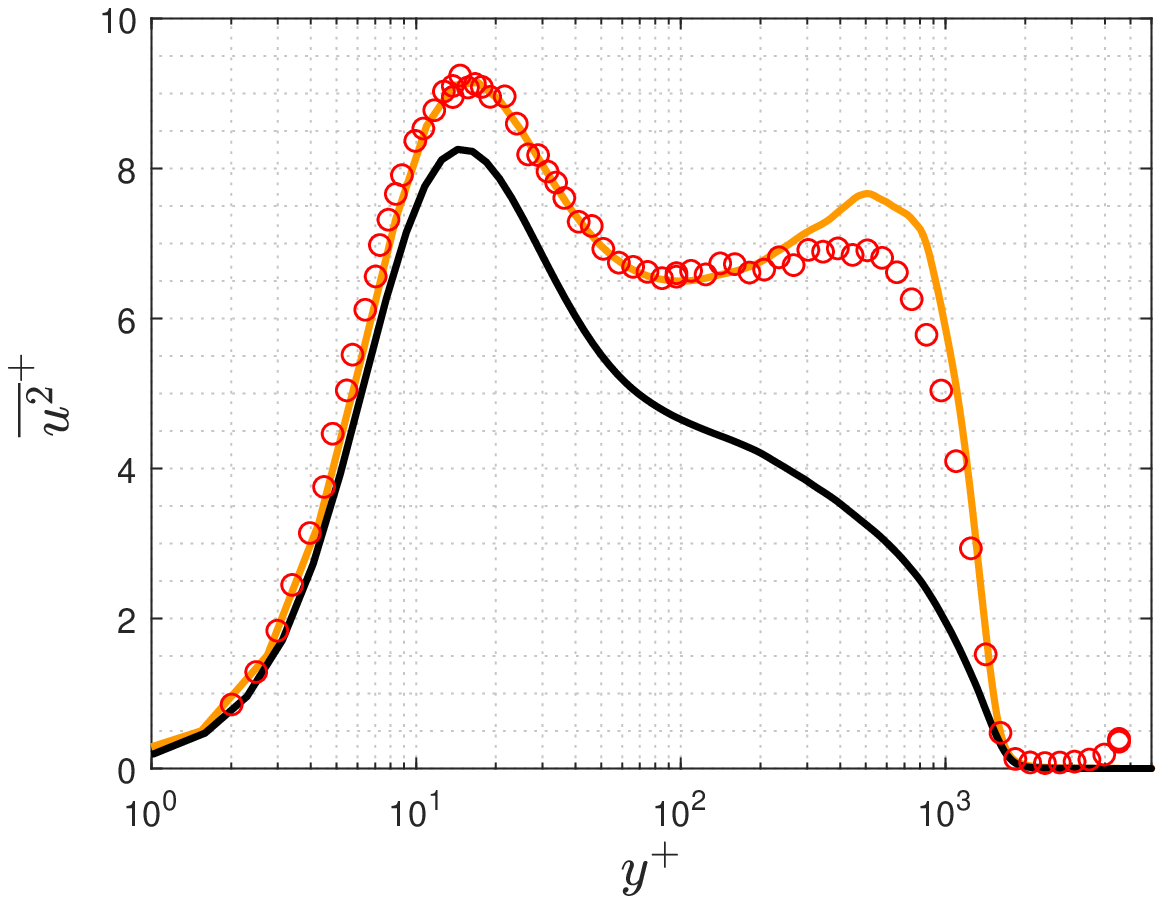}
\includegraphics[width=0.49\textwidth]{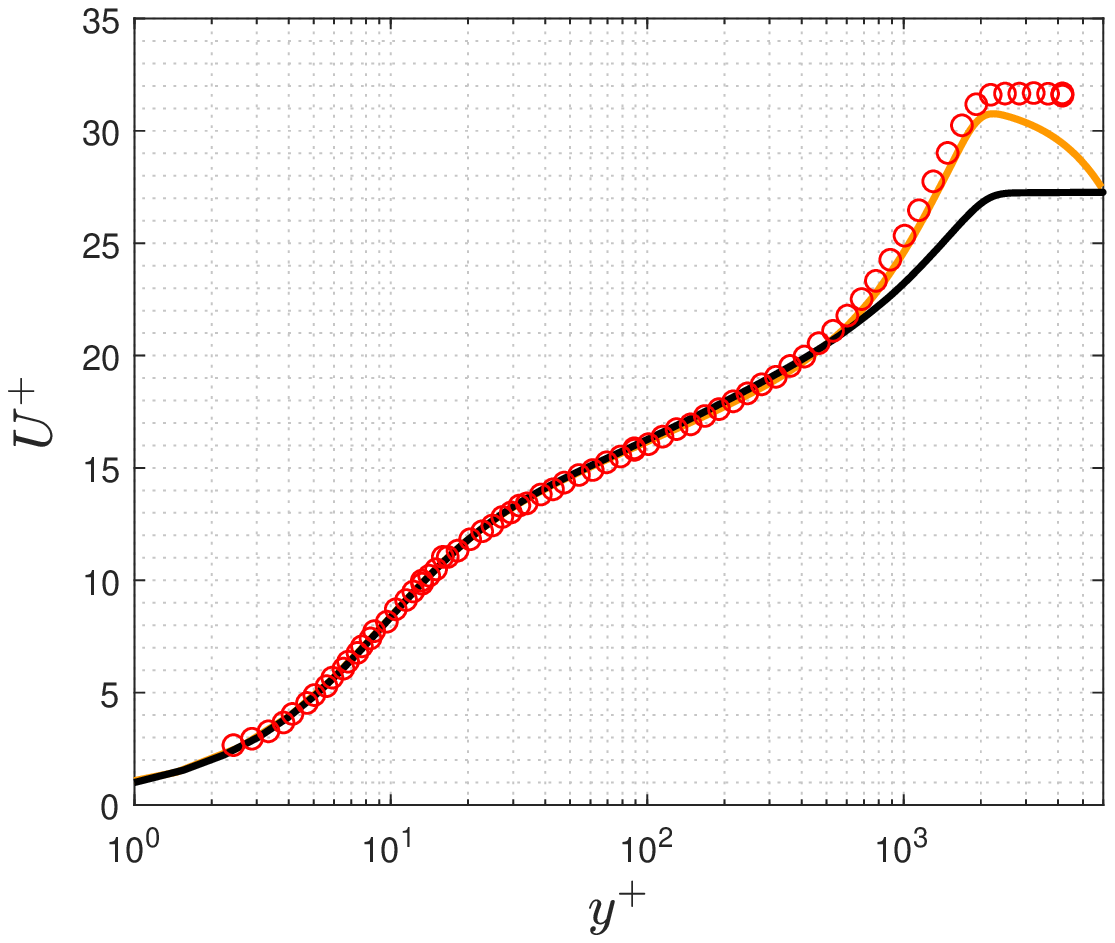}
\includegraphics[width=0.49\textwidth]{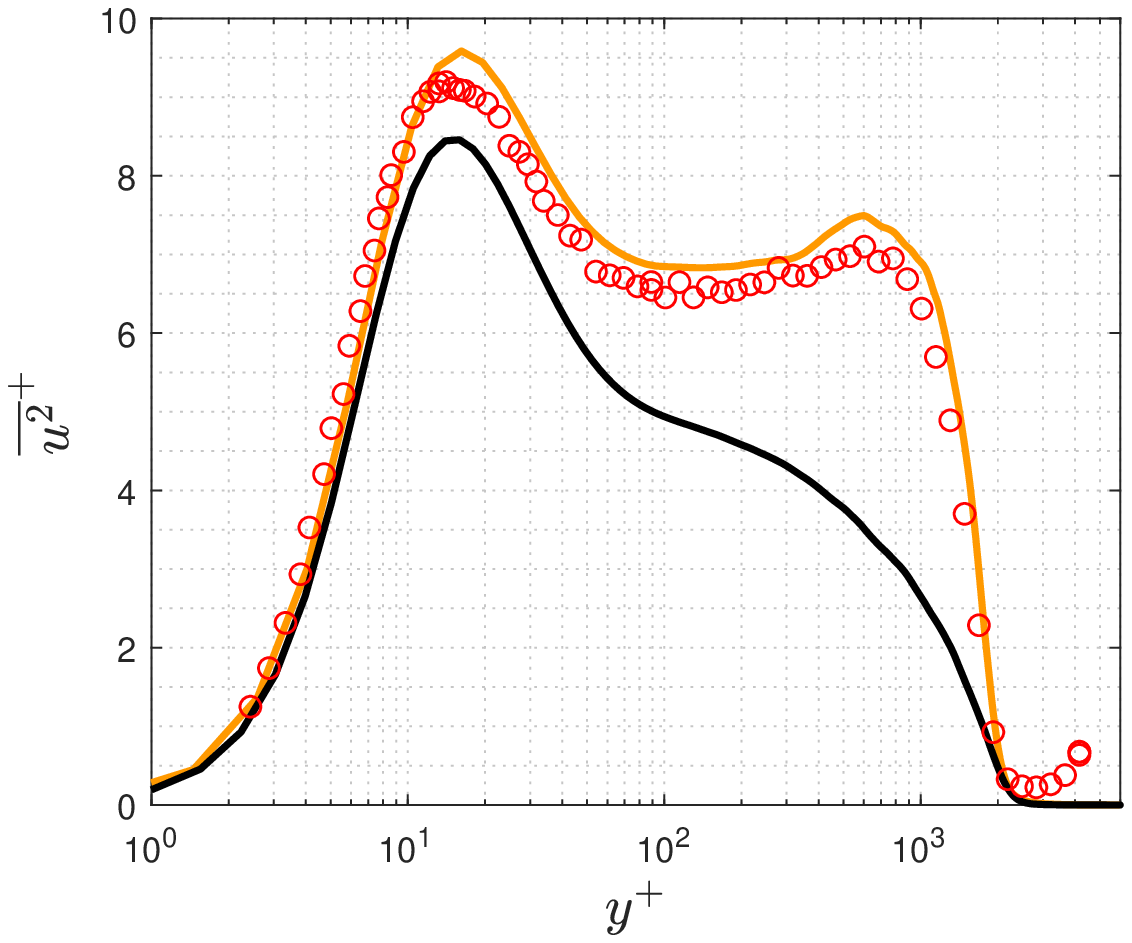}
\caption{Mean velocity (left column) and streamwise Reynolds stress (right column) scaled in viscous units as a function of the inner scaled wall-normal distance. The Reynolds numbers from top to bottom are $Re_{\tau}=\{1004, 1586, 2049\}$. The black solid line represents the ZPG by \cite{E-AmorZPG}, the orange line is the present b1.4 simulation and the red circles represent the experimental data by \cite{MTL_expSANMIGUEL}.}
\label{fig:experimentsMTL}
\end{figure}
%  Colors: (\protect\blackline) ZPG; (\protect\orangeline) b1.4; (\protect\redcircle) experiments by \cite{MTL_expSANMIGUEL}.

Three different profiles have been chosen in figure \ref{fig:experimentsMTL} to show the mean streamwise velocity and the streamwise Reynolds stress from the simulation and the experiment at matching $\Rey_{\tau}$. In the first row ($\Rey_{\tau}=1004$) the experimental TBL has a very low $\beta=0.3$, and it is very close to the ZPG, with small differences in the near-wall peak of $\overline{u^2}^+$ and a growing energy in the outer region of the TBL.
Here the simulation has a larger value of $\beta=1.6$, and it exhibits the most relevant features of APGs, including a prominent outer peak in $\overline{u^2}^+$.
In the middle row ($\Rey_{\tau}=1586$) the experimental data and the numerical simulation b1.4 exhibit the same  $\beta \simeq 1.4$. While the mean velocity profiles and the near-wall region of $\overline{u^2}^+$ is in good agreement in both cases, the simulation exhibits a larger fluctuation peak. This is due to the effect of flow history \citep{bobke2017, tanarro_2020}, but it is interesting to note that while the simulation is subjected to a mildly decaying $\beta(x)$ curve, the APG is rapidly increasing in the experiment. This implies that the smaller scales adapt more quickly to the local pressure gradient, while the larger scales require a longer streamwise distance.  
Finally, the higher Reynolds-number profile ($\Rey_{\tau}=2049$), where both TBLs have a value of $\beta \simeq 1.1$, exhibits a better collapse in both inner and outer peaks of $\overline{u^2}^+$, as well as in the mean velocity $U^+$.
This implies that, despite the different flow histories upstream, the two TBLs have been exposed to a similar PG magnitude for a sufficiently long streamwise distance such that their local turbulence features converge. Several profiles have been compared for $\Rey_{\tau}>1586$, and the best agreement between simulation and experiment is obtained for $\Rey_{\tau}=2049$. Upstream of this position the outer-peak value of the experimental data is still developing towards the value of the b1.4 case. The streamwise distance between the profiles at $\Rey_{\tau}=1586$ and 2049, which are subjected to a similar $\beta$ history, is around $11 \overline{\delta_{99}}$ for both simulation and experiment, where $\overline{\delta_{99}}$ is the average boundary-layer thickness between those profiles. A similar albeit lower of around $7\overline{\delta_{99}} $ was reported by \cite{bobke2017} for m16 simulation to converge towards b2 simulation at $\Rey_{\tau}=786$. One possible explanation for the longer distance reported here may be the higher Reynolds number, in which the large scales may require longer streamwise lengths to adapt to a particular pressure-gradient condition.

The APG in the simulation is achieved through a free-flow boundary condition, the turbulence is achieved through a tripping, therefore, the turbulence is confined inside the boundary layer with the Reynolds stresses being negligible outside of it. The experiments are performed inside a wind tunnel, the APG is imposed through changes in the geometry of the upper wall where another turbulent boundary layer develops. Even though the methods to establish the two APG TBLs are different, the results are remarkably similar. 
While the region outside of the boundary layer in the wind tunnel can be seen as the flow in the core of a channel where the mean $U$ does not exhibit a significant change, the free-flow APG condition has a negative $\partial{U}/\partial{y}$. This gradient, even if small, is present over a large distance in $y$, and when the mean profile is represented in a logarithmic scale it gives the impression of a drastic reduction of the mean velocity for $y>\delta_{99}$.
This comparison between numerical simulation and experiments shows a satisfactory collapse, and thus validates the high-Reynolds-number region of this numerical simulation.

\section{Scaling considerations} \label{sec:Scalings}

One important question of wall-bounded turbulence research is to find a scaling that produces a collapse of statistical quantities such as the mean velocity or the Reynolds stresses in different parts of the TBL. Ideally, the scaling parameters should include the information of the forces/boundary conditions that could affect the flow, {\it i.e.} pressure gradients, temperature gradients, friction forces, flow history, etc.
Here we have an incompressible simulation of a flat-plate turbulent boundary layer where outside of the boundary layer there is an initial zero-pressure-gradient condition, which after a certain streamwise distance evolves into an adverse-pressure-gradient condition. The scaling parameters should include the effects of the friction at the wall ($\tau_w$), the pressure gradient, the Reynolds number and the flow history.
In the literature there are many studies of self-similarity (or self-preservation) focused on the equation of momentum conservation in $x$ under some ansatz and simplifications \citep{rotta1950theorie, Townsend_1956_structure, Castillo_2004, Kitsios2016, Gibis2019}. Most of those studies find parameters that should be kept constant to achieve self-similarity. Another question regarding self-similarity is whether it can be achieved across the whole boundary layer or whether the boundary layer can only be self-similar in different regions, with various scales.

%\subsection{Self-similarity analysis and scaling of the outer region}

\begin{figure}
\includegraphics[width=0.49\textwidth]{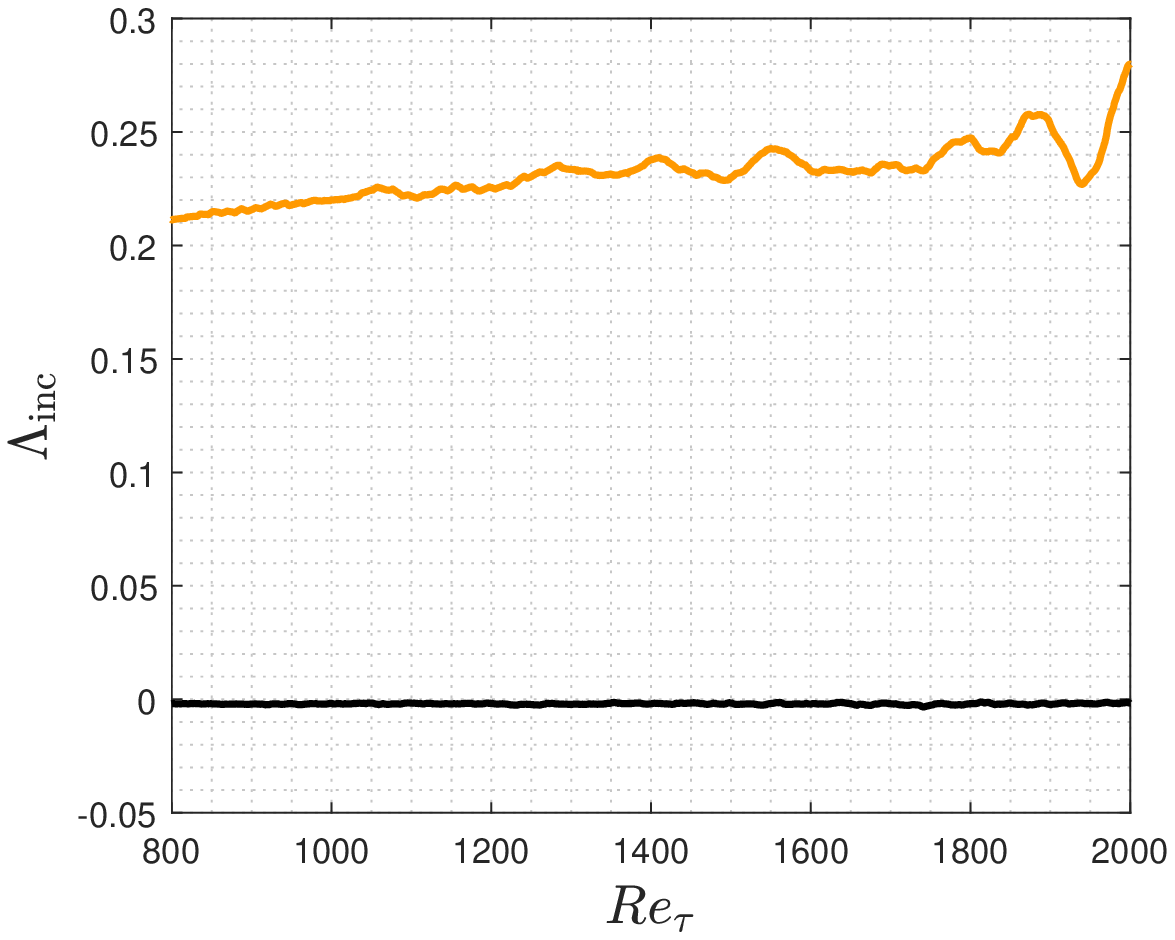}
\includegraphics[width=0.49\textwidth]{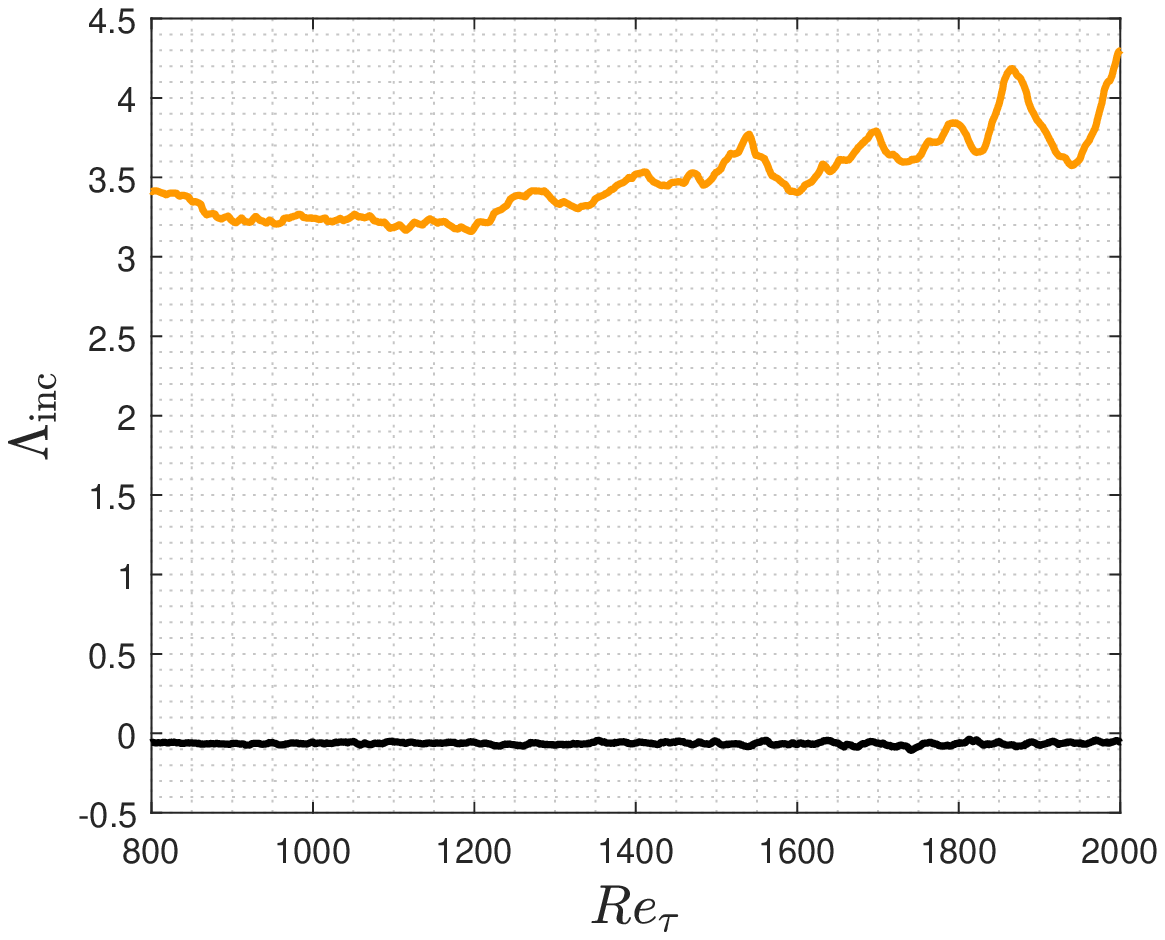}
\includegraphics[width=0.49\textwidth]{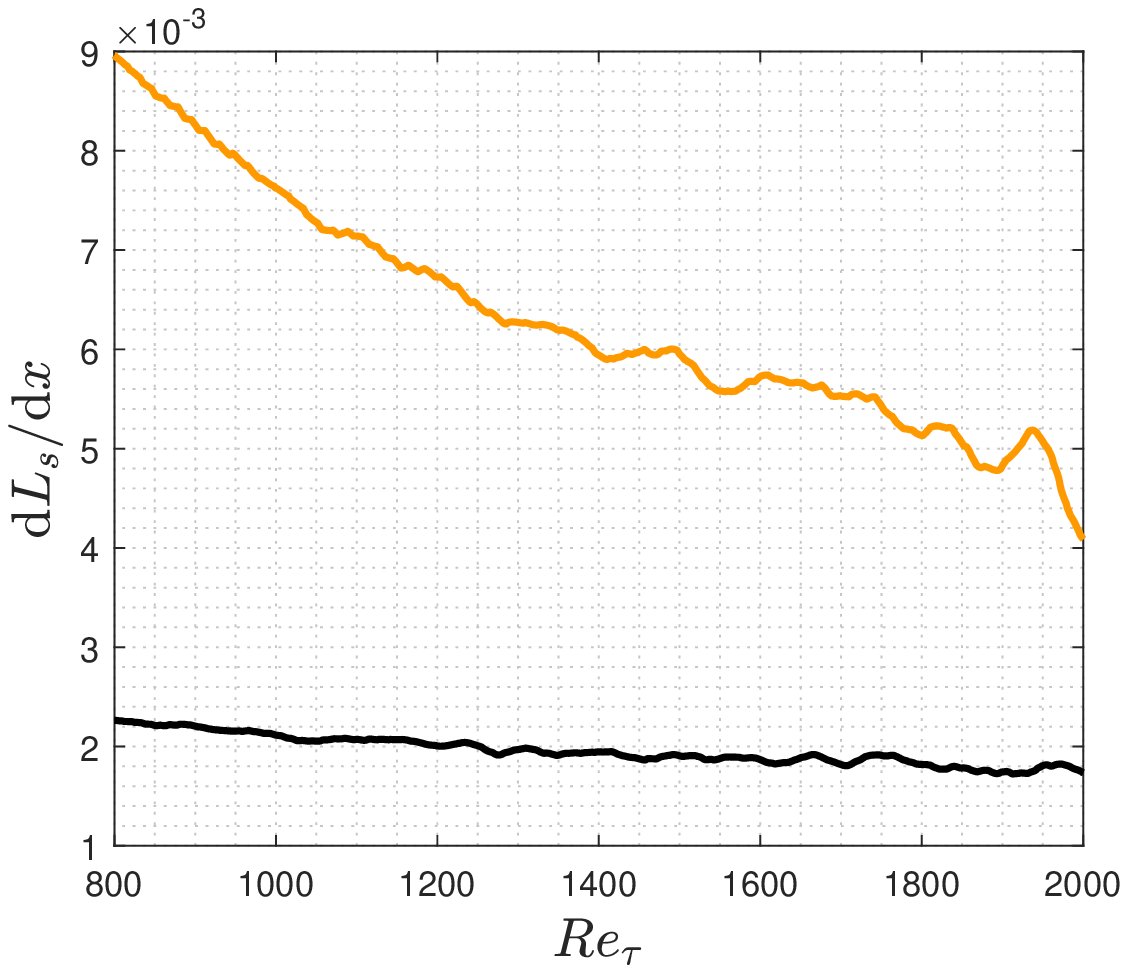}
\includegraphics[width=0.49\textwidth]{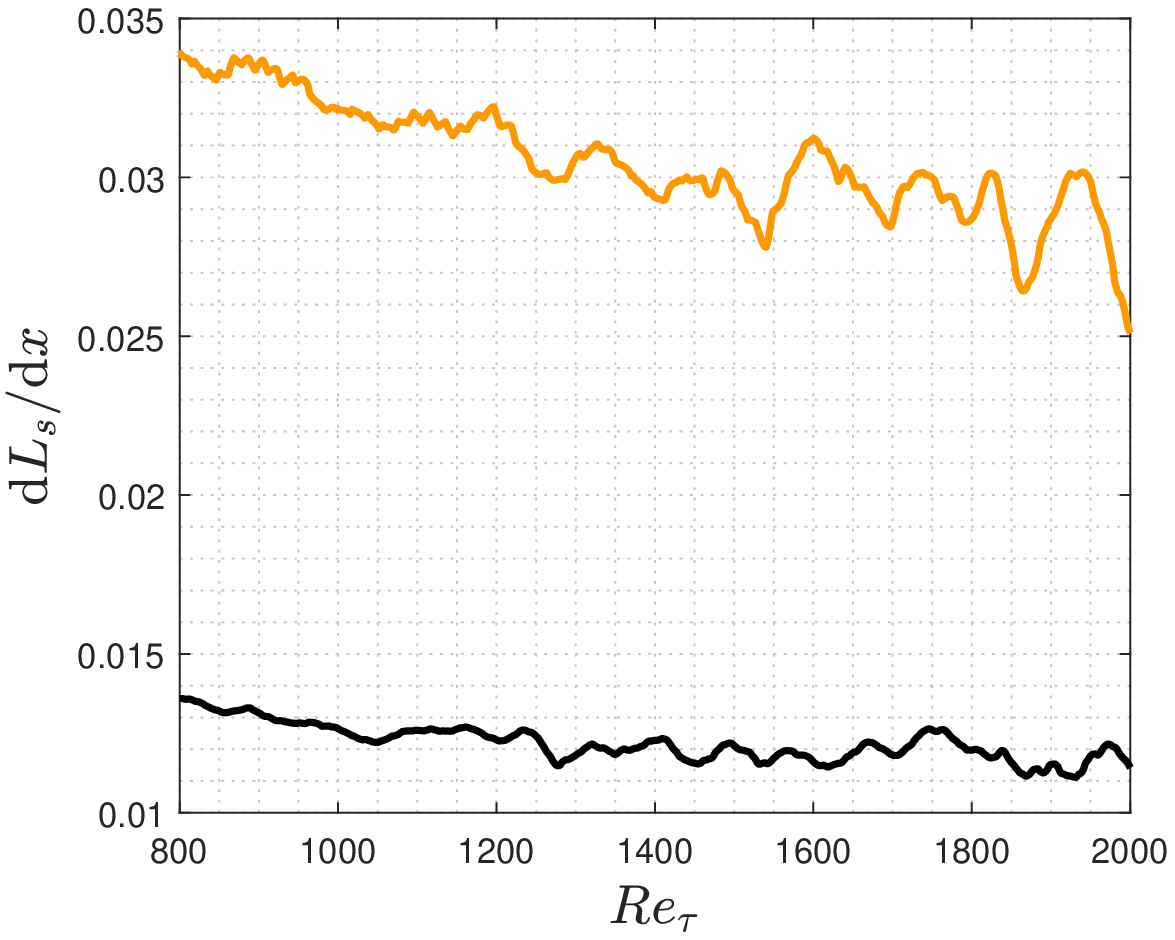}
\caption{Different pressure-gradient parameters based on the self-similarity analysis for the outer layer performed in \citet{Gibis2019}. Left column represents the edge scaling, where $L_s=\delta^*$ and $U_s=U_e$. Right column shows the Zagarola--Smits scaling with $L_s=\delta_{99}$ and $U_s=U_{e}\delta^*/\delta_{99}$. A Savitzky--Golay filter has been applied to $\textrm{d} L_s/\textrm{d}x$ as in \cite{Gibis2019}. The black solid line represents the ZPG by \cite{E-AmorZPG} and the orange line is the present b1.4 simulation.}
\label{fig:pg_parameters}
\end{figure}

Following the considerations about self-similarity in \citet{Gibis2019}, two sets of scalings will be considered in the outer region for the b1.4 and ZPG databases: the edge and the Zagarola--Smits (ZS) scalings as discussed below. The equilibrium character of the boundary layer will be assessed using the Rotta--Clauser pressure-gradient parameter $\beta$ together with the pressure-gradient boundary-layer growth parameter $\Lambda_{\rm inc}$ (\ref{eq:PG_Lambda}), for the chosen outer scalings.

\begin{equation} \label{eq:PG_Lambda}
    \Lambda_{\rm inc}=\frac{L_s}{\rho U_s^2 (\mathrm{d}L_s/\mathrm{d} x)} \frac{\mathrm{d}p}{\mathrm{d}x}.
\end{equation}
In the parameter $\Lambda_{\rm inc}$, $L_s$ represents the chosen length scale and $U_s$ the velocity length scale. If $L_s=\delta^*$ and $U_s=u_{\tau}$, then the difference with respect to the parameter $\beta$ will be given by the streamwise derivative of the length scale $\textrm{d} L_s/\textrm{d}x$, which contains some information of the flow history linking the local profile with the surrounding flow.
The different scalings with their respective velocity and length scales are summarized in table \ref{tab:Scaling_parameters}.
In figure \ref{fig:pg_parameters} the pressure-gradient parameter $\Lambda_{\rm inc}$ is shown for simulation b1.4 in the first row. While the edge scaling (left) shows a monotonically rising $\Lambda_{\rm inc}$ from the beginning, the ZS scaling (right) exhibits an initial peak, then a plateau, and finally a slowly-increasing region. On the other hand, $\beta$ (figure \ref{fig:beta}) exhibits a maximum and a slow monotonic decrease throughout the domain. Recalling that the ROI starts at $\Rey_{\tau}=800$, the range of variation of the PG parameter within the ROI is $[0.2, 0.25]$ for the edge scaling, $[3.2, 4]$ for the ZS scaling and $[1.65, 1.2]$ for $\beta$. The rates of variation of the PG parameters and $\textrm{d} L_s/\textrm{d}x$ over the ROI, although not constant, are relatively small, which is a requirement for similarity according to the various studies.

%*********************************************************************************
% Tobias Gibis JFM2019  "Self-Similar compressible turbulent...outer layer."
%*********************************************************************************
\begin{table}
  \begin{center}
\def~{\hphantom{0}}
    \begin{tabular}{ l c  c  l}
    Scaling         &     $L_s$        & $U_s$                          & Reference  \\[3pt]
    Rotta--Clauser  &  $U_e^+\delta^*$ & $u_{\tau}$                     & \cite{Clauser_1956}   \\
    Edge            &  $\delta^*$      & $U_e$                          & \cite{Kitsios2016}    \\
    Zagarola--Smits &  $\delta_{99}$   & $U_{e}(\delta^*/\delta_{99})$  & \cite{zagarola_smits} \\
    \end{tabular}
  \caption{Parameters for different scalings, where $L_s$ and $U_s$ correspond to the length and velocity scales, respectively.}
  \label{tab:Scaling_parameters}
  \end{center}
\end{table}

\begin{figure} 
\includegraphics[width=0.49\textwidth]{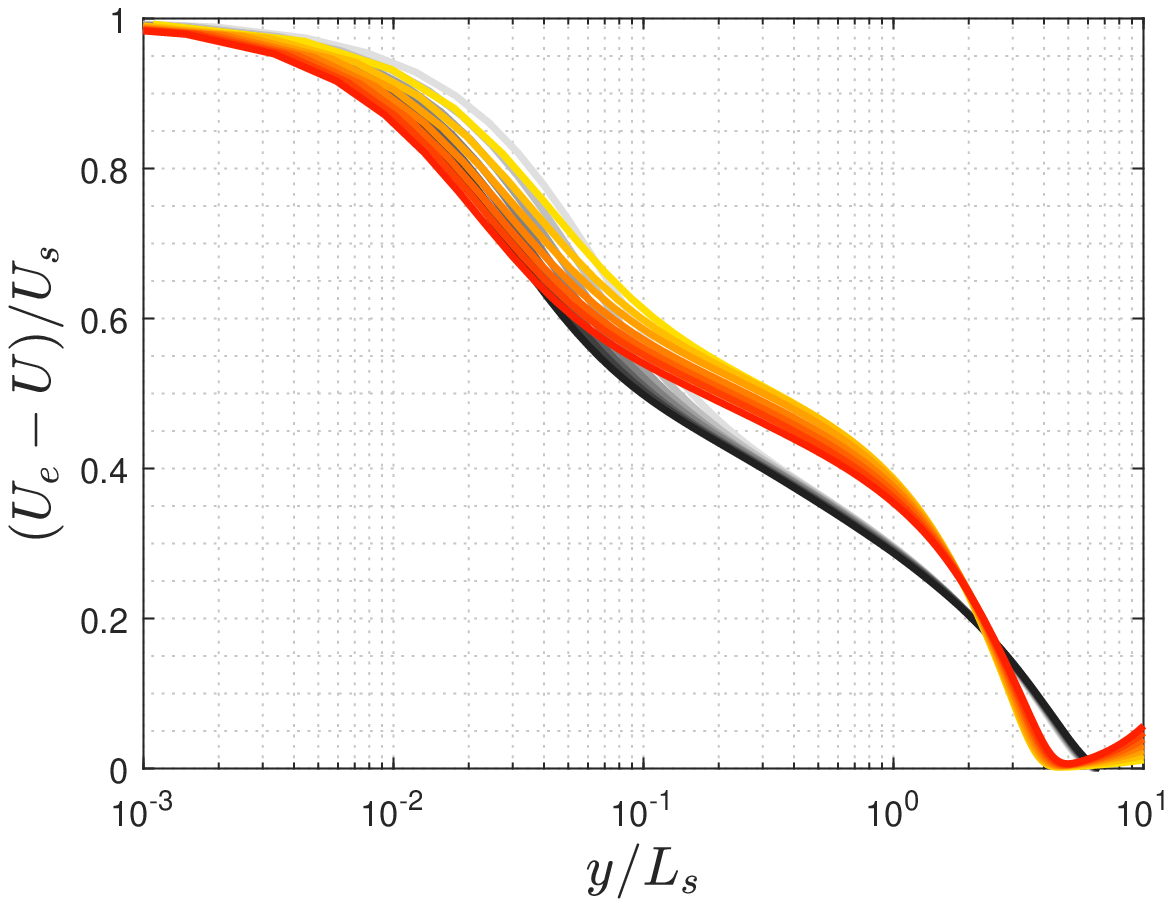}
\includegraphics[width=0.49\textwidth]{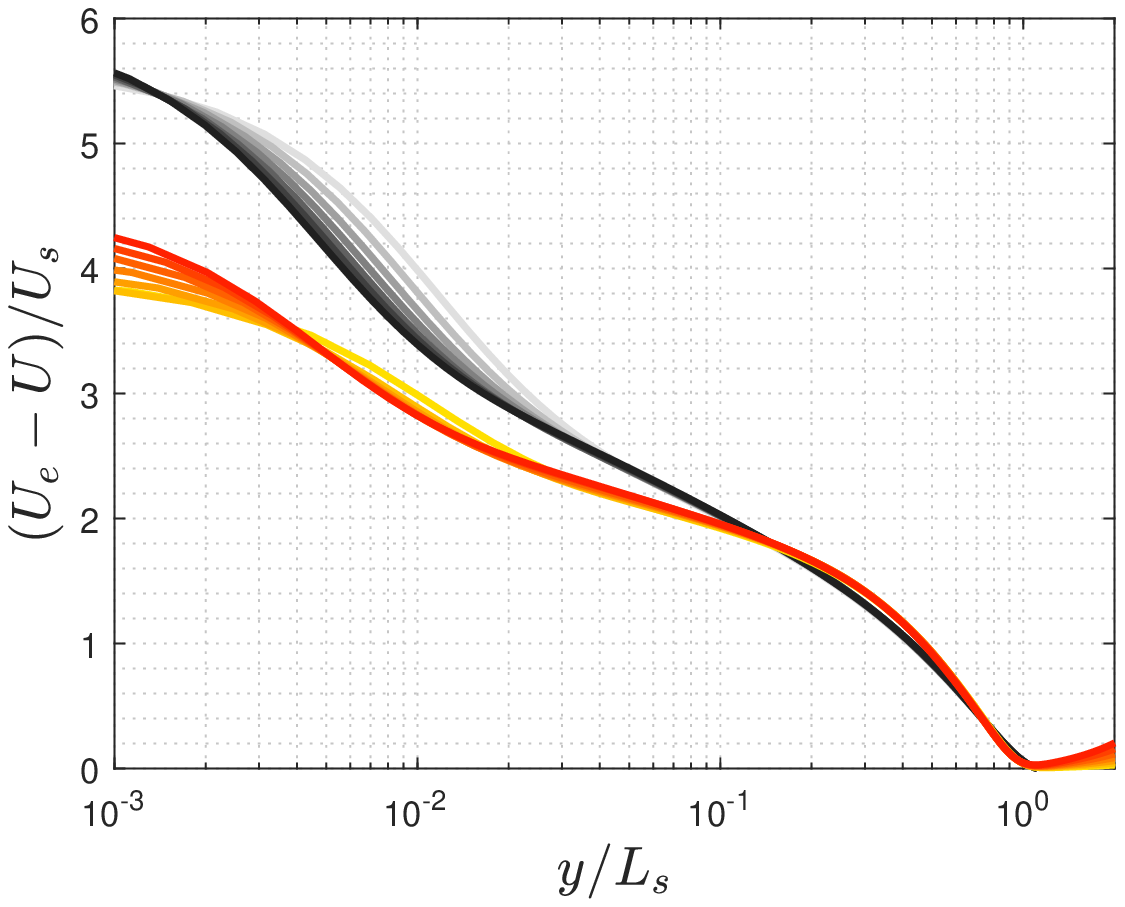}
\includegraphics[width=0.49\textwidth]{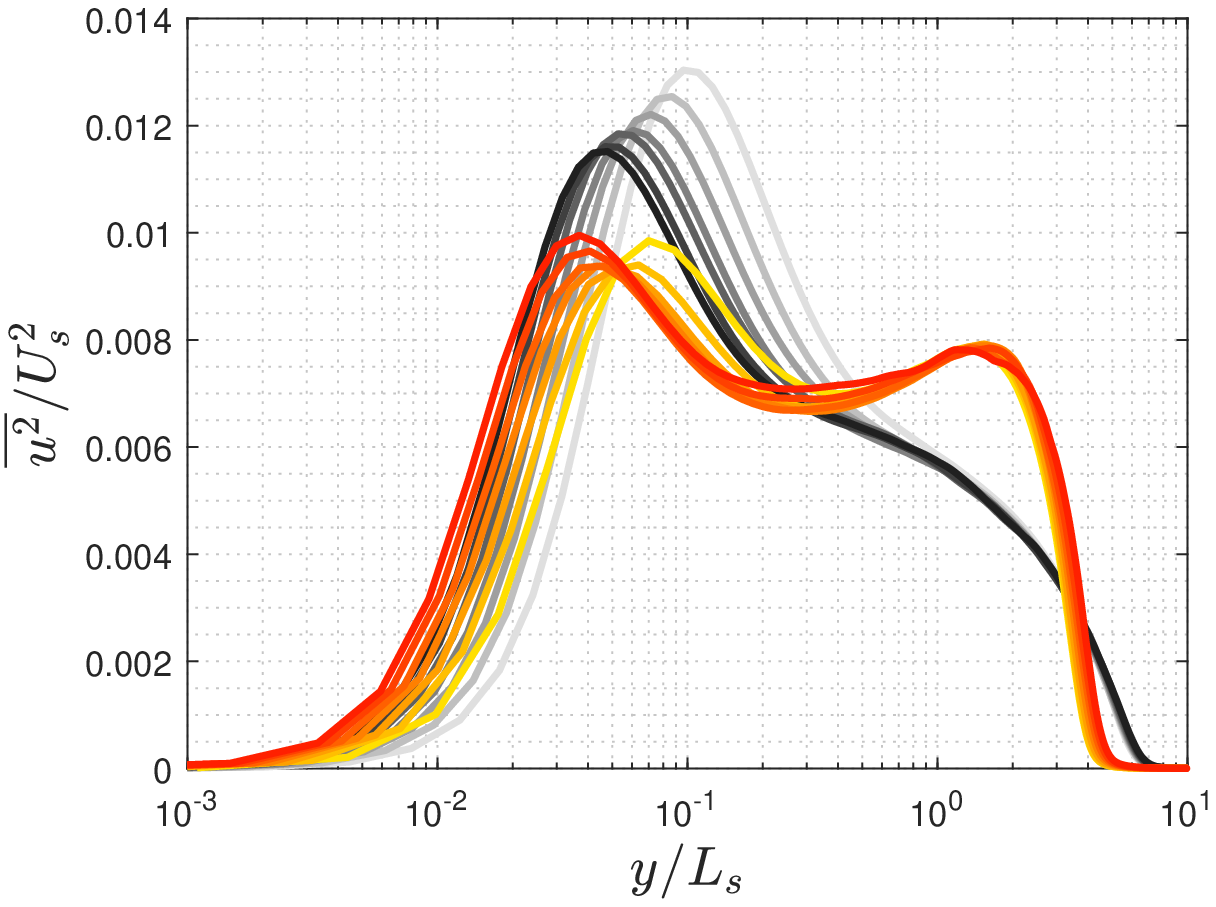}
\includegraphics[width=0.49\textwidth]{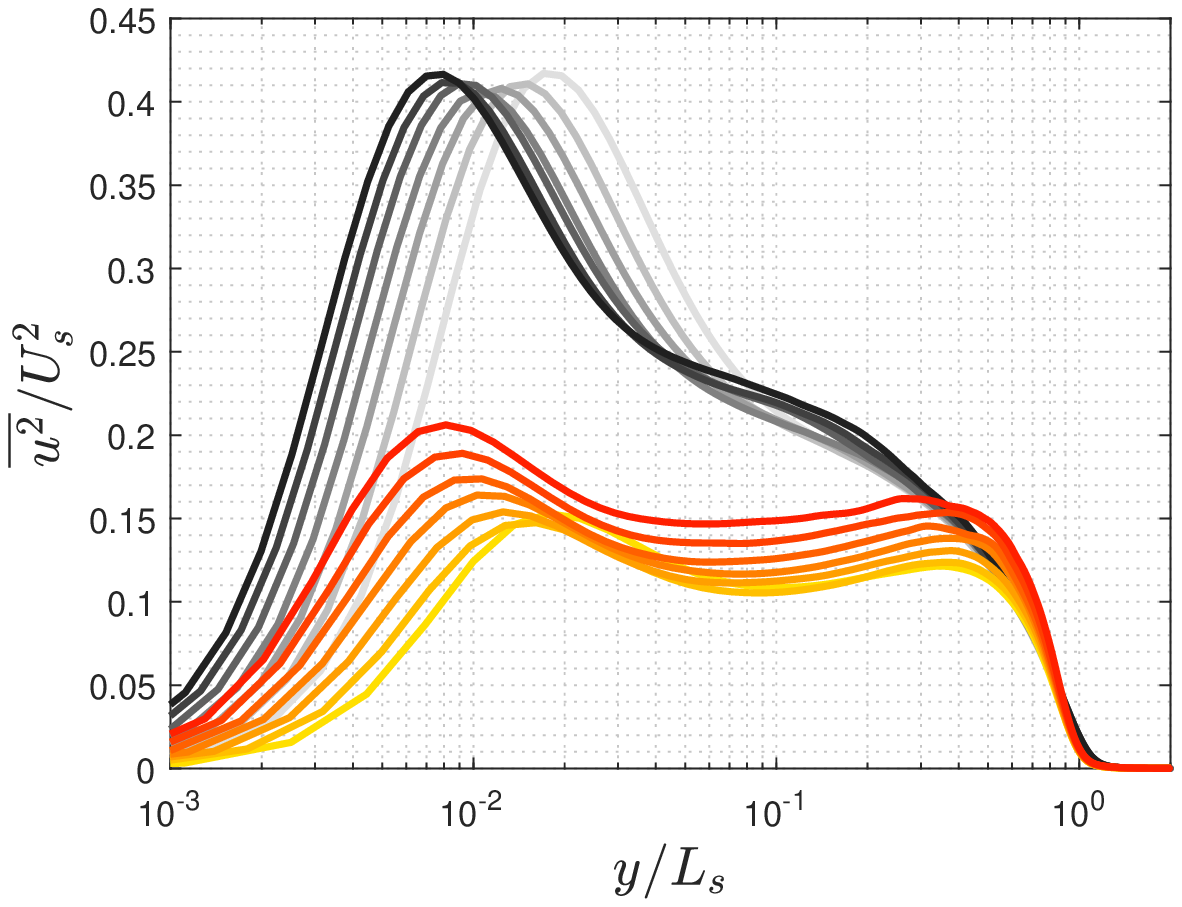}
\caption{(Top) Mean velocity defect and (bottom) streamwise Reynolds stress scaled with: edge (left) and Zagarola--Smits (right) scalings. Profiles from $Re_{\tau}=800$ to $Re_{\tau}=2000$. Lines in gray scale represent ZPG data \citep{E-AmorZPG}, increasing the Reynolds number from white to black. APG data from the b1.4 simulation increases Reynolds number from yellow to red.}
\label{fig:def_U_uu}
\end{figure}

In figure \ref{fig:def_U_uu} we use the edge and ZS scalings for the streamwise mean velocity defect and the streamwise RS for the profiles within the ROI.
The first observation is the common good agreement of both scalings from the logarithmic region all the way to the edge of the TBL in the ZPG.

Equation~(\ref{eq:defect_scaled_2}) shows the definition of $\delta^*$. For the ZS and edge scalings, the velocity and length scales are just a combination of the parameters present in equation~(\ref{eq:defect_scaled_2}) ($U_e$, $\delta^*$, $\delta_{99}$), therefore we can divide both sides by the term $U_e \delta^*$ and using the length and velocity scales in table \ref{tab:Scaling_parameters} it is possible to rewrite the integral in a common form (right-hand side of equation~(\ref{eq:defect_scaled_2})) for both ZS and edge scalings.

\begin{equation}\label{eq:defect_scaled_2}
    \int_{0}^{\delta_{99}} (U_e-U) \textrm{d} y = U_e \delta^*  \Rightarrow \int_{0}^{\delta_{99}/L_s} \frac{U_e-U}{U_s} \textrm{d}(y/L_s)=1. 
\end{equation}

Using this form we can directly relate the integral with the mean defect velocity curves in figure \ref{fig:def_U_uu}. 
 The value of the normalized integral of the mean velocity defect is the same for both scalings, where the integrands are the various curves and the only difference would be the upper limit of the integration. That upper limit in the ZS scaling does not change with the Reynolds number: it is 1 since $L_s=\delta_{99}$. The upper limit for the edge scaling is variable with Reynolds number because $L_s=\delta^*$. 
The functional form in the edge scaling fixes all the profiles to start from the same point, thus the differences between profiles increase from the wall. The functional form of the ZS scaling makes the profiles start from different values for each Reynolds number, and since the value of the integral is the same and the upper limit is also the same for all profiles, the differences are concentrated close to the wall, allowing for a better collapse in the outer region of the TBL.
This can be observed in figure \ref{fig:def_U_uu} (top right), where the ZS scaling exhibits a better collapse even in the overlap region. 
The edge scaling, figure \ref{fig:def_U_uu} (top left), exhibits a good collapse in the wake region, with differences in the overlap region.

We have previously discussed that the inner peak of $\overline{u^2}$ has a location $y^+ \simeq 15$. As shown in figure \ref{fig:def_U_uu} (bottom), and also in figure \ref{fig:uupeaks_loc}b) and c), the outer-peak location for the edge scaling is at $y/\delta^*=1.4$ and around $y/\delta_{99}=0.35$ for the ZS scaling.
While there is a better collapse in the outer region of the mean defect profiles using the ZS scaling, the $\overline{u^2}$ profiles collapse in the outer region when using the edge scaling.
Our results suggests that it is not possible to collapse the wall-normal profiles at all positions 
with a single length scale. In the mean flow, the inner region collapses in inner scaling and the outer region using the ZS scaling. Regarding the streamwise RS, the location of the near-wall peak slightly varies around $y^+\simeq 15$, it slowly grows with the Reynolds number and the slope is larger with higher $\beta$, whereas both the magnitude and location of the outer peak are fixed using the edge scaling. 
Since the inner and outer scales are related by $\Rey_{\tau}$, to achieve self-similarity throughout the whole profile would require that $\Rey_{\tau}$ remains constant in $x$, which is not the case even in the ZPG. 

\begin{figure}
\includegraphics[width=0.49\textwidth]{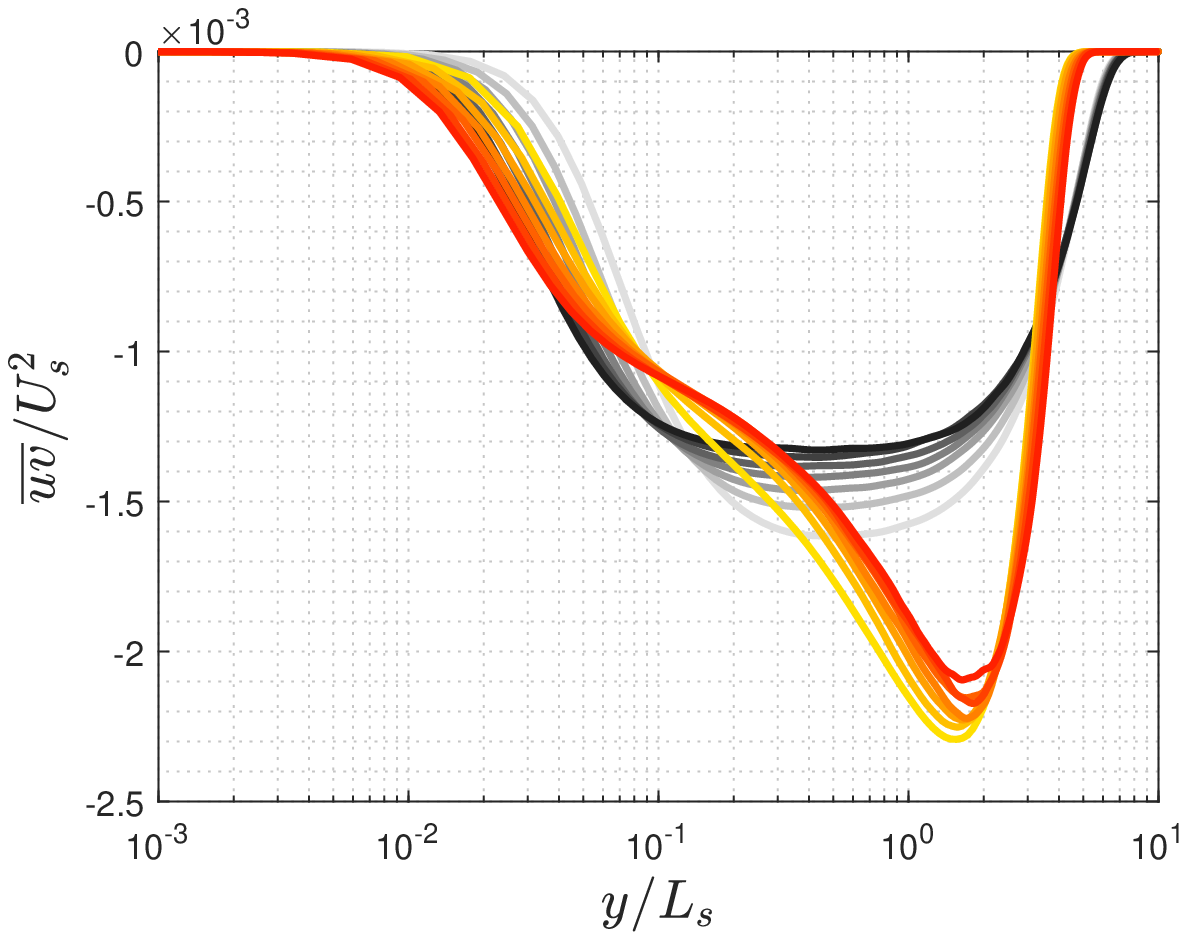}
\includegraphics[width=0.49\textwidth]{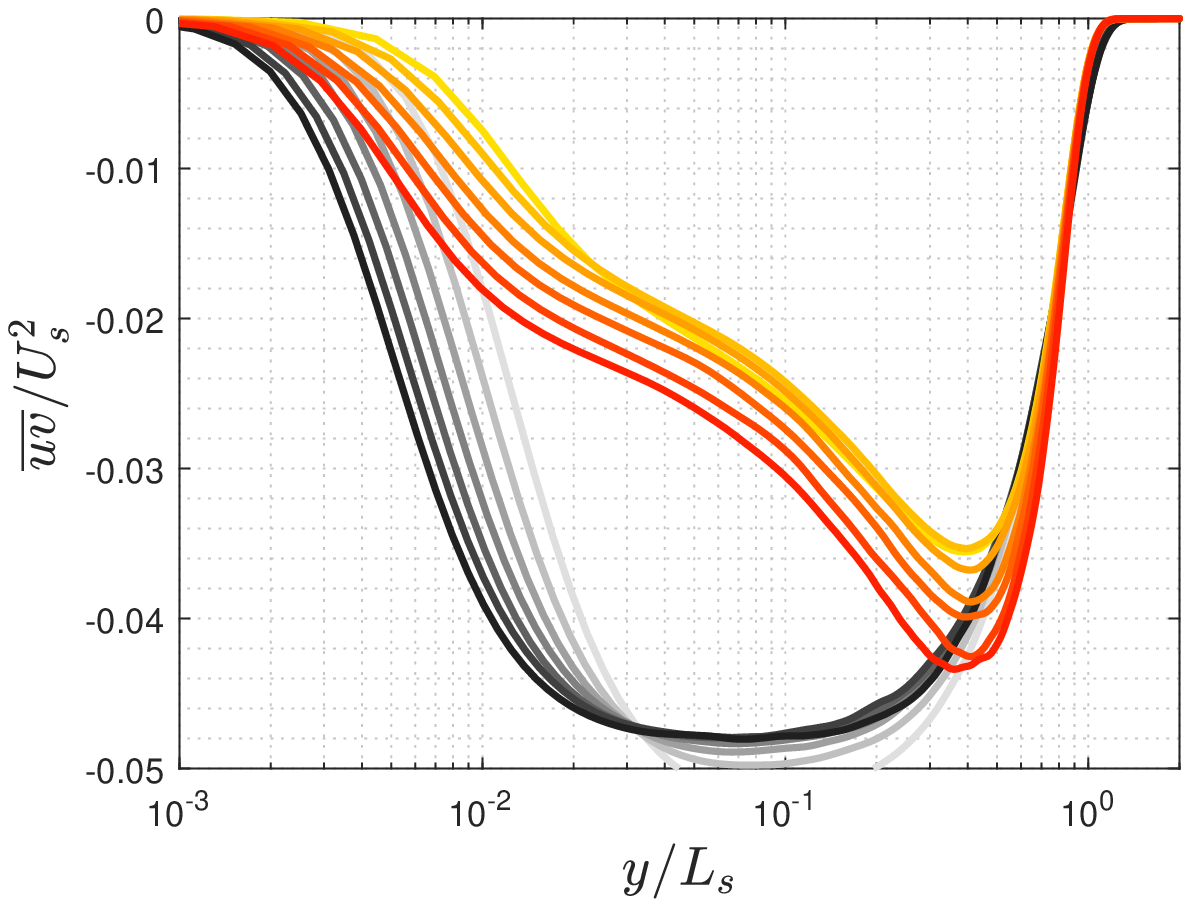}
\includegraphics[width=0.49\textwidth]{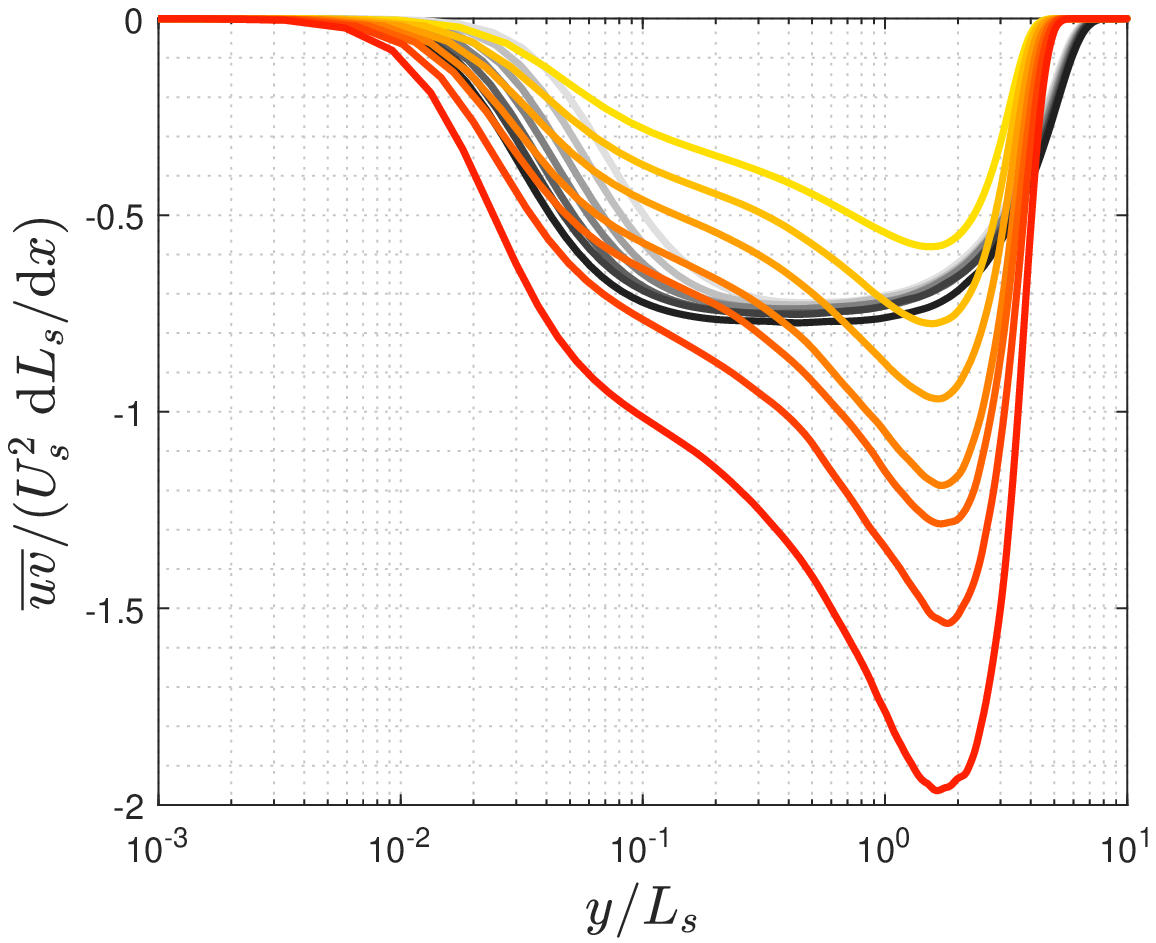}
\includegraphics[width=0.49\textwidth]{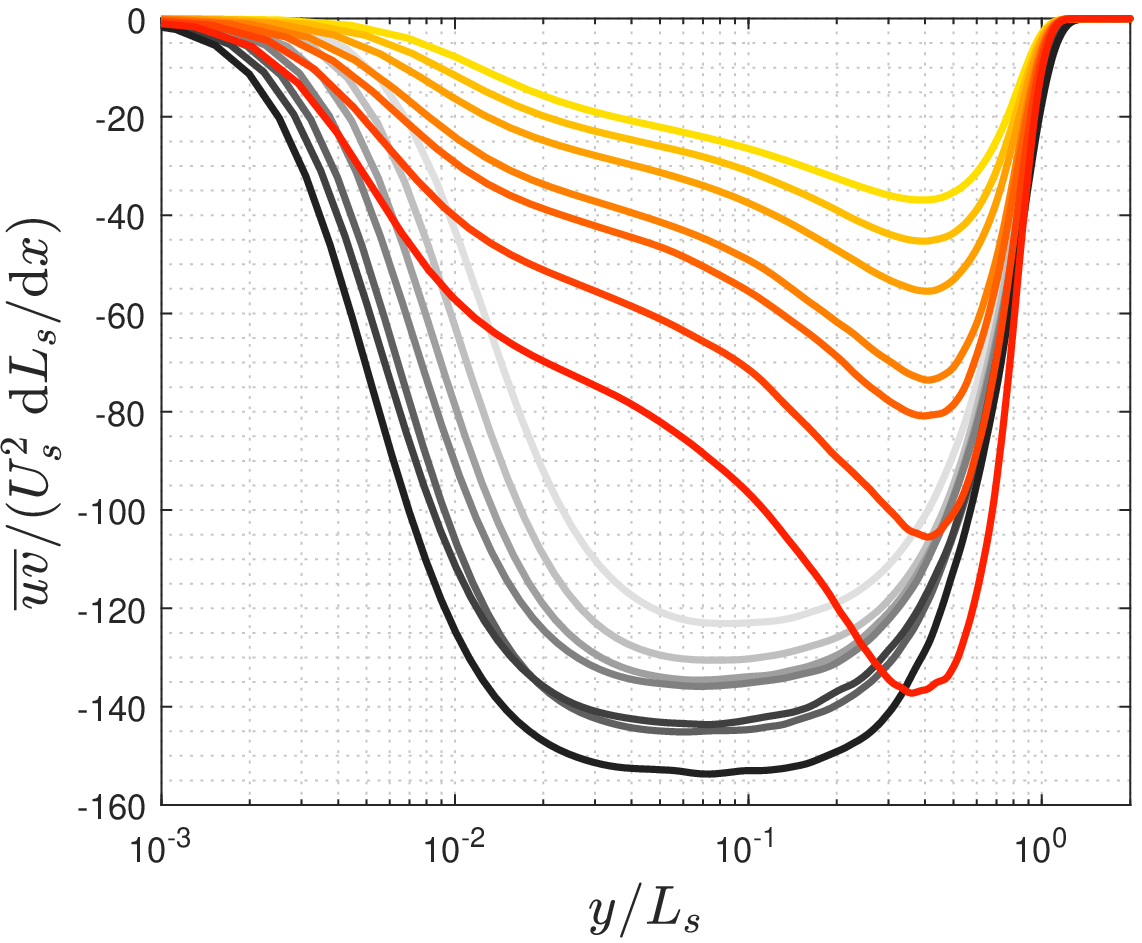}
\caption{Reynolds shear stress $\overline{uv}$ scaled with: edge (left) and Zagarola--Smits (right) scalings. The second row shows the effect of the evolution of the characteristic length scale $\textrm{d} L_s/\textrm{d}x$. Profiles from $Re_{\tau}=800$ to $Re_{\tau}=2000$. Lines in gray scale represent ZPG data \citep{E-AmorZPG}, increasing the Reynolds number from white to black. APG data from the b1.4 simulation increases Reynolds number from yellow to red.}
\label{fig:uv_scalings}
\end{figure}

In the next similarity analysis we consider the Reynolds shear stress, where in figure \ref{fig:uv_scalings} (top) we consider $U_s^2$, and in figure \ref{fig:uv_scalings} (bottom) we use $U_s^2 \textrm{d} L_s/\textrm{d}x$ as in \cite{Kitsios2016, Gibis2019}.
The edge scaling leads to a moderate collapse of the $\overline{uv}$ profiles in the outer region, although the collapse is not as good as the one observed for $\overline{u^2}$.
The additional term $\textrm{d} L_s/\textrm{d}x$ applied to the Reynolds shear stress, worsens the collapse for the APG in both scalings, while it improves the collapse for the ZPG in the edge scaling.

To summarize, the edge scaling yields a good scaling of the APG profiles in the outer region, and given the good collapse with viscous units close to the wall, it can be stated that the APG TBL is in near-equilibrium  \citep{Marusic_PoF_2010, bobke2017} conditions in the ROI.
The ZS scaling leads to a better collapse in the outer region of the mean defect profiles.

\begin{figure}
\includegraphics[width=0.32\textwidth]{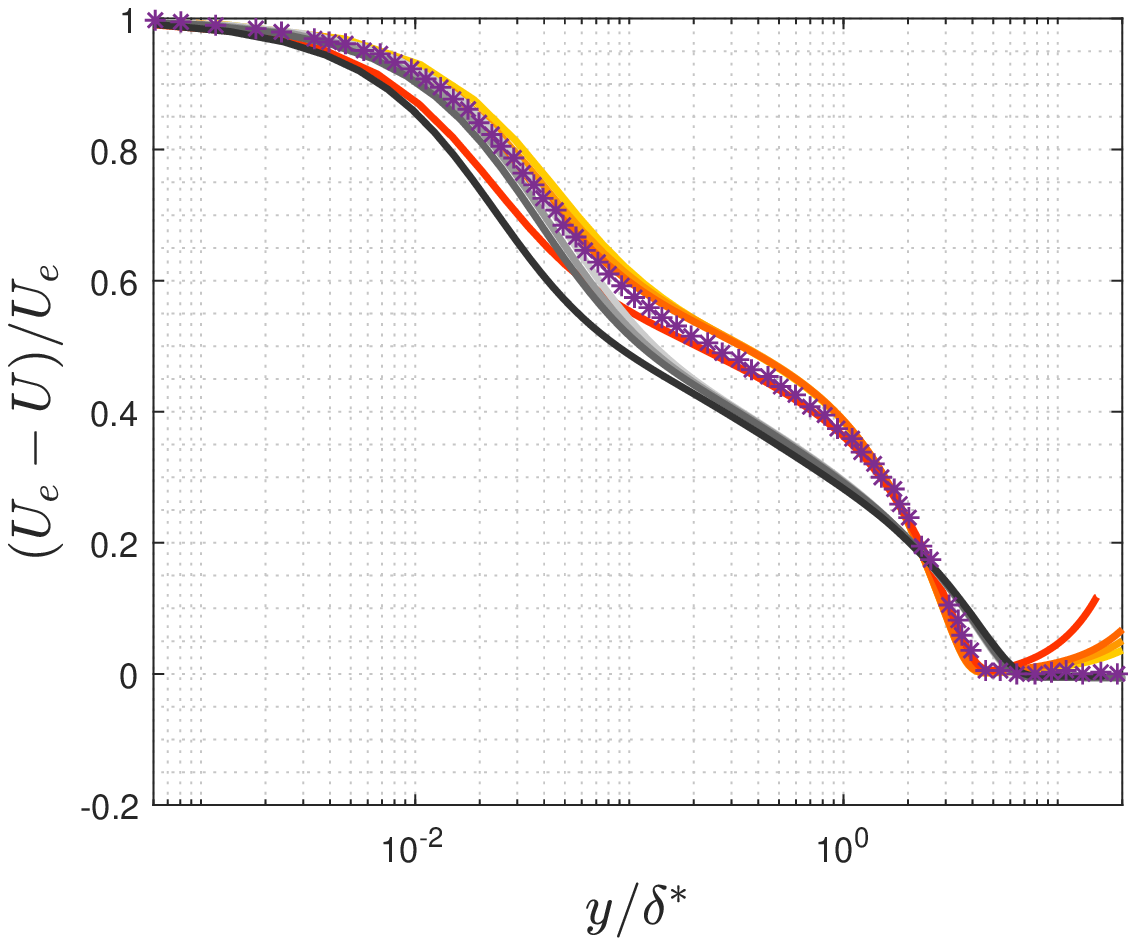}
\includegraphics[width=0.32\textwidth]{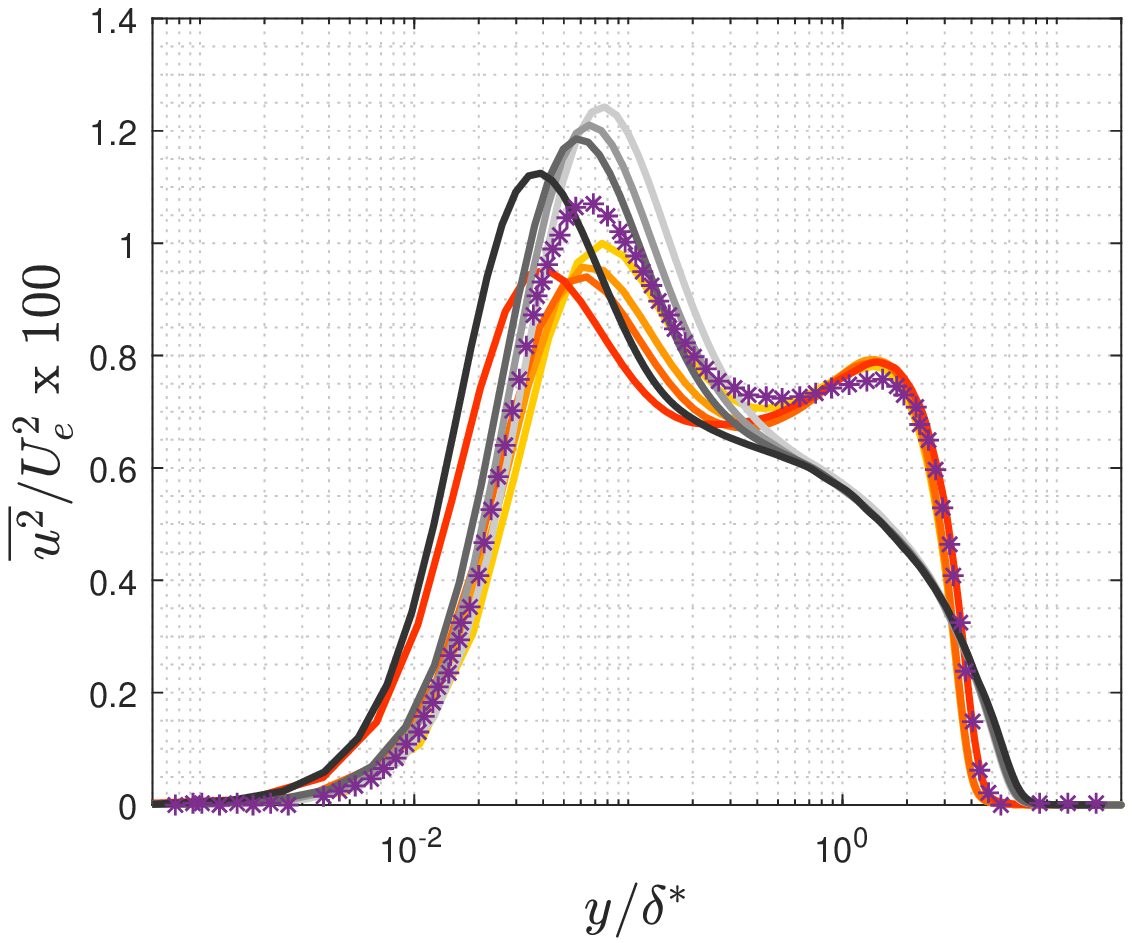}
\includegraphics[width=0.32\textwidth]{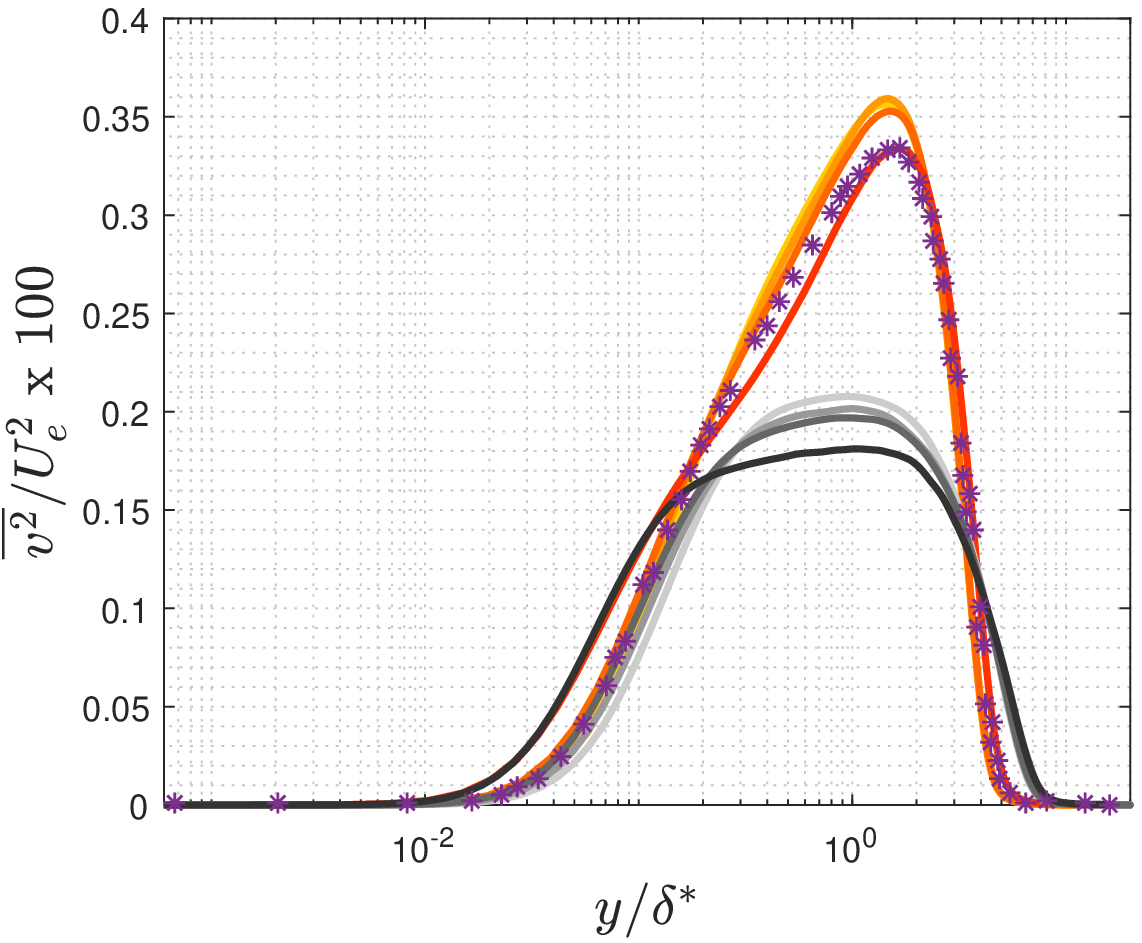}
\includegraphics[width=0.32\textwidth]{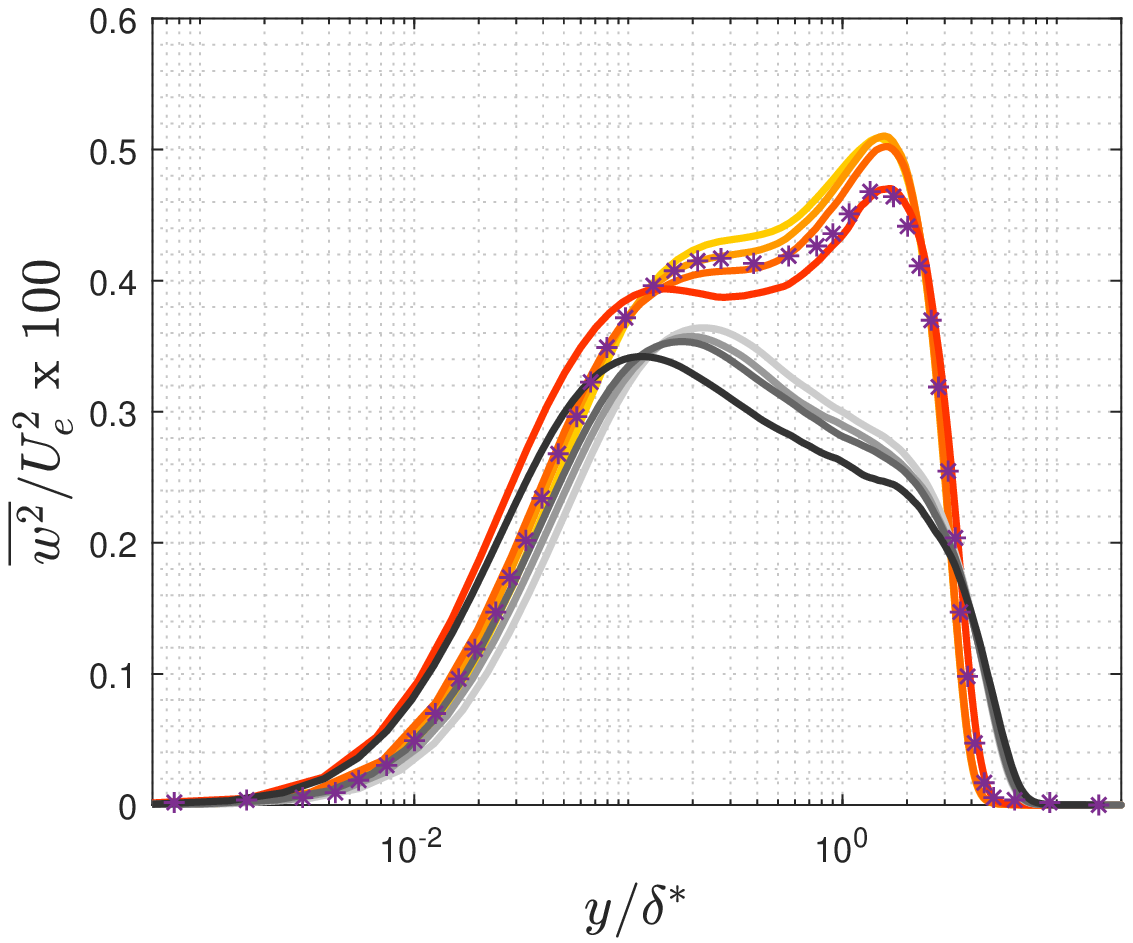}
\includegraphics[width=0.32\textwidth]{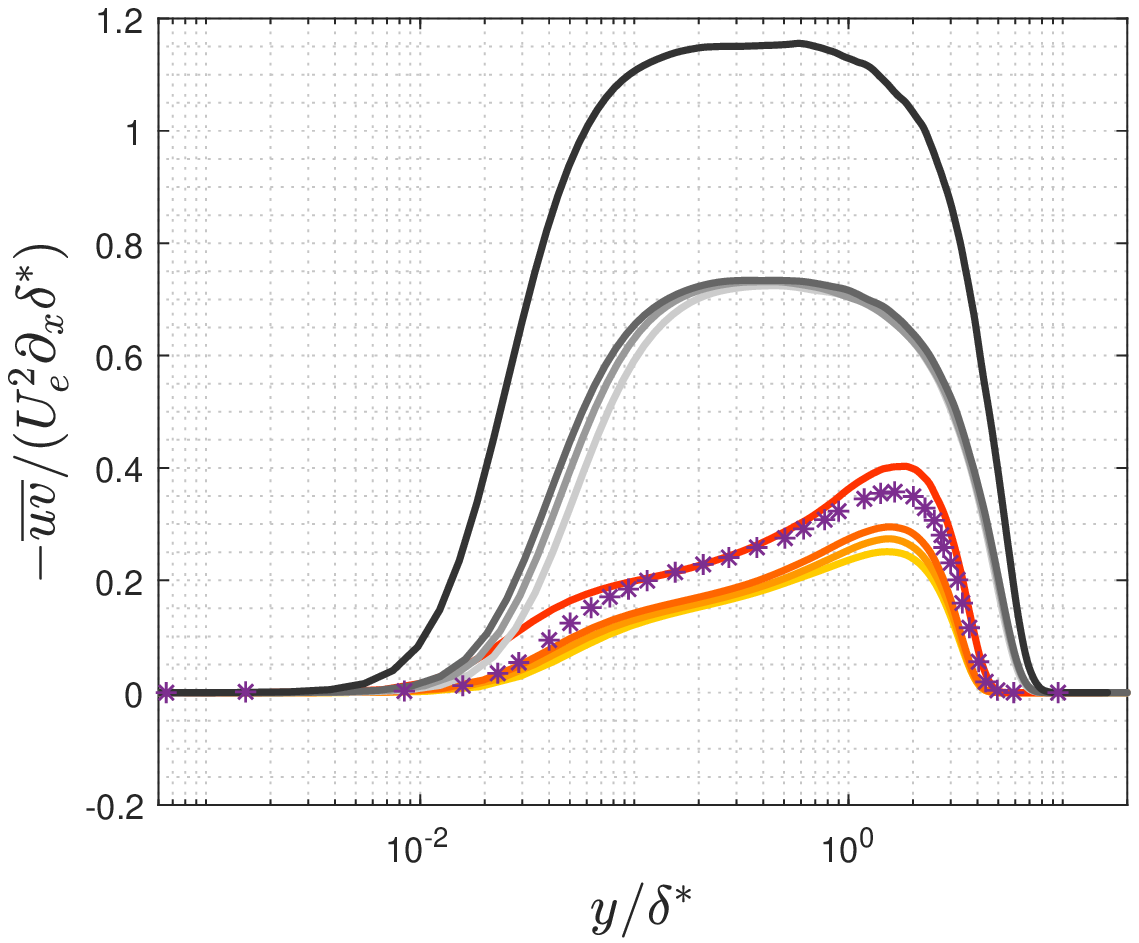}
\caption{Mean streamwise velocity defect and Reynolds-stress tensor components $\overline{u^2}$, $\overline{v^2}$, $\overline{w^2}$, $\overline{uv}$ scaled using the edge scaling as in \cite{Kitsios2016}, and using the streamwise derivative of the length scale $\partial_x \delta^*$ in the case of $\overline{uv}$. The purple asterisks are used for the collapsed data by \cite{Kitsios2016}. The profiles have been taken at $\Rey_{\theta}=\{3500, 4150, 4800, 8200\}$, where the first three are in the same range as \cite{Kitsios2016} and the last at one is the highest $\Rey_{\theta}$ available in ZPG and b1.4 cases. Gray lines show the ZPG data growing in $\Rey$ from light to dark. The b1.4 lines show increase in Reynolds number from yellow to red.}
\label{fig:kitsios_scalings}
\end{figure}

As observed in figure \ref{fig:kitsios_scalings},
comparing the first three profiles for both the ZPG and b1.4 cases a clear lack of collapse 
can be observed, and as discussed in figure \ref{fig:def_U_uu} the profiles only collapse in the inner or outer regions using the adequate scaling. We argue that any claims of self-similarity throughout the complete boundary layer may be based on limited regions of near-equilibrium conditions, which may hide the existing Reynolds-number trends. 
With these considerations we remark the importance of considering long regions of near equilibrium, where it is possible to clearly observe the trends in the inner and outer layers and to assess how the scales separate as the TBL develops. 

\section{Spectral analysis} \label{sec:Spectra}

The turbulent fluctuations are due to the interactions of a wide range of coherent structures of different sizes. Each of these structures will have a characteristic length and energy content. Depending on the approach used to decompose the energy content of the Reynolds-stress components in space/time we can have different types of representations.
The spectral analysis used here is based on a Fourier decomposition of the spanwise two-point correlations. The velocity correlation between two points along the spanwise direction provides an idea of lengths at which the fluctuations are highly correlated, and this gives an indication of the presence of a certain structure or pattern. In a multi-scale phenomenon such as turbulence at high Reynolds numbers, the two-point correlations will contain a mix of all the different scales and it will be difficult to obtain meaningful information from it. 
Since the spanwise direction is periodic, a Fourier decomposition of the two-point correlation is possible, and the result is a spectral decomposition of the energy content in different wavenumbers $k_z$ associated with their corresponding wavelengths $\lambda_z=2\pi/k_z$. 
In the next sections we show the premultiplied power-spectral density of the different Reynolds-stress components at matched values of $\Rey_{\tau}=500, 1000,1500,2000$. For reference, an additional contour has been added in gray for the maximum $\Rey_{\tau}=2386$ in the ZPG simulation.

%-------------------- Spec1D APG-ZPG ------------------------------------------------------------------------
\subsection{One-dimensional power-spectral density in z}

\begin{figure}
\includegraphics[width=0.49\textwidth]{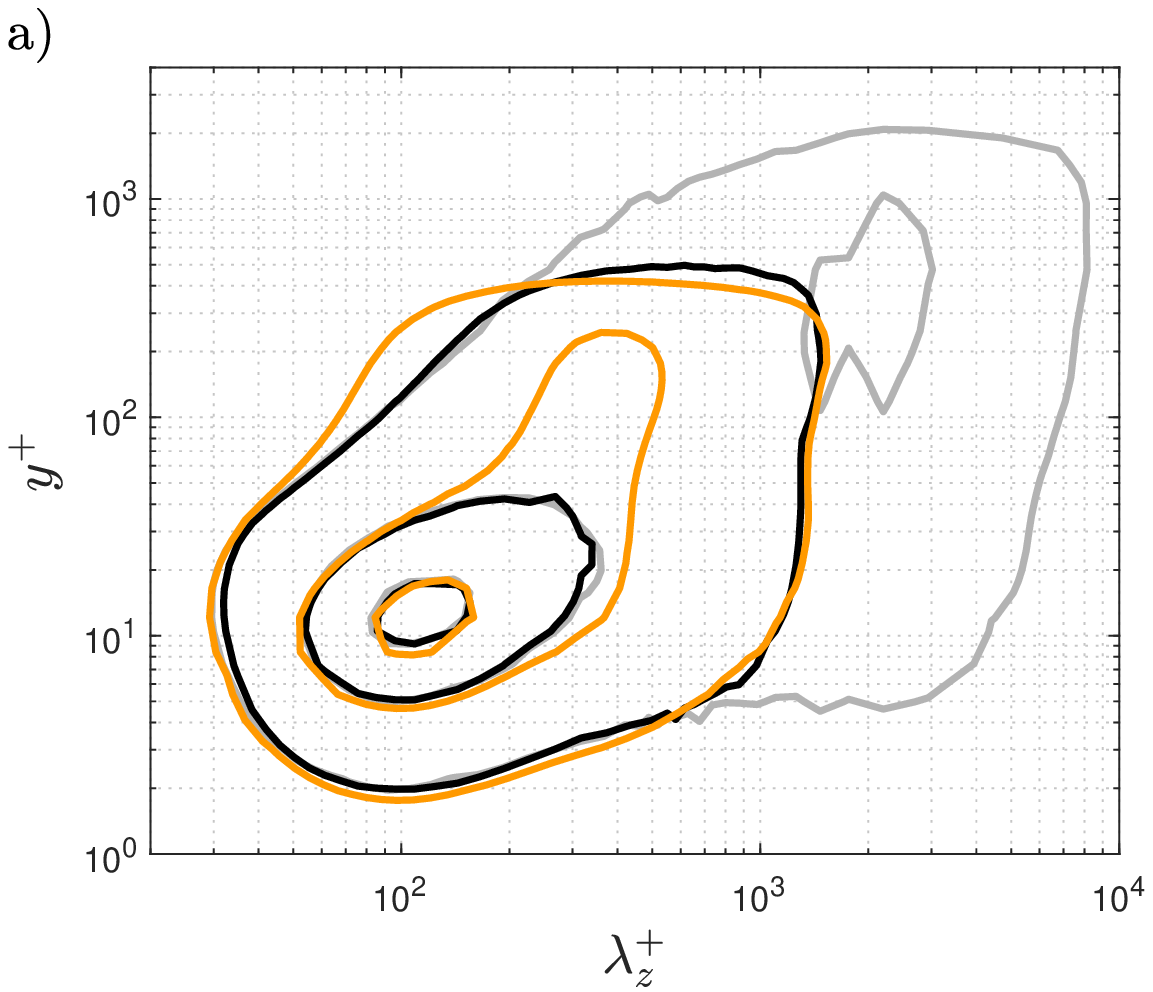}
\includegraphics[width=0.49\textwidth]{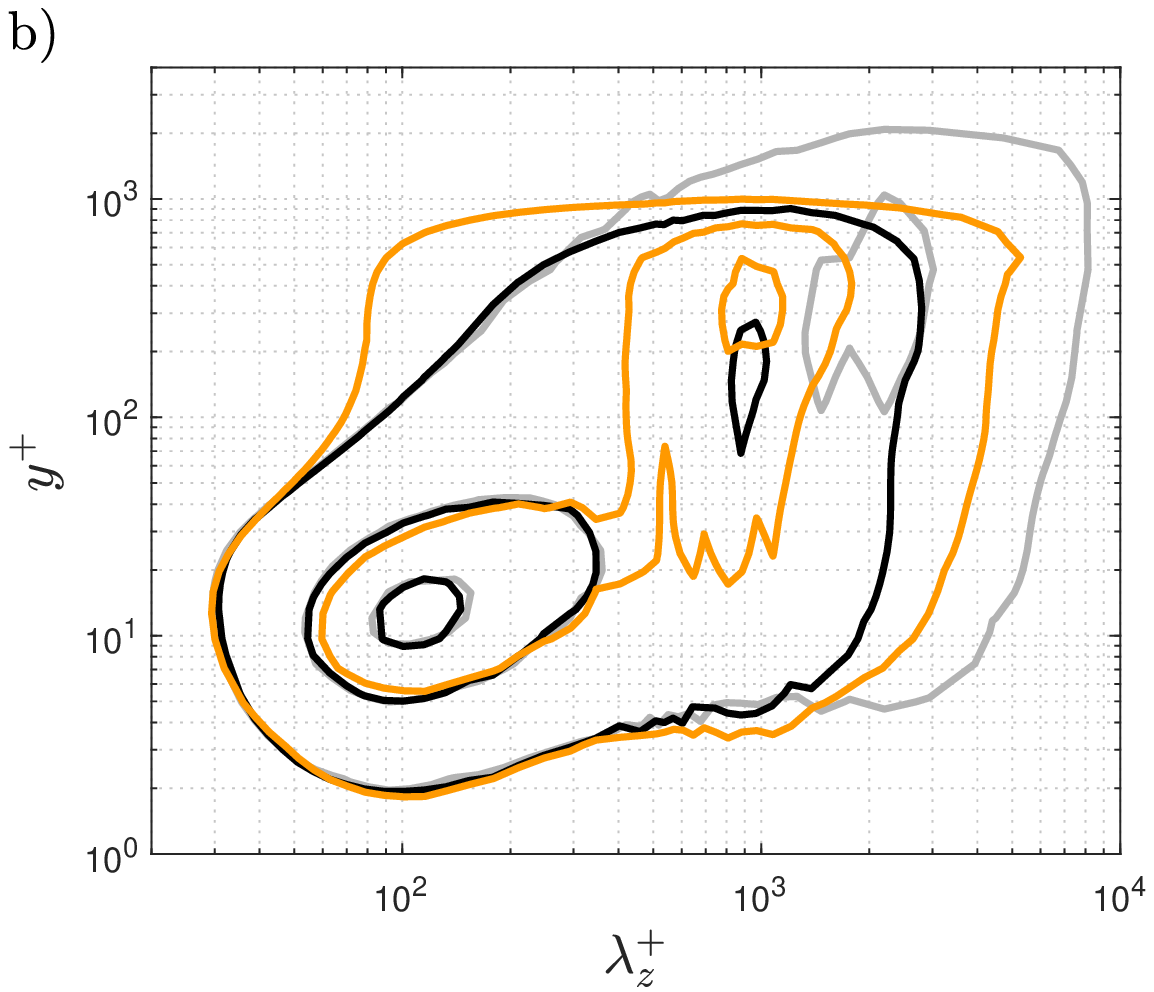}
\includegraphics[width=0.49\textwidth]{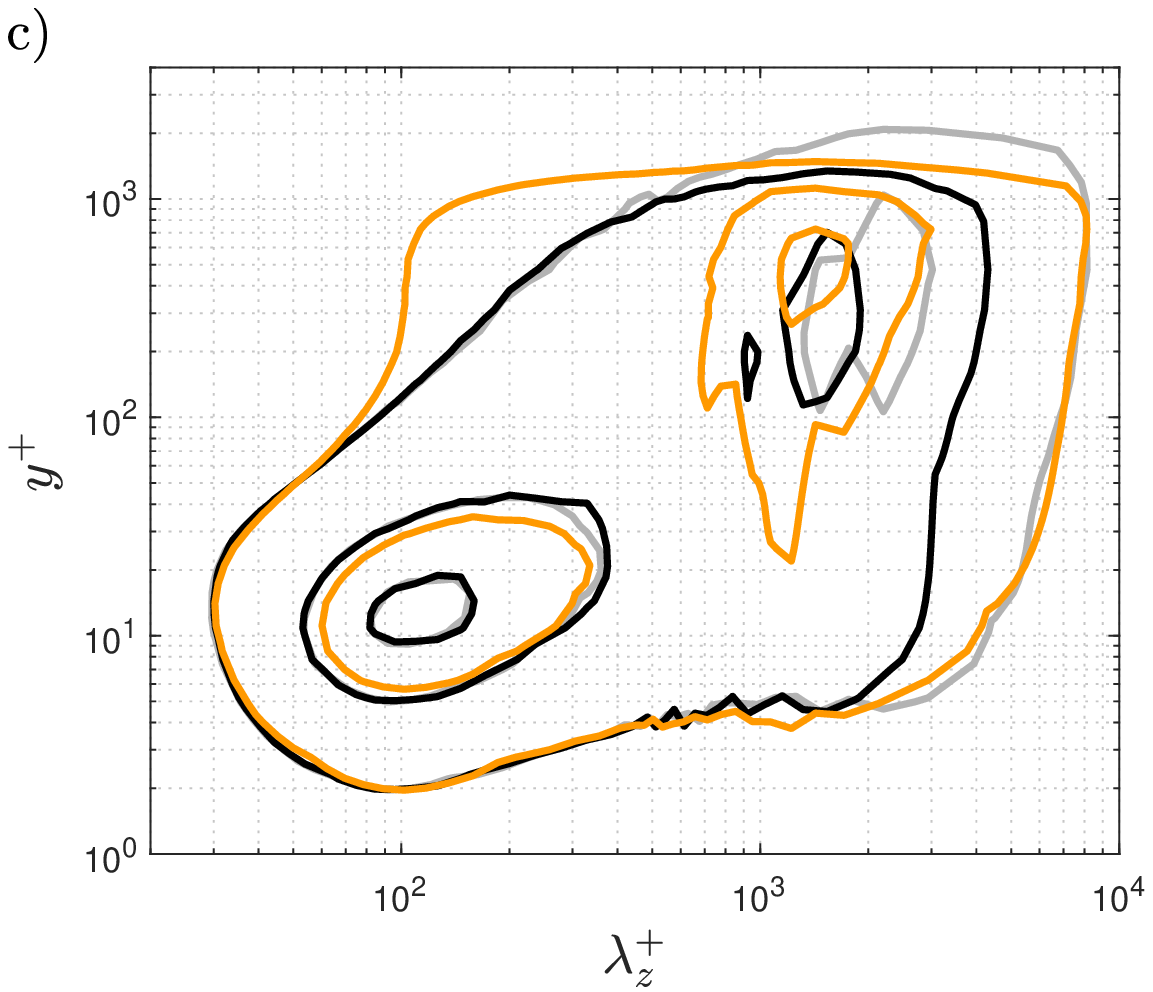}
\includegraphics[width=0.49\textwidth]{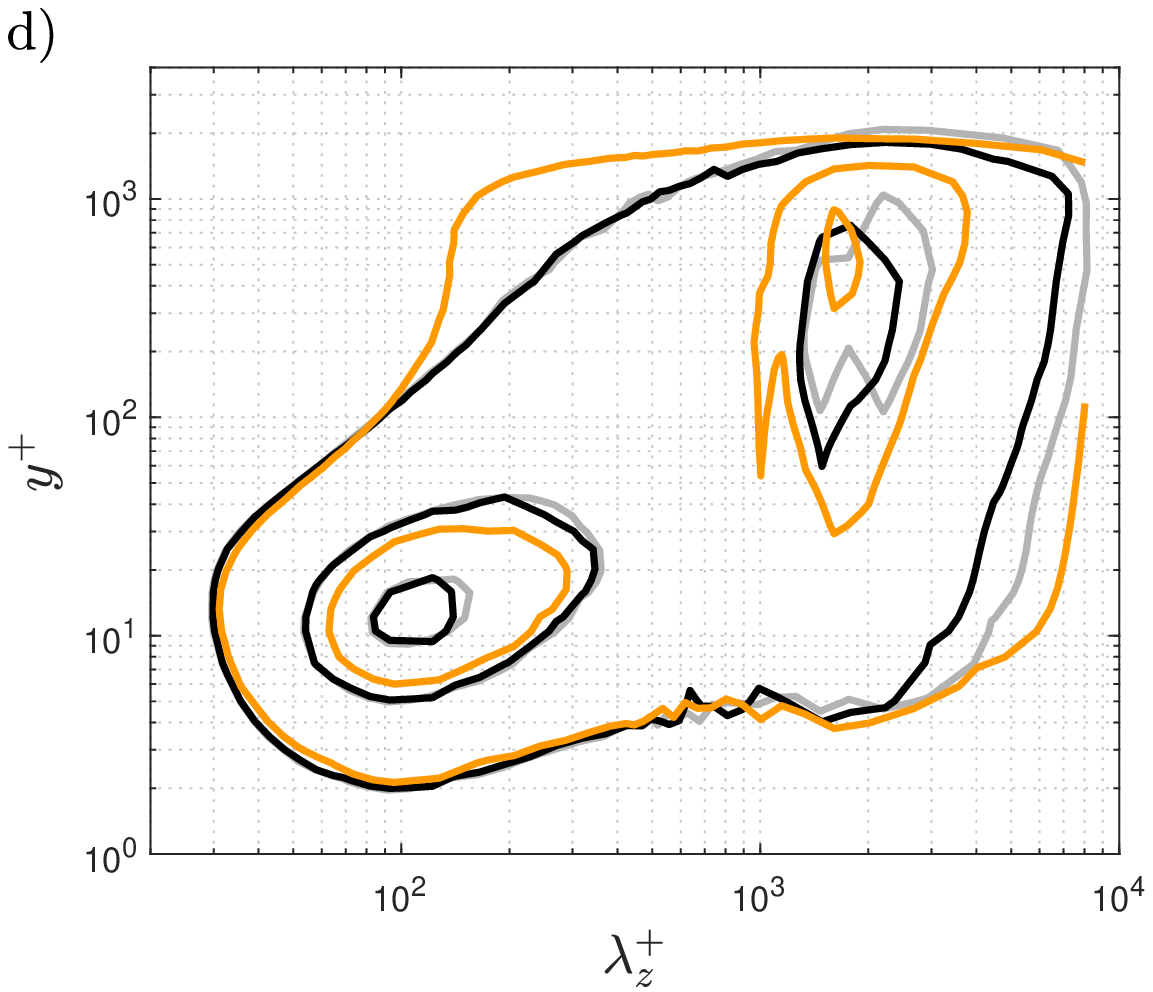}
  \caption{Premultiplied spanwise power-spectral density $k_z |\phi_{uu}|$ scaled with the local maximum for the b1.4 and ZPG cases at matched $\Rey_{\tau}$. Contours taken at $10\%$, $50\%$, $90\%$ of the maximum value. Reference contour in gray colour: ZPG at $Re_{\tau}=2386$. Contours with (\protect\blackline) for ZPG and (\protect\orangeline) for b1.4. (Top-left) $Re_{\tau}=500$, (top-right) $Re_{\tau}=1000$, (bottom-left) $Re_{\tau}=1500$, (bottom-right) $Re_{\tau}=2000$.}
\label{fig:spec1DUU}
\end{figure}

The premultiplied power-spectral energy density of the streamwise velocity fluctuations is shown in figure \ref{fig:spec1DUU}; note that this corresponds to the highest energetic components of the turbulent kinetic energy.
This figure shows that the low-Reynolds-number case $Re_{\tau}=500$ exhibits similarities between the b1.4 and the ZPG TBLs. The APG starts to show its effects by lifting small scales with $\lambda_z^+ \approx 100$ from the wall to the outer region, and increasing the energy of the scales with $\lambda_z^+ \approx 400$ in the outer region. The near-wall spectral peak located at $y^+\approx 15$ and $\lambda_z^+ \approx 100$ remains similar in the APG and ZPG cases.
For increasing Reynolds number, the near-wall peak magnitude does not exhibit significant changes in the APG, but the maximum value shifts to the outer region at $\Rey_{\tau} \approx 700$, and is associated with scales of wavelength $\lambda_z=\delta_{99}$.
The inner and outer peaks are separated by a region of lower energy content, and the lowest-energy contour shows the characteristic rising of small scales by the APG in the outer region \citep{tanarro_2020, VINUESA2018}. The rest of the spectra exhibit similar features between the ZPG and the APG, except for a wider range of $\lambda_z^+$ across the boundary layer at the same $\Rey_{\tau}$, which is an effect of the footprint of large scales residing in the logarithmic region, similar to what was reported by \cite{Hoyas_PoF2006} for channel flows.
Note that at these Reynolds numbers the outer spectral peak of the APG has a magnitude similar to that of the near-wall peak in the ZPG.

\begin{figure}
\includegraphics[width=0.49\textwidth]{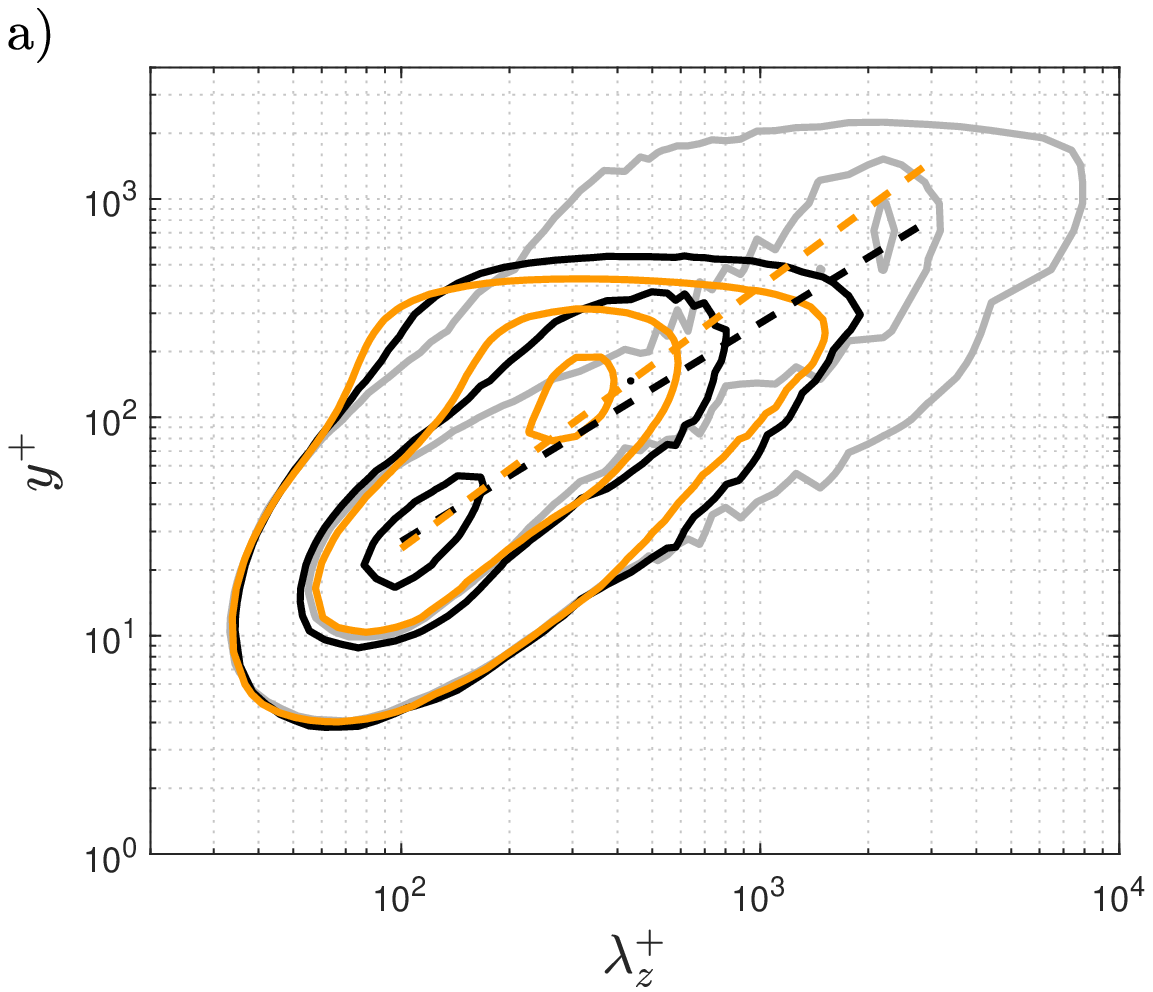}
\includegraphics[width=0.49\textwidth]{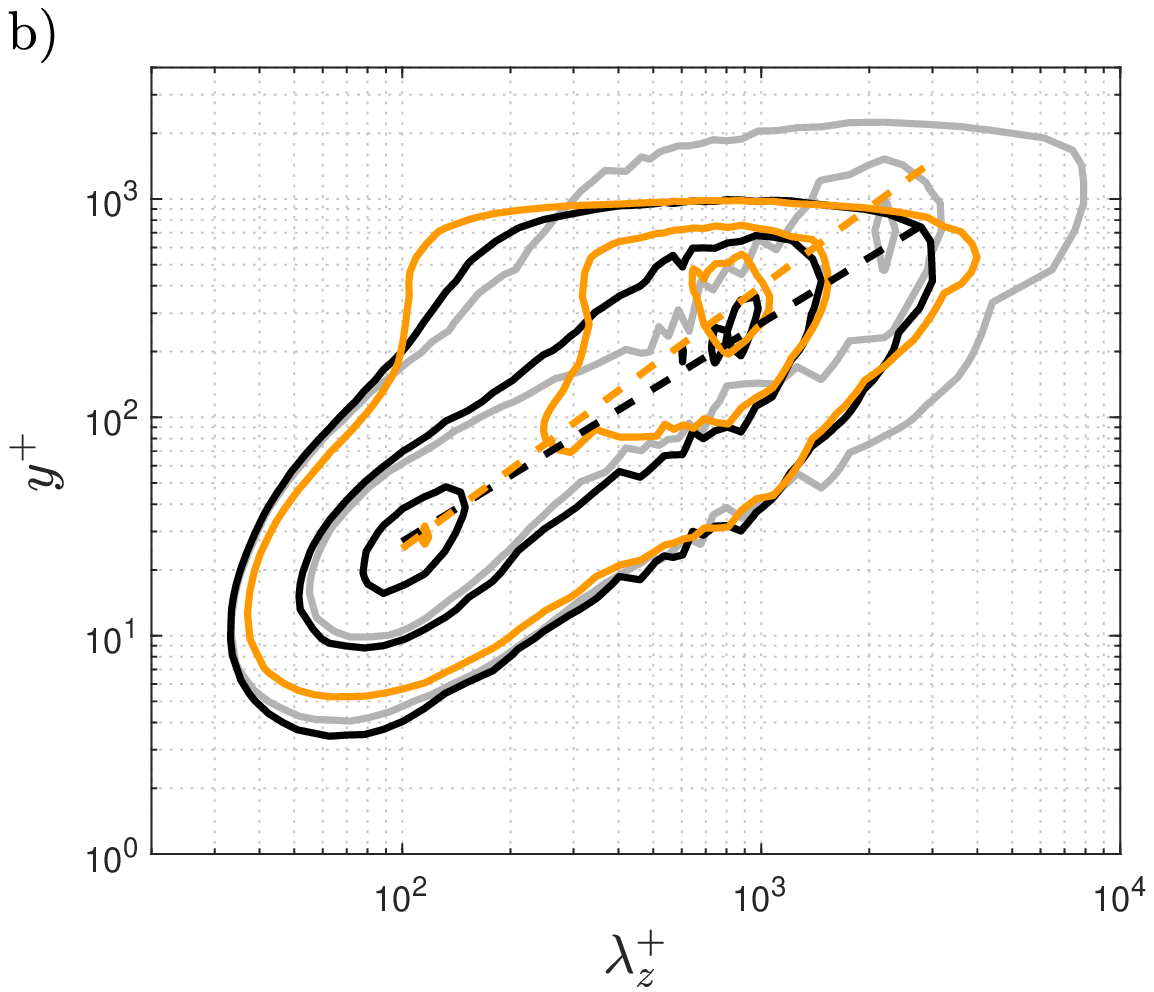}
\includegraphics[width=0.49\textwidth]{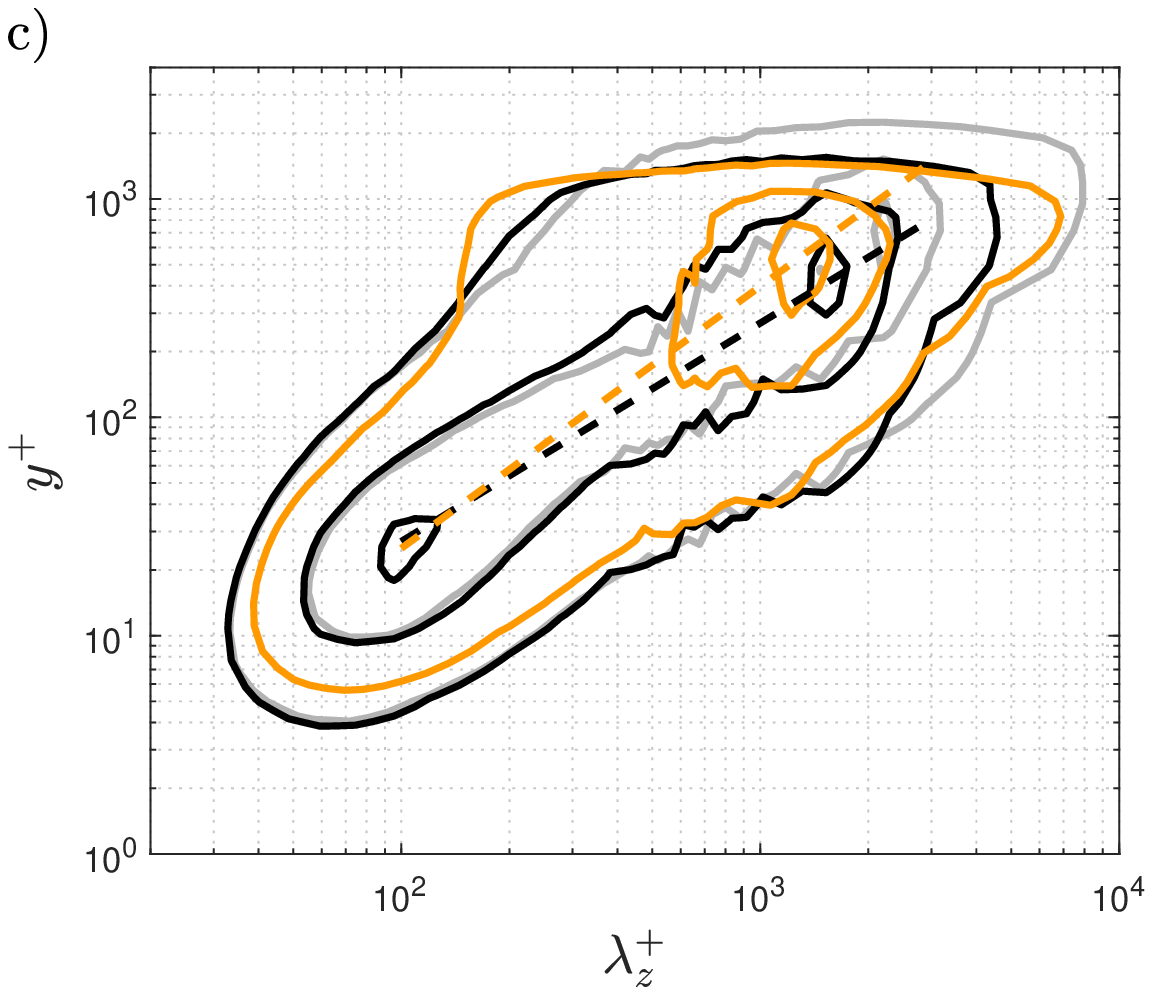}
\includegraphics[width=0.49\textwidth]{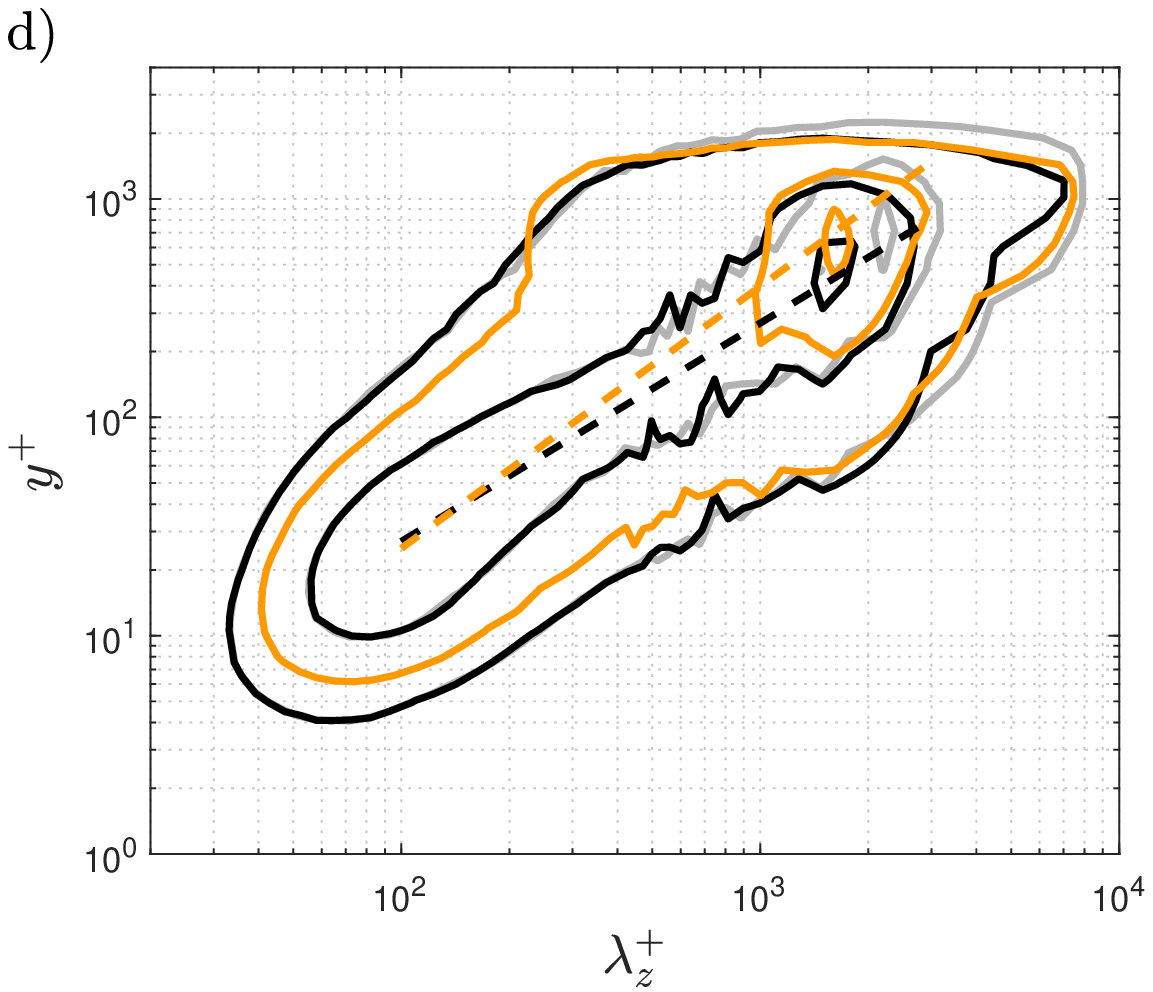}
  \caption{ Premultiplied cospectra $k_z |\phi_{uv}|$ scaled with the local maximum for the b1.4 and ZPG cases at matched $\Rey_{\tau}$. Contours taken at $10\%$, $50\%$, $90\%$ of the maximum value. Reference contour in gray colour: ZPG at $Re_{\tau}=2386$. Dashed black lines show the curve $y^+=0.27 \lambda_z^+$ as in \cite{giovanetti2016}, while the orange dashed lines represent $y^+=0.1 (\lambda_z^+)^{1.2}$, which is the ridge for the b1.4 case. Colors: (\protect\blackline) ZPG; (\protect\orangeline) b1.4. (Top-left) $Re_{\tau}=500$, (top-right) $Re_{\tau}=1000$, (bottom-left) $Re_{\tau}=1500$, (bottom-right) $Re_{\tau}=2000$.}
\label{fig:spec1DUV}
\end{figure}

In figure \ref{fig:spec1DUV} we show the premultiplied cospectra $k_z |\phi_{uv}|$ at the same matched $\Rey_{\tau}$ as in figure \ref{fig:spec1DUU}. 
At the lowest $\Rey_{\tau}$ the contours in the APG and the ZPG are similar, with the difference that the maximum value is in the outer region in the former ($y^+=120$ and $\lambda_z^+=307$) and closer to the wall ($y^+=26$ and $\lambda_z^+=109$) in the latter.
In the $10\%$ and $50\%$ contours, it is possible to see a small contribution of the small scales in the outer region compared to the ZPG contours.
At higher Reynolds numbers, the ZPG  develops a plateau of energy with contour levels following a straight line of slope $C$, which in a logarithmic plot corresponds to a power law of the form: $y^+=(\lambda_z^+)^C$.
This line connects the near-wall peak  with the outer region. Furthermore, the plateau indicates a progressive growth of the region containing this level of energy following the previous power law, which at $\Rey_{\tau}=1000$ develops an outer peak with a magnitude similar to that of the near-wall peak. Note that at higher Reynolds numbers the outer peak progressively rises over the magnitude of the near-wall peak.
The effect of the APG is to displace the near-wall energy to the outer region, which becomes dominant in the premultiplied cospectra of the Reynolds shear stress. At higher Reynolds numbers, the premultiplied cospectra exhibit a peak at $\lambda_z \simeq \delta_{99}$, which implies that this peak scales in outer units.
The constant $C$ which defines the slope of the black dashed lines in figure \ref{fig:spec1DUV} was reported to be $\approx 1$ by \cite{giovanetti2016}. The APG exhibits $10\%$ contours similar to those of the ZPG, indicating the presence of some energy in the near-wall region. If a line from the inner peak of the ZPG is drawn towards the APG peak in the outer region (orange dashed lines in figure \ref{fig:spec1DUV}), its slope is larger than that of the ZPG. If a self-similar hierarchy of motions in the log-layer is suggested by the black dashed line, in connection with the attached-eddy hypothesis \citep{Townsend_1976, deshpande_2021}, then the APG either rises the slope of that hierarchy of scales or follows the same hierarchy of motions as in the ZPG with an additional contribution in the outer region by small scales risen from the wall by the wall-normal convection of the APG and by the more energetic large scales.

\begin{figure}
\includegraphics[width=0.245\textwidth]{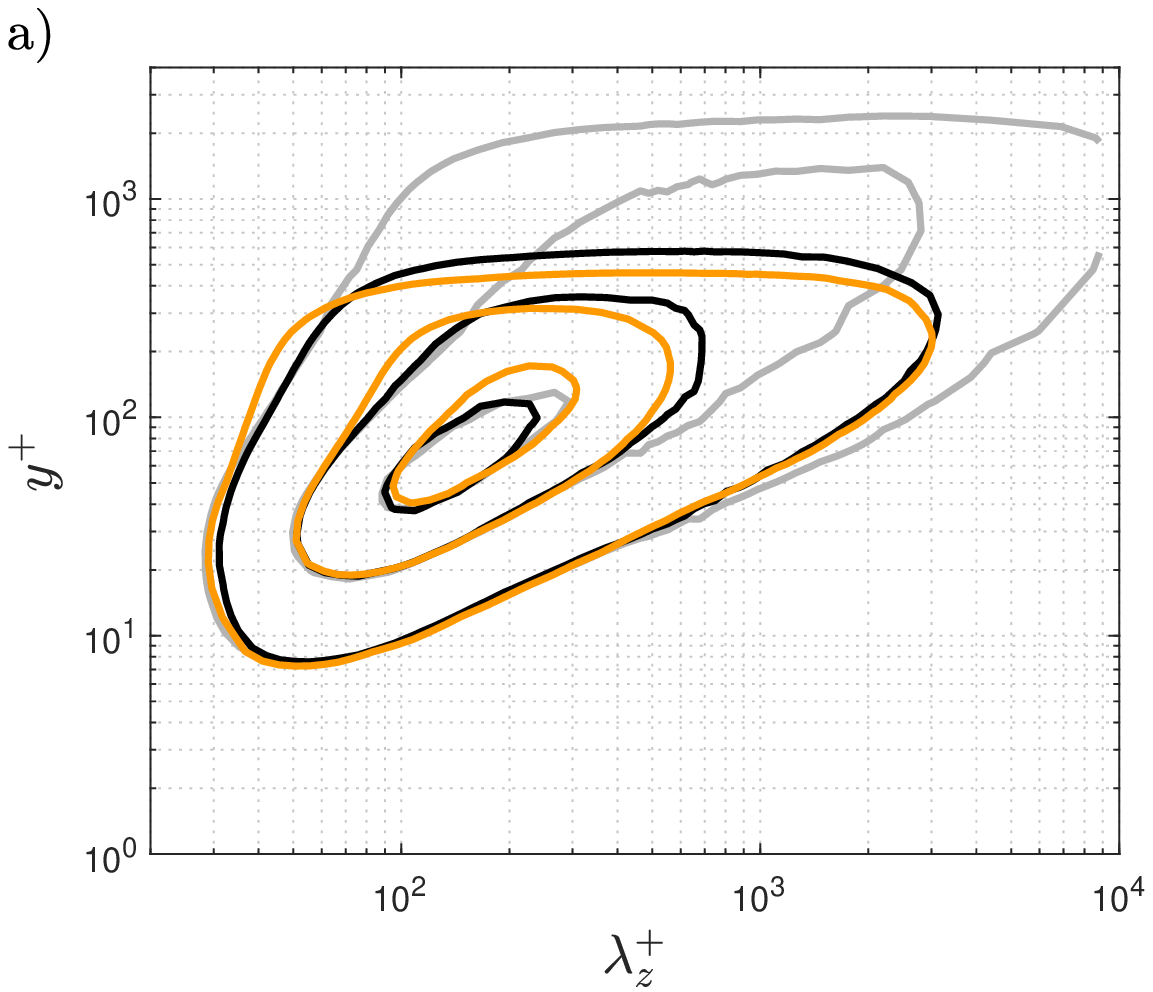}
\includegraphics[width=0.245\textwidth]{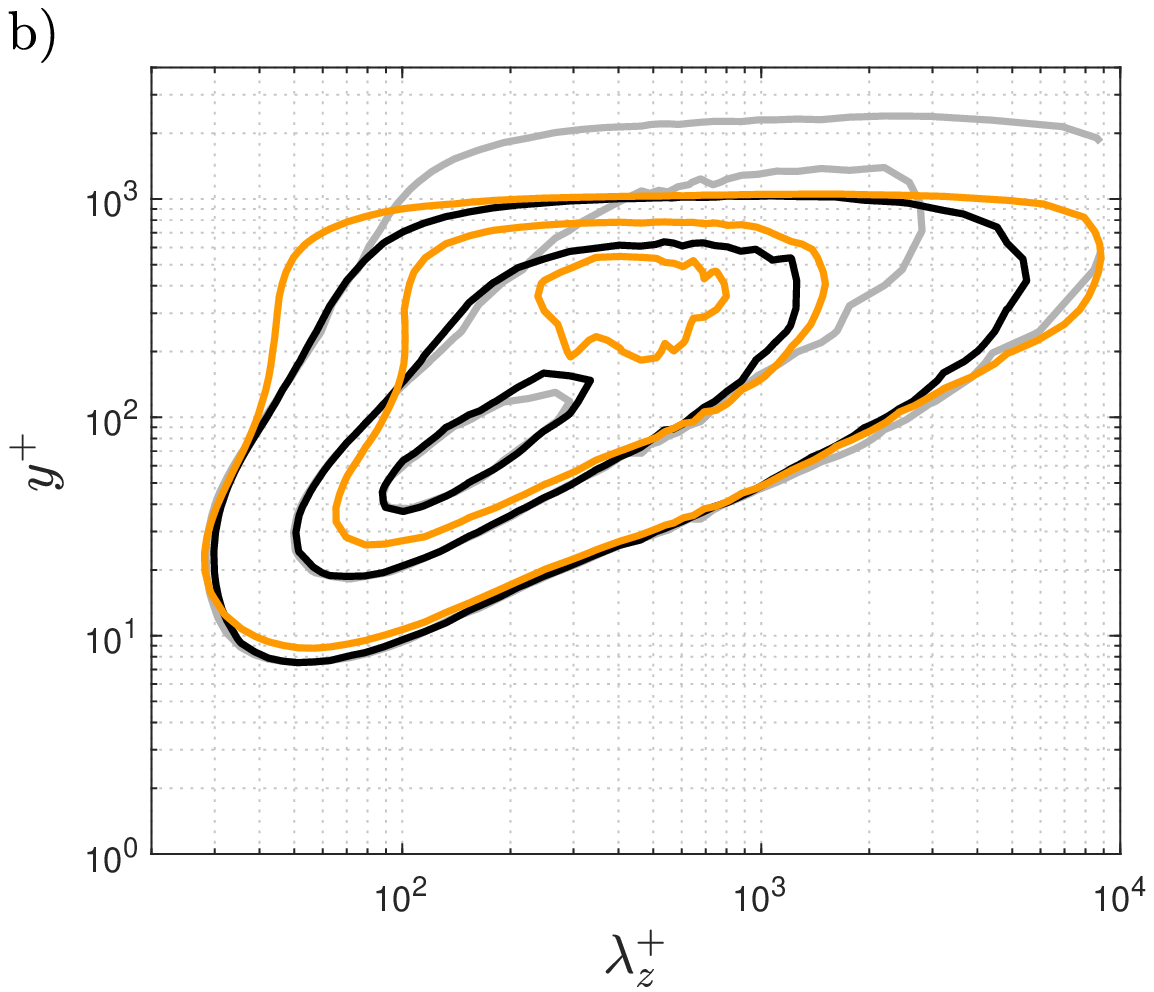}
\includegraphics[width=0.245\textwidth]{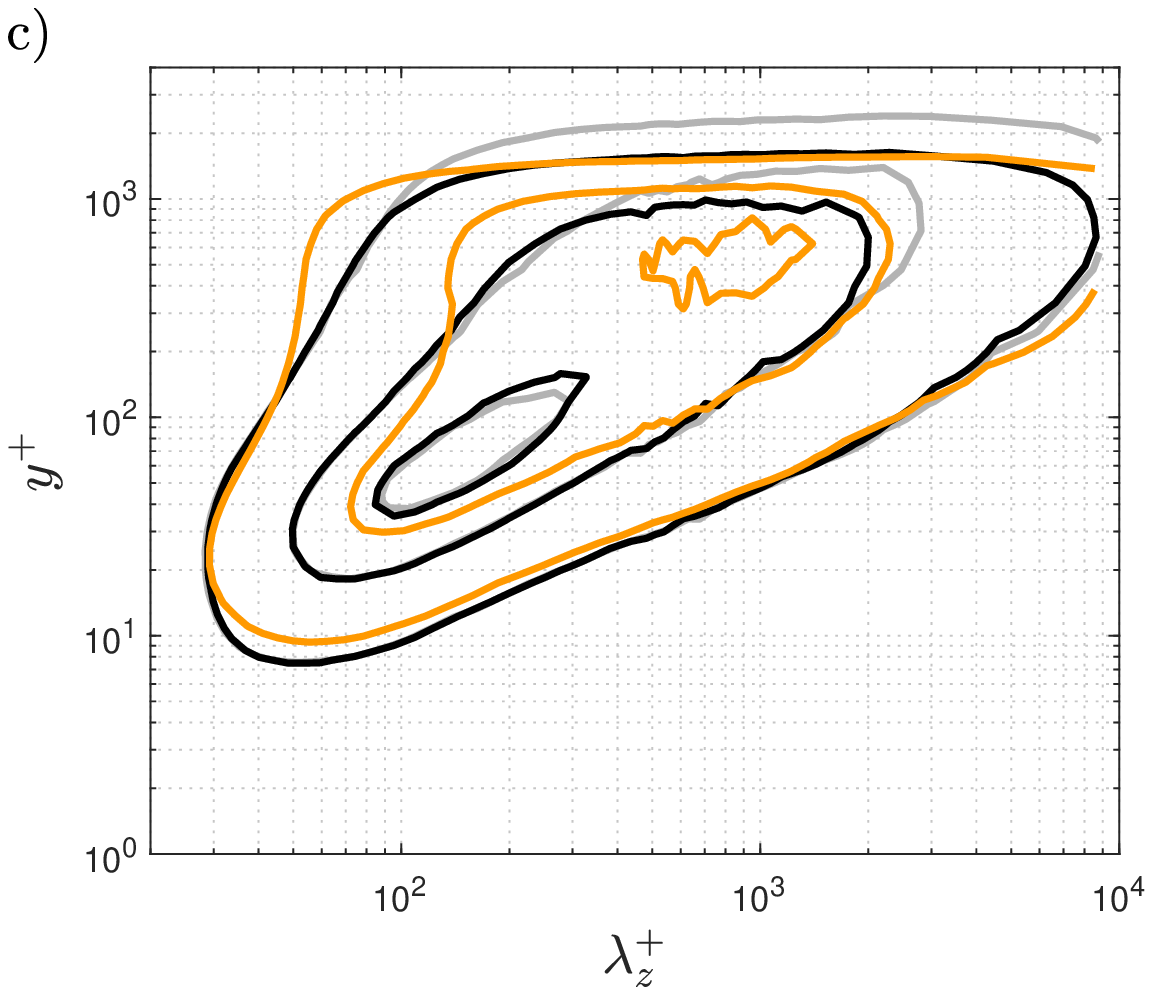}
\includegraphics[width=0.245\textwidth]{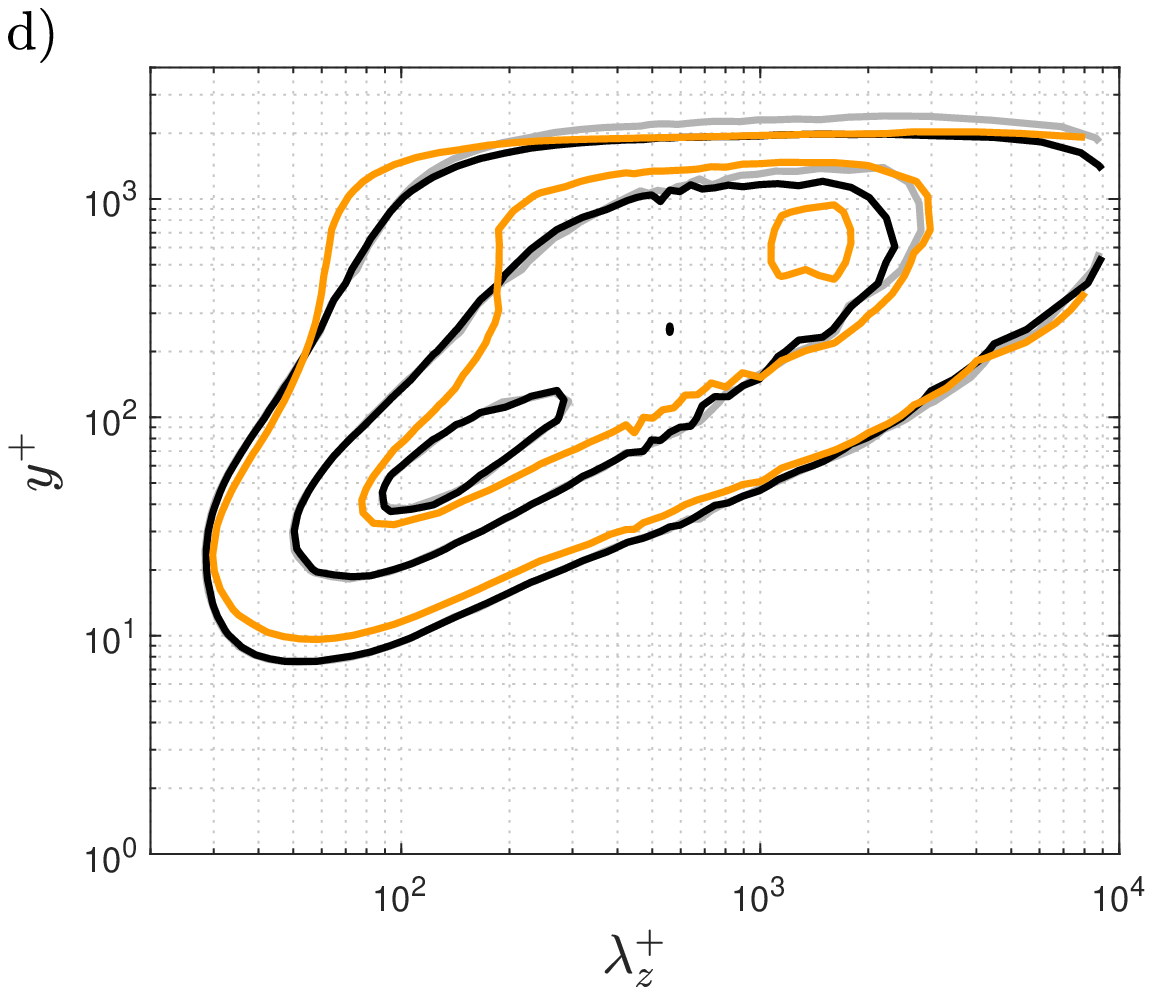}
\includegraphics[width=0.245\textwidth]{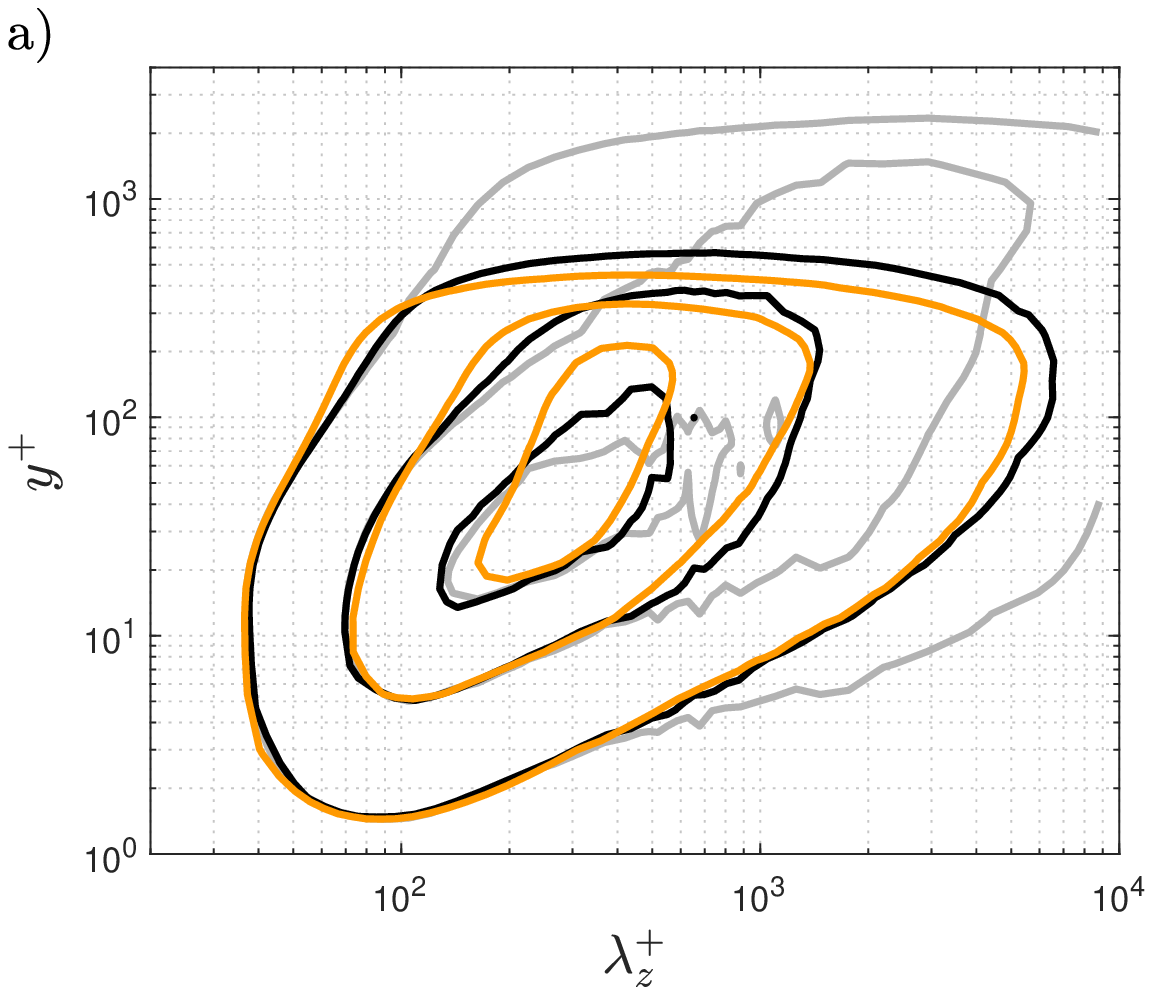}
\includegraphics[width=0.245\textwidth]{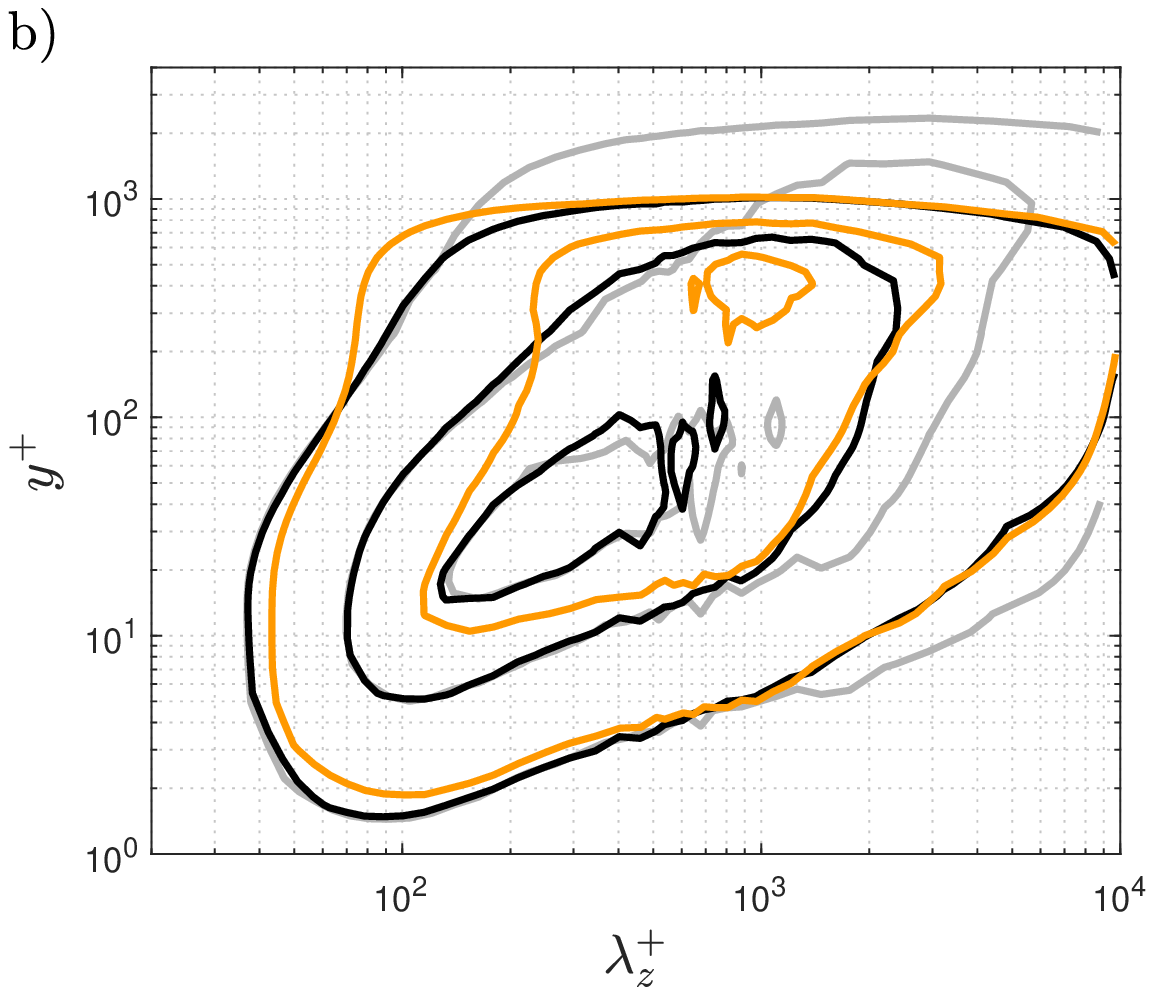}
\includegraphics[width=0.245\textwidth]{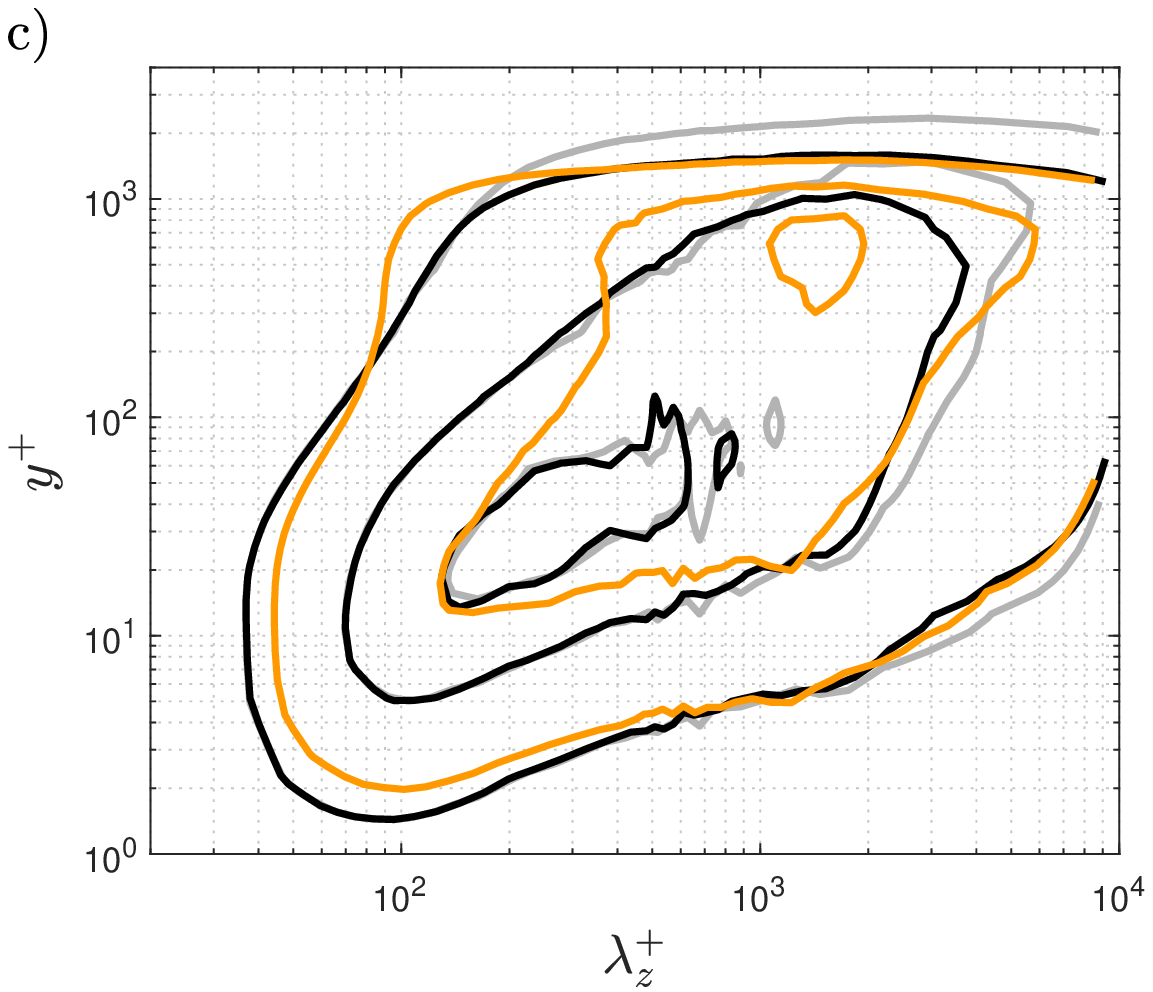}
\includegraphics[width=0.245\textwidth]{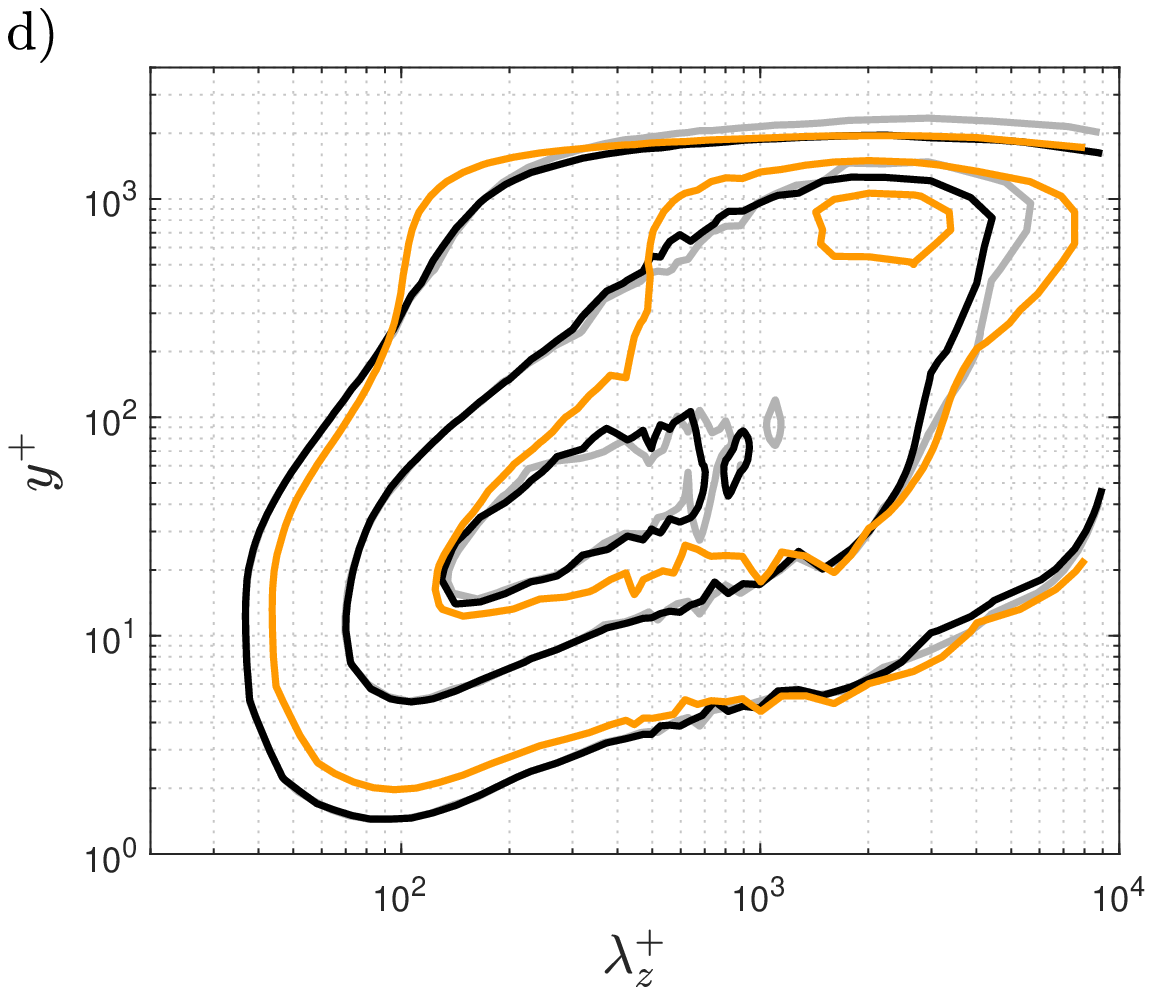}
  \caption{Premultiplied spanwise power-spectral density of $k_z |\phi_{vv}|$ (first row) and $k_z |\phi_{ww}|$ (second row) scaled with the local maximum for the b1.4 and ZPG cases at matched $\Rey_{\tau}$. Contours taken at $10\%$, $50\%$, $90\%$ of the maximum value. Reference contour in gray colour: ZPG at $Re_{\tau}=2386$. Contours with (\protect\blackline) for ZPG and (\protect\orangeline) for b1.4. From left to right: $Re_{\tau}=500$, $Re_{\tau}=1000$, $Re_{\tau}=1500$ and $Re_{\tau}=2000$.}
\label{fig:spec1D_VV_WW}
\end{figure}

The premultiplied spectra for the wall-normal fluctuations is shown in the first row of figure \ref{fig:spec1D_VV_WW}. It exhibits features similar to those of the cospectra of the Reynolds shear stress: small-scale energy in the outer region due to the APG and a different location of the maximum power-spectral density in the ZPG and the APG. While the ZPG exhibits an elongated $90\%$ contour around $\lambda_z^+ \approx 150$ at $y^+\approx 90$ for the different Reynolds numbers, the APG starts to stretch the peak at the lowest $\Rey_{\tau}=500$, and at $\Rey_{\tau}=1000$ the peak is located in the outer region with scales of the order of $\lambda_z \simeq \delta_{99}$.
In the ZPG the $10\%$ contours in the wall-normal spectra exhibit a shape similar to that of the $50\%$ contours in the cospectra. Approximating the $10\%$ contour by an ellipse, the major axis also follows a trend $y^+=(\lambda_z^+)^C$ with $C=1$, as in the cospectra. The $50\%$ contours also exhibit linear regions, but in the wall-normal spectra the slopes are different. This could still be related to a hierarchy of motions in the logarithmic layer, just indicating that the range of wall-normal scales grows with the wall-normal location and as before, the APG adds an extra contribution from the small and highly-energetic large scales in the outer part of the logarithmic layer. The $10\%$ contours of the APG follow the trend dictated by $\Rey_{\tau}$ and far from the trend marked by $y^+=(\lambda_z^+)^C$; for this reason, we have not included those linear trends.
As can be observed in the second row of figure \ref{fig:spec1D_VV_WW}, the spectra of the spanwise fluctuations exhibit effects similar to those shown in the wall-normal spectra in the first row for the ZPG and the APG. In particular we identify the small-scale contribution to the outer region and a stable $90\%$ contour for the ZPG expanding a long range of scales from $\lambda_z^+\approx 15$ to $\lambda_z^+\approx 70$ in a region between $y^+=15$ and $y^+=100$, which is inside the overlap region. This $90\%$ contour is displaced by the APG towards regions farther than $y^+=300$, already in the wake region. The $\lambda_z^+$ of this peak also scales with the Reynolds number.

%-----------------Spectra 2D------------------------------------------------------------
%-----uu--------------------

\subsection{Two-dimensional power-spectral density}

The two-dimensional power-spectral energy $E_{u_iu_j}(k_z,k_t,y)$ is obtained using temporal series of the velocities in all the spanwise grid points at selected streamwise and wall-normal positions. Once the mean in time and spanwise direction is substracted, the turbulent components are transformed to Fourier space in the spanwise wavenumbers. To obtain the power spectra in time, Welch's method is used with 8 independent subdivisions in time overlapped with 7 subdivisions for a total of 15 bins. The window function is a Hamming window.
The spectral energy is then divided by $\Delta k_z \Delta k_t$ to obtain the two-dimensional power-spectral density $\phi(k_z,k_t,y)$. As above, the figures used to illustrate the effects of the Reynolds number and the APG in the spectral-density content will be premultiplied, in this case, with the factor $k_z k_t$.
It has been verified that the addition of the spectral energy for all the wavenumbers in time yields the one-dimensional power-spectral energy in the spanwise direction $E_{u_i u_j}(k_z,y)$.

\subsubsection{Two-dimensional power-spectral density in the near-wall region}

\begin{figure}
\includegraphics[width=0.49\textwidth]{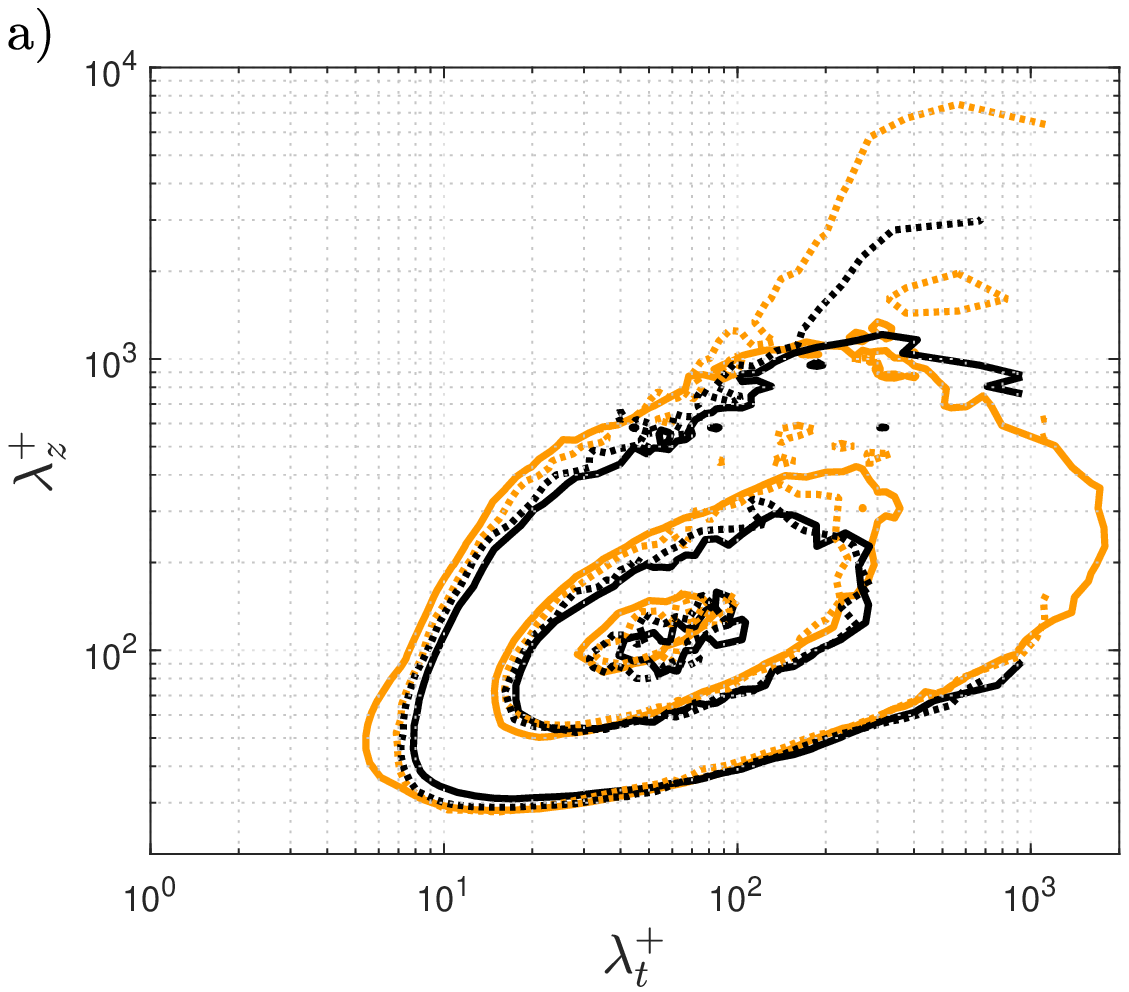}
\includegraphics[width=0.49\textwidth]{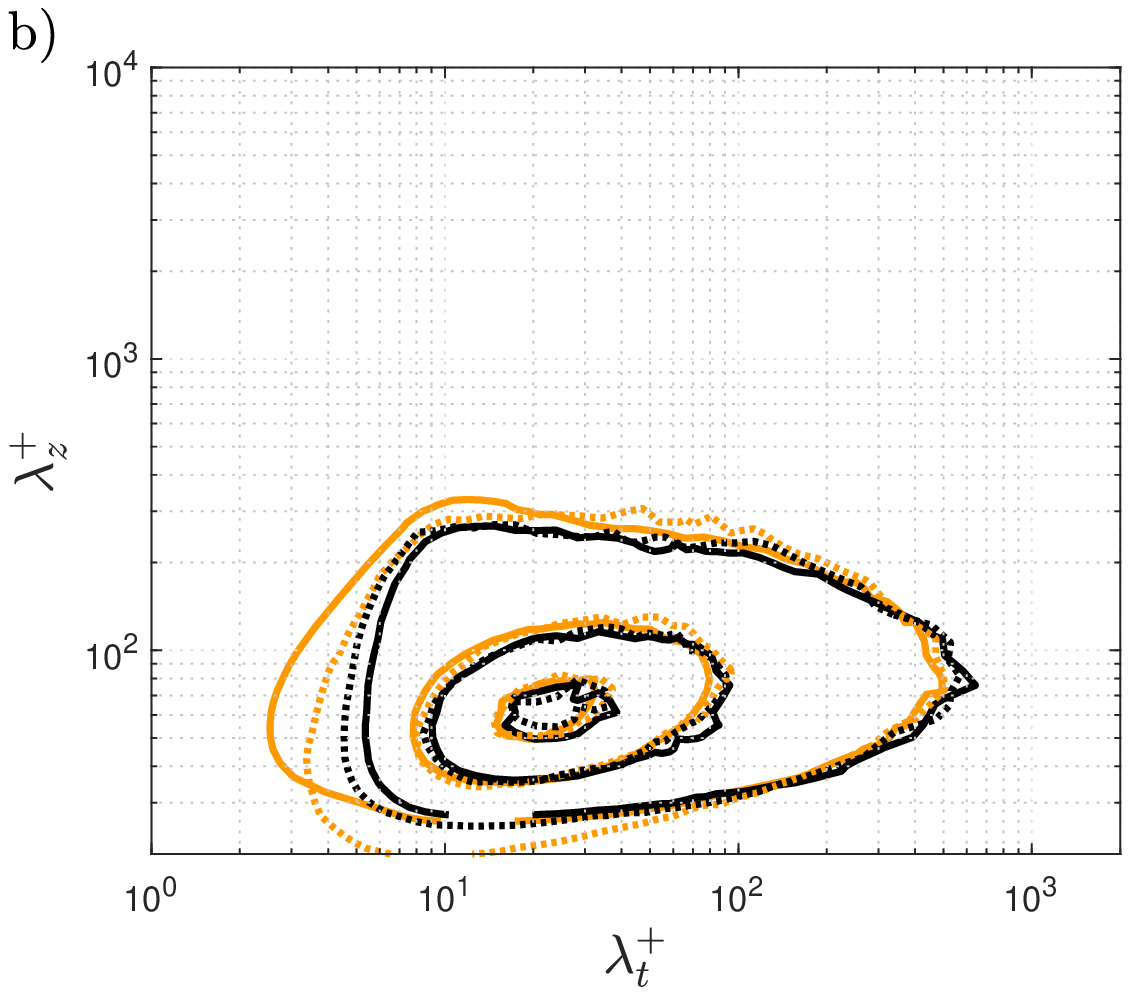}
\includegraphics[width=0.49\textwidth]{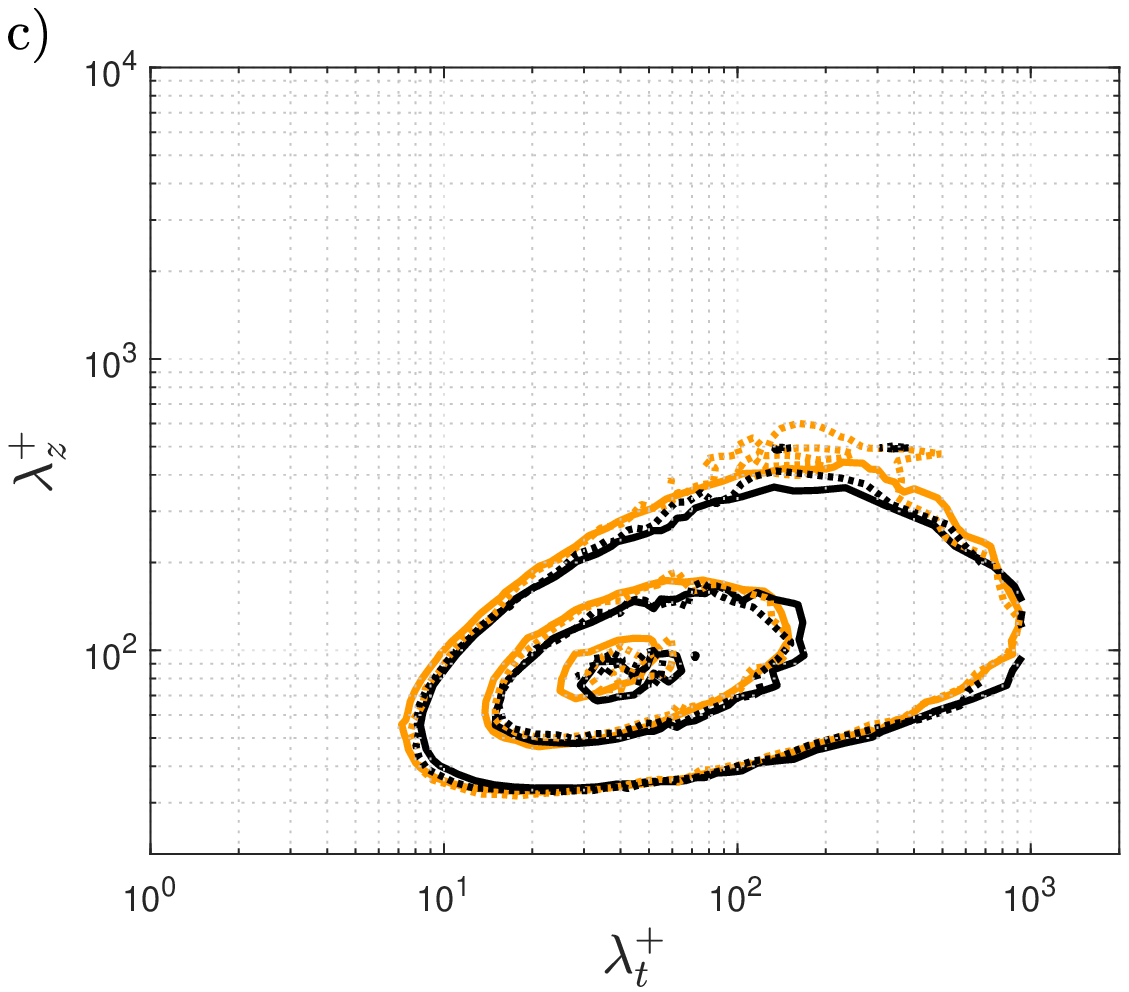}
\includegraphics[width=0.49\textwidth]{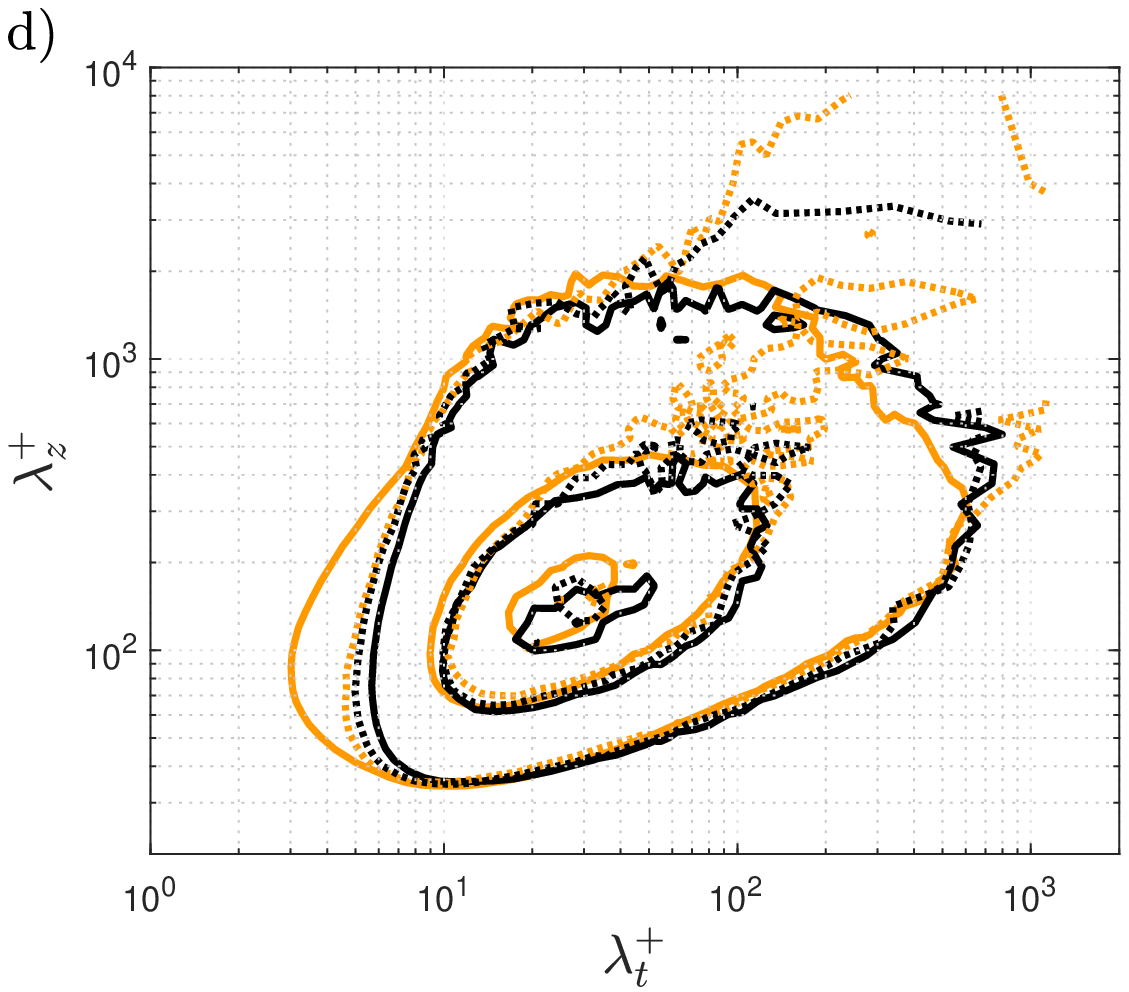}
  \caption{Evolution with the Reynolds number of the two-dimensional premultiplied power-spectral density in time and $z$ for the various Reynolds-stress components $k_z k_t |\phi_{u_iu_j}|$ at $y^+=15$ scaled with the local maximum. The panels show spectra of: a) $uu$, b) $vv$, c) $uv$, d) $ww$. Contours taken at $10\%$, $50\%$, $90\%$ of the maximum value. Solid lines for $Re_{\tau}=500$ and dotted lines for $Re_{\tau}=2000$. Colors: (\protect\blackline) for ZPG and (\protect\orangeline) for b1.4.}
\label{fig:spec2D_uiuj_y15}
\end{figure}

The premultiplied power-spectral density in time and the spanwise direction $k_zk_t\phi(\lambda_z, \lambda_t)$ is first analysed at $y^+=15$, a wall-normal location which shows the characteristics of the near-wall peak of the streamwise RS and the production of TKE.
\begin{figure}
\includegraphics[width=0.24\textwidth]{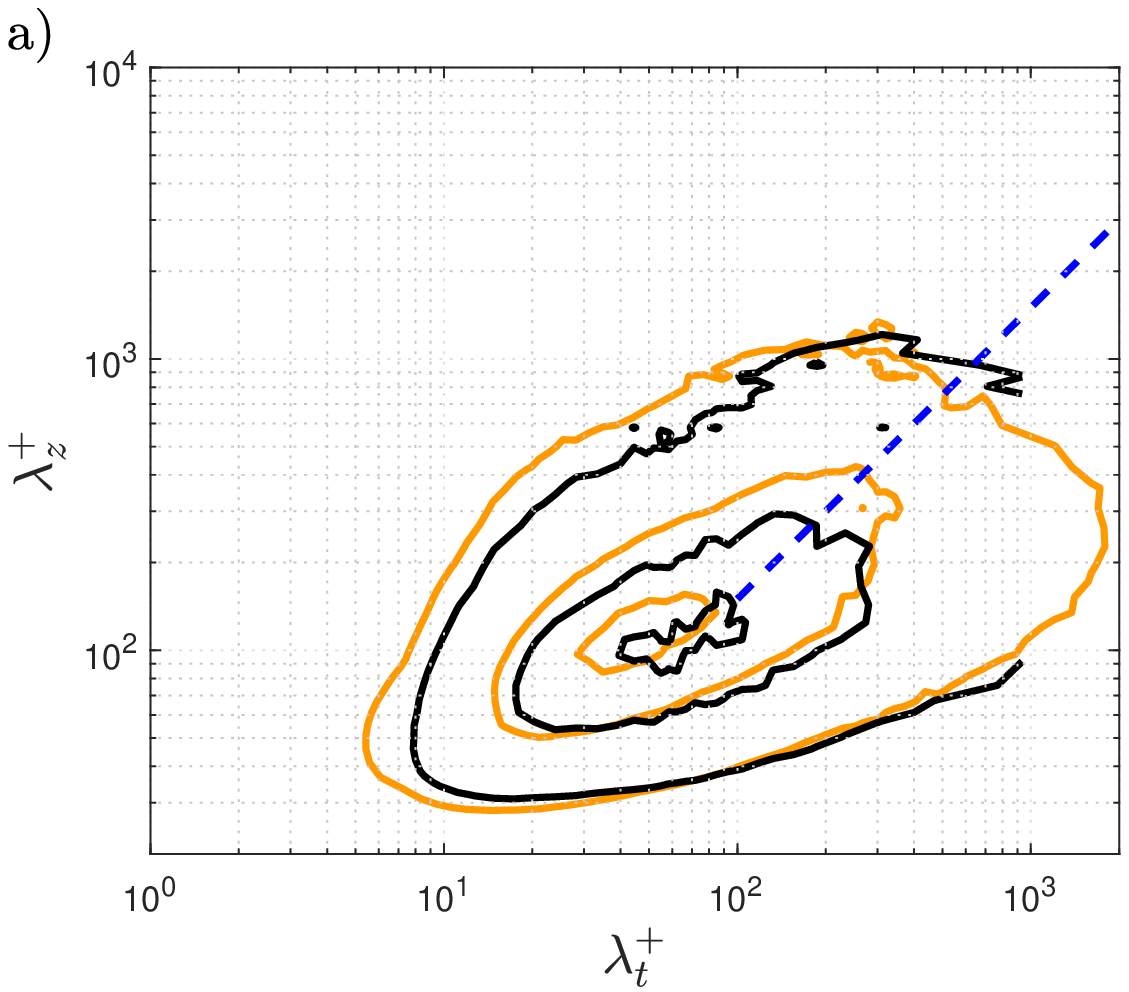}
\includegraphics[width=0.24\textwidth]{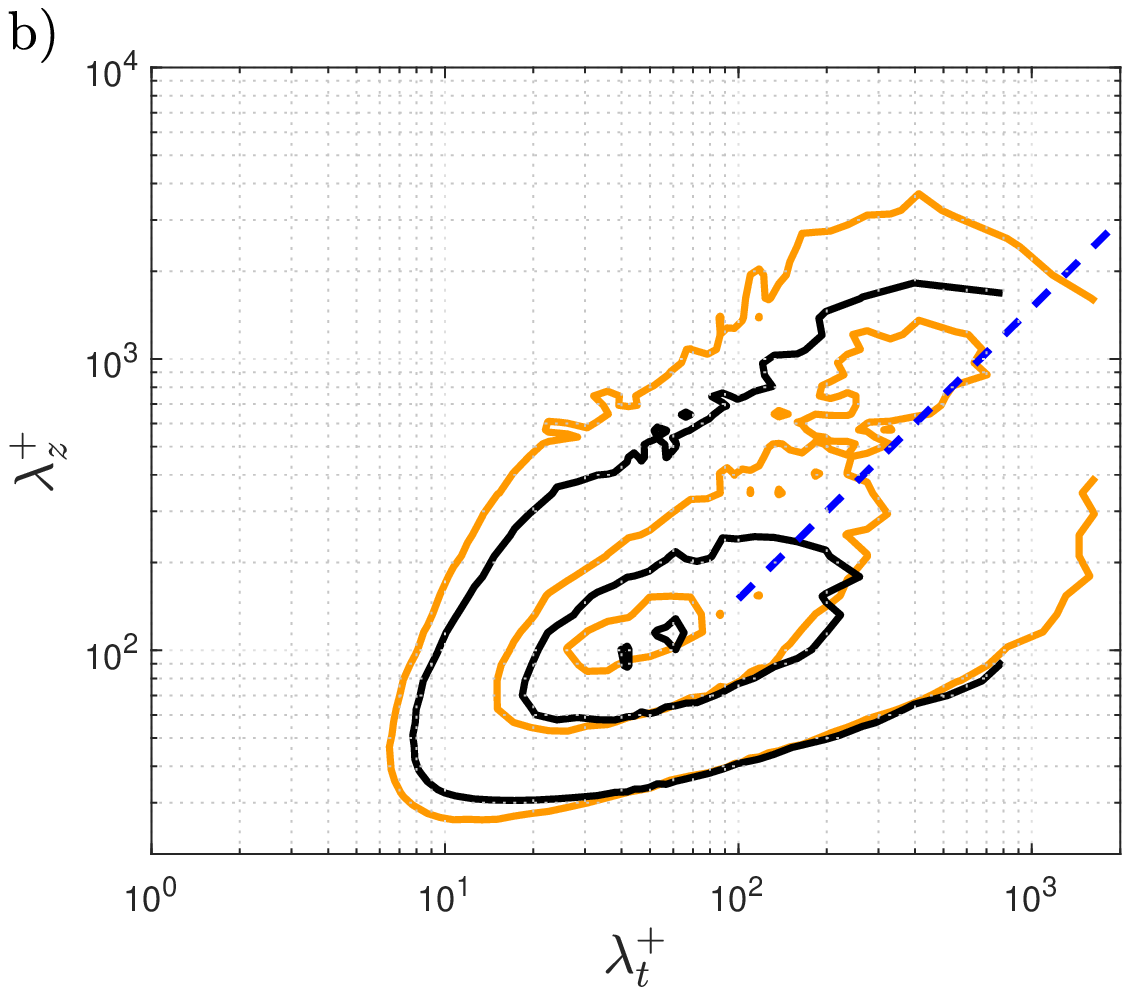}
\includegraphics[width=0.24\textwidth]{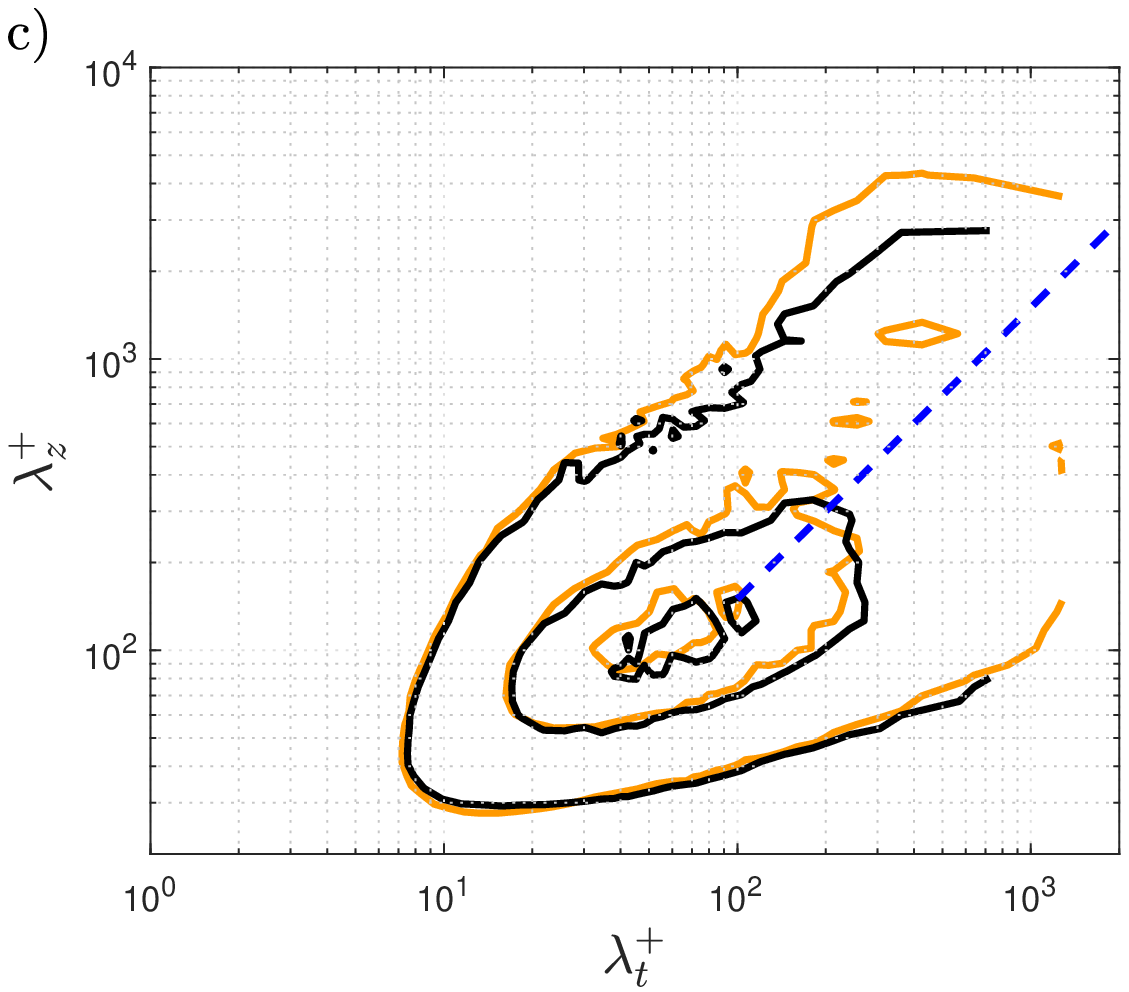}
\includegraphics[width=0.24\textwidth]{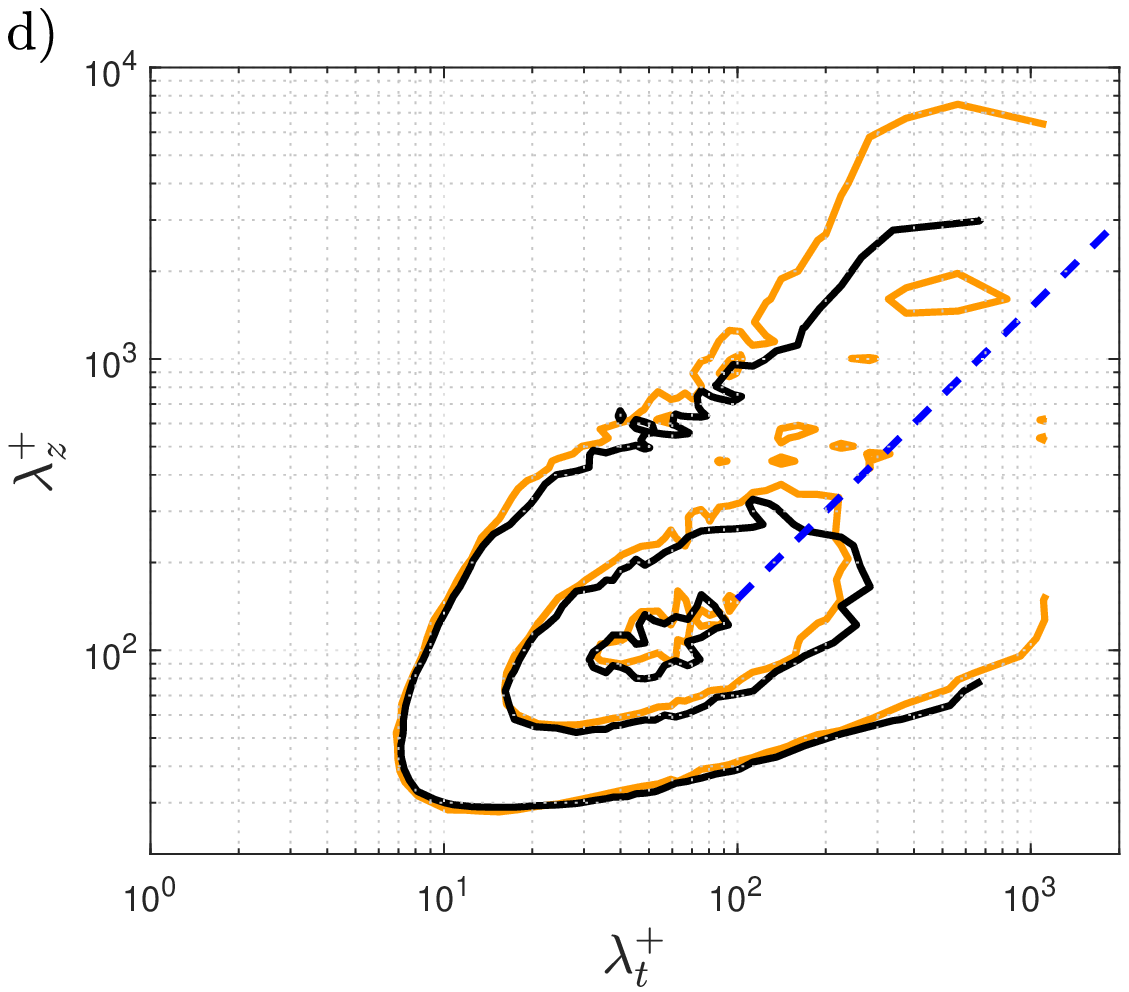}
  \caption{Two-dimensional premultiplied power-spectral density $k_z k_t |\phi_{uu}|$ at $y^+=15$ scaled with the local maximum. Contours taken at $10\%$, $50\%$, $90\%$ of the maximum value. From left to right: $Re_{\tau}=500$, $Re_{\tau}=1000$, $Re_{\tau}=1500$ and $Re_{\tau}=2000$. The dashed blue line represents $\lambda_z^+ = 1.5\lambda_t^+$. Contours with (\protect\blackline) for ZPG and (\protect\orangeline) for b1.4.}
\label{fig:spec2D_uu_y15}
\end{figure}
For all the Reynolds numbers shown in figure \ref{fig:spec2D_uu_y15} there is a near-wall spectral peak with $\lambda_z^+ \approx 100$ and $\lambda_t^+$ of the order of 100. The blue dashed line represents $\lambda_z^+ = 1.5\lambda_t^+$, which describes the evolution with $\Rey$ of the ridge in the upper-right part of the spectrum. This ridge was documented by \cite{Hoyas_PoF2006} for channels, \cite{Sillero_2011}, for ZPG TBLs and \cite{tanarro_2020} for APG TBLs.
The APG does not significantly modify the spectrum around the near-wall peak and the local viscous length and time scales $l_{\tau}$ and $t_{\tau}$ lead to the collapse of the inner region for the three displayed contour levels.
Some extra energy is seen in the APG with the $50\%$ contours in the region with larger $\lambda_z^+$ and $\lambda_t^+$.
The biggest difference in the inner region is seen for the lowest Reynolds number, which is upstream of the ROI where near-equilibrium is achieved. At higher $\Rey_{\tau}$ values of 1000, 1500 and 2000, the spectra around the near-wall peak are identical for each simulation and grow closer to the ZPG contours. In figure \ref{fig:spec2D_uiuj_y15}, only the contours at $\Rey_{\tau}=500$ and 2000 are represented to show the differences in the region around the near-wall peak due to $\Rey_{\tau}=500$ being outside of the near-equilibrium region and to show the growth of the largest spatio-temporal scales.

Surprisingly, the cospectra of the Reynolds shear stress shown in figure \ref{fig:spec2D_uiuj_y15}c) at $y^+=15$ does not exhibit significant differences between APG and ZPG nor with increasing $\Rey_{\tau}$. 
The behaviour of the streamwise and spanwise components is similar, since at higher Reynolds numbers the ZPG and the APG contours grow closer, both developing a region with larger values of $\lambda_z^+$, $\lambda_t^+$ with a higher energy content.
The wall-normal Reynolds stress does not develop a region with larger scales, and the lowest-density contour grows in the direction of smaller $\lambda_z^+$, $\lambda_t^+$.
The effect of $\Rey_{\tau}=500$ not being in near-equilibrium, as opposed to the higher-$\Rey_{\tau}$ profiles at 1000, 1500 and 2000 is reflected in a different slope of the $10\%$ contour in the lower region of the premultiplied spectra for the normal Reynolds stresses.

\subsubsection{Two-dimensional power-spectral density in the overlap region}
The analysis for the overlap region will be done at $y^+=150$ as in \cite{tanarro_2020} to compare the trends of $\lambda_{z}^+=f(\lambda_{t}^+)$ previously reported for channel flows in \cite{delAlamo_jfm_2004} and for boundary layers in \cite{chandran_jfm_rapids_2017}.
\begin{figure}
\includegraphics[width=0.31\textwidth]{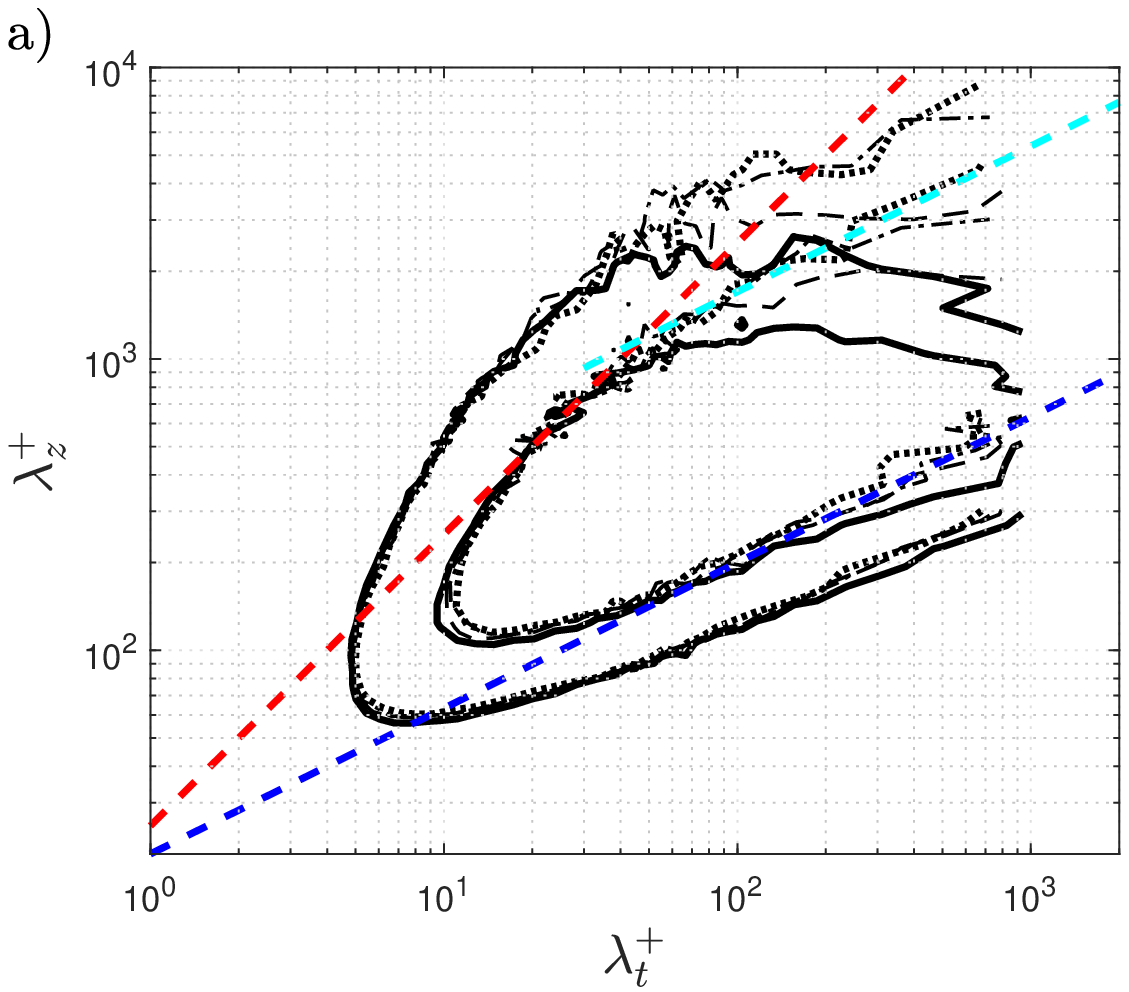}
\includegraphics[width=0.31\textwidth]{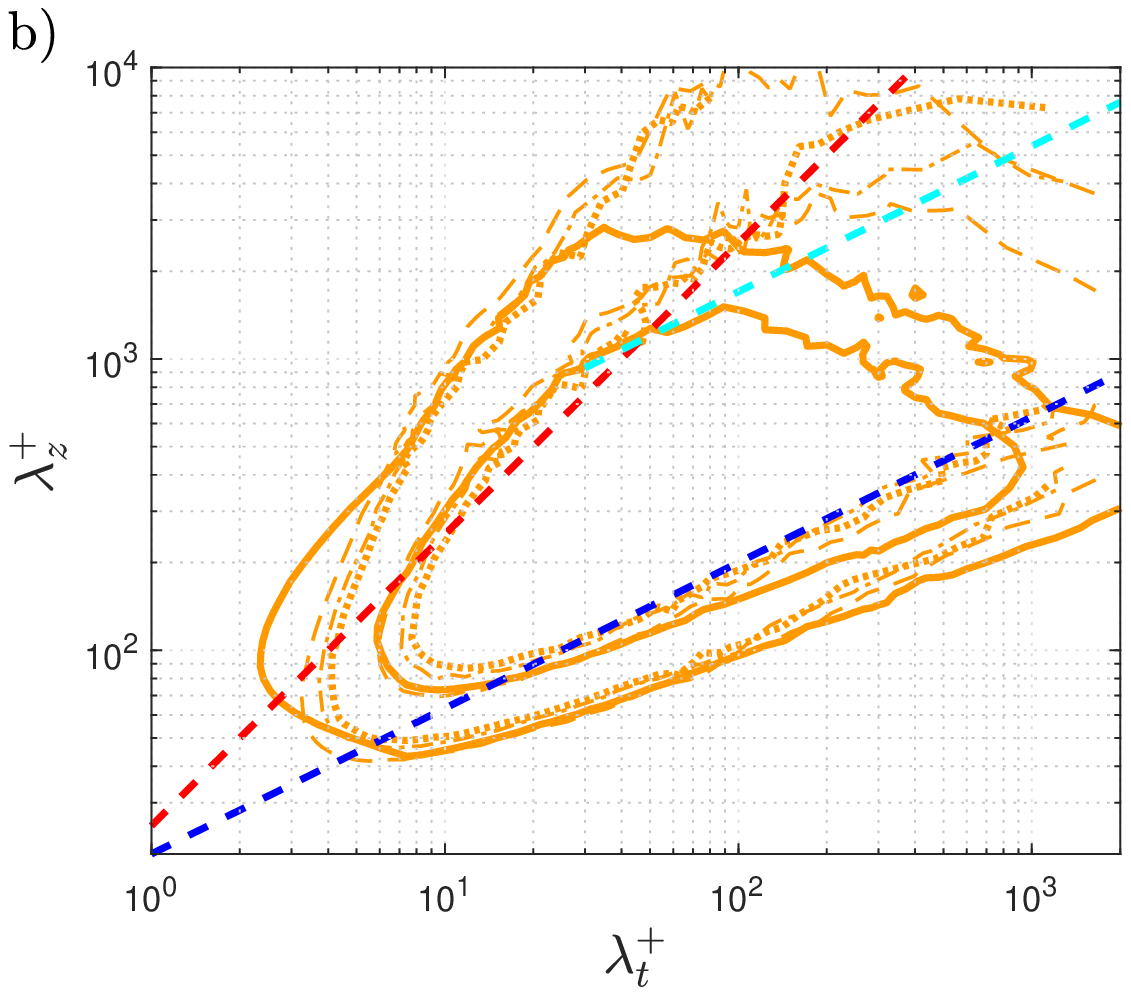}
\includegraphics[width=0.31\textwidth]{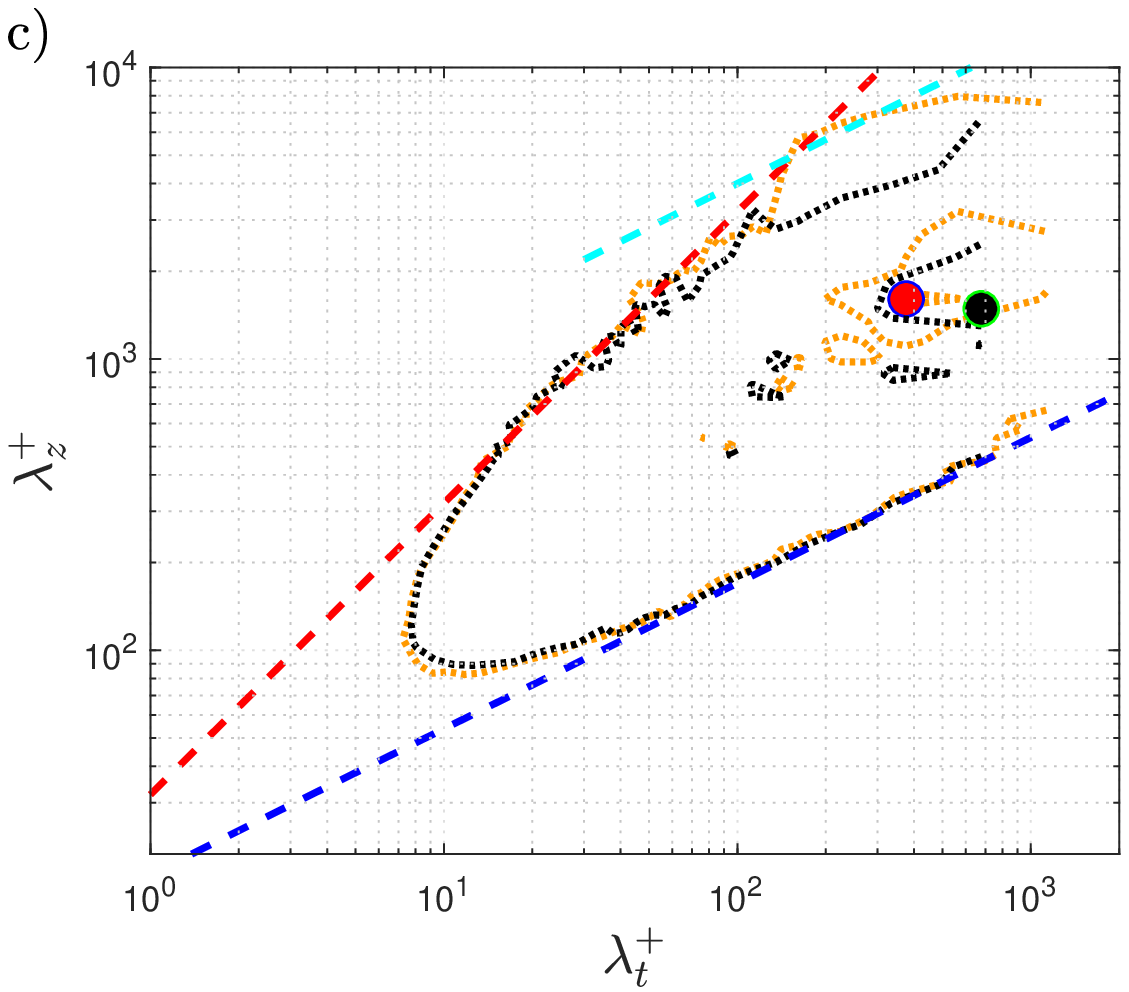}
  \caption{Two-dimensional premultiplied power-spectral density $k_z k_t |\phi_{uu}|$ at $y^+=150$. The line styles solid, dashed, dash-dotted and dotted correspond to $Re_{\tau}=500, 1000, 1500$ and 2000 respectively. a) and b) are scaled with inner units, $u_{\tau}^2$, and show the contour levels 0.05 and 0.15 of the ZPG and b1.4 respectively. In a) and b) the red, blue and cyan lines are tangent to the contour level 0.15 of the ZPG. c) Represents both ZPG and b1.4 scaled with the local maximum marked as a black dot for ZPG and as a red dot for b1.4. The contours are taken at $10\%$ and $50\%$ of the maximum value. In c) the red, blue and cyan lines are moved to be tangent to the $10\%$ contour. Blue and cyan lines follow $\lambda_z^+ \propto (\lambda_t ^{+})^{0.5} $, while red line represents $\lambda_z^+ \propto \lambda_t^+$. Colors: (\protect\blackline) for ZPG and (\protect\orangeline) for b1.4.}
\label{fig:spec2Duu_yp150}
\end{figure}
In figure \ref{fig:spec2Duu_yp150} we show for $y^+=150$ the contour levels of energy 0.05 and 0.15 for the ZPG and for b1.4. At this wall-normal location and for the energy level 0.15, it was reported in \cite{tanarro_2020} a lower bound for small time and spanwise scales following $\lambda_z^+ \propto (\lambda_t ^{+})^{0.5} $ (blue-dashed line) and an upper bound composed by the red-dashed line ($\lambda_z^+ \propto \lambda_t^+$) for shorter time scales, as well as the cyan-dashed line for longer time scales (same slope as the blue-dashed line).
For the ZPG it is possible to see that for short scales there is a good collapse for different Reynolds numbers, and the lower limit (blue line) serves as a good approximation in spite of a small tendency with higher Reynolds numbers (at higher $\lambda_t^+$) to go towards wider scales for the same time scale.
The lower-energetic level 0.05 also presents a good collapse for the different Reynolds numbers, however, the slope of the trend (0.4) is slightly smaller than the one for energy level 0.15 (0.5).
The upper part of the contours are more curved than the lower part, therefore it is harder to come up with a linear trend in a logarithmic plot (potential law). The red line approximates a short region, $\lambda_t^+ \in (20, 60)$, while the cyan line seems to be a good approximation for the larger scales at higher Reynolds numbers.
The tangent lines to the energetic contour 0.15 of the ZPG are presented also in figure \ref{fig:spec2Duu_yp150}b) as a way to compare with the same energetic contour in b1.4. It is clear that in inner-scaling, the same energy contour is bigger in the APG than in the ZPG, expanding over shorter and larger scales. There is also a bigger dispersion of the contours due to the increasing Reynolds numbers.
The lower limits exhibit a slightly different slope than the blue line, however, with an increasing Reynolds number (dotted orange line) the trend is closer to the blue-dashed line. Since this is a fixed location at $y^+=150$ (see figure \ref{fig:spec1DUU}) at higher Reynolds numbers the differences between APG and ZPG are shifted towards higher wall-normal positions and larger $\lambda_z^+$ scales.
The red line ($\lambda_z^+ \propto \lambda_t^+$) appears to be a better approximation for the upper limit that will expand over a longer region , {\it i.e.} $\lambda_t^+ \in (10, 200)$ for $\Rey_{\tau}=2000$ compared to the case of the ZPG. The cyan line also seems to be a good approximation for the larger scales.
For $\Rey_{\tau}=2000$ the location of the local maximum for APG and ZPG is very similar. If the power-spectral density is non-dimensionalized by the local maximum as in figure \ref{fig:spec2Duu_yp150}c) the $10\%$ and $50\%$ contours of the ZPG and APG cases appears to collapse, except for some extra energy in the APG in the largest $\lambda_z^+$ and $\lambda_t^+$. The trends marked by the blue, red and cyan dashed lines are a better approximation at the higher $\Rey$.

\begin{figure}
\includegraphics[width=0.49\textwidth]{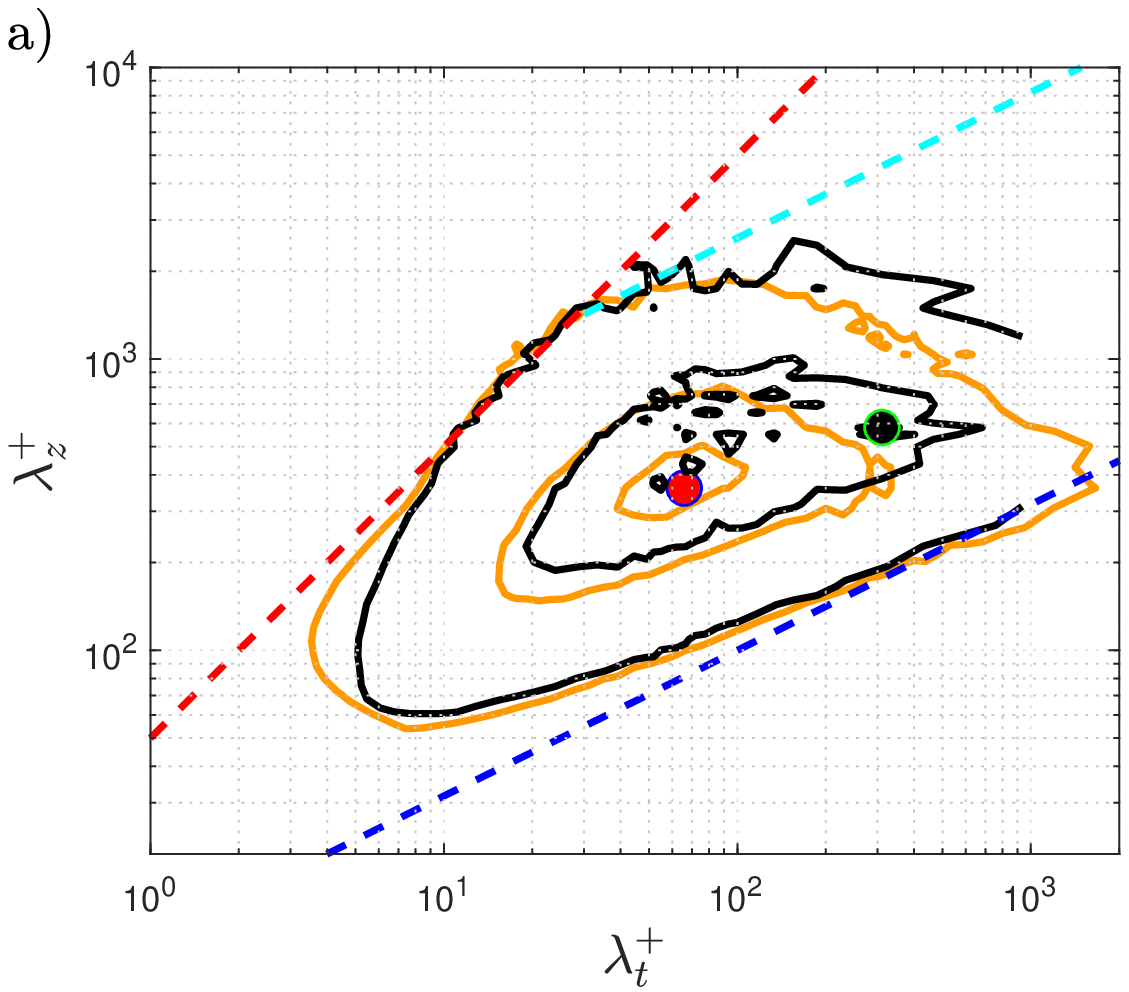}
\includegraphics[width=0.49\textwidth]{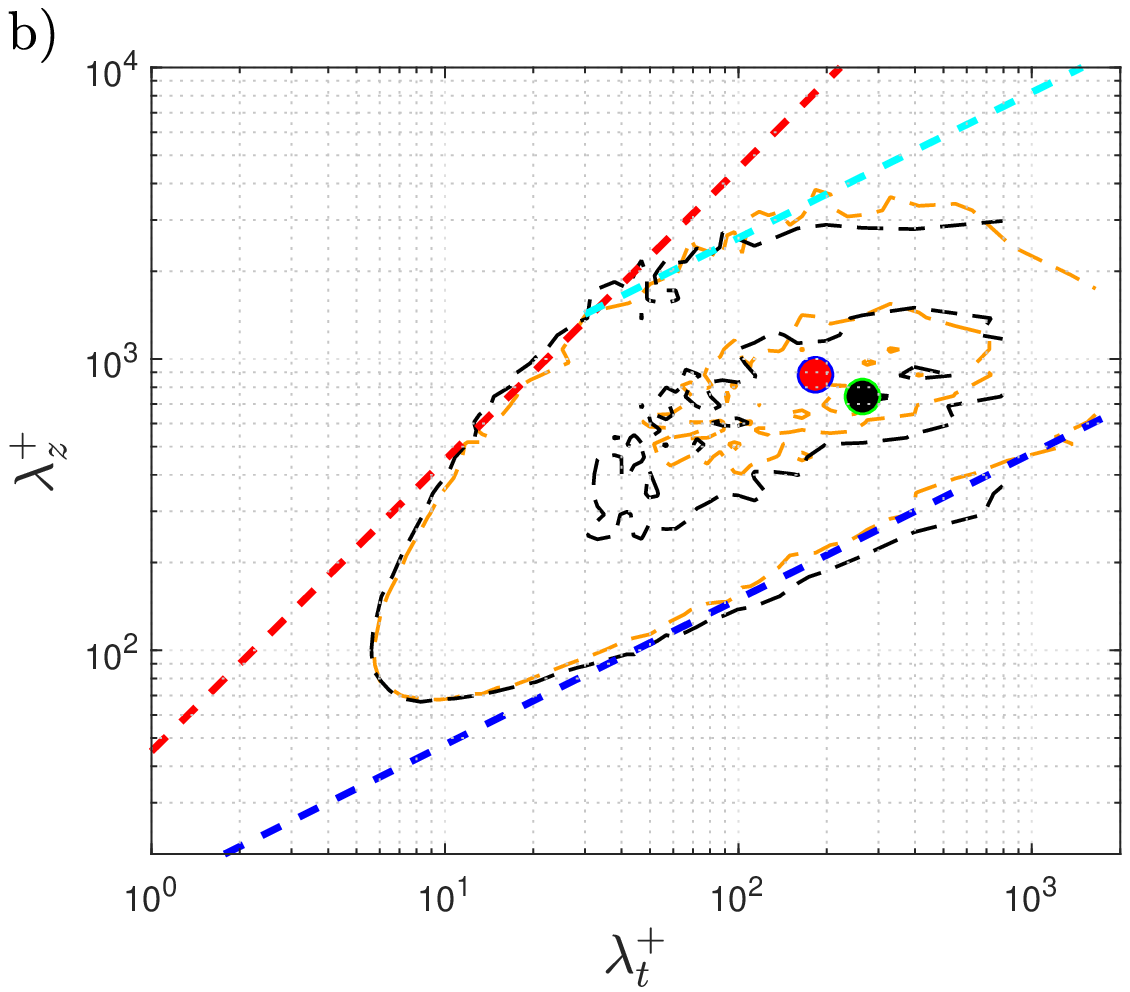}
\includegraphics[width=0.49\textwidth]{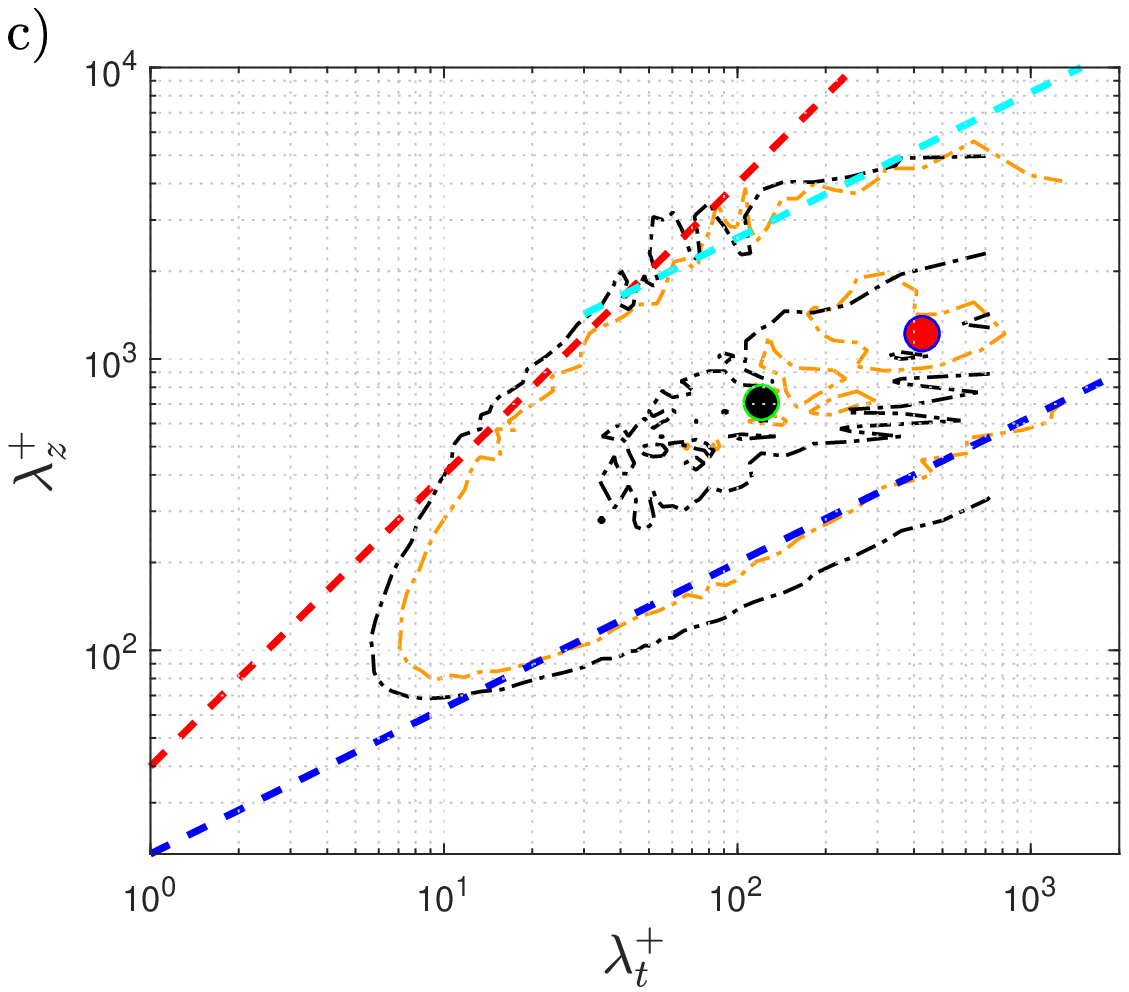}
\includegraphics[width=0.49\textwidth]{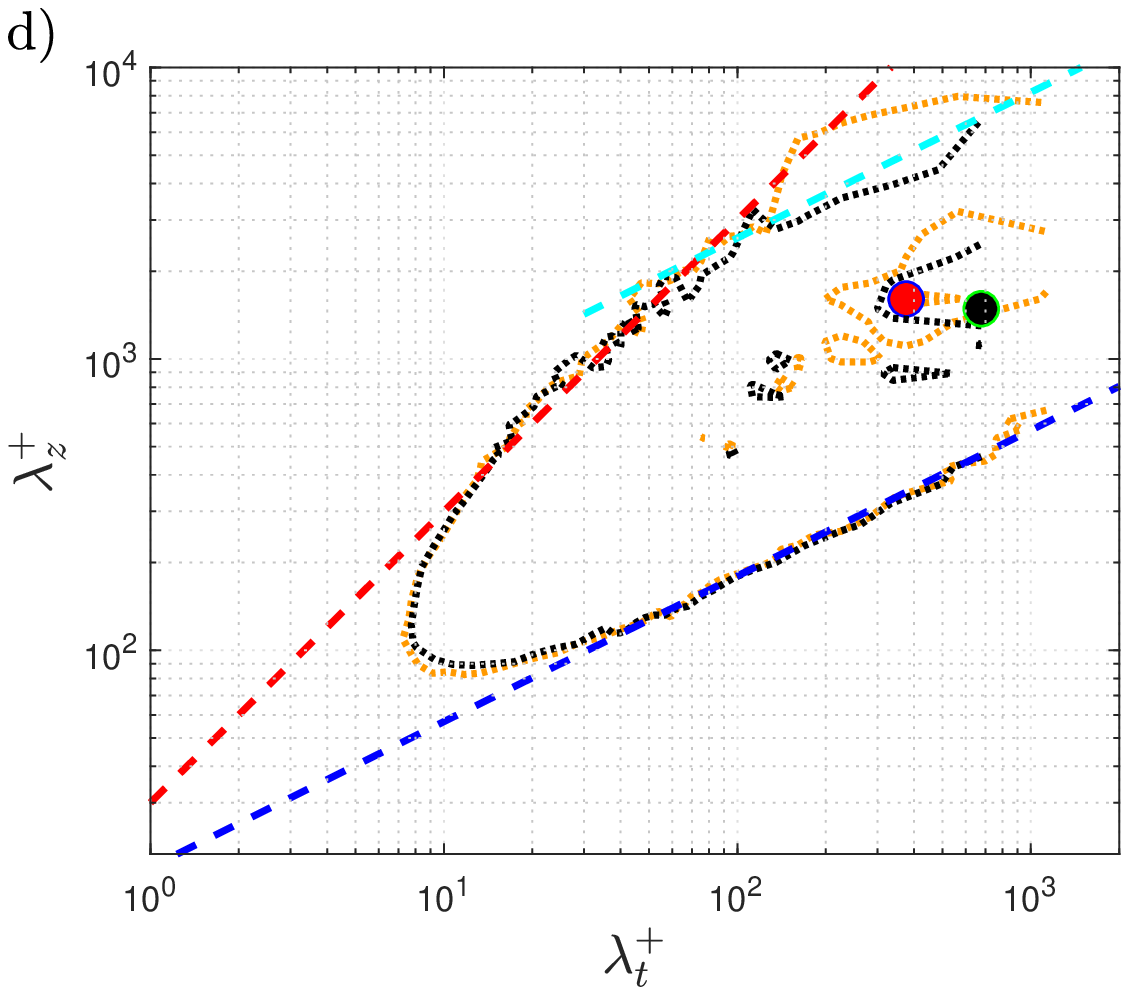}
  \caption{Two-dimensional premultiplied power-spectral density $k_z k_t |\phi_{uu}|$ at $y^+=150$ scaled with the local maximum. Contours taken at $10\%$, $50\%$, $90\%$ of the maximum value. From left to right: a) $Re_{\tau}=500$, b) $Re_{\tau}=1000$, c) $Re_{\tau}=1500$ and d) $Re_{\tau}=2000$. The dashed blue and cyan lines represent $\lambda_z^+ \approx (\lambda_t^+)^{1/2}$, the dashed red line represents $\lambda_z^+ \approx \lambda_t^+$. The red and black dots mark the position of the local maximum for b1.4 and ZPG respectively. Colors: (\protect\blackline) for ZPG and (\protect\orangeline) for b1.4.}
\label{fig:spec2Duu_yp150_max}
\end{figure}

For the lowest $\Rey_{\tau}$ in figure \ref{fig:spec2Duu_yp150_max} there is not a clear separation of scales. For each wall-normal plane, the local maximum of $k_z k_t |\phi_{uu}|$ evolves from the characteristic inner-peak wavelengths of $\lambda_z^+=100$ at $y^+=15$ towards wider scales $\lambda_z$ of the order of $\delta_{99}$ in the outer region.
At a higher Reynolds number as $\Rey_{\tau}=1000$ the location of the maxima in these premultiplied plots changes radically from scales $\lambda_z^+ \approx \mathcal{O} (100)$ and $\lambda_t^+ \approx \mathcal{O} (\lambda_z^+/2)$ to scales $\lambda_z \approx \mathcal{O}(\delta_{99})$. This change appears at $y^+\approx 23$ for b1.4 and $y^+\approx 40$ for the ZPG.  

\begin{figure}
\includegraphics[width=0.49\textwidth]{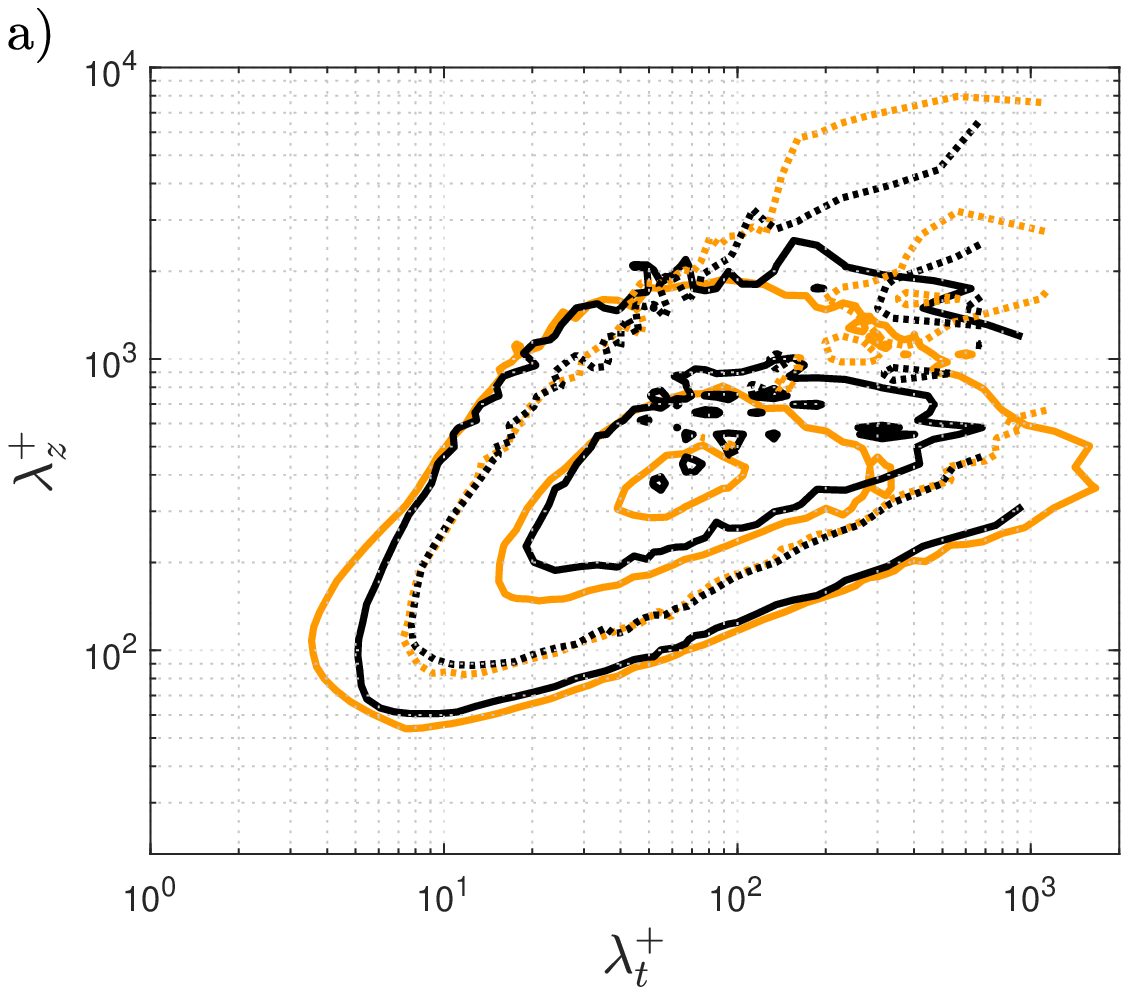}
\includegraphics[width=0.49\textwidth]{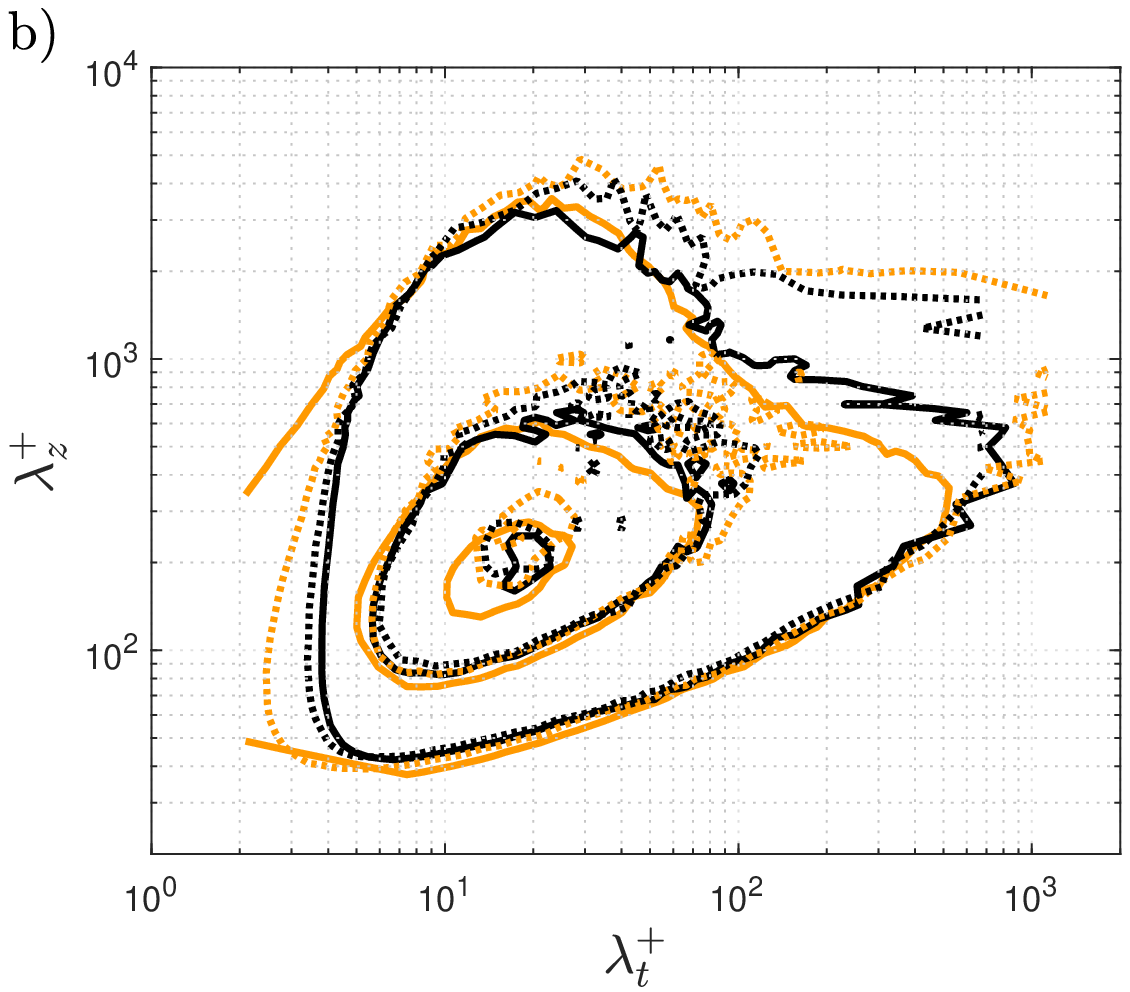}
\includegraphics[width=0.49\textwidth]{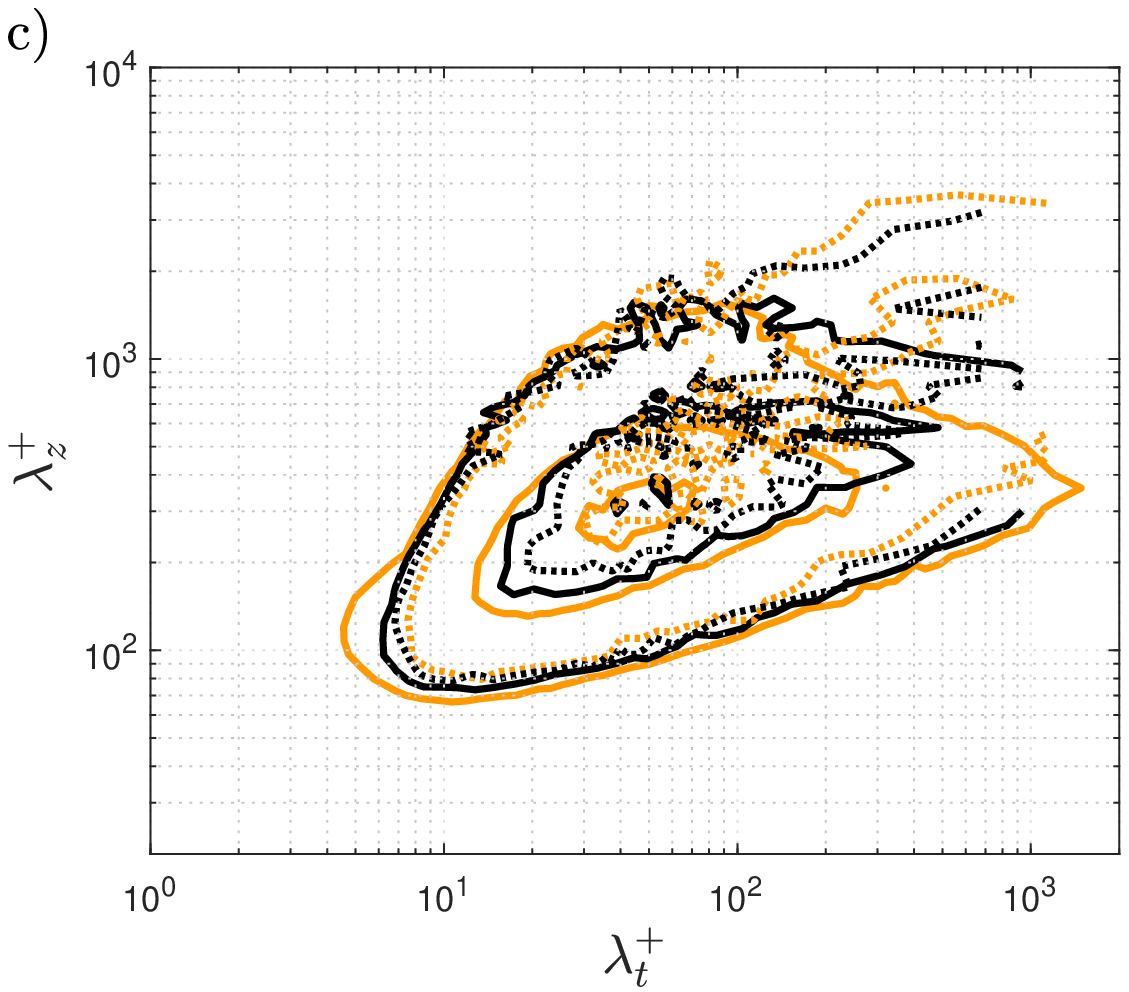}
\includegraphics[width=0.49\textwidth]{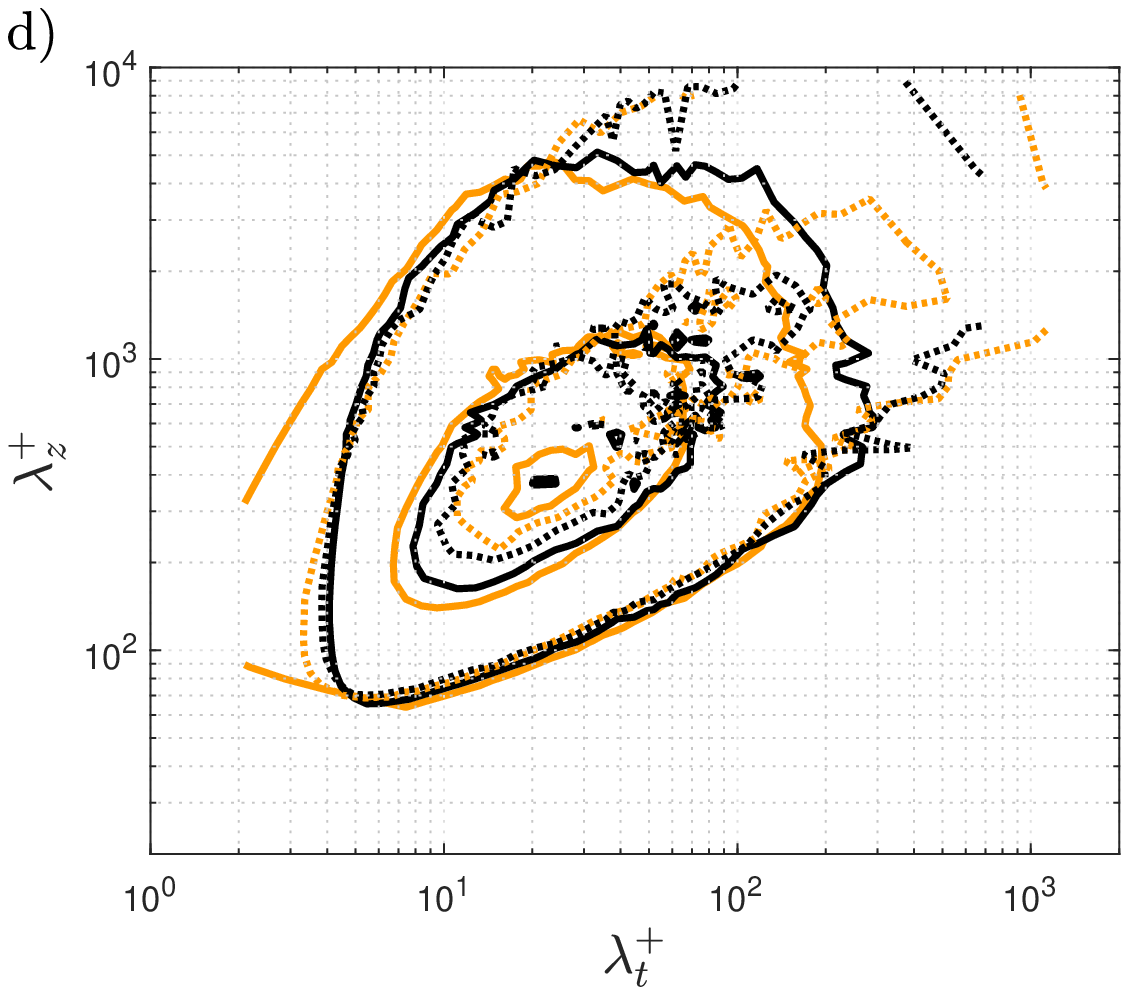}
  \caption{Evolution with the Reynolds number of the two-dimensional premultiplied power-spectral density of the Reynolds-stress components $k_z k_t |\phi_{u_iu_j}|$ at $y^+=150$ scaled with the local maximum. The panels show spectra of: a) $uu$, b) $vv$, c) $uv$, d) $ww$. Contours taken at $10\%$, $50\%$, $90\%$ of the maximum value. Solid lines for $Re_{\tau}=500$ and dotted lines for $Re_{\tau}=2000$. Colors: (\protect\blackline) for ZPG and (\protect\orangeline) for b1.4.}
\label{fig:spec2Duiuj_150}
\end{figure}

At the position $y^+=150$ the peak of $k_z k_t |\phi_{uu}(k_z,k_t)|$ is already in scales $\lambda_z \approx \delta_{99}$ as it can be seen in figure \ref{fig:spec2Duu_yp150_max}c).
The other components of the Reynolds stress tensor are shown in figure \ref{fig:spec2Duiuj_150}.
As it was seen near the wall, at the lowest $\Rey_{\tau}=500$ (solid lines), the $10\%$ contour shows an additional content of energy for b1.4 in the smaller $\lambda_z^+$ scales with lower $\lambda_t^+$. 
The APG effects are manifested as a deviation of the local peak position and as it was seen near the wall, in the presence of small scales with shorter $\lambda_t^+$ in the wall-normal component, see figure \ref{fig:spec2Duiuj_150}b).
Previously, near the wall, the spectra containing wall-normal velocities in figure \ref{fig:spec2Duiuj_150}b) and c) was practically unaffected by an increase in the Reynolds number. On the other hand, in the outer region, both components develop larger spatio-temporal scales when the Reynolds number is increased. 
As observed in figure \ref{fig:spec2Duu_yp150_max}d) (and also in figure \ref{fig:spec2Duu_yp150}c)), the red-dashed line which approximates the contours for scales up to $\lambda_t^+ =100$ in the ZPG, reaches scales up to $\lambda_t^+ =200$ in the APG \citep{chandran_jfm_rapids_2017}.
The extra energy in the small scales with lower $\lambda_t^+$ is only seen in the lower $\Rey_{\tau}=500$ profiles of b1.4; for $\Rey_{\tau}=1000$, 1500 and 2000 that region is exhibits collapse.
In the 2D spectra shown in \cite{tanarro_2020}, the profiles were taken at different levels of the premultiplied spectral density scaled either in viscous or outer units. They reported that at the same contour level the APG effects extend over a wider range of scales compared to the ZPG contours at the same low $\Rey_{\tau}=305$. However, note that in their case the Reynolds number was low, the APG very strong and their boundary layers were not in near-equilibrium, exhibiting a rapid change in the $\beta(x)$ curve. Even if they achieved a value of $\beta$ larger than in the b1.4 simulation and an increase of the energy in the outer regions of the $\overline{u^2}^+$ TKE production profiles, the separation of scales in their study was not enough to observe the effects exhibited by the b1.4 simulation.

\section{Summary and conclusions} \label{sec:Conclusions}

A new well-resolved LES of a TBL developing on a flat plate subjected to an adverse pressure gradient is presented. The relevance of this simulation lies in the high Reynolds number we achieve starting from a laminar flow under similar conditions to those in experiments, and obtaining a long region where the TBL is in near-equilibrium. 
To the authors’ knowledge, this is the first TBL under approximately-constant APG magnitude  ($\beta\approx 1.4$) over a long near-equilibrium region up to $\Rey_{\theta}=8700$. The characteristics of this simulation have enabled a direct comparison with experimental data in a similar range of $\beta$ and $\Rey_{\tau}$, as well as the comparison with other near-equilibrium databases at lower Reynolds number as discussed below. 
The data obtained from the simulation consists of two-dimensional turbulence statistics for all the grid points in the $xy$ plane, together with time series and two-point correlations at 20 streamwise locations (containing all the spanwise grid points), with 10 of those profiles in the near-equilibrium ROI.

The turbulence statistics were compared with lower-Reynolds-number simulations of different APGs \citep{bobke2017}, with another high-Reynolds-number well-resolved LES of a ZPG \citep{E-AmorZPG} and also with high-Reynolds-number experiments \citep{MTL_expSANMIGUEL} with a similar $\beta(x)$ development. 
These comparisons highlight the need of more simulations with long near-equilibrium regions to be able to distinguish the effects of the APG and the effects of the Reynolds number. The near-equilibrium features have been analysed with the Rotta-Clauser pressure-gradient parameter $\beta$ and the parameter $\Lambda_{\rm inc}$ for different sets of velocity and length scales, {\it i.e.} the edge and the Zagarola--Smits scalings. Near-equilibrium conditions were obtained in the region from $\Rey_{\tau}=800$ to around 2000. 
The results were also compared to another constant-$\beta$ database \citep{Kitsios2016}, and we showed that any self-similarity analysis has to be performed along regions of near-equilibrium at high Reynolds numbers  to be able to study the collapse of the different regions in the TBL.
For the APGs under study here, the viscous scaling collapses the near-wall region for the streamwise mean velocity and although the magnitude of the near-wall peak of the streamwise RS increases in inner scaling, its location remains close to $y^+ \simeq 15$.
Furthermore, the outer scaling shows a good collapse of the outer region.

The large scales associated with the APG have an effect in the inner and outer regions, similarly to the large scales present in the flow at high Reynolds numbers in simpler geometries, such as channel flow \citep{Hoyas_PoF2006}. This was assessed for the various terms of the Reynolds-stress tensor using the spanwise one-dimensional power-spectral density and the two-dimensional power-spectral density in spanwise spatial scales and the temporal scales. 
This study shows that the displacement of small scales with $\lambda_z^+ \simeq 100$ from the inner to the outer region, documented in APGs at low Reynolds numbers by \cite{tanarro_2020, VINUESA2018}, is also observed in high-$\Rey$ APGs.
The present analysis at higher Reynolds numbers shows that there is an energization associated not only with the small scales but also with longer spatial and temporal scales in the streamwise and spanwise Reynolds stresses near the wall and in the overlap region.
The current database enables spatio-temporal analysis over a wide range of Reynolds numbers in near-equilibrium conditions due to the long near-equilibrium region at high $\Rey$. 
This type of high-quality APG TBL in near equilibrium is necessary to deepen our insight into the similar but different effects of Reynolds number and APG, and to further understand the role of flow history on the local features of wall-bounded turbulence

\section*{Declaration of Interests}
The authors report no conflict of interest.

\section*{Acknowledgements}
RV acknowledges the financial support provided by the Swedish Research Council (VR).
The computations and data handling were enabled by resources provided by the Swedish National Infrastructure for Computing (SNIC), partially funded by the Swedish Research Council.

% \clearpage
\FloatBarrier

% \begin{appendices}
\section*{  Appendix \hypertarget{AppA}{A}} \label{sec:AppendixA}

For the sake of completeness and documentation, the usual time- and spanwise-averaged statistics are shown for the b1.4 simulation. They are compared with those of the ZPG case to analyse the APG effects for a wide range of Reynolds numbers. The lower-$\Rey$ APG simulations b1 and b2 are used to compare the effects of different APG intensities at low Reynolds numbers.

% -------- Retau & Retheta --------------------------------
\begin{figure}
\includegraphics[width=0.49\textwidth]{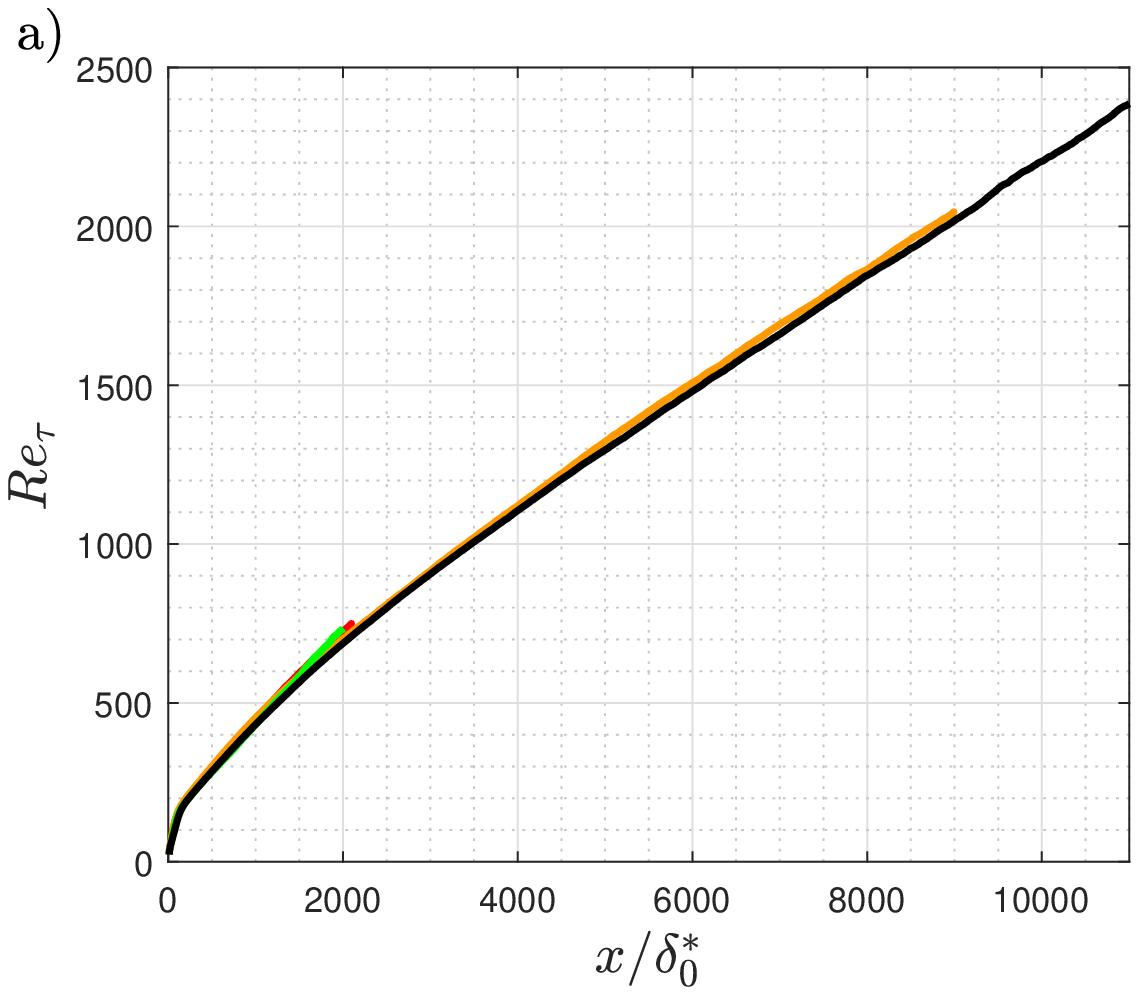}
\includegraphics[width=0.49\textwidth]{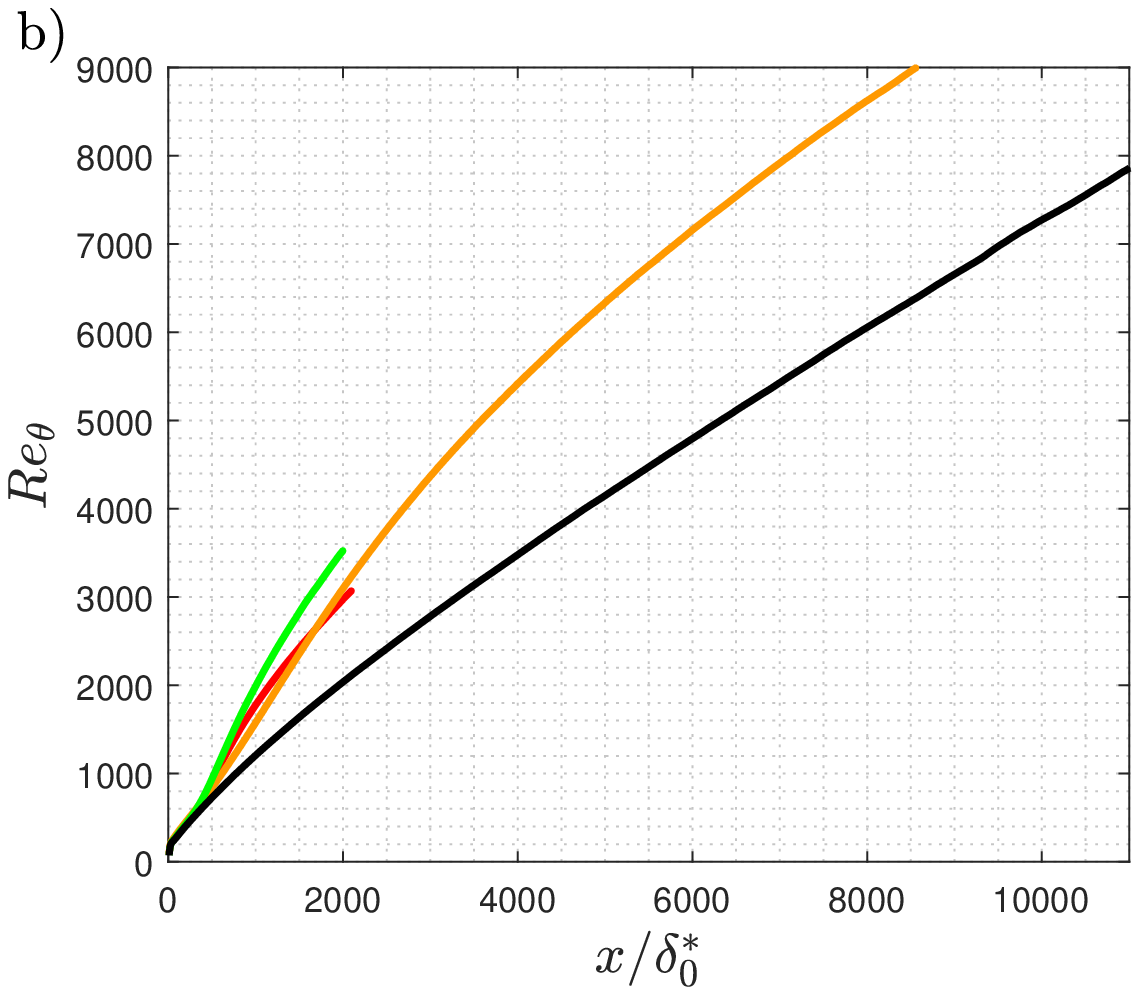}
  \caption{Streamwise development of the a) friction Reynolds number $Re_{\tau}$ and b) Reynolds number based on momentum thickness $Re_{\theta}$ as a function of the streamwise coordinate $x/\delta^{*}_{0}$.  Colors: (\protect\blackline) ZPG; (\protect\orangeline) b1.4; (\protect\redline) b1; (\protect\greenline) b2.}
%   Colors as in table \ref{tab:param}.}
\label{fig:RetauRetheta}
\end{figure}

In figure \ref{fig:RetauRetheta} we show the development of the Reynolds number based on friction velocity ($\Rey_{\tau}=u_{\tau}\delta_{99}/\nu$) and momentum thickness ($\Rey_{\theta}=U_{e} \theta / \nu$) along the streamwise coordinate $x/\delta_0^*$.
The evolution of $\Rey_{\tau}$, which corresponds to the development of the boundary-layer thickness in viscous units $\delta_{99}^+$, is very similar for all the simulations. 
The effects of the APG are more noticeable in the evolution of $\Rey_{\theta}$ where the APGs starts to diverge from the ZPG around $x/\delta_0^*\approx 200$ ($\Rey_{\theta}\approx 400$) and among the APGs the divergence is seen at $x/\delta_0^*\approx 400$ ($\Rey_{\theta}\approx 700$). 

%   --------Cf & H12----------------------------------------
 \begin{figure}
\includegraphics[width=0.49\textwidth]{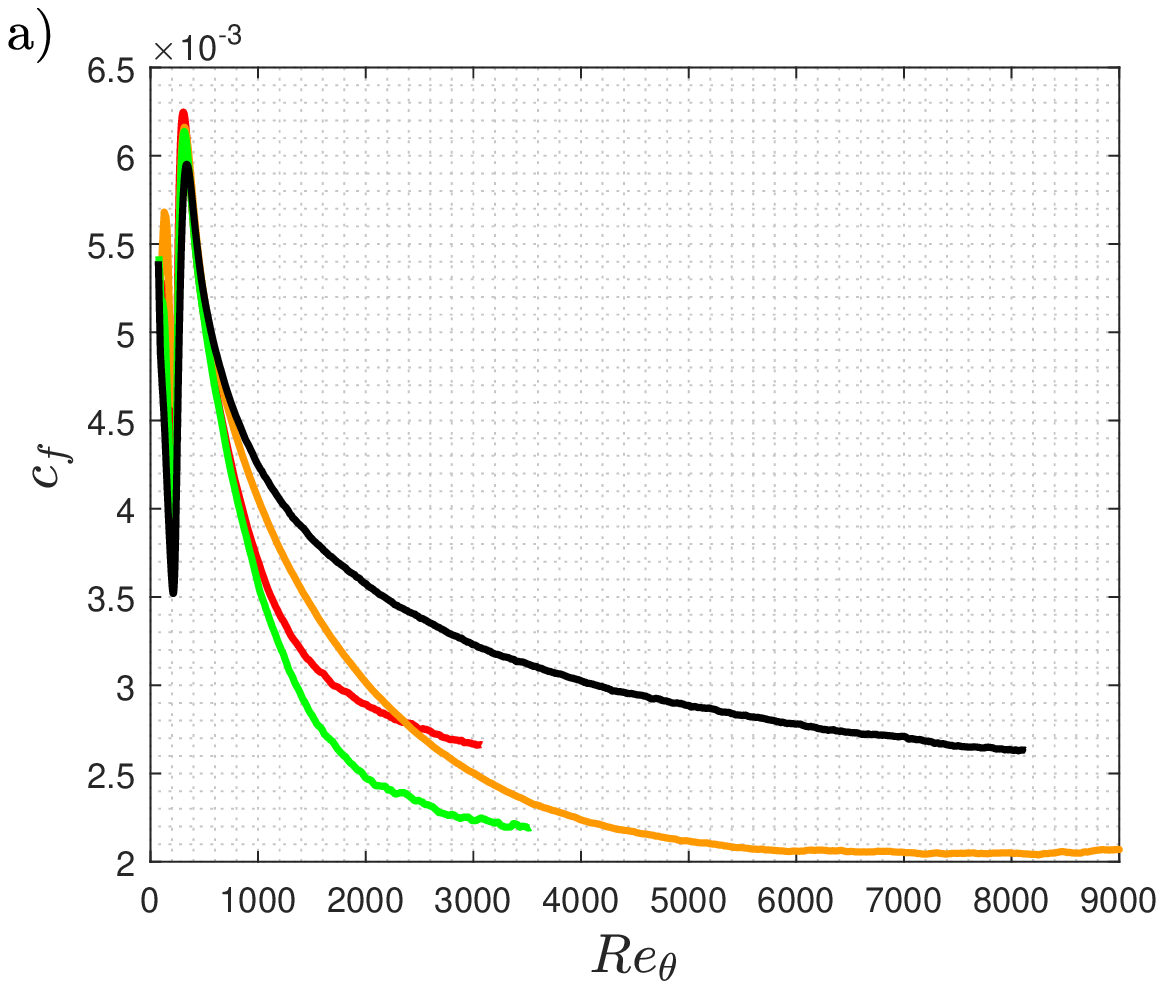}
\includegraphics[width=0.49\textwidth]{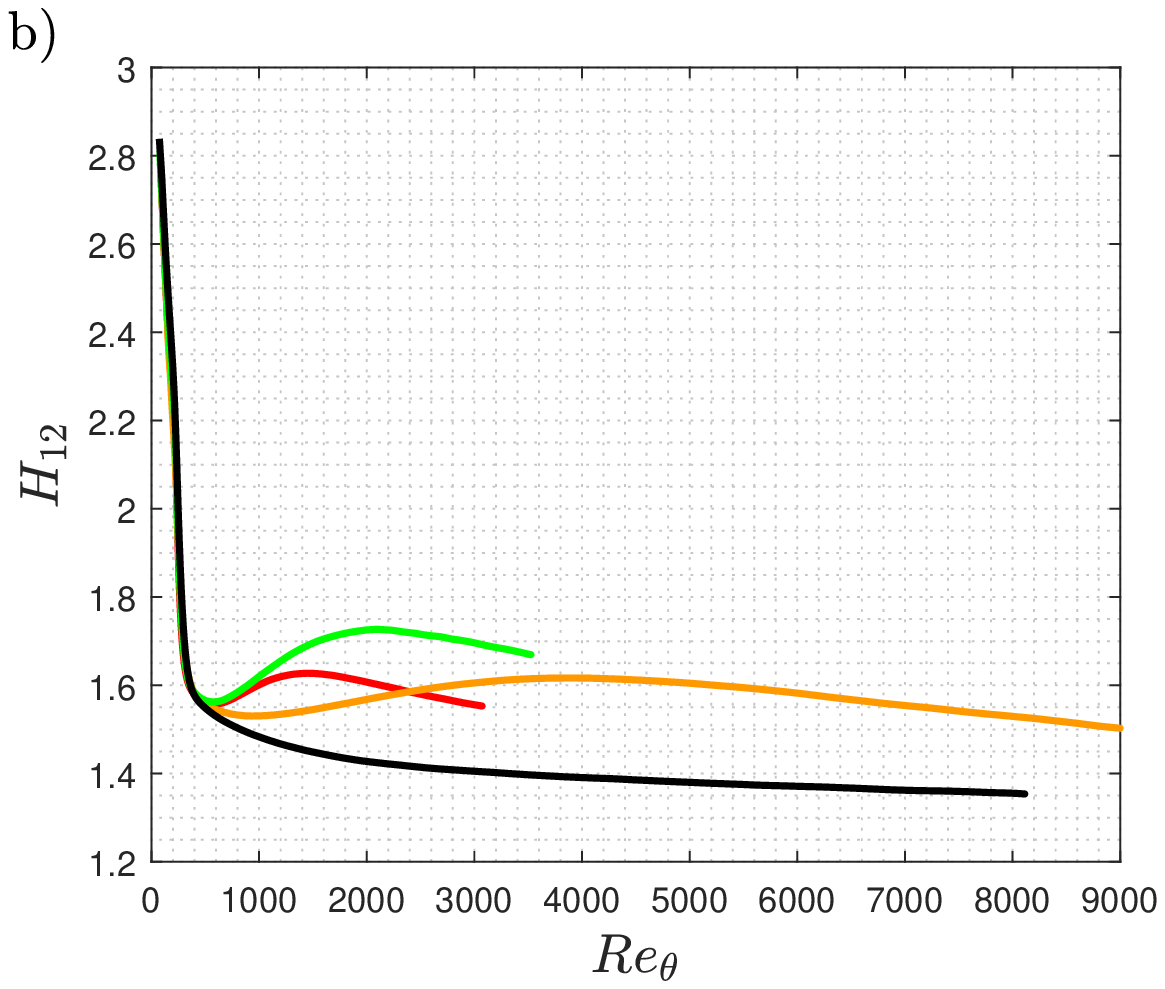}
  \caption{Evolution of a) the skin-friction coefficient $c_f$ and b) the shape factor $H_{12}$ as a function of the momentum-thickness-based Reynolds number $Re_{\theta}$.  Colors: (\protect\blackline) ZPG; (\protect\orangeline) b1.4; (\protect\redline) b1; (\protect\greenline) b2.}
%   Colors as in table \ref{tab:param}.}
\label{fig:cfH12}
\end{figure}
In figure \ref{fig:cfH12} we show the skin-friction coefficient $c_f=2(u_\tau/U_{e})^2$ and the shape factor $H_{12}$ as a function of $\Rey_{\theta}$ (plotted against $\Rey_{\tau}$ would show similar trends). For all the simulations the data was trimmed close to the fringe region, where there was a clear growing tendency in $c_f$. The APG simulations by \cite{bobke2017} and the ZPG case exhibit a decreasing trend in $c_f$ for increasing $\Rey$, while the b1.4 simulation suggest an asymptotic behaviour of $c_f$ for growing $\Rey$, where the value of the asymptote may be set by the PG magnitude and the flow history.
The APGs exhibit local minima in $H_{12}$ after the transition region ($\Rey_{\theta} \approx 600$), and local maxima located very close to the respective maxima in $\beta$. Note that this behavior is not observed for ZPGs.

 \subsection*{ Statistics in the wall-normal direction.}
%  ----------------- Mean velocity and RS ------------------------------------------
\begin{figure}
\includegraphics[width=0.49\textwidth]{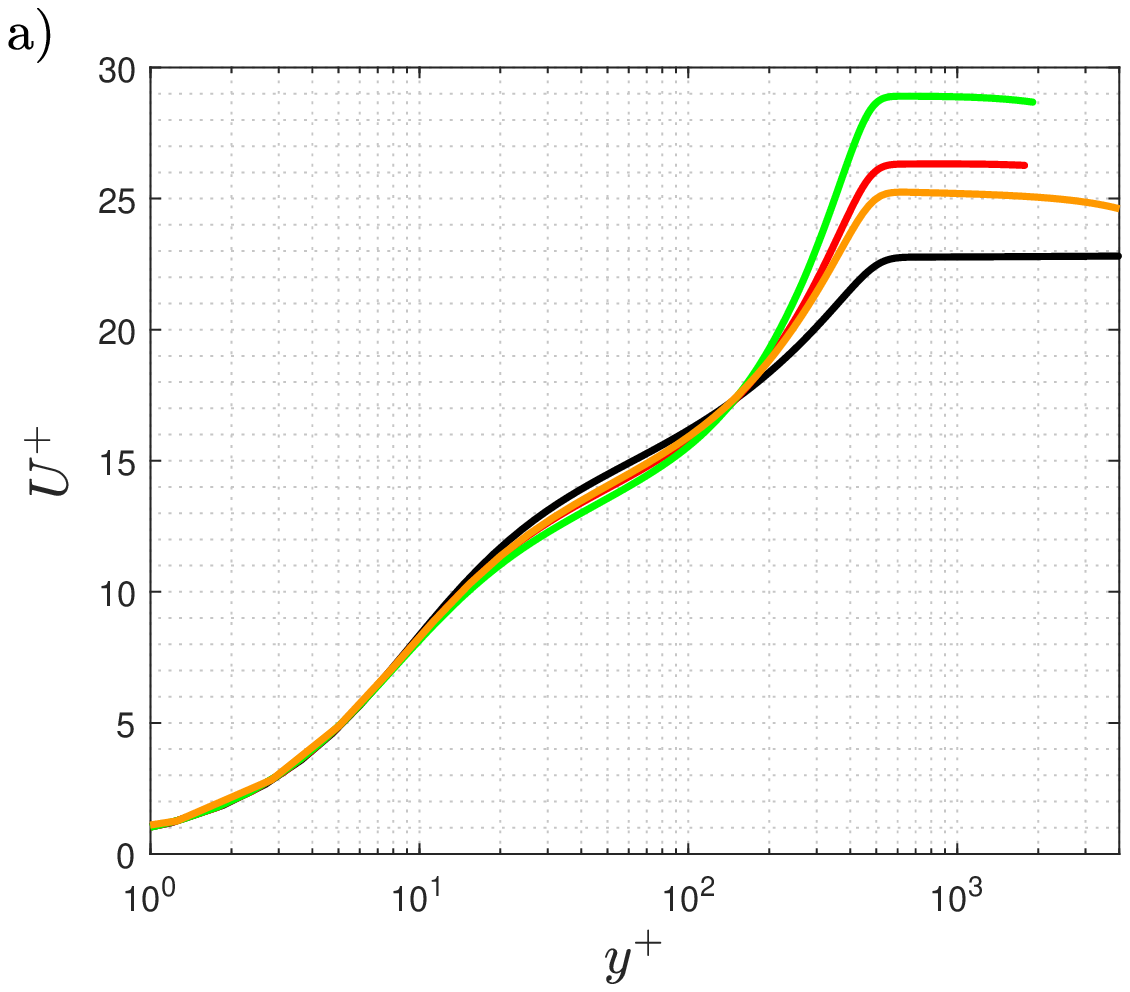}
\includegraphics[width=0.49\textwidth]{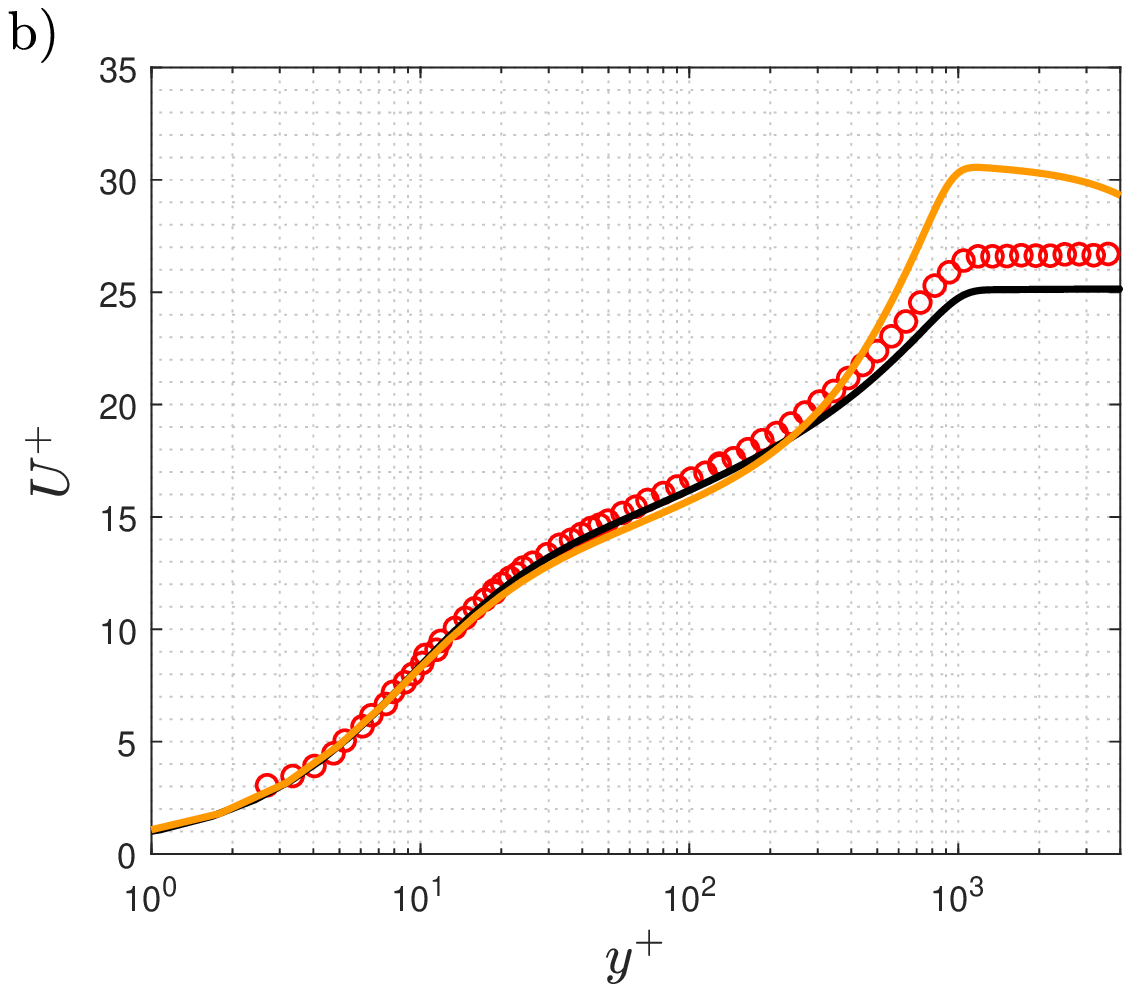}
\includegraphics[width=0.49\textwidth]{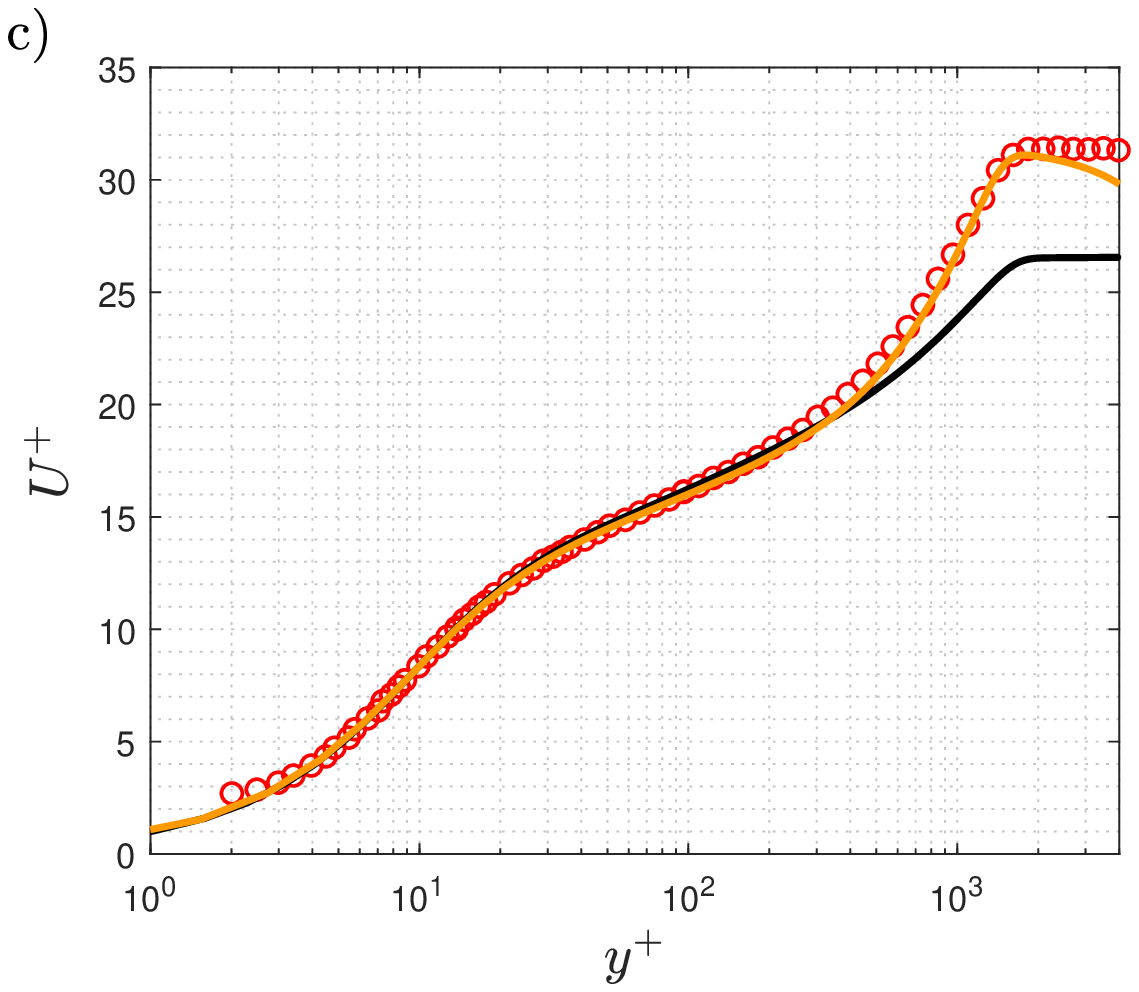}
\includegraphics[width=0.49\textwidth]{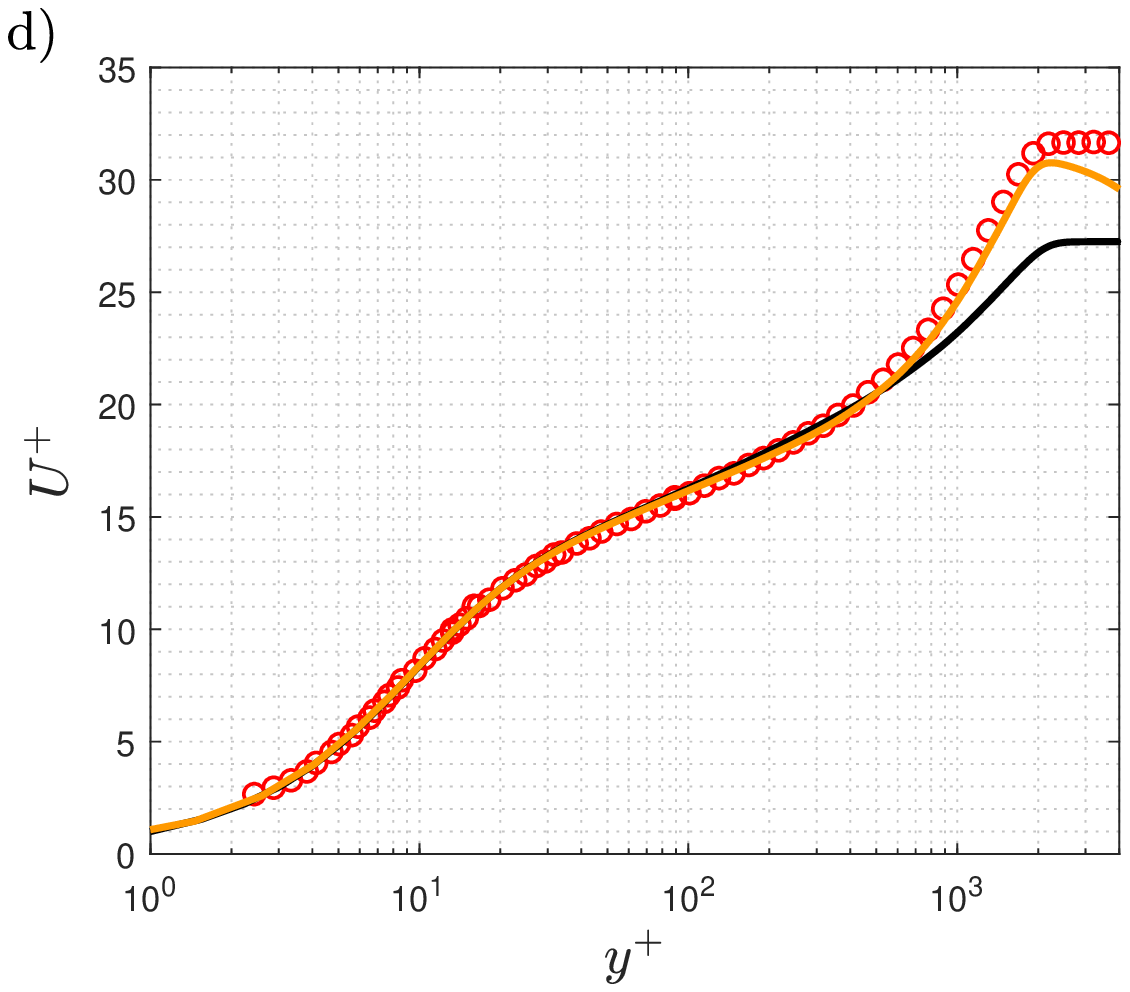}
  \caption{ Inner-scaled streamwise mean velocity at different friction Reynolds numbers: a) $Re_{\tau}=500$ where $\beta(Re_{\tau})$ intersects for the simulations b1 and b1.4 ; b) $Re_{\tau}=1004$ ; c) $Re_{\tau}=1586$ ; d) $Re_{\tau}=2049$. Colors and symbols: (\protect\blackline) ZPG; (\protect\redline) b1; (\protect\orangeline) b1.4; (\protect\greenline) b2; (\protect\redcircle) exp; as in table \ref{tab:param}. }
\label{fig:meanU}
\end{figure}
 The mean streamwise velocity $U$ is represented in viscous scaling in figure \ref{fig:meanU}. It is possible to see the collapse for all the simulations for $y^+ \leq 10$. For the lowest $\Rey_{\tau}=500$, there is a difference in the buffer and logarithmic region, where an increasing APG is reflected in a lower $U^+$. Although b1 and b1.4 have the same $\beta \simeq 1.2$ at this $\Rey_{\tau}$, the b1.4 case has been exposed to lower values of $\beta$ in the upstream region than the b1 case. This is manifested in the buffer region, which exhibits a lower accumulated PG effect with its values being closer to the ZPG than those of the b1 APG. If we compared both curves at $\Rey_{\tau}=587$ where $c_f$ is the same for b1 and b1.4, we would be matching $U_e^+$ and both curves would show a better collapse; note that at this position the PG of b1.4 is higher and in the buffer region we could see that b1 is closer to the ZPG. This is also an example that even matching local $\Rey_{\tau}$ and $\beta$ does not imply a collapse of profiles as in \cite{tanarro_2020}, and the effects of the flow history need to be taken into account, not only the local states. At around $y^+=200$ the overlap region ends at these Reynolds numbers, and the curves diverge in the wake region, showing different values of $U_e^+$ since the $c_f$ are different.
Increasing the Reynolds number we can see a better collapse of the b1.4 and ZPG simulations along the inner region including the overlap layer. The APG effects are then confined to the wake region for the mean streamwise velocity.

% Reynolds stress in outer units
\begin{figure}
\includegraphics[width=0.49\textwidth]{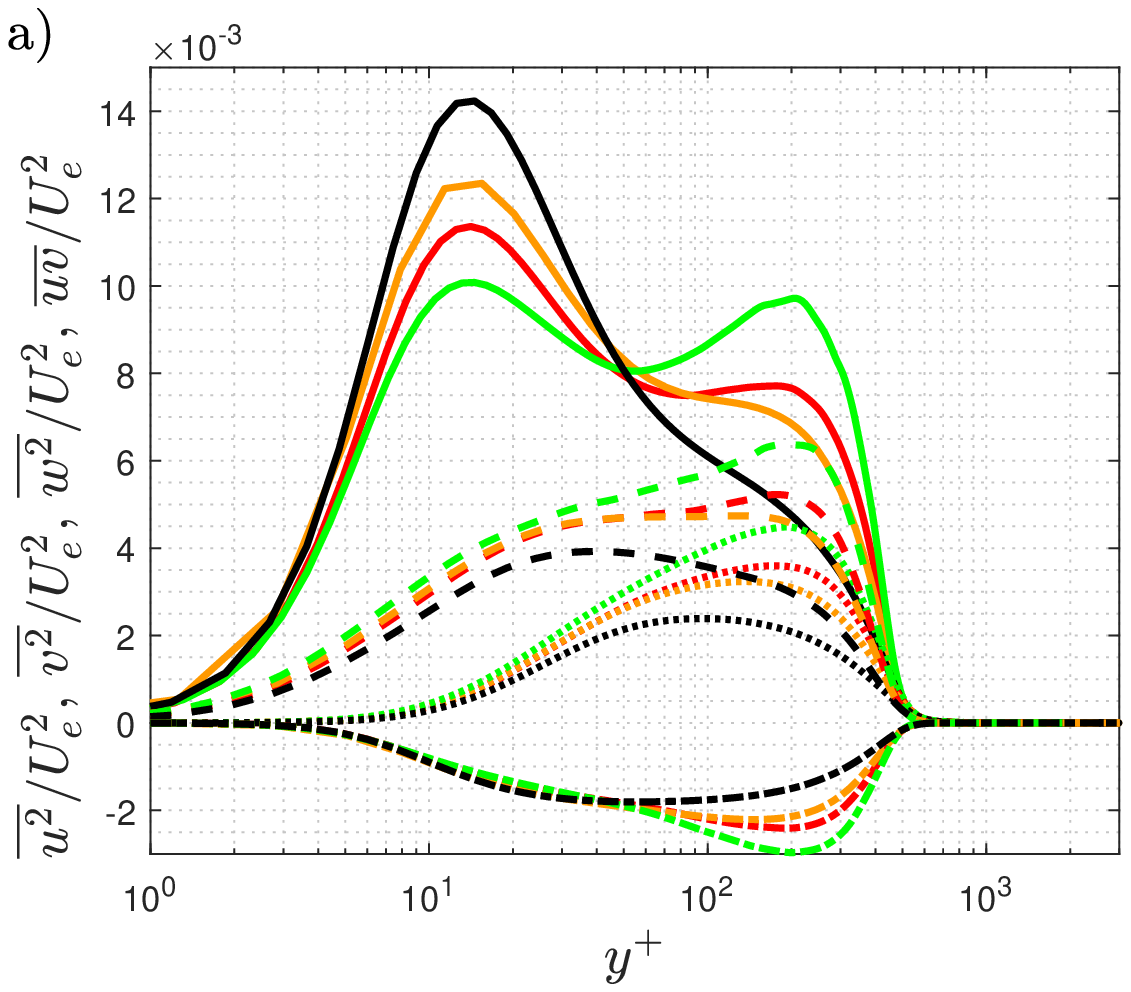}
\includegraphics[width=0.49\textwidth]{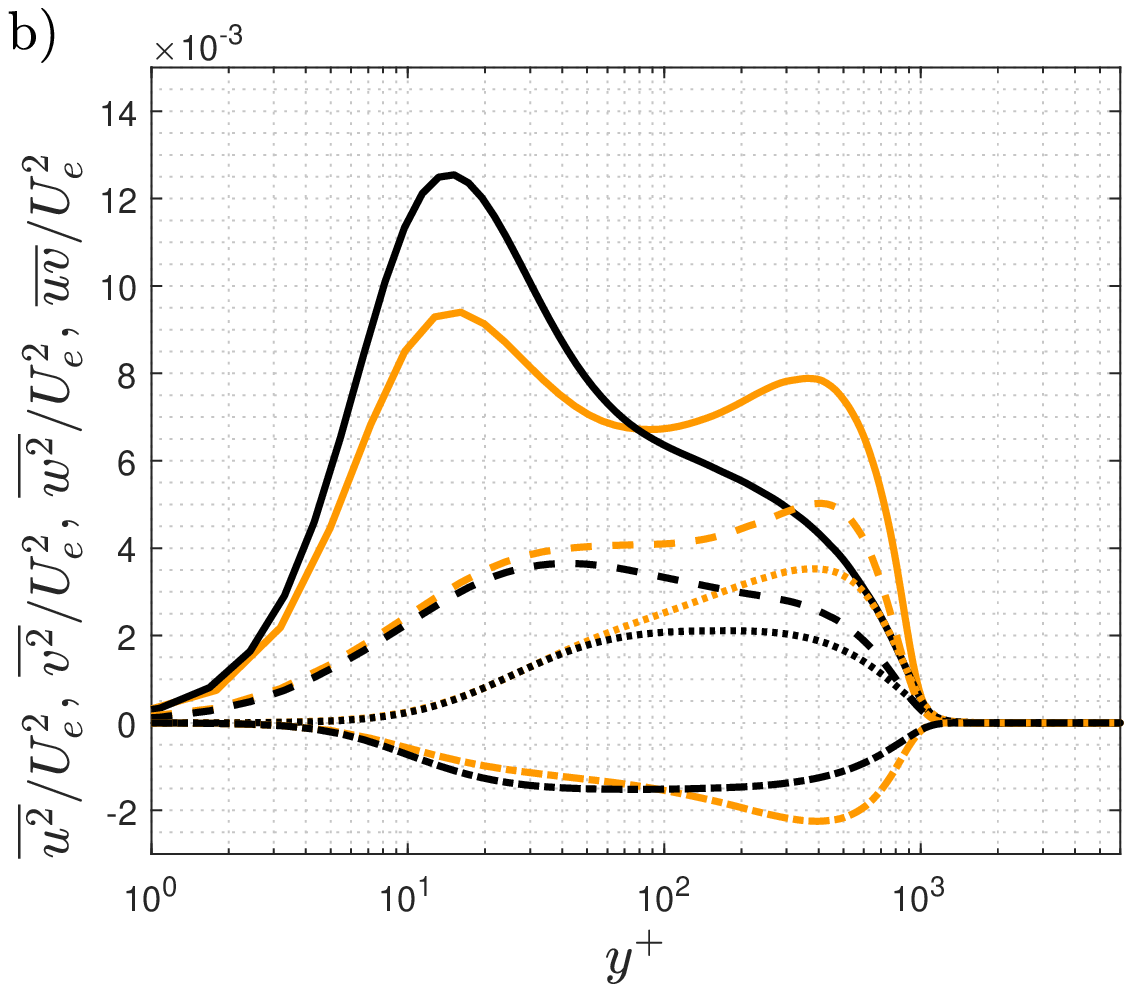}
\includegraphics[width=0.49\textwidth]{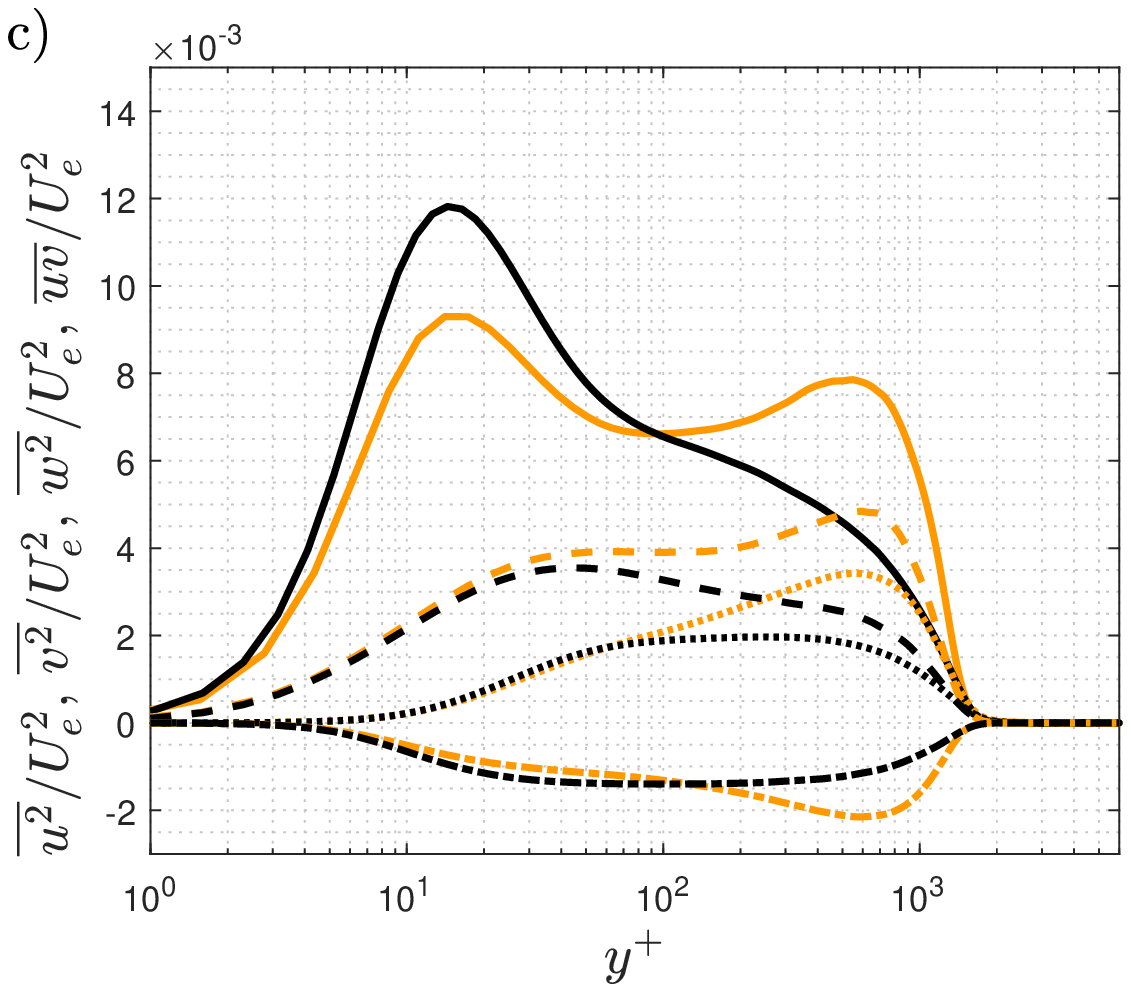}
\includegraphics[width=0.49\textwidth]{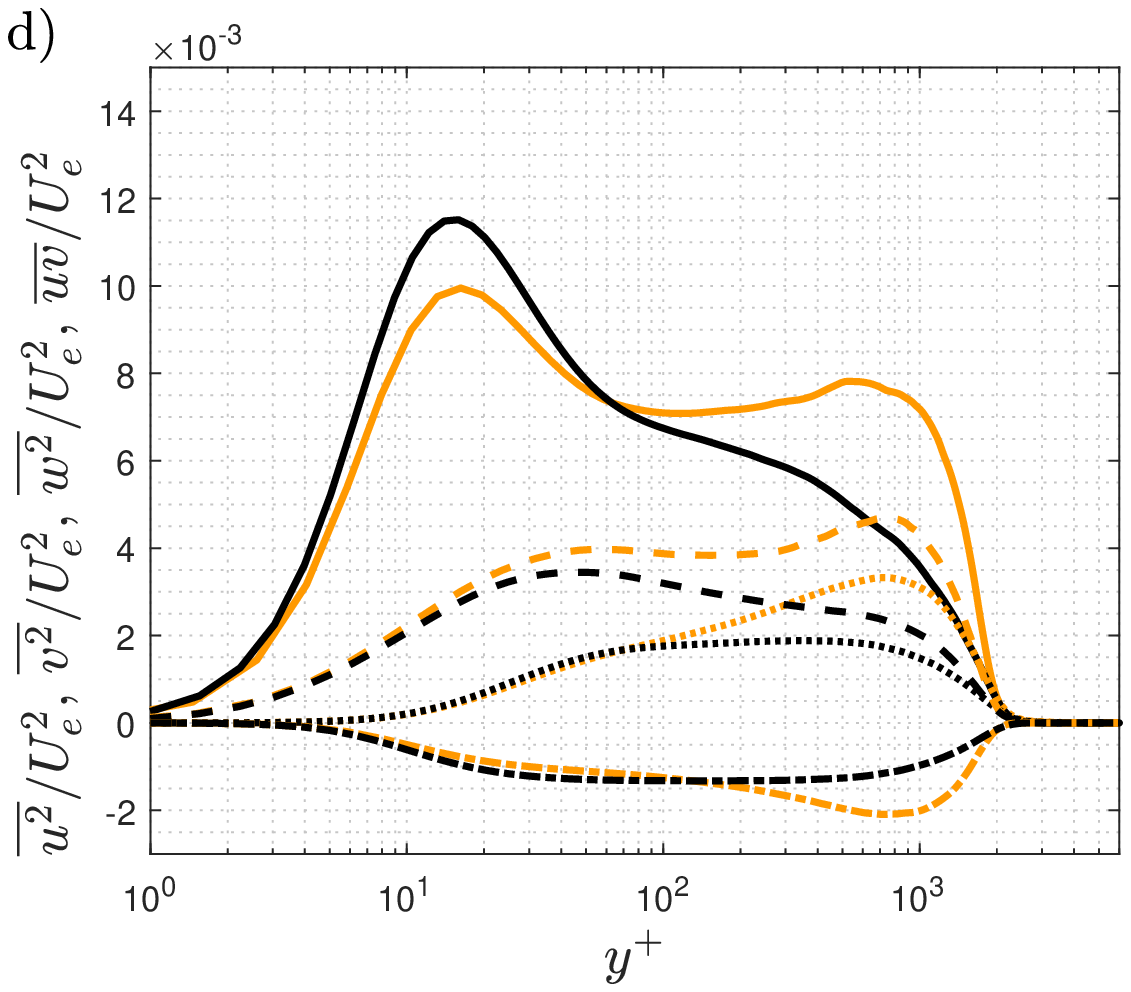}
  \caption{ Reynolds stresses scaled with the edge velocity $U_e$ at various matched $Re_{\tau}$: a) $Re_{\tau}=500$ where $\beta(Re_{\tau})$ intersects for the simulations b1 and b1.4; b) $Re_{\tau}=1000$ ; c) $Re_{\tau}=1500$ ; d) $Re_{\tau}=2000$. Symbols: (\protect\blackline) $\overline{u^2}/U_e^2$; (\protect\blackdotted) $\overline{v^2}/U_e^2$; (\protect\blackdash) $\overline{w^2}/U_e^2$; (\protect\blackdashdot) $\overline{uv}/U_e^2$.  Colors as in table \ref{tab:param}. Colors and symbols: (\protect\blackline) ZPG; (\protect\redline) b1; (\protect\orangeline) b1.4; (\protect\greenline) b2; (\protect\redcircle) exp; as in table \ref{tab:param}. }
\label{fig:RSouter}
\end{figure}

\subsection*{  TKE-budget equations}
% ----------------------------------  TKE --------------------------
The transport equation of the TKE, defined as $\overline{k}=1/2( \overline{u^2} + \overline{v^2} + \overline{w^2} ) $, is decomposed into the following terms:

\begin{equation*}
     \frac{\partial }{\partial t} \overline{k} = P^k + \varepsilon^k + D^k + T^k  + \Pi^k + C^k ,
\label{eq:tke_bud}
\end{equation*}
 where the production term is computed as $P^k= - \overline{u_i u_j}({\partial U_i}/{\partial x_j})$, the dissipation as $\varepsilon^{k} = -\nu \overline{ ({\partial u_i}/{\partial x_j})^2}$ or $-2\nu (\overline{s_{ij}s_{ij}})$, (where $s_{ij}$ is the fluctuating strain rate), the viscous diffusion is defined as $D^k=(\nu/2)({\partial^2 \overline{u_iu_i}}/{\partial x^2_j})$, the velocity-pressure-gradient correlation $\Pi^k = -(1/\rho) ({\partial \overline{pu_i}}/{\partial x_i})$, the turbulent transport $T^k=-(1/2){\partial \overline{u_iu_iu_j}}/{\partial x_j})$ and the convection is $C^k=-(1/2)U_j({\partial \overline{u_iu_i}}/{\partial x_j})$.

% TKE budget in inner units
\begin{figure}
\includegraphics[width=0.49\textwidth]{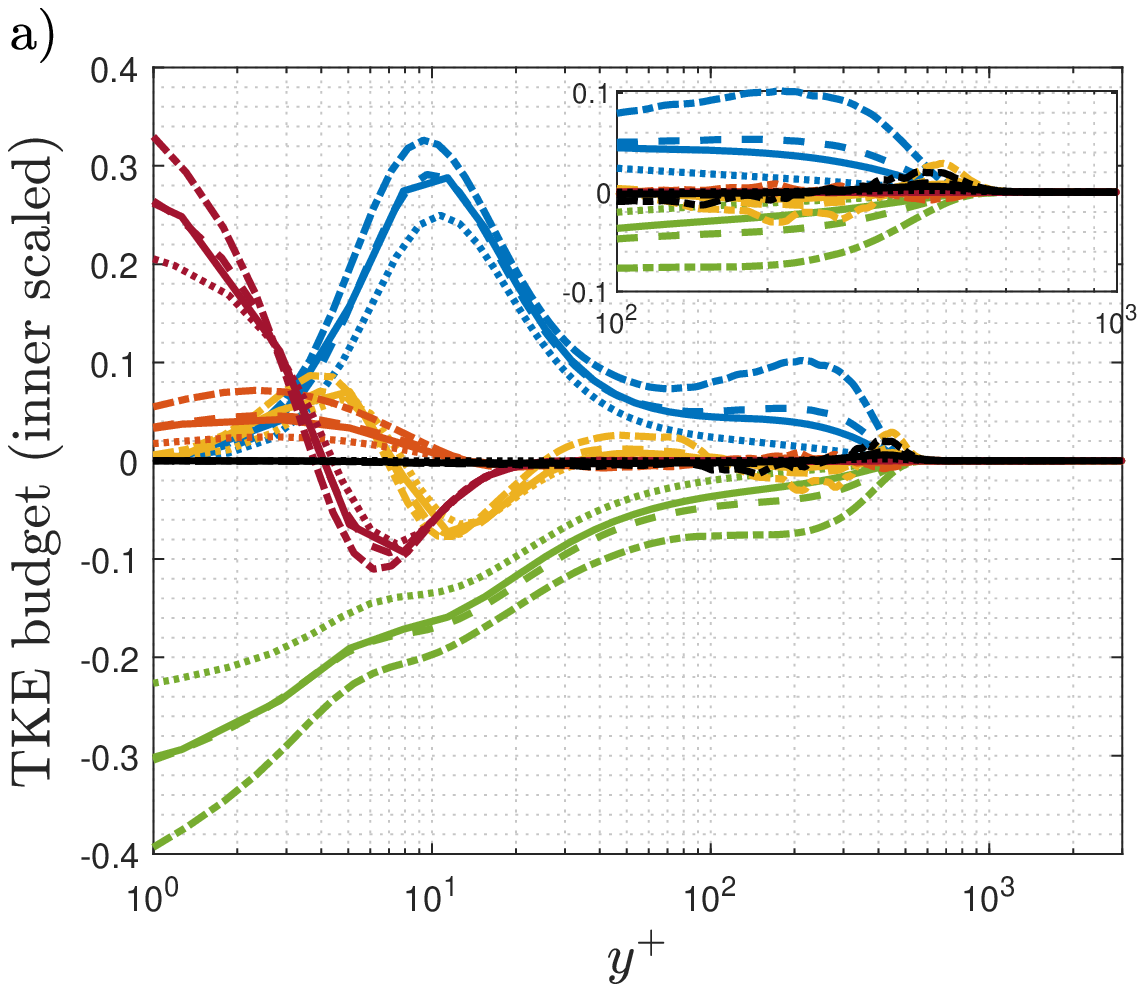}
\includegraphics[width=0.49\textwidth]{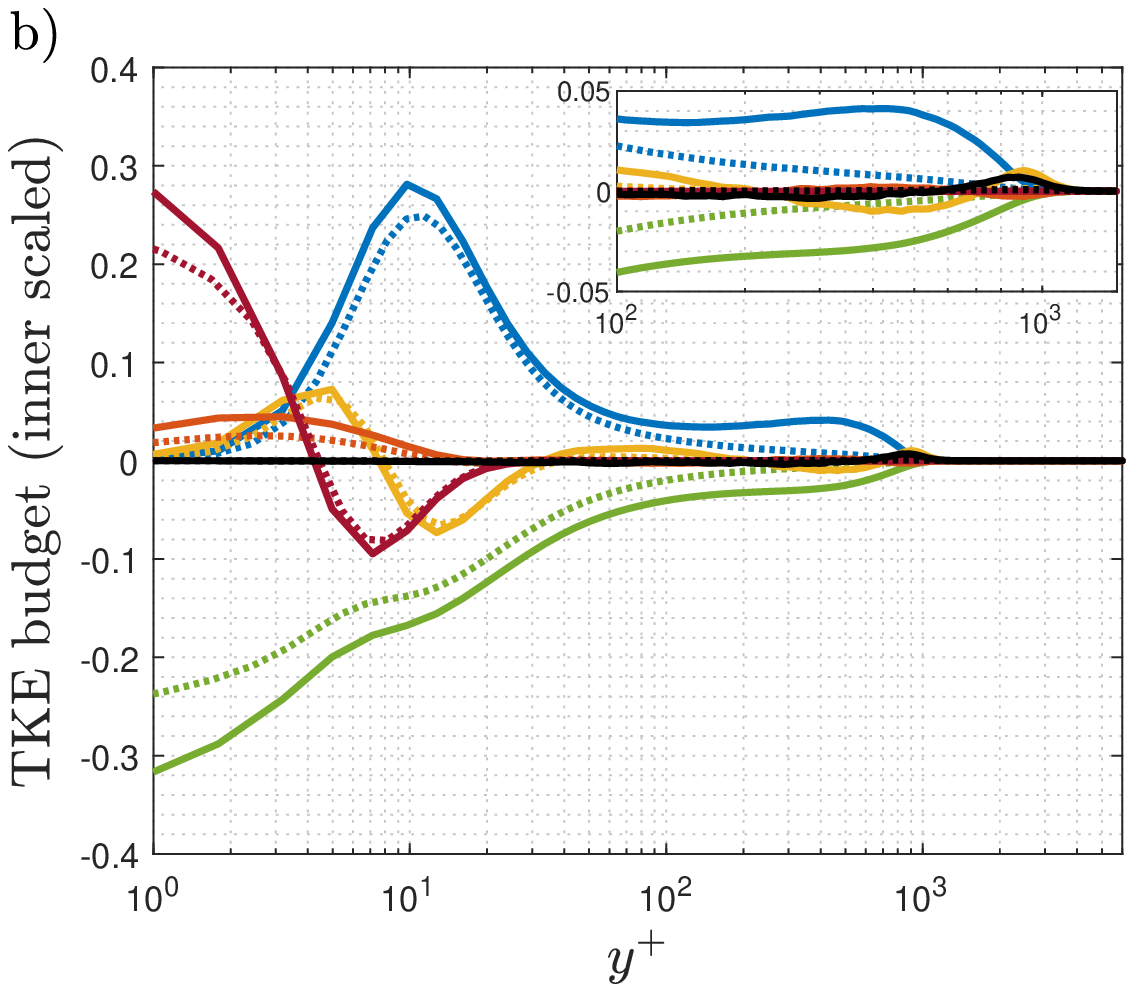}
\includegraphics[width=0.49\textwidth]{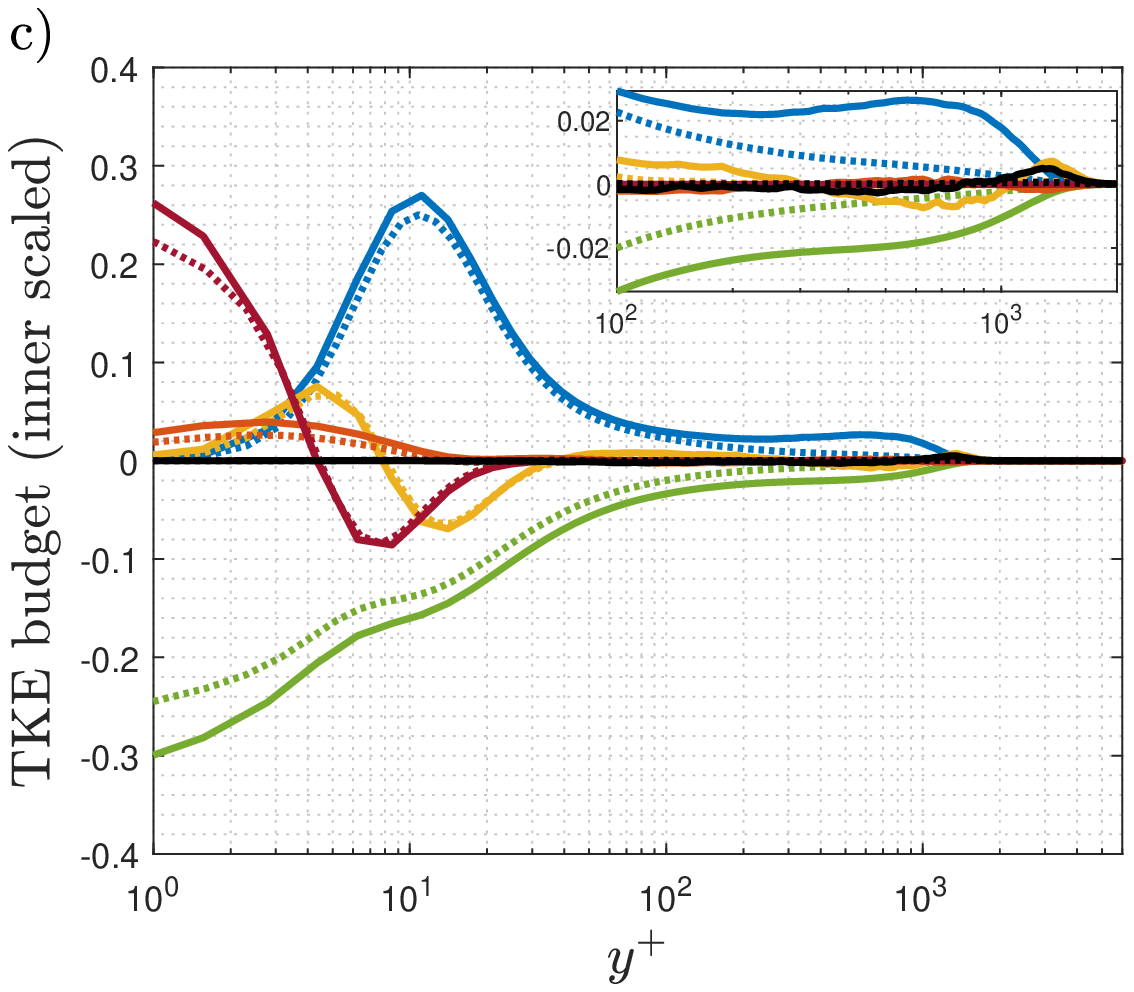}
\includegraphics[width=0.49\textwidth]{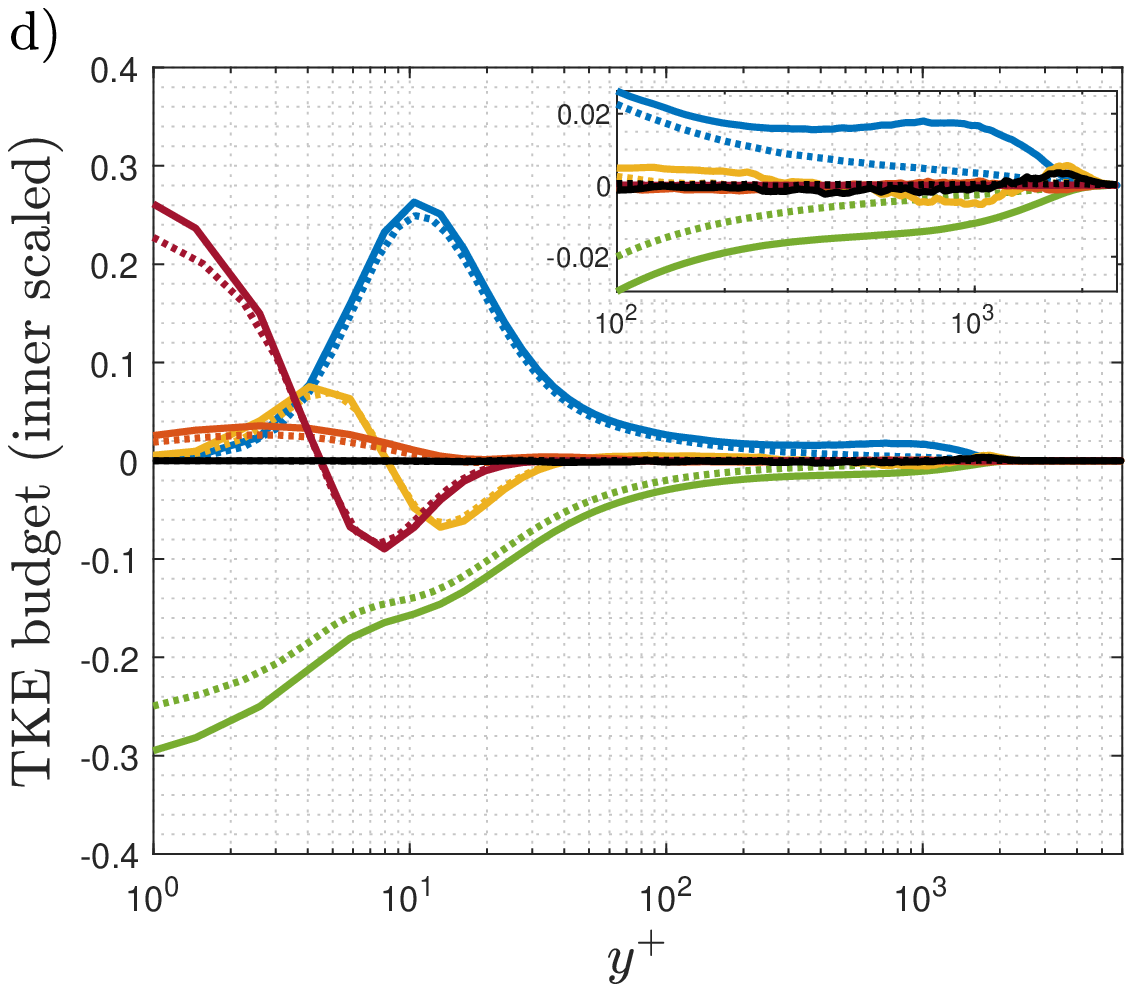}
  \caption{ Inner-scaled turbulent-kinetic-energy budget at different $Re_{\tau}$: (Top-left) $Re_{\tau}=500$ where $\beta(Re_{\tau})$ intersects for the simulations b1 and b1.4 . (Top-right) $Re_{\tau}=1000$, (bottom-left) $Re_{\tau}=1500$, (bottom-right) $Re_{\tau}=2000$. Symbols: (\protect\blackline) b1.4; (\protect\blackdotted) ZPG; (\protect\blackdash) b1 and (\protect\blackdashdot) b2. The colours correspond to the following terms of the TKE budget: production (\protect\blueline), dissipation (\protect\greenline), turbulent transport (\protect\yellowline), velocity-pressure-gradient correlation (\protect\orangeline), viscous diffusion (\protect\magentaline) and convection (\protect\blackline).}
\label{fig:TKEinner}
\end{figure}

% TKE budget in outer units
\begin{figure}
\includegraphics[width=0.49\textwidth]{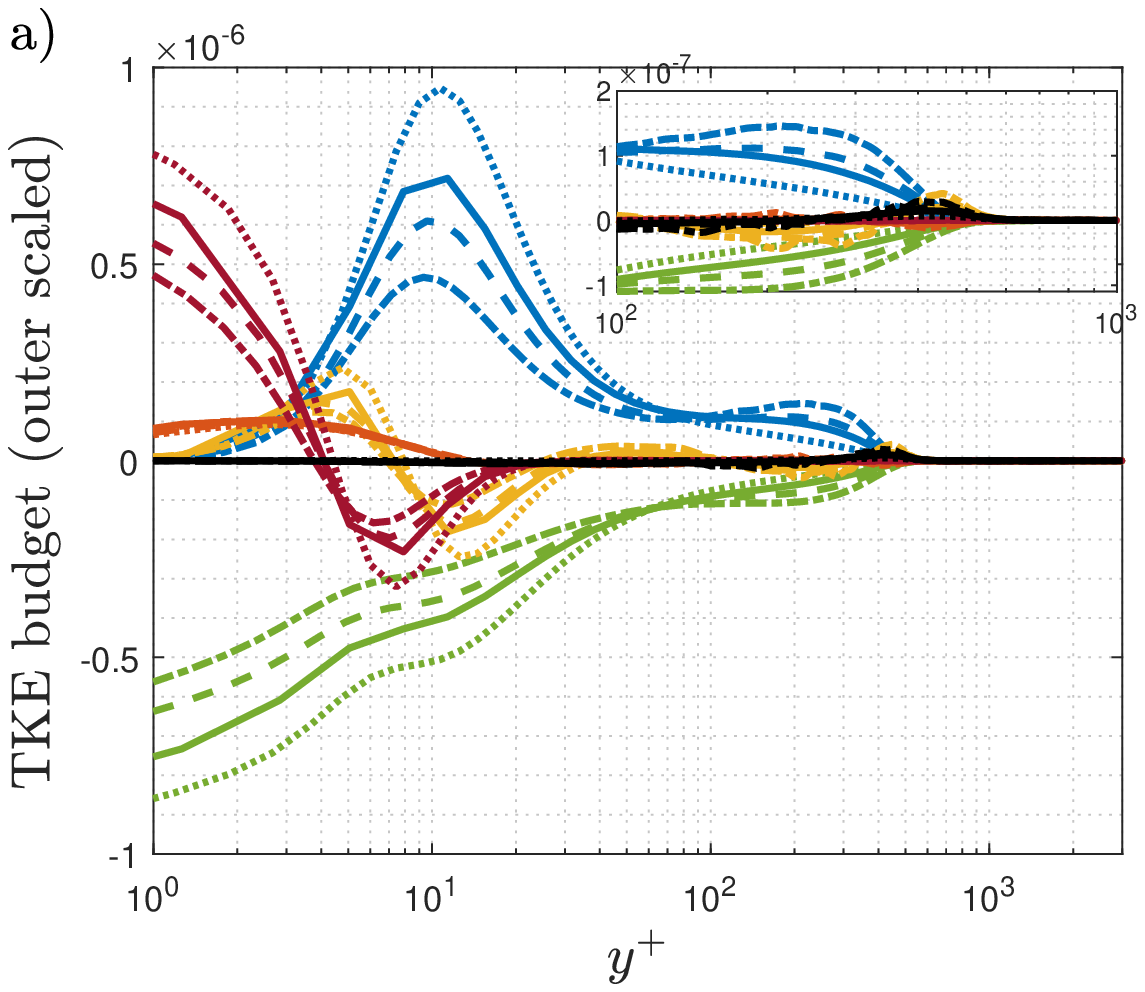}
\includegraphics[width=0.49\textwidth]{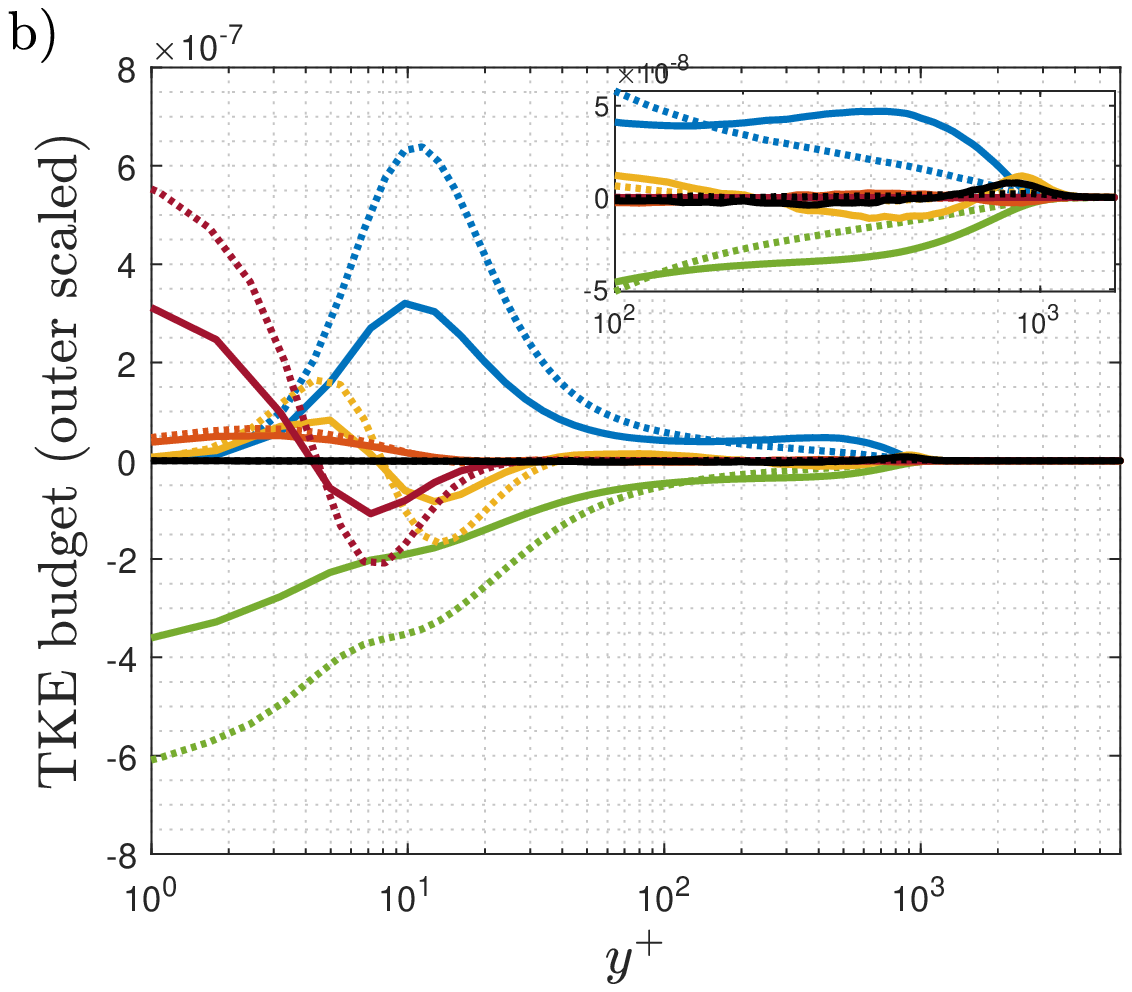}
\includegraphics[width=0.49\textwidth]{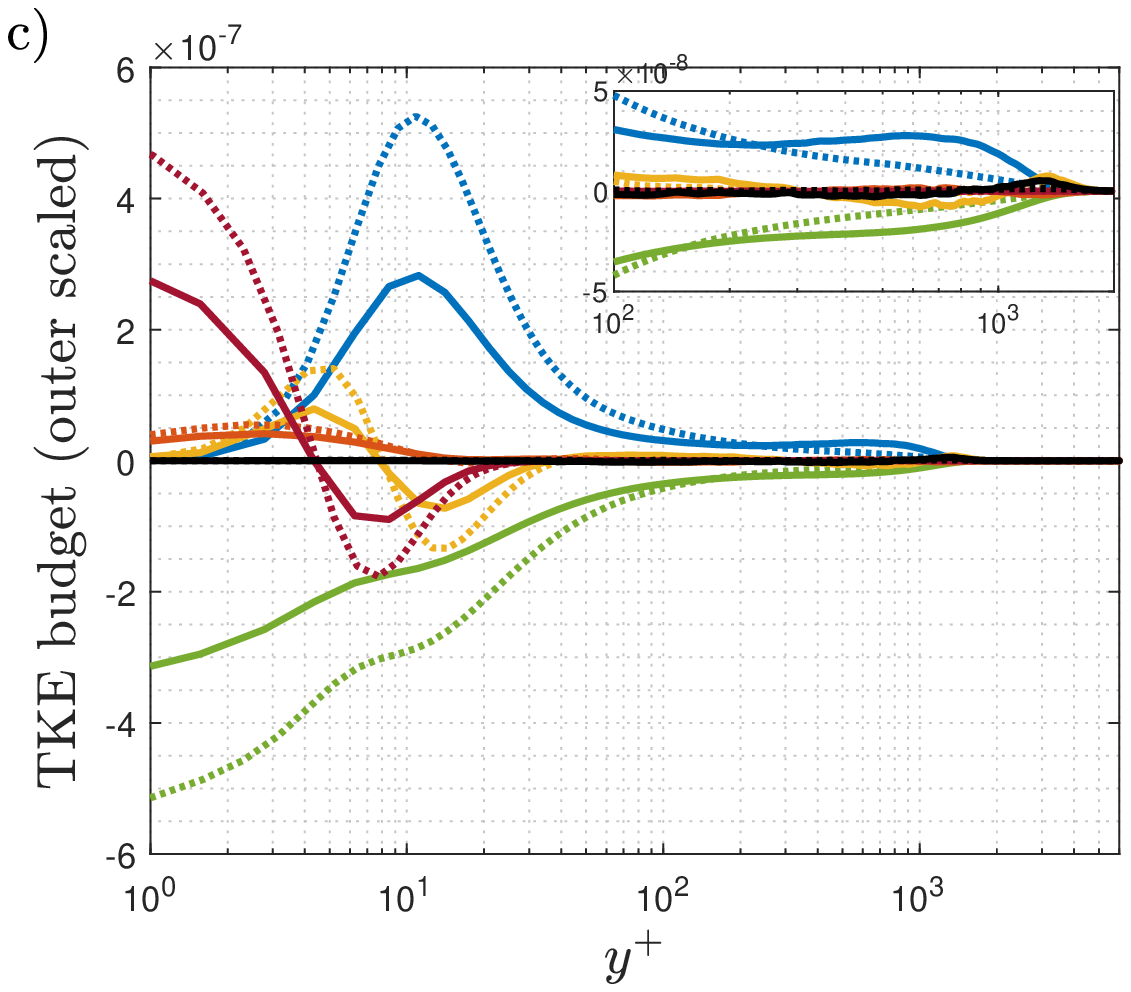}
\includegraphics[width=0.49\textwidth]{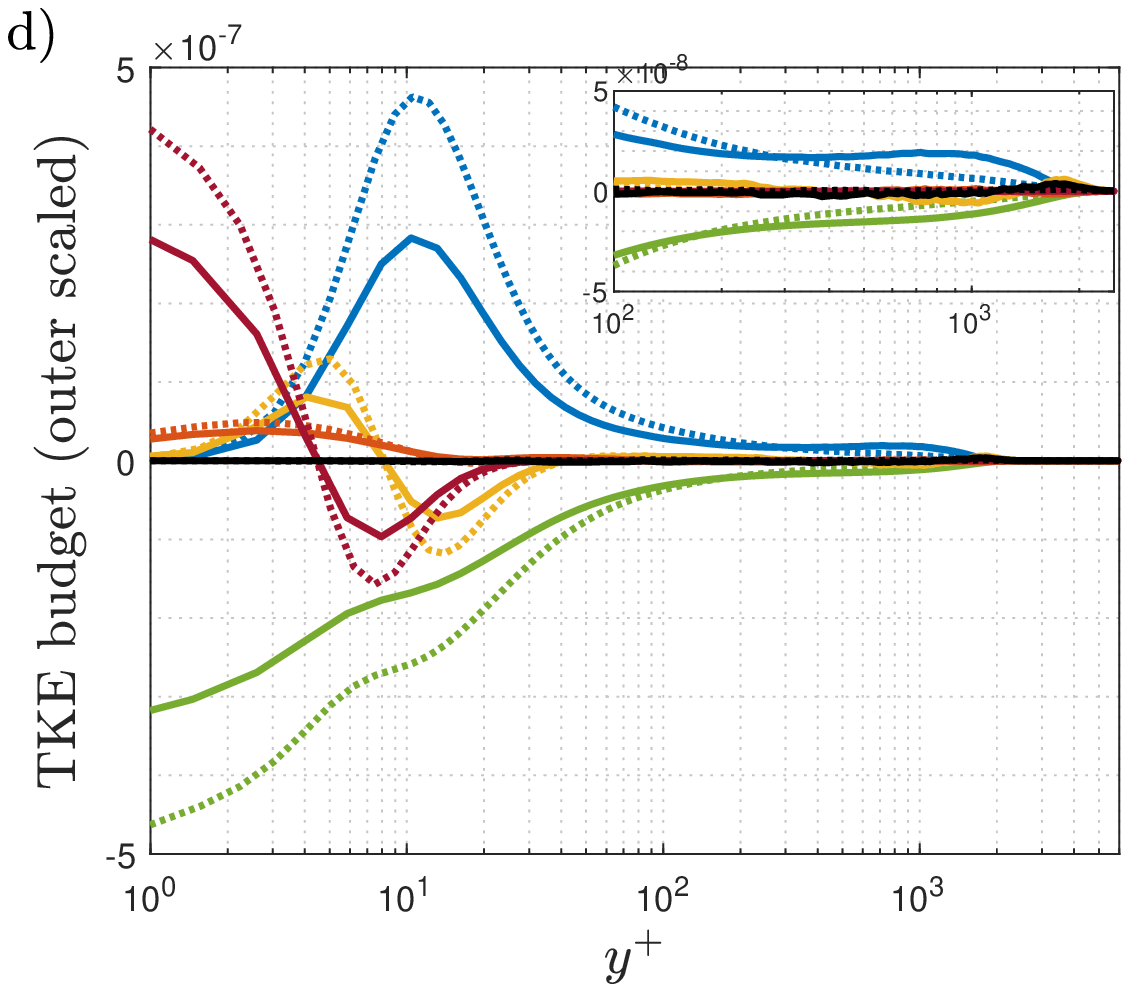}
  \caption{ Outer-scaled turbulent-kinetic-energy budget at different $Re_{\tau}$: (Top-left) $Re_{\tau}=500$ where $\beta(Re_{\tau})$ intersects for the simulations b1 and b1.4 . (Top-right) $Re_{\tau}=1000$, (bottom-left) $Re_{\tau}=1500$, (bottom-right) $Re_{\tau}=2000$. Symbols: (\protect\blackline) b1.4; (\protect\blackdotted) ZPG; (\protect\blackdash) b1 and (\protect\blackdashdot) b2. The colours correspond to the following terms of the TKE budget: production (\protect\blueline), dissipation (\protect\greenline), turbulent transport (\protect\yellowline), velocity-pressure-gradient correlation (\protect\orangeline), viscous diffusion (\protect\magentaline) and convection (\protect\blackline).}
\label{fig:TKEouter}
\end{figure}

The turbulent-kinetic-energy (TKE) budgets are shown in inner scale in figure~\ref{fig:TKEinner} and in outer scale in figure~\ref{fig:TKEouter}. As seen before in the inner region of the streamwise RS, the magnitude of the various budget terms increases in inner scaling with the APG, as opposed to the behaviour in outer scale where the APG produces a reduction of these magnitudes. 

Figure \ref{fig:TKEinner} shows that, as the distance from the wall is increased, differences in the velocity-pressure-gradient correlation are observed.

As the Reynolds number is increased, the general magnitude of each TKE-budget term as well as the relative differences between the various simulations are reduced. The most prominent effects of the APG are seen in the viscous sub-layer for the pair viscous-diffusion/dissipation and in the wake region for production/dissipation.
The outer scaling leads to an excellent collapse of the velocity-pressure-gradient correlation for all the simulations in all the regions of the TBL.
The phenomena discussed above for the Reynolds-stress terms where the APG leads to the development of an outer peak in all the components of the tensor, is manifested here as a second peak in the TKE production and dissipation in the outer region. Note that while the near-wall production peak becomes progressively smaller for higher $\beta$ when scaled in outer units, the outer-production peak exhibits the opposite behavior and grows with $\beta$. The value of $\beta$ decreases slowly in b1.4 with Reynolds number, and the outer peak in the TKE budget terms is also reduced at higher $\Rey$, approaching the ZPG values in inner scale.

The equations for the different terms of the TKE budgets are the same for ZPG and APG, but the magnitude of the streamwise gradients ($\partial / \partial x$) is larger in the APG compared to the ZPG.
We will analyze the major differences seen for the production $P^k=-\overline{u^2}\partial U/ \partial x -\overline{v^2}\partial V/ \partial y -\overline{uv} (\partial U/ \partial y + \partial V/ \partial x)$ in inner units.
   
The largest contribution to $P^k$ is given by the third term $-\overline{uv} \partial U/ \partial y$, which is much larger than the other three.

In the inner region the main differences come from the first two terms, since the magnitude of $\partial U/ \partial x$ for APGs is greater than for the ZPG and as a result of the continuity equation, the magnitude of $\partial V/ \partial y$ is also larger for APG than for ZPG. As discussed above, a larger $\beta$ leads to a larger inner-scaled near-wall peak of $\overline{u^2}$ for the APG than for the ZPG, therefore the first term of $P^k$ is also larger. In the outer region of the APG, all the terms in the RS tensor exhibit an outer peak, and these outer peaks, together with the gradients of $U$ and $V$ being larger than in the ZPG, produce an outer peak in the production term. Note from figure \ref{fig:meanU} that the slope of $U$, which is $\partial U/ \partial y$, is larger for APG than for the ZPG.

 % Appendix with statistics
  \section*{Appendix \hypertarget{AppB}{B}}\label{sec:AppendixB}

Turbulent flows are characterised by fluctuations of the flow variables, which may have different amplitudes and frequencies, and are linked with a wide range of eddy structures of different spatial and temporal scales.

As the TBL develops and the Reynolds number increases, the range of scales of the eddies grows, and a finer grid is needed to properly resolve all the scales. In this work, we implement a well-resolved LES which resolves the larger eddies in a suitable grid and a subgrid-scale (SGS) model is used to take into account the effects of the smaller scales. The SGS model used in this study is the approximate deconvolution model with a relaxation-term (ADM-RT), which is further documented in \cite{Schlatter_2004}. 
This model does not involve an eddy viscosity, and it is based on filters applied on the equidistant grids in Fourier space and on the non-equidistant wall-normal direction. As stated in \cite{Schlatter_2004}, the model does not disturb the flow development as long as it is still sufficiently well resolved, otherwise it adds the additional necessary dissipation.
The relaxation term can be seen as an SGS force of the form:
\begin{equation*}
    \frac{\partial \tau_{ij}}{\partial x_j} = \chi H_N * \overline{u}_i,
\end{equation*}
which is added to the right-hand side of the filtered Navier--Stokes equations.
The coefficient $\chi=0.2$ is proportional to the inverse of the time step of the integration. The high-order filter $H_N$ uses a cutoff frequency $\omega_c \in (0, \pi]$ (in this simulation $\omega_c=2\pi/3$), which only affects the smallest scales. The filter is applied through a convolution (denoted by the symbol $*$) to the velocities $\overline{u}_i$. The velocities are marked with an overline to indicate that they are implicitly filtered because of the lower resolution of the LES grid.

This SGS model has been compared with DNS simulations on multiple occasions: for transitional channel flow in \cite{Schlatter_2004}, for ZPG TBL in \cite{E-AmorZPG} and for turbulent wings in \cite{NEGI_2018}. The filter has been used in the APG cases by \cite{Bobke_2016, bobke2017}.
\cite{E-AmorZPG} reports for $\Rey_{\tau}=3600$ , $87.2\%$ of the dissipation of the DNS being resolved by the LES, while the addition of the SGS contributed to recover $99.8\%$.

 % Appendix with LES method
% \end{appendices}

% \newpage
% \clearpage
\FloatBarrier
\bibliographystyle{jfm}
\bibliography{jfm-instructions}

\begin{thebibliography}{51}
\expandafter\ifx\csname natexlab\endcsname\relax\def\natexlab#1{#1}\fi
\def\au#1{#1} \def\ed#1{#1} \def\yr#1{#1}\def\at#1{#1}\def\jt#1{\textit{#1}}
  \def\bt#1{#1}\def\bvol#1{\textbf{#1}} \def\vol#1{#1} \def\pg#1{#1}
  \def\publ#1{#1}\def\arxiv#1{#1}\def\org#1{#1}\def\st#1{\textit{#1}}

\bibitem[Araya {\em et~al.\/}(2015)Araya, Castillo \& Hussain]{FPG_araya2015}
{\sc \au{Araya, G.}, \au{Castillo, L.} \& \au{Hussain, F.}} \yr{2015}  \at{The
  log behaviour of the {Reynolds} shear stress in accelerating turbulent
  boundary layers}.  \jt{J.~Fluid Mech.}  \bvol{775},  \pg{189–200}.

\bibitem[Bailey {\em et~al.\/}(2013)Bailey, Hultmark, Monty, Alfredsson, Chong,
  Duncan, Fransson, Hutchins, Marusic, McKeon \& et~al.]{bailey_2013_JFM}
{\sc \au{Bailey, S. C.~C.}, \au{Hultmark, M.}, \au{Monty, J.~P.},
  \au{Alfredsson, P.~H.}, \au{Chong, M.~S.}, \au{Duncan, R.~D.}, \au{Fransson,
  J. H.~M.}, \au{Hutchins, N.}, \au{Marusic, I.}, \au{McKeon, B.~J.} \&
  \au{et~al.}} \yr{2013}  \at{Obtaining accurate mean velocity measurements in
  high reynolds number turbulent boundary layers using pitot tubes}.
  \jt{J.~Fluid Mech.}  \bvol{715},  \pg{642–670}.

\bibitem[Bobke {\em et~al.\/}(2016)Bobke, Vinuesa, Örlü \&
  Schlatter]{Bobke_2016}
{\sc \au{Bobke, A.}, \au{Vinuesa, R.}, \au{Örlü, R.} \& \au{Schlatter, P}}
  \yr{2016}  \at{Large-eddy simulations of adverse pressure gradient turbulent
  boundary layers}.  \jt{J. Phys. Conf. Ser.}  \bvol{708},  \pg{012012}.

\bibitem[Bobke {\em et~al.\/}(2017)Bobke, Vinuesa, Örlü \&
  Schlatter]{bobke2017}
{\sc \au{Bobke, A.}, \au{Vinuesa, R.}, \au{Örlü, R.} \& \au{Schlatter, P.}}
  \yr{2017}  \at{History effects and near equilibrium in
  adverse-pressure-gradient turbulent boundary layers}.  \jt{J.~Fluid Mech.}
  \bvol{820},  \pg{667–692}.

\bibitem[Castillo {\em et~al.\/}(2004)Castillo, Wang \& George]{Castillo_2004}
{\sc \au{Castillo, L.}, \au{Wang, X.} \& \au{George, W.~K.}} \yr{2004}
  \at{Separation criterion for turbulent boundary layers via similarity
  analysis.}  \jt{J. Fluids Eng.}  \bvol{126}~(3),  \pg{297--304}.

\bibitem[Chandran {\em et~al.\/}(2017)Chandran, Baidya, Monty \&
  Marusic]{chandran_jfm_rapids_2017}
{\sc \au{Chandran, D.}, \au{Baidya, R.}, \au{Monty, J.~P.} \& \au{Marusic, I.}}
  \yr{2017}  \at{Two-dimensional energy spectra in high-reynolds-number
  turbulent boundary layers}.  \jt{J.~Fluid Mech.}  \bvol{826},  \pg{R1}.

\bibitem[Chevalier {\em et~al.\/}(2007)Chevalier, Schlatter, Lundbladh \&
  Henningson]{simson_techrep}
{\sc \au{Chevalier, M.}, \au{Schlatter, P.}, \au{Lundbladh, A.} \&
  \au{Henningson, D.~S.}} \yr{2007}  \bt{A pseudo-spectral solver for
  incompressible boundary layer flows.} {\em Tech. Rep.\/}.  \org{TRITA-MEK.
  Stockholm}.

\bibitem[Clauser(1954)]{Clauser_1954_exp}
{\sc \au{Clauser, F.~H.}} \yr{1954}  \at{Turbulent boundary layers in adverse
  pressure gradients}.  \jt{J. Aeronaut. Sci.}  \bvol{21}~(2),  \pg{91--108}.

\bibitem[Clauser(1956)]{Clauser_1956}
{\sc \au{Clauser, F.~H.}} \yr{1956}  \at{The turbulent boundary layer}.
  \jt{Adv. Appl. Mech.}  \bvol{4}~(2),  \pg{1--51}.

\bibitem[Deshpande {\em et~al.\/}(2021)Deshpande, Monty \&
  Marusic]{deshpande_2021}
{\sc \au{Deshpande, R.}, \au{Monty, J.~P.} \& \au{Marusic, I.}} \yr{2021}
  \at{Active and inactive components of the streamwise velocity in wall-bounded
  turbulence}.  \jt{J.~Fluid Mech.}  \bvol{914},  \pg{A5}.

\bibitem[Eitel-Amor {\em et~al.\/}(2014)Eitel-Amor, Örlü \&
  Schlatter]{E-AmorZPG}
{\sc \au{Eitel-Amor, G.}, \au{Örlü, R} \& \au{Schlatter, P.}} \yr{2014}
  \at{Simulation and validation of a spatially evolving turbulent boundary
  layer up to ${Re}_{\theta}=8300$}.  \jt{Int. J. Heat Fluid Flow}  \bvol{47},
  \pg{57--69}.

\bibitem[Gibis {\em et~al.\/}(2019)Gibis, Wenzel, Kloker \& Rist]{Gibis2019}
{\sc \au{Gibis, T.}, \au{Wenzel, C.}, \au{Kloker, M.} \& \au{Rist, U.}}
  \yr{2019}  \at{Self-similar compressible turbulent boundary layers with
  pressure gradients. p 2. {Self-similarity} analysis of the outer layer.}
  \jt{J.~Fluid Mech.}  \bvol{880},  \pg{284--325}.

\bibitem[de~Giovanetti {\em et~al.\/}(2016)de~Giovanetti, Hwang \&
  Choi]{giovanetti2016}
{\sc \au{de~Giovanetti, M.}, \au{Hwang, Y.} \& \au{Choi, H.}} \yr{2016}
  \at{Skin-friction generation by attached eddies in turbulent channel flow}.
  \jt{J.~Fluid Mech.}  \bvol{808},  \pg{511–538}.

\bibitem[Griffin {\em et~al.\/}(2021)Griffin, Fu \&
  Moin]{d99_determination_2020}
{\sc \au{Griffin, K.~P.}, \au{Fu, L.} \& \au{Moin, P.}} \yr{2021}  \at{General
  method for determining the boundary layer thickness in nonequilibrium flows}.
   \jt{Phys. Rev. Fluids}  \bvol{6},  \pg{024608}.

\bibitem[Gungor {\em et~al.\/}(2017)Gungor, Maciel, Simens \&
  Gungor]{Gungor_DNS_2017}
{\sc \au{Gungor, A.~G.}, \au{Maciel, Y.}, \au{Simens, M.~P.} \& \au{Gungor,
  T.}} \yr{2017}  \bt{Direct numerical simulation of a non-equilibrium adverse
  pressure gradient boundary layer up to ${Re}_{\theta}=8000$}.

\bibitem[Harun {\em et~al.\/}(2013)Harun, Monty, Mathis \&
  Marusic]{harun_monty_2013}
{\sc \au{Harun, Z.}, \au{Monty, J.~P.}, \au{Mathis, R.} \& \au{Marusic, I.}}
  \yr{2013}  \at{Pressure gradient effects on the large-scale structure of
  turbulent boundary layers}.  \jt{J.~Fluid Mech.}  \bvol{715},
  \pg{477–498}.

\bibitem[Hiroyuki(2019)]{Abe_2019}
{\sc \au{Hiroyuki, A.}} \yr{2019}  \at{Direct numerical simulation of a
  turbulent boundary layer with separation and reattachment over a range of
  reynolds numbers}.  \jt{Fluid Dyn. Res.}  \bvol{51}~(1),  \pg{011409}.

\bibitem[Hoyas \& Jiménez(2006)]{Hoyas_PoF2006}
{\sc \au{Hoyas, S.} \& \au{Jiménez, J.}} \yr{2006}  \at{Scaling of the
  velocity fluctuations in turbulent channels up to ${Re}_{\tau}=2003$}.
  \jt{Phys. Fluids}  \bvol{18}~(1),  \pg{011702}.

\bibitem[Kitsios {\em et~al.\/}(2016)Kitsios, Atkinson, Sillero, Borrell,
  Gungor, Jiménez \& Soria]{Kitsios2016}
{\sc \au{Kitsios, V.}, \au{Atkinson, C.}, \au{Sillero, J.A.}, \au{Borrell, G.},
  \au{Gungor, A.G.}, \au{Jiménez, J.} \& \au{Soria, J.}} \yr{2016}  \at{Direct
  numerical simulation of a self-similar adverse pressure gradient turbulent
  boundary layer}.  \jt{Int. J. Heat Fluid Flow}  \bvol{61},  \pg{129--136}.

\bibitem[Kitsios {\em et~al.\/}(2017)Kitsios, Sekimoto, Atkinson, Sillero,
  Borrell, Gungor, Jiménez \& Soria]{Kitsios2017}
{\sc \au{Kitsios, V.}, \au{Sekimoto, A.}, \au{Atkinson, C.}, \au{Sillero,
  J.~A.}, \au{Borrell, G.}, \au{Gungor, A.~G.}, \au{Jiménez, J.} \& \au{Soria,
  J.}} \yr{2017}  \at{Direct numerical simulation of a self-similar adverse
  pressure gradient turbulent boundary layer at the verge of separation}.
  \jt{J.~Fluid Mech.}  \bvol{829},  \pg{392–419}.

\bibitem[Marusic {\em et~al.\/}(2015)Marusic, Chauhan, Kulandaivelu \&
  Hutchins]{marusic_2015}
{\sc \au{Marusic, I.}, \au{Chauhan, K.~A.}, \au{Kulandaivelu, V.} \&
  \au{Hutchins, N.}} \yr{2015}  \at{Evolution of zero-pressure-gradient
  boundary layers from different tripping conditions}.  \jt{J.~Fluid Mech.}
  \bvol{783},  \pg{379–411}.

\bibitem[Marusic {\em et~al.\/}(2010)Marusic, McKeon, Monkewitz, Nagib, Smits
  \& Sreenivasan]{Marusic_PoF_2010}
{\sc \au{Marusic, I.}, \au{McKeon, B.~J.}, \au{Monkewitz, P.~A.}, \au{Nagib,
  H.~M.}, \au{Smits, A.~J.} \& \au{Sreenivasan, K.~R.}} \yr{2010}
  \at{Wall-bounded turbulent flows at high {Reynolds} numbers: {Recent}
  advances and key issues}.  \jt{Phys. Fluids}  \bvol{22}~(6),  \pg{065103}.

\bibitem[Marusic {\em et~al.\/}(1997)Marusic, Uddin \& Perry]{Marusic_1997_ZPG}
{\sc \au{Marusic, I.}, \au{Uddin, A. K.~M.} \& \au{Perry, A.~E.}} \yr{1997}
  \at{Similarity law for the streamwise turbulence intensity in
  zero-pressure-gradient turbulent boundary layers}.  \jt{Physics of Fluids}
  \bvol{9}~(12),  \pg{3718--3726},  \arxiv{arXiv:
  https://doi.org/10.1063/1.869509}.

\bibitem[Mellor \& Gibson(1966)]{mellor_gibson_1966}
{\sc \au{Mellor, G.~L.} \& \au{Gibson, D.~M.}} \yr{1966}  \at{Equilibrium
  turbulent boundary layers}.  \jt{J.~Fluid Mech.}  \bvol{24}~(2),
  \pg{225–253}.

\bibitem[Monty {\em et~al.\/}(2011)Monty, Harun \& Marusic]{MONTY2011}
{\sc \au{Monty, J.P.}, \au{Harun, Z.} \& \au{Marusic, I.}} \yr{2011}  \at{A
  parametric study of adverse pressure gradient turbulent boundary layers}.
  \jt{Int. J. Heat Fluid Flow}  \bvol{32}~(3),  \pg{575--585}.

\bibitem[Narasimha \& Sreenivasan(1973)]{narasimha_sreenivasan_1973}
{\sc \au{Narasimha, R.} \& \au{Sreenivasan, K.~R.}} \yr{1973}
  \at{Relaminarization in highly accelerated turbulent boundary layers}.
  \jt{J.~Fluid Mech.}  \bvol{61}~(3),  \pg{417–447}.

\bibitem[Negi {\em et~al.\/}(2018)Negi, Vinuesa, Hanifi, Schlatter \&
  Henningson]{NEGI_2018}
{\sc \au{Negi, P.S.}, \au{Vinuesa, R.}, \au{Hanifi, A.}, \au{Schlatter, P.} \&
  \au{Henningson, D.S.}} \yr{2018}  \at{Unsteady aerodynamic effects in
  small-amplitude pitch oscillations of an airfoil}.  \jt{"Int. J. Heat Fluid
  Flow"}  \bvol{71},  \pg{378--391}.

\bibitem[Nordström {\em et~al.\/}(1999)Nordström, Nordin \&
  Henningson]{SIMSON_fringe}
{\sc \au{Nordström, J.}, \au{Nordin, N.} \& \au{Henningson, D.}} \yr{1999}
  \at{The fringe region technique and the {Fourier} method used in the direct
  numerical simulation of spatially evolving viscous flows}.  \jt{SIAM J. Sci.
  Comput.}  \bvol{20},  \pg{1365--1393}.

\bibitem[Rotta(1950)]{rotta1950theorie}
{\sc \au{Rotta, J.C.}} \yr{1950} {\em {\"U}ber die Theorie der turbulenten
  Grenzschichten\/}. {\em Mitteilungen aus dem Max-Planck-Institut f{\"u}r
  Str{\"o}mungsforschung\/} .  \publ{Max-Planck-Institut f.
  Str{\"o}mungsforschung}.

\bibitem[Rotta(1962)]{rotta1962turbulent}
{\sc \au{Rotta, J.C.}} \yr{1962} {\em Turbulent boundary layers in
  incompressible flow\/}.  \publ{Pergamon Press}.

\bibitem[Sanmiguel~Vila {\em et~al.\/}(2020)Sanmiguel~Vila, Vinuesa, Discetti,
  Ianiro, Schlatter \& \"Orl\"u]{Sanmiguel_PRF}
{\sc \au{Sanmiguel~Vila, C.}, \au{Vinuesa, R.}, \au{Discetti, S.}, \au{Ianiro,
  A.}, \au{Schlatter, P.} \& \au{\"Orl\"u, R.}} \yr{2020}  \at{Separating
  adverse-pressure-gradient and reynolds-number effects in turbulent boundary
  layers}.  \jt{Phys. Rev. Fluids}  \bvol{5},  \pg{064609}.

\bibitem[{Sanmiguel Vila} {\em et~al.\/}(2020){Sanmiguel Vila}, Vinuesa,
  Discetti, Ianiro, Schlatter \& Örlü]{MTL_expSANMIGUEL}
{\sc \au{{Sanmiguel Vila}, C.}, \au{Vinuesa, R.}, \au{Discetti, S.},
  \au{Ianiro, A.}, \au{Schlatter, P.} \& \au{Örlü, R.}} \yr{2020}
  \at{Experimental realisation of near-equilibrium adverse-pressure-gradient
  turbulent boundary layers}.  \jt{Exp. Therm. Fluid Sci.}  \bvol{112},
  \pg{109975}.

\bibitem[Schlatter {\em et~al.\/}(2010)Schlatter, Li, Brethouwer, Johansson \&
  Henningson]{Schlatter_etAl_LES_2010}
{\sc \au{Schlatter, P.}, \au{Li, Q.}, \au{Brethouwer, G.}, \au{Johansson,
  A.~V.} \& \au{Henningson, D.~S.}} \yr{2010}  \at{Simulations of spatially
  evolving turbulent boundary layers up to ${Re}_{\theta}=4300$}.  \jt{Int. J.
  Heat Fluid Flow}  \bvol{31}~(3),  \pg{251--261}, sixth International
  Symposium on Turbulence and Shear Flow Phenomena.

\bibitem[Schlatter {\em et~al.\/}(2004)Schlatter, Stolz \&
  Kleiser]{Schlatter_2004}
{\sc \au{Schlatter, P.}, \au{Stolz, S.} \& \au{Kleiser, L.}} \yr{2004}
  \at{{LES} of transitional flows using the approximate deconvolution model}.
  \jt{Int. J. Heat Fluid Flow}  \bvol{25}~(3),  \pg{549--558}.

\bibitem[Schlatter \& Örlü(2010)]{schlatter_orlu_2010}
{\sc \au{Schlatter, P.} \& \au{Örlü, R.}} \yr{2010}  \at{Assessment of direct
  numerical simulation data of turbulent boundary layers}.  \jt{J.~Fluid Mech.}
   \bvol{659},  \pg{116–126}.

\bibitem[Schlatter \& Örlü(2012)]{schlatter_orlu_2012}
{\sc \au{Schlatter, P.} \& \au{Örlü, R.}} \yr{2012}  \at{Turbulent boundary
  layers at moderate {Reynolds} numbers: inflow length and tripping effects}.
  \jt{J.~Fluid Mech.}  \bvol{710},  \pg{5–34}.

\bibitem[Sillero {\em et~al.\/}(2011)Sillero, Jim{\'{e}}nez, Moser \&
  Malaya]{Sillero_2011}
{\sc \au{Sillero, J.}, \au{Jim{\'{e}}nez, J.}, \au{Moser, R.~D.} \& \au{Malaya,
  N.~P.}} \yr{2011}  \at{Direct simulation of a zero-pressure-gradient
  turbulent boundary layer up to ${Re}_{\theta}= 6650$}.  \jt{J. Phys. Conf.
  Ser.}  \bvol{318}~(2),  \pg{022023}.

\bibitem[Sillero {\em et~al.\/}(2013)Sillero, Jiménez \&
  Moser]{Sillero_2013pof}
{\sc \au{Sillero, J.~A.}, \au{Jiménez, J.} \& \au{Moser, Robert~D.}} \yr{2013}
   \at{One-point statistics for turbulent wall-bounded flows at reynolds
  numbers up to $\delta^+ \approx 2000$}.  \jt{Phys. Fluids}  \bvol{25}~(10),
  \pg{105102}.

\bibitem[Skåre \& Krogstad(1994)]{skare_krogstad_1994}
{\sc \au{Skåre, P.~E.} \& \au{Krogstad, P.}} \yr{1994}  \at{A turbulent
  equilibrium boundary layer near separation}.  \jt{J.~Fluid Mech.}
  \bvol{272},  \pg{319–348}.

\bibitem[Tanarro {\em et~al.\/}(2020)Tanarro, Vinuesa \&
  Schlatter]{tanarro_2020}
{\sc \au{Tanarro, Á.}, \au{Vinuesa, R.} \& \au{Schlatter, P.}} \yr{2020}
  \at{Effect of adverse pressure gradients on turbulent wing boundary layers}.
  \jt{J.~Fluid Mech.}  \bvol{883},  \pg{A8}.

\bibitem[Townsend(1956{\natexlab{{\em a\/}}})]{townsend_1956_eqBL}
{\sc \au{Townsend, A.~A.}} \yr{1956{\natexlab{{\em a\/}}}}  \at{The properties
  of equilibrium boundary layers}.  \jt{J.~Fluid Mech.}  \bvol{1}~(6),
  \pg{561–573}.

\bibitem[Townsend(1956{\natexlab{{\em b\/}}})]{Townsend_1956_structure}
{\sc \au{Townsend, A.~A.}} \yr{1956{\natexlab{{\em b\/}}}}  \at{The structure
  of turbulent shear flow.}  \jt{J.~Fluid Mech.}  \bvol{1}~(5),
  \pg{554–560}.

\bibitem[Townsend(1961)]{townsend_1961}
{\sc \au{Townsend, A.~A.}} \yr{1961}  \at{Equilibrium layers and wall
  turbulence}.  \jt{J.~Fluid Mech.}  \bvol{11}~(1),  \pg{97–120}.

\bibitem[Townsend(1976)]{Townsend_1976}
{\sc \au{Townsend, A.~A.}} \yr{1976} {\em The Structure of Turbulent Shear
  Flow. 2nd Edition.\/}.  \publ{Cambridge University Press}.

\bibitem[Vinuesa {\em et~al.\/}(2016{\natexlab{{\em a\/}}})Vinuesa, Bobke,
  Örlü \& Schlatter]{diagnostic_Vinuesa}
{\sc \au{Vinuesa, R.}, \au{Bobke, A.}, \au{Örlü, R.} \& \au{Schlatter, P.}}
  \yr{2016{\natexlab{{\em a\/}}}}  \at{On determining characteristic length
  scales in pressure-gradient turbulent boundary layers}.  \jt{Phys. Fluids}
  \bvol{28}~(5),  \pg{055101}.

\bibitem[Vinuesa {\em et~al.\/}(2018)Vinuesa, Negi, Atzori, Hanifi, Henningson
  \& Schlatter]{VINUESA2018}
{\sc \au{Vinuesa, R.}, \au{Negi, P.S.}, \au{Atzori, M.}, \au{Hanifi, A.},
  \au{Henningson, D.S.} \& \au{Schlatter, P.}} \yr{2018}  \at{Turbulent
  boundary layers around wing sections up to ${Re_c}=1,000,000$}.  \jt{Int. J.
  Heat Fluid Flow}  \bvol{72},  \pg{86--99}.

\bibitem[Vinuesa {\em et~al.\/}(2017)Vinuesa, \"Orl\"u, Sanmiguel~Vila, Ianiro,
  Discetti \& Schlatter]{Vinuesa_2017}
{\sc \au{Vinuesa, R.}, \au{\"Orl\"u, R.}, \au{Sanmiguel~Vila, C.}, \au{Ianiro,
  A.}, \au{Discetti, S.} \& \au{Schlatter, P.}} \yr{2017}  \at{Revisiting
  history effects in adverse-pressure-gradient turbulent boundary layers}.
  \jt{Flow Turbul. Combust.}  \bvol{99},  \pg{565--587}.

\bibitem[Vinuesa {\em et~al.\/}(2016{\natexlab{{\em b\/}}})Vinuesa, Prus,
  Schlatter \& Nagib]{vinuesa2016_mec}
{\sc \au{Vinuesa, R.}, \au{Prus, C.}, \au{Schlatter, P.} \& \au{Nagib, H.~M.}}
  \yr{2016{\natexlab{{\em b\/}}}}  \at{Convergence of numerical simulations of
  turbulent wall-bounded flows and mean cross-flow structure of rectangular
  ducts}.  \jt{Meccanica}  \bvol{51},  \pg{3025--3042}.

\bibitem[Zagarola \& Smits(1998)]{zagarola_smits}
{\sc \au{Zagarola, M.~V.} \& \au{Smits, A.~J.}} \yr{1998} A new mean velocity
  scaling for turbulent boundary layers.  \bt{In {\em Proc. of FEDSM'98\/}},
  \pg{pp. 1--6}.  \publ{Washington, DC: ASME}.

\bibitem[del Álamo {\em et~al.\/}(2004)del Álamo, Jiménez, Zandonade \&
  Moser]{delAlamo_jfm_2004}
{\sc \au{del Álamo, J.~C.}, \au{Jiménez, J.}, \au{Zandonade, P.} \&
  \au{Moser, R.~D.}} \yr{2004}  \at{Scaling of the energy spectra of turbulent
  channels}.  \jt{J.~Fluid Mech.}  \bvol{500},  \pg{135–144}.

\bibitem[Örlü \& Schlatter(2013)]{Orlu_Schlatter_exp2013}
{\sc \au{Örlü, R.} \& \au{Schlatter, P.}} \yr{2013}  \at{Comparison of
  experiments and simulations for zero pressure gradient turbulent boundary
  layers at moderate {Reynolds} numbers.}  \jt{Exp. Fluids}  \bvol{54},
  \pg{1547}.

\end{thebibliography}

\end{document}